АЖМУХАМЕДОВ И.М.

# РЕШЕНИЕ ЗАДАЧ ОБЕСПЕЧЕНИЯ ИНФОРМАЦИОННОЙ БЕЗОПАСНОСТИ НА ОСНОВЕ СИСТЕМНОГО АНАЛИЗА И НЕЧЕТКОГО КОГНИТИВНОГО МОДЕЛИРОВАНИЯ

МОНОГРАФИЯ

Астрахань 2012

# ОГЛАВЛЕНИЕ

# СПИСОК СОКРАЩЕНИЙ.

СЗИ – система защиты информации

ЗИ – защита информации, защищаемая информация

КСЗИ – комплексная система защиты информации

КС – комплексная система

СС – слабоструктурированная система

КТ – коммерческая тайна

ДФ – дестабилизирующие факторы

НСД – несанкционированный доступ

ОИБ – обеспечение информационной безопасности

ОС – организационная система, организационная структура

ОУ – объект управления

СУ – субъект управления

ТЭО – технико-экономическое обоснование

ТЗ – техническое задание

ТП – технический проект

ТС – технические средства

СФС – структурно-функциональная схема

СБ – служба безопасности

АСОД – автоматизированная система обработки данных

ИС – информационная система

ГК – гражданский кодекс

ЛПР – лицо, принимающее решения

УР – управленческое решение

ВТ – вычислительная техника

ЧС – чрезвычайная ситуация

# ВВЕДЕНИЕ

В Доктрине информационной безопасности Российской Федерации настоящий этап развития общества характеризуется возрастающей ролью информационной сферы, представляющей собой совокупность информации, информационной инфраструктуры, субъектов, осуществляющих сбор, формирование, распространение и использование информации, а также системы регулирования [1].

Интенсивное развитие и использование современных информационных технологий уже в настоящее время привели к серьезным качественным изменениям во всех сферах общественной жизни. Современное общество вступает в постиндустриальный период своего развития, который по всеобщему мнению можно назвать информационным.

Принятая в 2008 г. стратегия развития информационного общества в России так характеризует его отличительные черты:

- существенный рост доли в валовом внутреннем продукте отраслей экономики, связанных с производством знаний, с созданием и внедрением наукоемких, в том числе информационных, технологий, других продуктов интеллектуальной деятельности, с оказанием услуг в области информатизации, образования, связи, а также в области поиска, передачи, получения и распространения информации;

- радикальное ускорение технического прогресса, превращение научных знаний в реальный фактор производства, повышения качества жизни человека и общества;

- участие значительной части трудоспособного населения в производственной деятельности, связанной с созданием и использованием информационных технологий, информации и знаний;

- глобализация экономической, политической и духовной сфер жизни общества.

В этих условиях на передний план экономического и социального развития выходят проблемы совершенствования систем информационного обеспечения всех сфер деятельности общества.

Эти проблемы считаются одними из наиболее актуальных и неотложных задач общества. Для их решения в последние годы ведутся весьма интенсивные и крупномасштабные исследования и разработки.

Вместе с тем развитие информационного общества, помимо расширения созидательных возможностей, приводит и к росту угроз национальной безопасности, связанных с нарушением установленных режимов использования информационных и коммуникационных систем, ущемлением конституционных прав граждан, распространением вредоносных программ, а также с использованием возможностей современных информационных технологий для враждебных, террористических и других преступных действий [2].

Отечественная статистика в настоящее время не в состоянии ответить на вопрос: какие финансовые потери несут отечественные организации от компьютерных преступлений? Поэтому обратимся к зарубежной статистике.

Так, по данным ФБР США, в течение последних двух лет 78% опрошенных компаний понесли финансовые убытки, связанные с недостаточной информационной безопасностью. Основные причины потерь:

• 76% компаний, понесших убытки, пострадали от компьютерных вирусов (убытки американского бизнеса от эпидемий вирусов оцениваются в $7.6 миллиардов долларов);

• 42% - от атак изнутри;

• 25% - от атак извне;

• 70% компаний пострадали от ошибок, связанных с невнимательностью;

• 10% - от промышленного шпионажа.

Примерно 18 % опрошенных из этого числа заявляют, что потеряли более миллиона долларов в ходе нападений, более 66 процентов потерпели убытки в размере 50 тыс. долларов. Свыше 22% атак были нацелены на промышленные секреты или документы, представляющие интерес, прежде всего, для конкурентов. По зарубежным оценкам в США и Западной Европе экономические потери от компьютерных преступлений только в банковской сфере составляют ежегодно около 130 млрд. долларов.

В связи с этим особую актуальность приобретает проблема обеспечения информационной безопасности, и прежде всего надежной защиты информации (предупреждения ее искажения или уничтожения, несанкционированной модификации, злоумышленного получения и использования).

Обострению проблемы ОИБ способствует также появление в огромных количествах дешевых персональных компьютеров и построенных на их основе локальных и глобальных национальных и транснациональных сетей ЭВМ, использующих спутниковые каналы связи, повсеместная и массовая компьютеризация информационных процессов, создание высокоэффективных систем разведки и добычи информации.

Для надежной защиты информационных активов необходимо, прежде всего, решение следующих задач:

- организация практических работ по защите информации и управление ими на государственном, ведомственном, региональном и объектовом уровнях;
- проведение научных исследований и разработок всех аспектов рассматриваемой проблемы;
- разработка, производство и распространение средств защиты;
- подготовка кадров по защите информации.

При этом специфическими особенностями создания систем обеспечения информационной безопасности являются [3]:

- неполнота и неопределенность исходной информации о составе и характере угроз;
- многокритериальность задачи, связанная с необходимостью учета большого числа частных показателей;
- наличие как количественных, так и качественных показателей, которые необходимо учитывать при решении задач разработки и внедрения систем защиты;

- невозможность применения классических методов оптимизации.

В связи с тем, что процессы защиты информации подвержены сильному влиянию случайных факторов, методы классической теории систем оказываются практически непригодными для использования в качестве основы научно-методологического базиса решения проблем безопасности.

Таким образом, при формировании теории защиты возникает актуальная задача расширения арсенала классической теории систем за счет использования методов нечетких множеств, лингвистических переменных, неформального оценивания и поиска оптимальных решений. Причем, исключительно важное значение для решения современных проблем защиты приобретают методы экспертных оценок, эвристического программирования и «мозгового штурма» [4].

Основой стратегии защиты информации (ЗИ) является унифицированная концепция, которая фактически представляет собой последовательную цепь методологий оценки уязвимости информации, выработки требований по ее защите, определения набора концептуальных решений по защите и оценке факторов, влияющих на требуемый уровень защиты [5-6].

Все эти методологии, начиная с формирования полного множества угроз, также представляют собой ярко выраженные плохо формализуемые и слабоструктурированные проблемы.

Для решения широкого круга задач, связанных с моделированием плохо формализованных процессов, их прогнозированием и поддержкой принятия решений часто используются нечеткие когнитивные модели. Неоспоримыми их достоинствами по сравнению с другими методами являются возможность формализации численно неизмеримых факторов, использования неполной, нечеткой и даже противоречивой информации [7].

Поэтому возникла необходимость обобщить имеющийся опыт теоретических исследований и практического решения задач защиты информации на основе нечеткого когнитивного моделирования, позволяющего унифицировать подходы к управлению комплексной безопасностью систем, что и стало целью написания данной книги.

Основное содержание книги основано на материалах различных литературных источников, результатах исследований отечественных и зарубежных ученых и специалистов в области обеспечения информационной безопасности, авторских разработках в этой сфере.

Первая глава посвящена рассмотрению основных понятий в области информационной безопасности, построению онтологической модели задачи обеспечения ИБ. Рассмотрены классификация угроз и уязвимостей, модели нарушителя и типы атак на информационные ресурсы. Сформулированы основные принципы обеспечения ИБ.

Во второй главе рассматриваются научно-методологические основы оценки эффективности систем комплексной информационной безопасности, современное состояние проблемы нечетко-множественного описания, анализ моделей и методов, основанных на процедурах нечеткой логики.

Третья глава посвящена рассмотрению когнитивных моделей и методов, основанных на математическом формализме теории нечетких множеств и процедур нечеткой логики.

В четвертой главе рассматриваются проблемы расширения арсенала классической теории систем за счет использования методов, позволяющих адекватно моделировать плохо формализуемые процессы, существенно зависящие от воздействия трудно предсказуемых факторов, и решать задачи анализа, т.е. оценки защищенности (уязвимости) информации, и синтеза, т.е. оптимизации распределения ресурсов, выделяемых на защиту.

Пятая глава посвящена построению динамической нечеткой когнитивной модели оценки безопасности информационных активов современного ВУЗа.

В шестой главе рассматриваются примеры применения когнитивной модели в задачах подготовки и подбора кадров в сфере информационной безопасности.

Седьмая глава посвящена рассмотрению особенностей оценки экономической эффективности мер по обеспечению информационной безопасности. Предложена методика определения оптимального комплекса мер по обеспечению информационной безопасности.

В восьмой главе рассмотрено решение задачи управление сетевым трафиком на основе когнитивной модели, основанной на нечетких правилах.

В приложении приведены работы, посвященные решению отдельных задач информационной безопасности и созданию механизмов и средств защиты информации.

Автор считает своим долгом особо отметить ценные замечания и предложения, сделанные д.т.н., профессором О.М.Проталинским в процессе работы над книгой, и выражает искреннюю благодарность рецензенту д.т.н., профессору Г.А.Попову.

# ГЛАВА 1.

# ОСНОВНЫЕ ПОНЯТИЯ В ОБЛАСТИ ИНФОРМАЦИОННОЙ БЕЗОПАСНОСТИ

## 1.1. Термины и определения

Мир находится на пороге глобальных изменений: новое информационное общество приходит на смену обществу индустриальному, в связи с чем новые информационные технологии все более и более проникают во все области деятельности человека, особенно в промышленность и общественную жизнь, ускоряя процессы глобализации и интеграции мировой экономики и мирового сообщества [8].

Процесс информатизации общества привел к тому, что компьютерная информация превратилась в основной товар, обладающий значительной ценностью, в своеобразный стратегический ресурс. Информационные системы и технологии, как компоненты информационной сферы непосредственно и активно влияют на состояние экономической, экологической, энергетической, транспортной, продовольственной, криминогенной, информационной и других составляющих комплексной безопасности РФ.

Таким образом, информационные системы являются одним из системообразующих факторов жизни современного общества, и влияние информационной безопасности и на все стороны жизни общества с течением времени будет только возрастать [9].

Для того чтобы повысить доверие к электронной торговле, к электронным банковским системам, к телемедицине, к электронному правительству, необходима общая сплоченность в вопросах информационной безопасности на международном уровне.

Наша страна вносит свой вклад в дело решения проблем развития культуры информационной безопасности (ИБ), противодействия рискам в этой сфере и росту глобального уровня знаний в сфере защиты информации (ЗИ).

В Российской Федерации массово издается специализированная литература по ЗИ. Высшая школа регулярно выпускает специалистов по информационной безопасности в соответствии с требованиями семи образовательных стандартов.

Однако при этом одной из актуальных проблем является отсутствие устоявшейся терминологической базы по ИБ. Первопричиной такой ситуации является отсутствие в стране юридически значимого базового определения информационной безопасности. В российских законах нет подобной дефиниции [10].

Доктрина информационной безопасности Российской Федерации определяет главные термины и понятия защиты государственной информации. В частности, под информационной безопасностью Российской Федерации в доктрине понимается: «состояние защищенности ее национальных интересов в информационной сфере, определяющихся совокупностью сбалансированных интересов личности, общества и государства» [1].

Данное определение, сформулированное для госсектора, получило широкое распространение (в том числе в учебной литературе) применительно к другим сферам деятельности без учета их особенностей, что противоречит положениям доктрины. Большинство авторов учебной литературы берут за основу положения, разработанные в прошлом веке только для защиты государственной тайны (ГТ), забывая, что страна перешла на специфические рыночные отношения, о чем свидетельствует появление соответствующих законов «О техническом регулировании», «О коммерческой тайне» и «О персональных данных» и др.

В Доктрине отмечено, что: «В каждой сфере жизнедеятельности общества и государства наряду с общими методами обеспечения информационной безопасности Российской Федерации могут использоваться частные методы и формы, обусловленные спецификой факторов, влияющих на состояние информационной безопасности».

Решая задачи обеспечения информационной безопасности бизнеса, необходимо помнить, что главной целью любого предпринимателя является прибыль, для получения которой он должен снижать издержки производства и реализации продукта. При этом весь бизнес-процесс должен сопровождаться расчетами, построенными на базе измерений и учета.

Слабым местом процесса расчетов в сфере ИБ является почти полное отсутствие количественных метрик. Частично решению обозначенной проблемы способствует Закон «О техническом регулировании» [11].

В статье 2 данного закона вводятся основные понятия, характерные для любого бизнеса. Например, «риск - вероятность причинения вреда... » и «безопасность - состояние, при котором отсутствует недопустимый риск, связанный с причинением вреда...».

Применяя данные определения для термина «информационная безопасность» можно получить следующую, ориентированную на бизнес, дефиницию: «информационная безопасность - состояние информации при допустимом риске ее уничтожения, изменения или раскрытия, связанном с причинением вреда владельцу или пользователю информации» [12].

При этом под термином *владелец информации* понимается субъект информационных отношений, обладающий правом владения, распоряжения и пользования информационным ресурсом по договору с собственником информации. *Пользователь информации* - субъект информационных отношений, обладающий правом пользования доверенным ему информационным ресурсом.

Достоинствами данного определения информационной безопасности являются гармонизация положений новых стандартов (ГОСТ Р ИСО/МЭК 15408-1-2002, 27001, 17799) и прежнего научно-технического задела, а также возможность получения через оценку рисков количественных метрик информационной безопасности [13].

В отечественных нормативно-методических документах по информационной безопасности до 2002 года отсутствовало понятие риск,

которое впервые появилось в стандарте ГОСТ Р ИСО/МЭК 15408-1-2002 «Информационная технология. Методы и средства обеспечения безопасности. Критерии оценки безопасности информационных технологий».

В настоящее время количество стандартов с этой дефиницией резко возросло [14-17]. Однако необходимо отметить, что пока в данных стандартах используются только качественные метрики риска.

На базе исследования и обобщения отечественной и зарубежной литературы, в том числе отечественных и международных стандартов в области защиты информации, определим значение наиболее значимых терминов из области ЗИ, а также терминов из смежных областей, которые необходимы для понимания сущности целенаправленной деятельности по обеспечению информационной безопасности.

*Информация* (в информационных технологиях) – это совокупность данных, обрабатываемых техническими средствами. Под *информацией в области защиты* понимаются сведения, раскрываемые через демаскирующие признаки объектов защиты или путем несанкционированного доступа к техническим средствам обработки информации.

*Информационная технология* - система технических средств и способов обработки информации.

*Информационный ресурс* - совокупность данных и программ, задействованных при обработке информации техническими средствами.

*Субъект информационных отношений* - физическое или юридическое лицо, обладающее определенным правом по отношению к информационному ресурсу. В зависимости от уровня полномочий субъект информационных отношений может быть источником, собственником, владельцем или пользователем информации.

*Источник информации* - материальный объект или субъект, способный накапливать, хранить, преобразовывать и выдавать информацию в виде сообщений или сигналов различной физической природы.

*Содержание информации* - конкретные сведения о данном объекте или явлении, определяющие совокупность элементов, сторон, связей, отношений между ними. По своему содержанию информация может иметь политический, военный, экономический, научно-технический, производственный или коммерческий характер и в зависимости от категории доступа к ней подразделяется на *общедоступную* информацию и *информацию ограниченного доступа.*

Ограничение доступа к информации устанавливается в целях защиты основ конституционного строя, нравственности, здоровья, прав и законных интересов других лиц, обеспечения обороны страны и безопасности государства.

Документированная информация с ограниченным доступом подразделяется на информацию, отнесенную к *государственной тайне* и информацию *конфиденциального* характера.

*Конфиденциальная информация* - информация, которая представляет собой коммерческую, служебную, профессиональную и другие виды тайн (всего около 40 видов), а также *персональные данные* (информация о гражданах).

Федеральными законами устанавливаются условия отнесения информации к сведениям, составляющим коммерческую, служебную и иную тайну, обязательность соблюдения конфиденциальности такой информации, а также ответственность за ее разглашение.

*Общедоступная информация* - информация, которая не представляет собой государственную, служебную, коммерческую или иную тайну и может быть опубликована в открытой печати. Такая информация не защищается от утечки, так как не содержит в себе каких-либо тайн, но в случае если она представлена в форме документов (библиотеки) или банка данных ЭВМ она может и должна защищаться от нарушений целостности и доступности.

К общедоступной информации относятся, например, сведения, доступ к которым не может быть ограничен [18]:

- нормативные правовые акты, затрагивающие права, свободы и обязанности человека и гражданина, а также устанавливающие правовое положение организаций и полномочия государственных органов, органов местного самоуправления;

- информация о состоянии окружающей среды;

- информация о деятельности государственных органов и органов местного самоуправления (за исключением сведений, составляющих государственную или служебную тайну);

- информация, накапливаемая в открытых фондах библиотек, музеев и архивов, а также в государственных, муниципальных и иных информационных системах, созданных для обеспечения граждан и организаций такой информацией;

- информация, недопустимость ограничения доступа к которой установлена федеральными законами.

По принадлежности информационные ресурсы подразделяются на государственные (принадлежащие органам государственной власти) и не государственные (принадлежащие юридическим и физическим лицам).

При этом обладатель информации вправе:

- разрешать или ограничивать доступ к информации, определять порядок и условия такого доступа;

- использовать информацию, в том числе распространять ее, по своему усмотрению;

- передавать информацию другим лицам по договору или на ином установленном законом основании;

- защищать установленными законом способами свои права в случае незаконного получения информации или ее незаконного использования иными лицами.

Законом РФ «О персональных данных» от 27 июля 2006 г. № 152-ФЗ регулируются отношения, связанные с обработкой персональных данных,

осуществляемой органами государственной власти, юридическими и физическими лицами.

*Персональные данные* - это любая информация, относящаяся к определенному физическому лицу, в том числе его фамилия, имя, отчество, год, месяц, дата и место рождения, адрес, семейное, социальное, имущественное положение, образование, профессия, доходы, другая информация.

В законе дано определение общедоступных и конфиденциальных персональных данных (ст. 3):

-*конфиденциальность персональных данных* - обязательное требование не допускать их распространение без согласия субъекта персональных данных или законного основания;

-*общедоступные персональные данные* - персональные данные, доступ неограниченного круга лиц к которым предоставлен с согласия субъекта персональных данных или в соответствии с федеральными законами.

В связи с этим положением закона, в целях информационного обеспечения могут создаваться общедоступные источники персональных данных (в том числе справочники, адресные книги). В них с письменного согласия субъекта персональных данных могут включаться его фамилия, имя, отчество, год и место рождения, адрес, абонентский номер, сведения о профессии и иные персональные данные, предоставленные субъектом персональных данных.

*Обработка персональных данных* осуществляется только с согласия в письменной форме субъекта персональных данных.

*Обработка специальных категорий* персональных данных, касающихся расовой, национальной принадлежности, политических взглядов, религиозных или философских убеждений, состояния здоровья, интимной жизни, не допускается, за исключением указанных в законе случаев (осуществление правосудия, угроза состоянию здоровья и т.п.).

Оператор при обработке персональных данных обязан принимать необходимые меры, в том числе криптографические средства, для защиты персональных данных от несанкционированного доступа (НСД), уничтожения, и иных неправомерных действий.

В случае достижения цели обработки персональных данных оператор обязан незамедлительно прекратить обработку и уничтожить соответствующие персональные данные в срок, не превышающий трех рабочих дней с даты достижения цели обработки персональных данных, если иное не предусмотрено федеральными законами.

*Государственная тайна* – это сведения, охраняемые государством, разглашение которых может оказать отрицательное воздействие на качественное состояние военно-экономического потенциала страны или повлечь другие тяжкие последствия для ее обороноспособности, государственной безопасности, экономических и политических интересов.

В зависимости от величины политического или экономического ущерба, который может быть нанесен интересам государства в случае разглашения такой информации, она может иметь гриф секретно, совершенно секретно или особой важности.

*Под защитой информации* понимают деятельность, направленную на сохранение государственной, служебной, коммерческой или иной тайны, а также на сохранение носителей информации любого содержания. В основе комплексной системы информационной защиты любого объекта лежат следующие методы:

- *правовые* (международные, государственные, местные, ведомственные и внутрифирменные правовые акты);
- *организационные* (создание служб безопасности и введение режима защиты информации, подготовка и переподготовка кадров, системы лицензирования и сертификации в области защиты информации)*;*
- *технические* (программные, аппаратные, криптографические средства, физические ограждения и препятствия).

*Система защиты информации* - комплекс организационных и технических мероприятий по защите информации, проведенный на объекте с применением необходимых технических средств и способов в соответствии с концепцией, целью и замыслом защиты.

*Концепция защиты информации* – это система взглядов и общих технических требований по защите информации.

*Цель защиты информации* - заранее намеченный уровень защищенности информации, получаемый в результате реализации системы защиты на объекте.

*Замысел защиты* - основная идея, раскрывающая состав, содержание, взаимосвязь и последовательность мероприятий, необходимых для достижения цели защиты информации на объекте.

*Техническое средство защиты информации* - техническое средство, предназначенное для устранения или ослабления демаскирующих признаков объекта, создания ложных (имитирующих) признаков, а также для создания помех техническим средством доступа информации.

*Способ защиты информации* - прием (метод), используемый для организации защиты информации.

*Технико-экономическое обоснование ЗИ* - определение оптимального объема организационных и технических мероприятий в составе системы защиты информации на объекте, необходимого для достижения цели защиты.

При проведении исследований по технико-экономическому обоснованию следует исходить из того, что стоимость затрат на создание системы защиты информации на объекте не должна превышать стоимость защищаемой информации. В противном случае защита информации становится нецелесообразной.

*Эффективность защиты информации* - это степень соответствия достигнутого уровня защищенности информации поставленной цели.

*Показатель эффективности защиты информации* - параметр технического признака объекта защиты, применительно к которому устанавливаются требования и/или нормы по эффективности защиты информации.

*Экспертиза системы защиты информации* - оценка соответствия представленных проектных материалов по защите информации (на объекте) поставленной цели, требованиям стандартов и других нормативных документов.

*Атака* - любое действие, нарушающее безопасность информационной системы. Более формально можно сказать, что атака - это действие или последовательность связанных между собой действий, использующих уязвимости данной информационной системы и приводящих к нарушению политики безопасности.

*Уязвимость* - слабое место в системе, с использованием которого может быть осуществлена атака.

*Риск* - вероятность того, что конкретная угроза будет реализована с использованием конкретной уязвимости. В конечном счете, каждая организация должна принять решение о допустимом для нее уровне риска. Это решение должно найти отражение в политике безопасности, принятой в организации.

*Сервис безопасности* - сервис, который обеспечивает задаваемую политикой безопасность систем и/или передаваемых данных, либо определяет осуществление атаки. Сервис использует один или более механизмов безопасности.

Основными сервисами безопасности являются:

*Конфиденциальность* - свойство информации при ее обработке техническими средствами, обеспечивающее предотвращение несанкционированного ознакомления с ней или несанкционированного документирования (снятия копий).

*Целостность* - свойство информации при ее обработке техническими средствами, обеспечивающее предотвращение ее несанкционированной модификации или несанкционированного уничтожения. Данный сервис гарантирует, что информация при хранении или передаче не изменилась. Может применяться к потоку сообщений, единственному сообщению или отдельным полям в сообщении, а также к хранимым файлам и отдельным записям файлов.

*Доступность* - свойство информации при ее обработке техническими средствами, обеспечивающее беспрепятственный доступ к ней для проведения санкционированных операций по ознакомлению, документированию, модификации и уничтожению.

*Аутентификация* - подтверждение того, что информация получена из законного источника, и получатель действительно является тем, за кого себя выдает. В случае передачи единственного сообщения аутентификация должна гарантировать, что получателем сообщения является тот, кто нужно, и сообщение получено из заявленного источника. В случае установления соединения имеют место два аспекта. Во-первых, при инициализации соединения сервис должен гарантировать, что оба участника являются требуемыми. Во-вторых, сервис должен гарантировать, что на соединение не воздействуют таким образом, что третья сторона сможет маскироваться под одну из легальных сторон уже после установления соединения.

*Невозможность отказа* - невозможность, как для получателя, так и для отправителя, отказаться от факта передачи. Таким образом, когда сообщение отправлено, получатель может убедиться, что это сделал легальный отправитель. Аналогично, когда сообщение пришло, отправитель может убедиться, что оно получено легальным получателем.

*Контроль доступа* - возможность ограничить и контролировать доступ к системам и приложениям по коммуникационным линиям.

*Механизм безопасности* - программное и/или аппаратное средство, которое определяет и/или предотвращает атаку.

Примерами механизмов безопасности являются:

*Алгоритмы симметричного шифрования* - алгоритмы шифрования, в которых для шифрования и дешифрования используется один и тот же ключ или ключ дешифрования легко может быть получен из ключа шифрования.

*Алгоритмы асимметричного шифрования* - алгоритмы шифрования, в которых для шифрования и дешифрования используются два разных ключа, называемые открытым и закрытым ключами, причем, зная один из ключей, вычислить другой невозможно.

*Хэш-функции* - функции, входным значением которых является сообщение произвольной длины, а выходным значением - сообщение фиксированной длины. Хэш-функции обладают рядом свойств, которые позволяют с высокой долей вероятности определять изменение входного сообщения.

*Электронная цифровая подпись* - используется для аутентификации текстов, передаваемых по телекоммуникационным каналам. Функционально она аналогична обычной рукописной подписи и обладает ее основными достоинствами.

Таким образом, с помощью шифрования обеспечиваются три состояния безопасности информации:

1. *Конфиденциальность.* Шифрование используется для сокрытия информации от неавторизованных пользователей при передаче или при хранении.
2. *Целостность.* Шифрование используется для предотвращения изменения информации при передаче или хранении.
3. *Идентифицируемость.* Шифрование используется для аутентификации источника информации и предотвращения отказа отправителя информации от того факта, что данные были отправлены именно им.

Кроме криптографических используются и другие механизмы защиты. Например, *протоколирование и аудит,* которые предусматривают отслеживание, регистрацию и анализ всех происходящих в системе событий, имеющих значение с точки зрения безопасности: сеансов пользователей различного уровня, изменения прав и меток безопасности, открытия и модификации критичных файлов и т.д. Этот механизм выполняет две основные функции: обнаружение атак и контроль за состоянием системы.

*Стеганография* представляет собой механизм защиты передаваемых данных путем маскировки самого факта передачи сообщения или существования канала связи, например, с использованием избыточности некоторых форматов данных (в основном, графических, звуковых и видеоданных) или нестандартной записи на определенные носители. Так конфиденциальное сообщение может быть "спрятано" в файле, который, будучи обработан стандартными средствами, выглядит как обычная картинка (фотография, схема, диаграмма и пр.)

Еще одним широко используемым механизмом является *физическая защита*. Наряду с традиционными механическими системами, функционирующими при доминирующем участии человека, разрабатываются и внедряются универсальные автоматизированные электронные средства физической защиты, такие как:

- скрытые портативные средства видеонаблюдения;
- специальные наклейки и детекторы, предотвращающие вынос документов, электронных носителей информации, приборов, узлов и блоков системы из помещения;
- экранирование рабочих помещений и/или маскировка побочных электромагнитных излучений средств вычислительной техники при помощи генераторов помех;
- специальное оборудование для радиомониторинга пространства, выявления и контроля электромагнитных излучений закладных устройств и т.д.

*Техническая разведка* - деятельность по получению разведывательной информации с помощью технических средств.

*Средство технической разведки* - аппаратура технической разведки, установленная и используемая на носителе.

*Аппаратура технической разведки* - совокупность технических устройств обнаружения, приема, регистрации, измерения и анализа, предназначенная для получения разведывательной информации.

В зависимости от параметра технического демаскирующего признака, используемого технической разведкой для получения интересующих ее сведений об объекте защиты, может быть использован очень большой арсенал различных видов разведывательной аппаратуры.

Для видеонаблюдения в закрытых помещениях используются такие современные оптико-электронные устройства как видеокамеры, эндоскопы и средства передачи видеосигнала на расстояние (видео передатчики, видеоприемники, видео трансляторы и т.д.).

Для ведения визуальной разведки на открытой местности могут использоваться как традиционные оптические приборы дальнего видения (бинокли, подзорные трубы, телескопы), так и сопряженные с ними оптико-электронные устройства.

Для документирования видеоинформации используются фото, киноаппараты и видеомагнитофоны, которые в сочетании с оптическими приборами дальнего видения, тепловизионными и инфракрасными приборами ночного видения обеспечивают визуальную разведку наблюдения и документирования разведывательных данных в любое время суток и в любых погодных условиях.

Для перехвата радиолокационных и радиотехнических сигналов, а также сигналов радиосвязи, используются специальные радиоприемные устройства соответствующего диапазона частот.

Для перехвата акустических сигналов, распространяющихся в различных средах, используется аудиоаппаратура, гидроакустические и сейсмические приемники. При необходимости акустический сигнал может быть записан на магнитофон.

Для перехвата лазерных излучений и лазерного зондирования (в том числе лазерного оконных стекол для прослушивания помещений) применяют лазерные устройства, работающие как правило в невидимом (ИК или УФ) спектре частот.

Для обнаружения изменений окружающей среды используются приборы радиационной и химической разведки, а также другие устройства и приспособления.

*Зона разведдоступности* - часть пространства вокруг объекта, в пределах которого реализуются возможности технической разведки.

*Технический канал утечки информации* - совокупность объекта технической разведки, физической среды и средства технической разведки, которыми добываются разведывательные данные.

*Объект защиты* - обобщающий термин для всех форм существования информации, требующих защиты от технических разведок. Информация и информационные технологии становятся основным рыночным товаром в обществе. Поэтому наиболее важными составляющими объекта защиты является информация и средства обработки и хранения информации. Поэтому целесообразно выделить особый объект защиты - информация совместно с системой ее обработки и организации (информационная инфраструктура). По своему составу объекты защиты могут быть единичными и групповыми.

*Единичный объект защиты* - конкретный носитель секретной или конфиденциальной информации, представляющий собой единое целое и предназначенный для выполнения определенных функций.

К единичным объектам (субъектам) защиты относятся люди, владеющие секретной или конфиденциальной информацией, секретные и конфиденциальные документы и изделия, в том числе технические средства обработки информации, выделенные здания и помещения, а также вспомогательные технические средства и системы, подверженные влиянию информационных физических полей.

*Групповой объект защиты* - структурное объединение единичных объектов как требующих так и не требующих защиты, предназначенных для совместного выполнения определенных функций.

К групповым объектам защиты относятся режимные предприятия и учреждения, военные и военно-промышленные объекты, а также конструкторское бюро, опытные производства, испытательные полигоны, базы, аэродромы и другие объекты, связанные с разработкой, испытанием и эксплуатацией секретных изделий.

*Комплексность защиты* - принцип защиты, предусматривающий мероприятия против всех опасных видов и средств технической разведки.

*Активность защиты* - принцип защиты, выражающийся в целенаправленном навязывании техническим разведкам ложного представления об объекте в соответствии с замыслом защиты, а также подавление возможностей технической разведки.

*Убедительность защиты* - принцип защиты, заключающийся в соответствии замысла защиты условиям обстановки, в которых он реализуется.

*Непрерывность защиты* - принцип защиты, заключающийся в организации защиты объекта на всех стадиях его жизненного цикла: в период разработки, изготовления (строительства), испытаний, эксплуатации и утилизации.

*Разнообразие защиты* - принцип защиты, предусматривающий исключение повторяемости в выборе путей реализации замысла защиты, в том числе с применением типовых решений.

*Способ защиты информации от технических разведок* - преднамеренное воздействие на технический канал утечки информации или на объект защиты для достижения целей защиты от технических разведок.

Основными способами защиты информации от технических разведок являются скрытие и дезинформации. Разновидностями дезинформации являются легендирование и имитации.

*Техническое средство обработки информации ТСОИ* - техническое средство, предназначенное для приема, хранения, поиска, преобразования, отображения и/или передачи информации по каналам связи.

К техническим средствам обработки информации относятся средства вычислительной техники, средства и системы связи, средства записи, усиления и воспроизведения звука, переговорные и телевизионные устройства, средства изготовления и размножения документов, кинопроекционная аппаратура и другие технические средства, связанные с приемом, накоплением, хранением, поиском, преобразованием, отображением и/или передачей информации по каналам связи.

*Средство вычислительной техники СВТ* - техническое средство обработки информации, в котором информация представлена в цифровом коде.

К средствам вычислительной техники относятся процессоры, каналы селективные и мультиплексные, внешние запоминающие устройства, устройства ввода и вывода данных, устройства непосредственной связи оператора с ЭВМ, устройства систем телеобработки данных, устройства повышения достоверности и т.д.

*Объект информатизации* - стационарный или подвижный объект, который представляет собой комплекс технических средств обработки информации, предназначенный для выполнения определенных функций. К объектам информатизации могут быть отнесены также отдельные технические средства обработки информации, выполняющие определенные функции обработки информации.

*Объект вычислительной техники (ОВТ)* - стационарный или подвижный объект, который представляет собой комплекс средств вычислительной техники, предназначенный для выполнения определенных функций обработки информации. К объектам вычислительной техники относятся автоматизированные системы (АС), автоматизированные рабочие места (АРМ), информационно - вычислительные центры (ИВЦ) и другие комплексы средств вычислительной техники (ГОСТ 34.003-90). К объектам вычислительной техники могут быть отнесены также отдельные средства вычислительной техники, выполняющие самостоятельные функции обработки информации.

*Вспомогательные технические средства и системы (ВТСС)* - технические средства и системы, которые непосредственно не задействованы для обработки информации, но находятся в электромагнитном поле побочных излучений технических средств обработки информации, в результате чего на них наводится опасный сигнал, который по токопроводящим коммуникациям может распространяться за пределы контролируемой зоны.

К ВТСС относятся средства и системы связи, пожарной и охранной сигнализации, электрочасофикации, радиофикации, электробытовые приборы и другие вспомогательные технические средства и системы. ВТСС играют роль так называемых "случайных антенн".

*Безопасность информации* – это состояние уровня защищенности информации при ее обработке техническими средствами, обеспечивающее сохранение таких ее качественных характеристик (свойств) как конфиденциальность, целостность и доступность.

Необходимо заметить, что термины "безопасность информации" и "защита информации" не являются синонимами. Термин "безопасность" включает в себя не только понятие защиты, но также и аутентификацию, аудит, обнаружение проникновения.

*Санкционированный доступ к информации* - доступ к информационному ресурсу, который осуществляется штатными техническими средствами в соответствии с установленными правилами.

*Несанкционированный доступ к информации (НСД)* - доступ к информации, осуществляемый штатными техническими средствами с нарушением установленных правил.

Несанкционированный доступ может создать угрозу любой из качественных характеристик безопасности информации: конфиденциальности, целостности или доступности.

*Компьютерное преступление* - осуществление несанкционированного доступа к информационному ресурсу, его модификация (подделка) или уничтожение с целью получения имущественных выгод для себя или для третьего лица, а также для нанесения имущественного ущерба своему конкуренту.

*Модель защиты информации от несанкционированного доступа* - абстрактное (формализованное или неформализованное) описание комплекса организационных мер и программно - аппаратных средств защиты от несанкционированного доступа штатными техническими средствами, являющееся основой для разработки системы защиты информации.

Основными способами защиты информации от НСД являются разграничение доступа, идентификация и аутентификация пользователя.

*Разграничение доступа* - наделение каждого пользователя (субъекта доступа) индивидуальными правами по доступу к информационному ресурсу и проведению операций по ознакомлению с информацией, ее документированию, модификации и уничтожению.

Разграничение доступа может осуществляться по различным моделям, построенным по тематическому признаку или по грифу секретности разрешенной к пользованию информации.

*Концепция доступа* - модель управления доступом, осуществляемая в абстрактной ЭВМ, которая посредничает при всех обращениях субъектов к информационным ресурсам.

Существуют следующие концепции доступа: дискреционное управление, мандатное управление, многоуровневая защита.

*Дискреционное управление доступом* - концепция (модель) доступа к объектам по тематическому признаку, при которой субъект доступа с определенным уровнем полномочий может передать свое право любому другому субъекту.

*Мандатное управление доступом* - концепция (модель) доступа субъектов к информационным ресурсам по грифу секретности разрешенной к пользованию информации, определяемому меткой секретности (конфиденциальности).

*Средство разграничения доступа* - программно - аппаратное средство, обеспечивающее разграничение доступа субъектов к информационным ресурсам в соответствии с принятой моделью.

Средствами разграничения доступа являются матрица доступа и метка секретности (конфиденциальности).

*Матрица доступа* - таблица, отображающая правила доступа субъектов к информационным ресурсам, данные о которых хранятся в диспетчере доступа.

*Метка секретности (конфиденциальности)* - элемент информации (бит), который характеризует степень секретности (конфиденциальности) информации, содержащейся в объекте.

*Идентификация доступа* - присвоение субъектам и объектам доступа идентификаторов и сравнение предъявленного идентификатора с утвержденным перечнем.

*Идентификатор* - средство идентификации доступа, представляющее собой отличительный признак субъекта или объекта доступа. Основным средством идентификации доступа для пользователей является пароль.

*Пароль* - средство идентификации доступа, представляющее собой кодовое слово в буквенной, цифровой или буквенно-цифровой форме, которое вводится в ЭВМ перед началом диалога с нею с клавиатуры терминала или при помощи идентификационной (кодовой) карты.

*Аутентификация пользователя* - подтверждение подлинности пользователя с помощью предъявляемого им аутентификатора.

*Аутентификатор* - средство аутентификации, представляющее отличительный признак пользователя.

Средствами аутентификации пользователя могут быть дополнительные кодовые слова, биометрические данные и другие отличительные признаки пользователя, которые вводятся в ЭВМ с клавиатуры дисплея, с идентификационной карты или при помощи специального устройства аутентификации по биометрическим данным.

*Биометрические данные* - средства аутентификации, представляющие собой такие личные отличительные признаки пользователя как тембр голоса, форма кисти руки, отпечатки пальцев и т.д., оригиналы которых в цифровом виде хранятся в памяти ЭВМ.

*Устройство регистрации доступа* пользователей - программно-аппаратное устройство, обеспечивающее регистрацию пользователей при всех их обращениях к вычислительной системе с указанием номера терминала, даты и времени обращения.

*Устройство прерывания программы пользователя* - программно-аппаратное устройство, обеспечивающее прерывание (блокирование) программы пользователя в случае попыток несанкционированного доступа.

*Устройство стирания данных* - программно-аппаратное устройство, обеспечивающее стирание оставшихся после обработки данных в ОЗУ путем записи нулей во все ячейки соответствующего блока памяти.

*Устройство повышения достоверности идентификации* - программно-аппаратное устройство, обеспечивающее корректировку ошибок идентификации, переданных с удаленных терминалов по каналам связи.

Для корректировки ошибок используются различные способы: обратная посылка сообщений на передающий конец для сравнения его с оригиналом, посылка одновременно с сообщением контрольных разрядов, использование избыточных кодов (код Хемминга, циклические коды) и т.д.

*"Выделенное" помещение* - специальное помещение, предназначенное для проведения собраний, совещаний, бесед и других мероприятий речевого характера по секретным или конфиденциальным вопросам.

Мероприятия речевого характера могут проводиться в выделенных помещениях как с использованием технических средств обработки речевой информации (ТСОИ), так и без них.

*Защита выделенного помещения* - проведение комплекса организационно - технических мероприятий по предотвращению утечки речевой секретной или конфиденциальной информации по техническим каналам за пределы выделенного помещения.

В общем случае комплекс мероприятий по защите выделенных помещений включает:

- защиту речевой информации, обрабатываемой техническими средствами от утечки за счет электромагнитных излучений и наводок (ПЭМИН);
- защиту речевой информации от утечки за счет эффекта электроакустического преобразования вспомогательных технических средств и систем (ВТСС);
- защиту речевой информации от утечки за счет лазерного зондирования стекол или стетоскопического прослушивания ограждающих конструкций;
- защиту речевой информации от утечки за счет несанкционированного доступа в помещение и скрытой установки в нем подслушивающих приборов (микрофонов, магнитофонов, радиопередатчиков и т.д.);
- акустическую защиту помещений.

*Аттестация выделенного помещения* - официальное подтверждение органом по аттестации (сертификации) или другим специально уполномоченным органом наличия необходимых и достаточных условий, обеспечивающих надежную акустическую защищенность выделенного помещения в соответствии с установленными нормами и требованиями.

Другие термины и понятия будут вводиться и поясняться в соответствующих разделах по мере возникновения необходимости.

## 1.2. Онтология задачи обеспечения информационной безопасности

Опираясь на введенные выше понятия, можно построить следующую онтологическую схему задачи обеспечения информационной безопасности (рис.1.1).

Рис. 1.1. Онтология задачи обеспечения безопасности информационных систем

*Субъекты информационных отношений (*источник, собственник, владелец или пользователь информации) определяют множество *информационных ресурсов*, которые должны быть защищены от различного рода *атак*. К активам ИС обычно относят: материальные ресурсы; информационные ресурсы (аналитическая, служебная, управляющая информация на всех этапах своего жизненного цикла: создание, обработка, хранение, передача, уничтожение); информационные технологические процессы жизненного цикла автоматизированных систем; предоставляемые информационные услуги и т.п. [9].

Атаки являются результатом реализации *угроз*, осуществляются через различные *уязвимости в защите*, и имеют вероятность реализации (*риск* атаки).

Основные *нарушения безопасности*: раскрытие информационных ценностей (*потеря конфиденциальности*), их неавторизованная модификация (*потеря целостности*) или неавторизованная потеря доступа к этим ценностям (*потеря доступности*).

В результате анализа уязвимостей защиты, свойств источников угроз (природы возникновения, характера, отношения к объектам ИС) и вероятностей их возможной реализации в конкретном окружении, определяются *риски* для данного набора информационных ресурсов. Это, в свою очередь, позволяет определить стратегию защиты, которая задается *политикой безопасности*.

Выработанная субъектом информационных отношений стратегия защиты может предусматривать для каждой из угроз одну из возможных линий поведения: попытку ликвидации источника угрозы, уклонение от угрозы, принятие угрозы, минимизация ущерба от атаки, вызванной данной угрозой, с помощью *сервисов* и *механизмов* безопасности. При этом следует учитывать, что отдельные уязвимости могут сохраниться и после применения мер защиты.

Политика безопасности определяет согласованную совокупность механизмов и сервисов безопасности, адекватную защищаемым ценностям и окружению, в котором они используются.

Процесс обеспечения безопасности информации должен носить комплексный характер и основываться на глубоком анализе возможных негативных последствий. Такой анализ предполагает обязательную идентификацию возможных источников угроз, факторов, способствующих их проявлению (уязвимостей) и, как следствие, определение актуальных угроз безопасности информации.

Исходя их данного принципа, моделирование и классификацию источников угроз и их проявлений, целесообразно проводить на основе анализа взаимодействия логической цепочки:

Источник угрозы → Угроза → Уязвимость→

→ Реализация угрозы (атака) → Последствия (ущерб).

При этом под термином угроза понимается возможная опасность (потенциальная или реально существующая) совершения какого-либо деяния (действия или бездействия), направленного против объекта защиты (информационных ресурсов), наносящего ущерб собственнику, владельцу или пользователю, проявляющегося в опасности искажения и/или потери информации.

### 1.3. Классификация угроз информационной безопасности

Обеспечение защиты информации невозможно без проведения системного анализа соответствующих угроз безопасности. Основу такого анализа должна составлять классификация угроз по определенным базовым признакам, дающая исследователю целостное представление о различных вариантах деструктивных воздействий и их последствиях.

В литературе предложен ряд классификаций угроз безопасности, отражающий те или иные аспекты рассматриваемой проблемы [19 - 23].

Однако, будучи разработанными для решения узкого круга специфических задач, они не могут являться основой общей систематизации угроз и выделения их наиболее значимых признаков для последующего обобщения и декомпозиции.

Необходимо разработать обобщенную классификацию, которая должна позволить рассматривать свойства угроз как предмет научного исследования и в последующем получить описания всех возможных их типов.

При этом классификацию факторов, влияющих на информационную безопасность, целесообразно осуществлять с учетом следующих требований:

- достаточности количества уровней классификации факторов, позволяющих формировать их полное множество;

- достаточной гибкости классификации, позволяющей расширять множества классифицируемых факторов, группировок и признаков, а также вносить необходимые изменения без нарушения структуры классификации.

Под угрозой безопасности, как отмечалось ранее, понимают ситуацию, при которой могут быть нарушены основные сервисы, например: целостность, конфиденциальность и доступность информации.

Морфологический анализ показывает, что можно выделить следующие основные составляющие угрозы информационной безопасности: источник воздействия на информационную систему, способ воздействия, информационные объекты воздействия, а также результат воздействия (причинённый ущерб).

Данные элементы при разработке классификации могут быть выбраны в качестве базовых классификационных признаков для последующей их декомпозиции.

Согласно ГОСТ Р 51275-2006 факторы, влияющие на информационную безопасность, можно подразделить по признаку отношения к природе их возникновения на *объективные* и *субъективные*, по отношению к объектам информационной системы - на *внутренние* и *внешние*.

Деление источников на субъективные и объективные оправдано исходя из соображений по определению вины за причинение ущерба информации. А деление на внутренние и внешние источники оправдано потому, что для одной и той же угрозы методы парирования для внешних и внутренних источников могу быть разными.

Кроме этого, и внешние и внутренние источники могут носить как преднамеренный, так и непреднамеренный характер.

*Непреднамеренные угрозы* возникают независимо от воли и желания людей. Данный тип угроз связан чаще всего с прямым природным или техногенным физическим воздействием на элементы информационной системы и ведет к нарушению работы этой системы и/или физическому повреждению (уничтожению) носителей информации, средств обработки и передачи данных, телекоммуникационных каналов.

Причиной возникновения угроз непреднамеренного характера могут быть как сбои вследствие конкретных ошибок персонала и прямых действий иных лиц (например, повреждение кабельной линии связи при проведении строительных работ), так и случайные нарушения в работе системы (например, вследствие поломки оборудования, сбоя в работе программного обеспечения и т.д.)

*Преднамеренные угрозы*, в отличие от непреднамеренных, могут быть созданы только людьми, действующими целенаправленно с целью дезорганизовать работу информационной системы. Преднамеренные угрозы, в свою очередь, подразделяются на пассивные и активные.

*Пассивные угрозы* связаны с несанкционированным доступом к информации без каких-либо ее изменений.

*Активные угрозы* связаны с попытками изменения (перехвата, модификации, уничтожения) информации или попытками лишения доступа к информационным ресурсам легитимных пользователей.

Общая схема классификации угроз информационной безопасности показана на рис.1.2.

По данным исследования, проведенного Executive Information Network, в 80% случаев носителем угрозы является человек. Случайные угрозы (сбои аппаратуры и ПО, ошибки операторов) составляют до 55% от общего количества, умышленные угрозы занимают 25% и на долю стихийных бедствий приходится 20%.

Таким образом, основными источниками угроз являются угрозы, обусловленные действиями субъекта (антропогенные источники угроз), угрозы, обусловленные техническими средствами (техногенные источники угроз) и угрозы, обусловленные стихийными источниками (природные источники угроз).

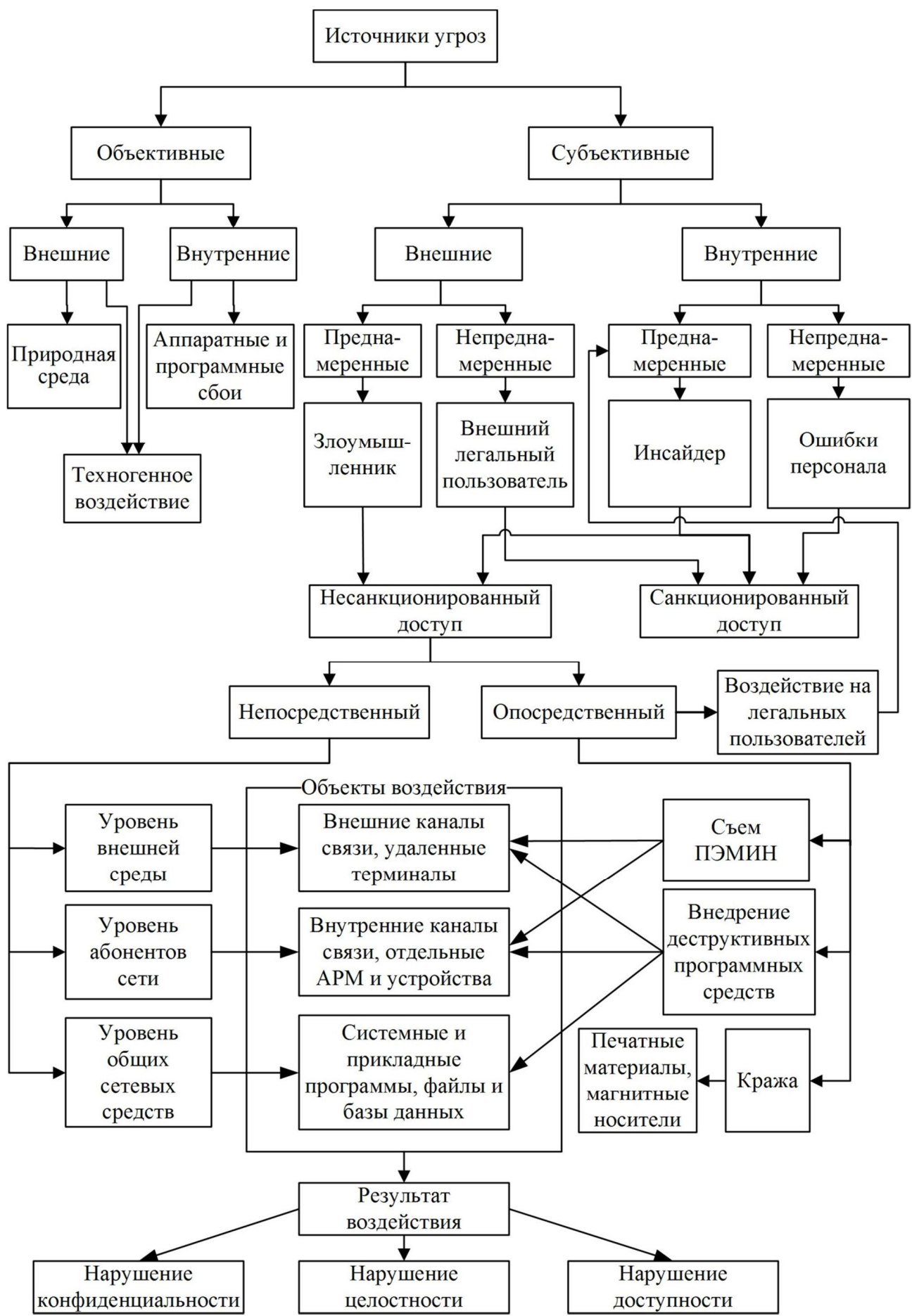

Рис.1.2. Классификация угроз информационным ресурсам.

*Антропогенные источники угроз.*

Антропогенными источниками угроз безопасности информации выступают субъекты, действия которых могут быть квалифицированы как умышленные или случайные нарушения. Эта группа наиболее обширна и представляет наибольший интерес с точки зрения организации защиты, так как действия субъекта можно оценить, спрогнозировать и принять адекватные меры защиты. Методы противодействия в этом случае управляемы и напрямую зависят от воли организаторов защиты информации.

В качестве антропогенного источника угроз можно рассматривать субъекта, имеющего доступ (санкционированный или несанкционированный) к работе со штатными средствами защищаемого объекта.

Субъекты (источники), действия которых могут привести к нарушению безопасности информации, могут быть как внешние, так и внутренние.

Внешние субъекты-источники угрозы, в свою очередь, могут быть случайными или преднамеренными и иметь разный уровень квалификации. К ним относятся:

- криминальные структуры;
- потенциальные преступники и хакеры;
- недобросовестные партнеры;
- технический персонал поставщиков телематических услуг;
- представители надзорных организаций и аварийных служб;
- представители силовых структур.

Внутренние субъекты (источники), как правило, представляют собой высококвалифицированных специалистов в области разработки и эксплуатации программного обеспечения и технических средств, знакомы со спецификой решаемых задач, структурой, основными функциями и принципами работы программно-аппаратных средств защиты информации, имеют возможность использования штатного оборудования. К ним относятся:

- основной персонал (пользователи, программисты, разработчики);

- представители службы защиты информации;
- вспомогательный персонал (уборщики, охрана);
- технический персонал (эксплуатация инженерных сетей).

В работе [24] приведен примерный список персонала и соответствующая степень опасности его действий:

- *наибольшая опасность*: администратор безопасности;
- *повышенный уровень опасности*: оператор системы, оператор ввода и подготовки данных, системный программист, управляющий по обработке данных;
- *средний уровень опасности*: ведущий инженер системы, ведущий инженер по программному обеспечению;
- *ограниченный уровень опасности*: прикладной программист, инженер или оператор по связи, администратор баз данных, инженер по оборудованию, оператор периферийного оборудования, пользователь-программист;
- *низкий уровень опасности*: инженер по периферийным средствам, библиотекарь прикладного программного обеспечения.

Таким образом, квалификация антропогенных источников угроз играет важную роль в оценке их влияния на безопасность ИС.

Особую группу внутренних антропогенных источников составляют лица с нарушенной психикой и специально внедренные и завербованные агенты, которые могут быть из числа основного, вспомогательного и технического персонала, а также представителей службы защиты информации. Данная группа рассматривается в составе перечисленных выше источников угроз, но методы парирования атак, исходящих из этой группы, могут иметь свои отличия.

*Техногенные источники угроз.*

Вторая группа содержит источники угроз, определяемые последствиями технократической деятельности человека, которые могут выйти из под его контроля и существовать сами по себе. Эти источники угроз менее прогнозируемые, напрямую зависят от свойств техники и поэтому требуют особого внимания.

Данный класс источников угроз безопасности информации особенно актуален в современных условиях, так как эксперты ожидают резкого роста числа техногенных катастроф, вызванных физическим и моральным устареванием используемого оборудования, а также отсутствием материальных средств на его обновление.

Технические средства, являющиеся источниками потенциальных угроз безопасности информации, так же могут быть внешними (средства связи, сети инженерных коммуникаций, транспорт) и внутренними (некачественные технические и/или программные средства обработки информации; вспомогательные средства охраны, сигнализации, телефонии; другие технические средства, применяемые в учреждении).

*Стихийные источники угроз*

Третья группа источников угроз объединяет обстоятельства, составляющие непреодолимую силу, то есть такие обстоятельства, которые носят объективный и абсолютный характер, распространяющийся на всех. К непреодолимой силе в законодательстве и договорной практике относят стихийные бедствия или иные обстоятельства, которые невозможно предусмотреть и/или предотвратить на современном уровне развития цивилизации [25].

Стихийные источники потенциальных угроз информационной безопасности, как правило, являются внешними по отношению к защищаемому объекту и под ними понимаются, прежде всего, природные катаклизмы: пожары; землетрясения; наводнения; ураганы; различные непредвиденные обстоятельства; необъяснимые явления и другие форс-мажорные обстоятельства [25].

Важно отметить, что угрозы безопасности различного типа на практике, как правило, демонстрируют сложную взаимосвязь. Так, например, маскарад (попытка выдать себя за авторизованного пользователя или процесс), будучи первичной угрозой, инициирующей утечку информации, может быть сам прямым следствием утечки, в частности, раскрытия пароля в результате, персональной неосторожности сотрудника.

Как следствие, наличие уязвимости системы защиты информации по отношению к какому-либо одному типу угрозы, может приводить к осуществлению целой комбинации угроз различного типа.

Таким образом, приведенная выше классификация описывает практически все основные характеристики угроз безопасности, которые могут иметь место в ИС при ее функционировании.

Следует заметить, что полные характеристики возможных угроз безопасности информации формируются на основе данной классификации декартовым произведением кортежей элементарных свойств, соответствующих базовым классификационным признакам.

При этом деструктивное воздействие на информацию осуществляется, как правило, через *несанкционированный доступ*, т. е. доступ к элементам ИС, выходящий за рамки разрешенных полномочий.

На рис.1.2 показаны схемы непосредственного и опосредованного доступа к хранимой, обрабатываемой и передаваемой информации.

Под непосредственным НСД понимается такой доступ, когда противник или персонал для реализации деструктивных воздействий на информацию использует в основном штатные средства ИС, реализуя те или иные встроенные документированные и недокументированные функции.

В отличие от внутреннего нарушителя внешнему противнику для достижения целей информационного воздействия необходимо преодолеть несколько уровней защиты. Как правило, это уровень внешней среды по отношению к объекту информатизации, уровень абонентов системы и уровень общих сетевых средств. Эффективность атаки внешнего противника напрямую связана с глубиной достигнутого им уровня.

Опосредованный доступ к информации связан с отсутствием у противника непосредственного контакта с элементами ИС. Данный тип доступа осуществляется только с использованием специальных технических и программных средств, путем кражи носителей информации или с помощью воздействия на легальных пользователей ИС.

По результатам исследования ФБР в 2010 году среди крупных американских компаний (в исследовании приняли участие 37% компаний-респондентов с годовым оборотом более 1 млрд. долларов) ущерб от умышленных или неумышленных действий сотрудников компании и хищения с помощью воздействия на персонал (например, хищение интеллектуальной собственности), составил более 50% от общих потерь. При этом для сравнения, ущерб от вирусов и вирусных эпидемий составил всего лишь около 30%.

Воздействие на легальных пользователей ИС может быть физическим или психологическим.

Физическое воздействие заключается в том, что злоумышленник прибегает к силовому воздействию на легальных пользователей ИС или его друзей и родственников с целью получить конфиденциальную информацию или нарушить функционирование каких-либо бизнес-процессов компании. Такие инциденты расследуются отделом физической безопасности компании и правоохранительными органами.

В рамках психологического воздействия широко распространены такие методы как шантаж, подкуп, социальная инженерия и многие другие. Рассмотрим более подробно социальную инженерию – наиболее интересный метод психологического воздействия с точки зрения реализации и способов защиты.

В научной литературе психологическое воздействие часто называется манипуляцией или манипулятивным воздействием. Манипуляция – это вид психологического воздействия, искусное исполнение которого ведет к скрытому возбуждению у другого человека намерений, не совпадающих с его актуально существующими желаниями [26].

Существует большое количество видов и методов манипулирования. Рассмотрим некоторые из них, используемые для получения доступа к информационным ресурсам.

*Психологическая атака* - метод активного воздействия на психику человека с целью отключения логического мышления: произведение положительного (отрицательного) впечатления или введение в состояние растерянности с последующим побуждением человека к нужной реакции.

Например, злоумышленник отправляет письмо сотруднику компании («жертве») на адрес электронной почты, в котором с помощью психологического воздействия заставляет его выполнить определенные действия (открыть веб-страницу, запустить прикрепленный файл и т.п.). Адрес электронной почты может быть получен из открытых источников, таких как поисковые системы, личные электронные дневники (блоги), форумы и др.

Примеры легенд атак на сотрудников компании и детали текстов писем приведены в [27].

*Пример GenoTree :*

Письмо-просьба, в котором отправитель (злоумышленник) рассказывает, что составляет генеалогическое дерево своего рода, причем у получателя (сотрудник компании, «жертва») и отправителя совпадают фамилии. Отправитель просит «жертву» посмотреть уже составленное им генеалогическое дерево; для просмотра дерева якобы используется специальная программа, которая также прикреплена к письму. На самом же деле к письму прикреплено троянское программное обеспечение:

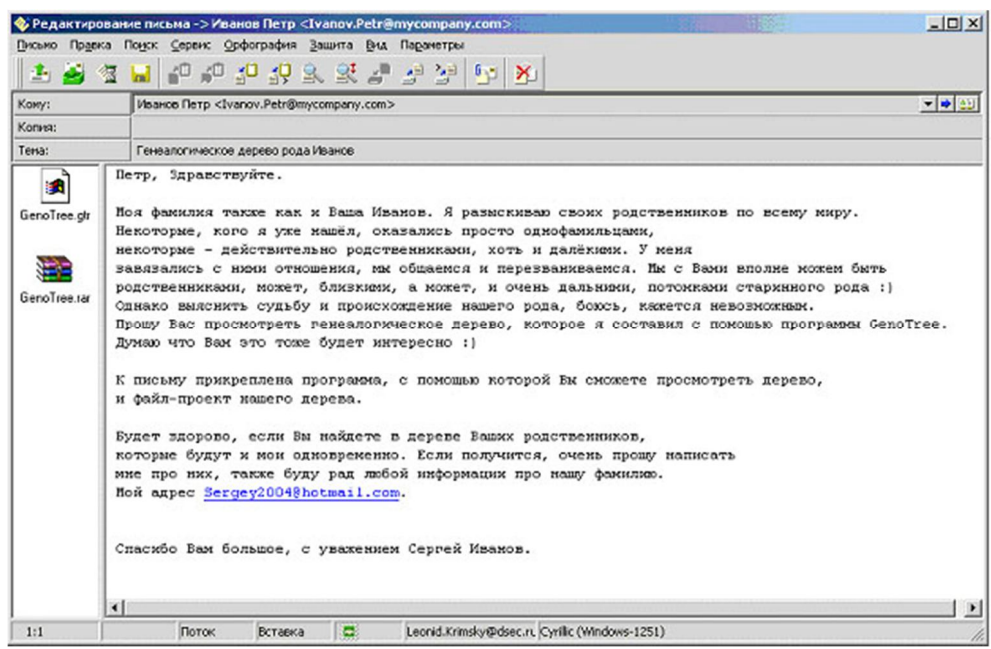

*Психологическое программирование* - метод однообразного или настойчивого воздействия на психику человека с целью выработки алгоритмов его поведения и образов мышления.

*Пример «Денежная пирамида»:*

Примером психологического программирования может быть регулярная рассылка различных сообщений одной и той же тематики (например, с предложением участия в традиционных денежных пирамидах). Сотрудник компании, получив сообщение с предложением легкого заработка и проигнорировав его, в следующий раз, получив подряд несколько аналогичных сообщений, возможно, задумается и через некоторое время выполнит указанные злоумышленником в инструкции действия.

*Психологическое манипулирование* - метод двойственного воздействия на психику человека, целью которого является поставить человека в положение выбора линии поведения между двумя альтернативами (между хорошим и плохим, добром и злом).

*Пример «Электронный платеж»:*

Письмо представляет собой фрагмент переписки двух человек, который якобы ошибочно был прислан сотруднику компании («жертве»). В переписке обсуждается возможность оплаты каких-либо услуг с помощью электронных денежных средств. Один из адресатов предлагает другому оплатить услуги с помощью электронного кошелька, при этом в письме приводятся номер электронного кошелька, пароль, инструкция по использованию. К письму прикреплен «плагин для совершения денежных операций» – троянское программное обеспечение. Таким образом, получатель попадает в положение выбора, находясь между двумя альтернативами: украсть или нет, воспользоваться чужими денежными средствами или нет.

*Психологическое давление* - метод внушительного воздействия на психику человека с целью принуждения его к определенным действиям.

*Пример «Критичное обновление»:*

Сотрудник компании («жертва») получает письмо якобы от имени другого сотрудника компании, обладающего более высоким статусом (большими полномочиями). Письмо представляет собой перечень действий (инструкцию), которые должен выполнить сотрудник компании. Для усиления психологического давления в письме могут быть приведены:

• основание для выполнения указанных в инструкции действий сотрудником, (например, вымышленный номер приказа);

• точные сроки выполнения указанных действий;

• пояснения к инструкции, ссылки на дополнительные источники информации;

• прочие атрибуты, усиливающие эффект доверия и подчинения.

Например, сотрудникам компании может быть разослано письмо якобы от имени руководителя отдела информационной безопасности компании, в котором сообщается о необходимости незамедлительно установить критичное обновление операционной системы, которое не вошло в репозитарий системы автоматического обновления по определенным причинам. К письму прикладывается детальная инструкция по установке обновления и само обновление (являющееся, естественно, троянским программным обеспечением).

Следует отметить, что рассмотренные методы психологического воздействия могут комбинироваться, образуя тем самым новые проявления воздействия на человека. Часто случается так, что «жертвы» психологической манипуляции не подозревают, что стали объектом атаки. В частности, это обстоятельство значительно затрудняет установление факта инцидента и его последующее расследование. Кроме этого, не всегда удается точно определить, были ли действия сотрудника умышленными или нет, и определить степень ответственности и меру взыскания, накладываемую на сотрудника.

Угрозы, связанные с воздействием на легальных пользователей ИС, легко реализуемы на практике, поэтому необходимо уделять им серьезное внимание. Для защиты информационной системы от таких угроз дополнительно следует:

• периодически проводить психологические тестирования сотрудников; предусмотреть возможность оказания психологической помощи (консультаций) сотрудникам компании.

• периодически проводить обучение пользователей вопросам информационной безопасности, в частности, детализируя методы, используемые злоумышленниками – социальными инженерами. Обучение рекомендуется завершать практическим тренингом и зачетом.

• использовать различные программные средства защиты информации, минимизирующие возможное влияние такого рода угроз.

## 1.4. Модель нарушителя

Как уже отмечалось выше, пользователи информационной системы и обслуживающий ее персонал, с одной стороны, являются составной частью, необходимым элементом ИС. С другой стороны, они же являются основной причиной и движущей силой нарушений в сфере безопасности.

Для достижения своих целей нарушитель прикладывает некоторые усилия, затрачивает определенные ресурсы. Исследовав личности нарушителей, суть применяемых ими приемов и средств, причины, толкнувшие их на нарушения, можно либо повлиять на сами эти причины (если это возможно), либо точнее определить требования к системе защиты от данного вида нарушений.

При этом модель нарушителя должна отражать его практические и теоретические возможности, априорные знания, время и место действия и т.п.

При разработке модели нарушителя определяются:
- категория лиц, к которым может принадлежать нарушитель;
- мотивы действий нарушителя (преследуемые нарушителем цели);
- квалификация нарушителя и его техническая оснащенность (используемые для совершения нарушения методы и средства);
- ограничения и предположения о характере возможных действий нарушителя.

Внутренним нарушителем может быть лицо из следующих категорий сотрудников:
- конечные пользователи (операторы) системы;
- персонал, обслуживающий технические средства (инженеры, техники);
- сотрудники отделов разработки и сопровождения ПО (прикладные и системные программисты);
- сотрудники службы безопасности ИС;
- руководители различных уровней.

Внешними нарушителями могут быть:
- технический персонал, обслуживающий здания (уборщики, электрики, сантехники и другие сотрудники, имеющие доступ в здания и помещения, где расположены компоненты ИС);

- клиенты (представители организаций, граждане);
- посетители (приглашенные по какому-либо поводу);
- представители организаций, взаимодействующих по вопросам обеспечения жизнедеятельности организации (энерго-, водо-, теплоснабжения и т.п.);
- представители конкурирующих организаций (иностранных спецслужб) или лица, действующие по их заданию;
- лица, случайно или умышленно нарушившие пропускной режим (без цели нарушить безопасность ИС);
- любые лица за пределами контролируемой зоны.

Можно выделить несколько основных мотивов нарушений:

- безответственность;
- самоутверждение;
- вандализм;
- принуждение;
- месть;
- корыстный интерес;
- идейные соображения.

При нарушениях, вызванных безответственностью, пользователь производит какие-либо разрушающие действия, не связанные, тем не менее, со злым умыслом. В большинстве случаев это следствие некомпетентности или небрежности.

Некоторые пользователи считают получение доступа к системным наборам данных крупным успехом, затевая своего рода игру "пользователь - против системы" ради самоутверждения либо в собственных глазах, либо в глазах коллег.

Нарушение безопасности ИС может быть связано, как указывалось ранее, с принуждением (шантаж), местью, идейными соображениями или корыстными интересами пользователя системы. В этом случае он будет целенаправленно пытаться преодолеть систему защиты для доступа к хранимой, передаваемой и обрабатываемой информации и другим ресурсам ИС.

По уровню знаний об ИС нарушителей можно классифицировать следующим образом:

- знает функциональные особенности ИС, основные закономерности формирования в ней массивов данных и потоков запросов к ним, умеет пользоваться штатными средствами;
- обладает высоким уровнем знаний и опытом работы с техническими средствами системы и их обслуживания;
- обладает высоким уровнем знаний в области программирования и вычислительной техники, проектирования и эксплуатации автоматизированных информационных систем;
- знает структуру, функции и механизм действия средств защиты, их сильные и слабые стороны.

По уровню возможностей (используемым методам и средствам):

- применяет только агентурные методы получения сведений;
- применяет пассивные средства (технические средства перехвата без модификации компонентов системы);
- использует только штатные средства и недостатки систем защиты для ее преодоления (несанкционированные действия с использованием разрешенных средств), а также компактные магнитные носители информации, которые могут быть скрытно пронесены через посты охраны;
- применяет методы и средства активного воздействия (модификация и подключение дополнительных технических средств, подключение к каналам передачи данных, внедрение программных закладок и использование специальных инструментальных и технологических программ).

По времени действия:

- в процессе функционирования ИС (во время работы компонентов системы);

- в период неактивности компонентов системы (в нерабочее время, во время плановых перерывов в ее работе, перерывов для обслуживания и ремонта и т.п.);
- как в процессе функционирования ИС, так и в период неактивности компонентов системы.

По месту действия:

- без доступа на контролируемую территорию организации;
- с контролируемой территории без доступа в здания и сооружения;
- внутри помещений, но без доступа к техническим средствам ИС;
- с рабочих мест конечных пользователей (операторов) ИС;
- с доступом в зону данных (серверов баз данных, архивов и т.п.);
- с доступом в зону управления средствами обеспечения безопасности ИС.

Могут учитываться, например, следующие ограничения и предположения о характере действий возможных нарушителей:

- работа по подбору кадров и специальные мероприятия затрудняют возможность создания коалиций нарушителей, т.е. объединения (сговора) для целенаправленных действий по преодолению подсистемы защиты двух и более нарушителей;
- нарушитель, планируя попытки НСД, скрывает свои несанкционированные действия от других сотрудников и т.п.

Определение конкретных значений характеристик возможных нарушителей в значительной степени субъективно.

Правильно построенная (адекватная реальности) модель нарушителя, в которой отражаются его практические и теоретические возможности, априорные знания, время и место действия и т.п. характеристики - важная составляющая успешного проведения анализа рисков и определения требований к составу и характеристикам системы защиты.

## 1.5. Классификация уязвимостей безопасности

Угрозы как возможные опасности совершения какого-либо действия, направленного против объекта защиты, проявляются не сами по себе, а через уязвимости, приводящие к нарушению безопасности информации на конкретном объекте информатизации (ОИ).

Уязвимости присущи объекту информатизации, неотделимы от него и обуславливаются недостатками процесса функционирования, свойствами архитектуры автоматизированных систем, протоколами обмена и интерфейсами, применяемыми программным обеспечением и аппаратной платформой, условиями эксплуатации и расположения и т.п.

Уязвимости могут присутствовать как в программно-аппаратном, так и организационно-правовом обеспечении ИБ.

Основная часть уязвимостей организационно-правового обеспечения обусловлена отсутствием на предприятиях нормативных документов, касающихся вопросов информационной безопасности. Примером уязвимости данного типа является отсутствие в организации утверждённой концепции или политики информационной безопасности, которая бы определяла требования к защите ИС, а также конкретные пути их реализации.

Уязвимости программно-аппаратного обеспечения могут присутствовать в программных или аппаратных компонентах рабочих станций пользователей ИС, серверов, а также коммуникационного оборудования и каналов связи ИС.

Источники угроз могут использовать уязвимости для нарушения безопасности информации, получения незаконной выгоды (нанесения ущерба собственнику, владельцу, пользователю информации). Кроме того, возможны незлонамеренные действия источников угроз по активизации тех или иных уязвимостей, способных нанести вред.

Каждой угрозе могут быть сопоставлены различные уязвимости. Устранение или существенное ослабление уязвимостей влияет на возможность реализации угроз безопасности информации (см. рис 1.1).

Существуют различные подходы к систематизации уязвимостей информационных систем и технологий.

В работе [28] для удобства анализа уязвимости разделены на классы, группы и подгруппы. Указано, что уязвимости безопасности информации могут быть:

- объективными;
- субъективными;
- случайными.

*Объективные уязвимости*

Объективные уязвимости зависят от особенностей построения и технических характеристик оборудования, применяемого на защищаемом объекте. Полное устранение этих уязвимостей невозможно, но они могут существенно ослабляться техническими и инженерно-техническими методами парирования угроз безопасности информации. К ним можно отнести:

- сопутствующие техническим средствам излучения (электромагнитные, электрические, звуковые);
- активизируемые (аппаратные и программные закладки);
- определяемые особенностями элементов (элементы, обладающие электроакустическими преобразователями или подверженные воздействию электромагнитного поля);
- определяемые особенностями защищаемого объекта (местоположением объекта, организацией каналов обмена информацией).

*Субъективные уязвимости*

Субъективные уязвимости зависят от действий сотрудников и, в основном, устраняются организационными и программно-аппаратными методами. К ним относятся:

- ошибки (при подготовке и использовании программного обеспечения, при управлении сложными системами, при эксплуатации технических средств);

- нарушения (режима охраны и защиты, режима эксплуатации технических средств, режима использования, режима конфиденциальности и т.д.).

*Случайные уязвимости*

Случайные уязвимости зависят от особенностей окружающей защищаемый объект среды и непредвиденных обстоятельств. Эти факторы, как правило, мало предсказуемы и их устранение возможно только при проведении комплекса организационных и инженерно-технических мероприятий по противодействию угрозам информационной безопасности. Это:

- сбои и отказы (отказы и неисправности технических средств, старение и размагничивание носителей информации, сбои программного обеспечения, сбои электроснабжения и др.);
- повреждения (жизнеобеспечивающих коммуникаций, ограждающих конструкций и т.п.).

В работе [29] приведен вариант классификации, основанный на этапах жизненного цикла информационных систем, и таким образом, в большей степени отвечающий задачам поиска и устранения уязвимостей (табл. 1.1).

Таблица 1.1.

Классификация уязвимостей в соответствии с этапами жизненного цикла ИС

| Этапы жизненного цикла ИС | Категории уязвимостей | Обнаружение | Устранение |
|---|---|---|---|
| 1. Проектирование | Уязвимости проектирования | Трудоемкий и длительный процесс | Трудоемкий и длительный процесс. Иногда устранение невозможно |
| 2. Реализация | Уязвимости реализации | Относительно трудно и долго | Несложно, но относительно долго |
| 3. Эксплуатация | Уязвимости конфигурации | Легко и быстро | Легко и быстро |

*Уязвимости проектирования.* Данный тип уязвимостей является наиболее серьезным и сложным для устранения, так как уязвимости этого типа свойственны основополагающим элементам проекта и алгоритмам реализации базовых функций. Как правило, уязвимости проектирования возникают как следствие недооценки требований безопасности при постановке задач и проявляются по мере возникновения новых угроз безопасности (например, уязвимости протоколов Интернет).

*Уязвимости реализации.* Эта категория уязвимостей появляется на этапе реализации в программном или аппаратном обеспечении корректного с точки зрения безопасности проекта. Обнаруживаются и устраняются подобного рода уязвимости относительно несложно, так как соответствующие изменения в программном или аппаратном обеспечении, как правило, не затрагивают принципиальных элементов системы.

*Уязвимости конфигурации.* Этот вид, наряду с уязвимостями реализации, является самой распространенной категорией уязвимостей. Локализовать и устранить такие уязвимости обычно проще всего. Проблема заключается лишь в том, чтобы определить само наличие уязвимости конфигурации. В лучшем случае это происходит на этапе тестирования системы, в худшем - индикатором наличия уязвимости конфигурации является успешно проведенная атака. Типичным источником уязвимостей конфигурации является широко применяемая большинством пользователей установка программного обеспечения со значениями "по умолчанию".

Характерные примеры уязвимостей этого типа:
- наличие слабых, не стойких к угадыванию паролей доступа к ресурсам информационной системы. При активизации этой уязвимости нарушитель может получить несанкционированный доступ к информации путём взлома пароля при помощи метода полного перебора или подбора по словарю;
- наличие в системе незаблокированных встроенных учётных записей пользователей, при помощи которых потенциальный нарушитель может собрать дополнительную информацию, необходимую для проведения атаки. Примерами таких учётных записей являются запись "Guest" в операционных системах или запись "Anonymous" на FTP-серверах;
- неправильным образом установленные права доступа пользователей к информационным ресурсам. В случае если в результате ошибки

администратора пользователи, работающие с системой, имеют больше прав доступа, чем это необходимо для выполнения их функциональных обязанностей, то это может привести к несанкционированному использованию дополнительных полномочий для проведения атак. Например, если пользователи будут иметь права доступа на чтение содержимого исходных текстов серверных сценариев, выполняемых на стороне Web-сервера, то этим может воспользоваться потенциальный нарушитель для изучения алгоритмов работы механизмов защиты Web-приложений и поиска в них уязвимых мест;

- наличие неиспользуемых, но потенциально опасных сетевых служб и программных компонентов. Так, например, большая часть сетевых серверных служб, таких как Web-серверы и серверы СУБД поставляются вместе с примерами программ, которые демонстрируют функциональные возможности этих продуктов. В некоторых случаях эти программы имеют высокий уровень привилегий в системе или содержат уязвимости, использование которых злоумышленником может привести к нарушению информационной безопасности системы. Примерами таких программ являются образцы CGI-модулей, которые поставляются вместе с Web-приложениями, а также примеры хранимых процедур в серверах СУБД;
- неправильная конфигурация средств защиты, приводящая к возможности проведения сетевых атак. Так, например, ошибки в настройке межсетевого экрана могут привести к тому, что злоумышленник сможет передавать через него пакеты данных.

## 1.6. Основные типы атак на информационные ресурсы.

В любой ИС существует информационный поток от отправителя (файл, пользователь, компьютер) к получателю (файл, пользователь, компьютер):

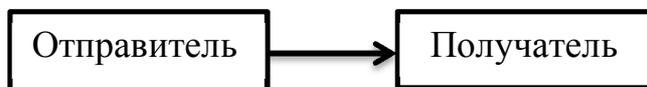

Рис. 1.3. Информационный поток

При этом все атаки можно разделить на два класса: пассивные и активные.

*Пассивная атака.*

Пассивной называется такая атака, при которой противник не имеет возможности модифицировать передаваемые сообщения и вставлять в информационный канал между отправителем и получателем свои сообщения. Целью пассивной атаки может быть только прослушивание передаваемых сообщений и анализ трафика.

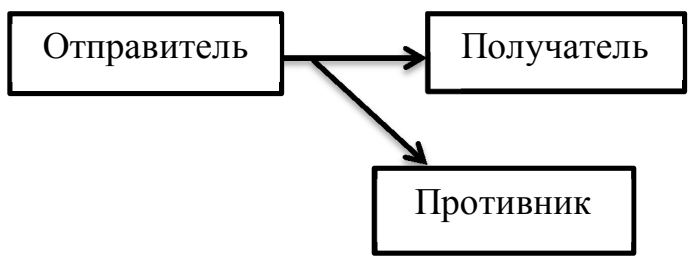

Рис. 1.4. Пассивная атака

*Активная атака.*

Активной называется такая атака, при которой противник имеет возможность модифицировать передаваемые сообщения и вставлять свои сообщения. Различают следующие типы активных атак:

*Отказ в обслуживании* - DoS-атака (Denial of Service)

Отказ в обслуживании нарушает нормальное функционирование сетевых сервисов. Противник может перехватывать все сообщения, направляемые определенному адресату. Другим примером подобной атаки является создание значительного трафика, в результате чего сетевой сервис не сможет обрабатывать запросы законных клиентов.

Классическим примером такой атаки в сетях TCP/IP является SYN-атака, при которой нарушитель посылает пакеты, инициирующие установление TCP-соединения, но не посылает пакеты, завершающие установление этого соединения. В результате может произойти переполнение памяти на сервере, и серверу не удастся установить соединение с законными пользователями.

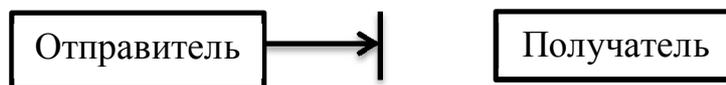

Рис. 1.5. DoS-атака

*Модификация потока данных* - атака "man in the middle"

Модификация потока данных означает либо изменение содержимого пересылаемого сообщения, либо изменение порядка сообщений.

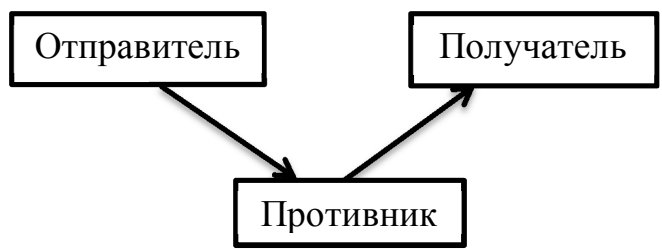

Рис. 1.6. Атака "man in the middle"

*Создание ложного потока* (фальсификация)

Фальсификация (нарушение аутентичности) означает попытку одного субъекта выдать себя за другого.

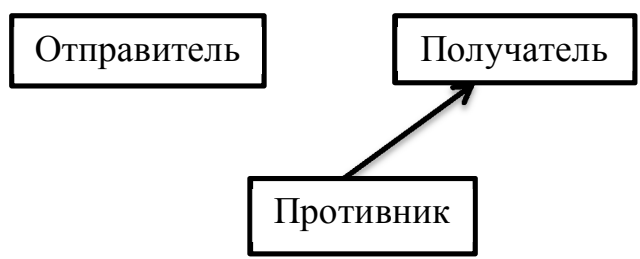

Рис. 1.7. Создание ложного потока

*Повторное использование* (replay-атака)

Повторное использование означает пассивный захват данных с последующей их пересылкой для получения несанкционированного доступа - это так называемая replay-атака.

На самом деле replay-атаки являются одним из вариантов фальсификации, но в силу того, что это один из наиболее распространенных вариантов атаки для получения несанкционированного доступа, его часто рассматривают как отдельный тип атаки.

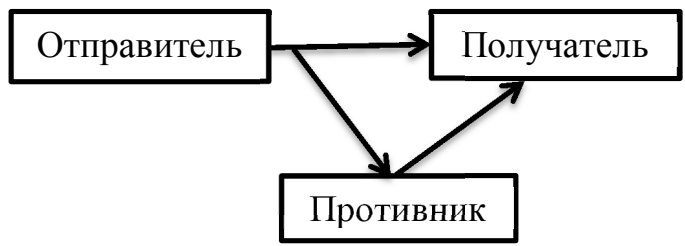

Рис. 1.8. Replay-атака

Перечисленные атаки могут существовать в любых типах сетей, а не только в сетях, использующих в качестве транспорта протоколы TCP/IP. Но в сетях, построенных на основе TCP/IP, атаки встречаются чаще всего, потому что, во-первых, Internet стал самой распространенной сетью, а во-вторых, при разработке протоколов TCP/IP требования безопасности никак не учитывались.

Атаки, реализуемые на локальном компьютере, можно рассматривать как частный случай информационного обмена между пользователем и локальной машиной.

Методы реализации атак можно разделить на следующие группы [30]:

- аналитические;
- технические;
- программные;
- социальные;
- организационные.

При этом необходимо учитывать, что само понятие «метод», применимо только при рассмотрении атак, реализуемых антропогенными источниками. Для техногенных и стихийных источников, это понятие трансформируется в понятие «предпосылка».

Классификация возможностей реализации угроз (атак), представляет собой совокупность возможных вариантов действий источника угроз определенными методами с использованием уязвимостей, которые приводят к реализации целей атаки.

Как было показано выше, существуют различные типа классификации атак. Например, деление на пассивные и активные, внешние и внутренние, умышленные и неумышленные атаки.

В [31] предложена несколько другая классификация:

- Удаленное проникновение (remote penetration). Атаки, которые позволяют реализовать удаленное управление компьютером через сеть. Для реализации такой атаки используются, например, программы NetBus или BackOrifice.

- Локальное проникновение (local penetration). Атака, которая приводит к получению несанкционированного доступа к узлу, на котором она запущена. Примером реализации такой атаки является программа GetAdmin.

- Удаленный отказ в обслуживании (remote denial of service). Атаки, которые позволяют нарушить функционирование или перегрузить компьютер через Internet. Для реализации такой атаки могут быть использованы, например, программы Teardrop или trin00.

- Локальный отказ в обслуживании (local denial of service). Атаки, которые позволяют нарушить функционирование или перегрузить компьютер, на котором они реализуются. Примером такой атаки является "враждебный" апплет, который загружает центральный процессор бесконечным циклом, что приводит к невозможности обработки запросов других приложений.

- Сетевые сканеры (network scanners). Программы, которые анализируют топологию сети и обнаруживают сервисы, доступные для атаки. Примером такой программы служит nmap.

- Сканеры уязвимостей (vulnerability scanners). Программы, которые ищут уязвимости на узлах сети и которые могут быть использованы для реализации атак. Пример такого сканера - системы SATAN или ShadowSecurityScanner.

- Взломщики паролей (password crackers). Программы, которые "подбирают" пароли пользователей. Примеры взломщика паролей: L0phtCrack для Windows или Crack для Unix.
- Анализаторы протоколов (sniffers). Программы, которые "прослушивают" сетевой трафик. При помощи этих программ можно автоматически искать такую информацию, как идентификаторы и пароли пользователей, информацию о кредитных картах и т.д. Таким анализатором протоколов являются Microsoft Network Monitor, NetXRay компании Network Associates или LanExplorer.

Компания Internet Security Systems, Inc. еще больше сократила число возможных категорий, доведя их до пяти:
- сбор информации;
- попытки несанкционированного доступа;
- отказ в обслуживании;
- подозрительная активность;
- системные атаки.

Первые 4 категории относятся к удаленным атакам, а последняя - к локальным, реализуемым на атакуемом узле. Можно заметить, что в данную классификацию не попал целый класс так называемых "пассивных" атак. Помимо "прослушивания" трафика, в эту категорию также попадают такие атаки, как "ложный DNS-сервер", "подмена ARP-сервера" и т.п.

Классификация, реализованная во многих системах обнаружения атак, не может быть категоричной. Например, атака, реализация которой для ОС Unix может иметь самые плачевные последствия (самый высокий приоритет), для ОС Windows NT может быть вообще не применима или иметь очень низкую степень риска (например, переполнение буфера statd). Кроме того, существует неразбериха и в самих названиях атак и уязвимостей. Например, одна и та же атака, может иметь совершенно различные наименования у разных производителей.

Для устранения неоднозначности в именованиях уязвимостей и атак в 1999 году компания MITRE Corporation предложила решение, независимое от производителя средств анализа защищенности, обнаружения атак и т.д.

Это решение было реализовано в виде базы данных CVE (Common Vulnerability Enumeration), которая затем была переименована в Common Vulnerabilities and Exposures. Это позволило всем специалистам и производителям разговаривать на одном языке.

В настоящее время CVE поддерживают такие компании как Cisco, Axent, BindView, IBM и другие.

Систематизация имеющихся данных об атаках и этапах их реализации дает необходимый базис для понимания технологий обнаружения атак и их дальнейшей нейтрализации. Как было показано, не будь уязвимостей в компонентах ИС, нельзя было бы реализовать многие атаки и, следовательно, традиционные системы защиты вполне эффективно справлялись бы с возможными атаками. Однако любая ИС имеет изъяны, вследствие чего и появляются уязвимости, которые используются злоумышленниками для реализации атак.

При этом цель атаки может не совпадать с целью реализации угрозы и может быть направлена на получения промежуточного результата, необходимого для достижения в дальнейшем цели реализации угрозы. В случае такого несовпадения атака рассматривается как этап подготовки к совершению действий, направленных на реализацию угрозы, то есть как «подготовка к совершению» противоправного действия.

Сама по себе атака на информационные ресурсы, не приведшая ни к каким последствиям, преступлением не является, и привлечь к ответственности за нее нельзя, на что не раз обращалось внимание на различных конференциях, в том числе представителями подразделений МВД по борьбе с преступлениями в сфере высоких технологий.

Результатом атаки являются последствия (ущерб), которые приводят к снижению уровня сервисов безопасности и/или способствуют повышению уязвимости ИС (рис.1.1).

**К атакам, связанным с непреднамеренными субъективными угрозами ИБ относятся:**

- неумышленные действия, приводящие к частичному или полному отказу системы или разрушению аппаратных, программных, информационных ресурсов системы (неумышленная порча оборудования, удаление, искажение файлов с важной информацией или программ, в том числе системных и т.п.);
- неправомерное отключение оборудования или изменение режимов работы устройств и программ;
- неумышленная порча носителей информации;
- запуск технологических программ, способных при некомпетентном использовании вызывать потерю работоспособности системы (зависания или зацикливания) или осуществляющих необратимые изменения в системе (форматирование или реструктуризацию носителей информации, удаление данных и т.п.);
- нелегальное внедрение и использование неучтенных программ (игровых, обучающих, технологических и др., не являющихся необходимыми для выполнения нарушителем своих служебных обязанностей) с последующим необоснованным расходованием ресурсов (загрузка процессора, захват оперативной памяти и памяти на внешних носителях);
- заражение компьютера вирусами;
- неосторожные действия, приводящие к разглашению конфиденциальной информации, или делающие ее общедоступной;
- разглашение, передача или утрата атрибутов разграничения доступа (паролей, ключей шифрования, идентификационных карточек, пропусков и т.п.);

- проектирование архитектуры системы, технологии обработки данных, разработка прикладных программ, с возможностями, представляющими опасность для работоспособности системы и безопасности информации;
- игнорирование организационных ограничений (установленных правил) при работе в системе;
- вход в систему в обход средств защиты (загрузка посторонней операционной системы со сменных магнитных носителей и т.п.);
- некомпетентное использование, настройка или неправомерное отключение средств защиты персоналом службы безопасности;
- пересылка данных по ошибочному адресу абонента (устройства);
- ввод ошибочных данных;
- неумышленное повреждение каналов связи.

**Атаки, связанные с преднамеренными субъективными угрозами:**

- физическое разрушение системы (путем взрыва, поджога и т.п.) или вывод из строя отдельных наиболее важных компонентов системы (устройств, носителей важной системной информации, лиц из числа персонала и т.п.);
- отключение или вывод из строя подсистем обеспечения функционирования вычислительных систем (электропитания, охлаждения и вентиляции, линий связи и т.п.);
- действия по дезорганизации функционирования системы (изменение режимов работы устройств или программ, забастовка, саботаж персонала, постановка мощных активных радиопомех на частотах работы устройств системы и т.п.);
- внедрение агентов в персонал системы (в том числе, возможно, и в административную группу, отвечающую за безопасность);
- вербовка (путем подкупа, шантажа и т.п.) персонала или отдельных пользователей, имеющих определенные полномочия;
- применение подслушивающих устройств, дистанционная фото- и видео-съемка и т.п.;

- перехват побочных электромагнитных, акустических и других излучений устройств и линий связи, а также наводка активных излучений на вспомогательные технические средства, непосредственно не участвующие в обработке информации (телефонные линии, сети питания, отопления и т.п.);
- перехват данных, передаваемых по каналам связи, их анализ с целью выяснения протоколов обмена, правил вхождения в сеть и авторизации пользователя, и последующая попытка имитации для проникновения в систему;
- хищение носителей информации (магнитных дисков, лент, микросхем памяти, запоминающих устройств и целиком ПЭВМ);
- несанкционированное копирование носителей информации;
- хищение производственных отходов (распечаток, записей, списанных носителей информации и т.п.);
- чтение остаточной информации из оперативной памяти и с внешних запоминающих устройств;
- чтение информации из областей оперативной памяти, используемых операционной системой (в том числе подсистемой защиты) или другими пользователями, в асинхронном режиме, используя недостатки мульти задачных операционных систем и систем программирования;
- незаконное получение паролей и других реквизитов разграничения доступа (агентурным путем, используя халатность пользователей, путем подбора, путем имитации интерфейса системы и т.д.) с последующей маскировкой под зарегистрированного пользователя ("маскарад");
- несанкционированное использование терминалов пользователей, имеющих уникальные физические характеристики, такие как номер рабочей станции в сети, физический адрес, адрес в системе связи, аппаратный блок шифрования и т.п.;
- вскрытие шифров криптозащиты информации;

- внедрение аппаратных "спецвложений", программных "закладок" и "вирусов" ("троянских коней" и "жучков"), то есть таких участков программ, которые не нужны для осуществления заявленных функций, но позволяющих преодолевать систему защиты, скрытно и незаконно осуществлять доступ к системным ресурсам с целью регистрации и передачи критической информации или дезорганизации функционирования системы;
- незаконное подключение к линиям связи с целью работы "между строк", с использованием пауз в действиях законного пользователя от его имени с последующим вводом ложных сообщений или модификацией передаваемых сообщений;
- незаконное подключение к линиям связи с целью прямой подмены законного пользователя путем его физического отключения после входа в систему и успешной аутентификации с последующим вводом дезинформации и навязыванием ложных сообщений.

Следует заметить, что чаще всего для достижения поставленной цели злоумышленник использует не одну, а некоторую совокупность из перечисленных выше атак.

## 1.7. Ущерб от реализации атак на информационные ресурсы

Важной характеристикой при рассмотрении задачи обеспечения информационной безопасности является величина ущерба, причиненного информационным ресурсам в результате реализации угрозы (атаки).

Для описания ущерба представляется целесообразным выделить значимость и тип цели, на которую нацелена та или иная атаки, и определить степень достижения этой цели.

Принято выделять четыре уровня значимости информации:

1. Жизненно важная – незаменимая информация, наличие которой необходимо для функционирования организации.

2. Важная — информация, которая может быть заменена или восстановлена, но процесс восстановления очень труден и требует больших затрат.

3. Полезная – информация, которая полезна и которую трудно восстановить, однако организация может функционировать и без нее.

4. Несущественная – информация, которая практически не нужна организации.

Тип цели можно классифицировать по нарушению основных характеристик безопасности (целостности, конфиденциальности, доступности), а степень ее достижения может представлять собой, например, количественный или качественный показатель, характеризующий ухудшение уровня основных сервисов безопасности.

Угроза, как следует из определения, это опасность причинения ущерба. В этом определении проявляется жесткая связь технических проблем с юридической категорией, каковой является «ущерб».

С юридической точки зрения «ущерб» - это фактические расходы, понесенные субъектом в результате нарушения его прав (например, разглашения или использования нарушителем конфиденциальной информации), утраты или повреждения имущества, а также расходы, которые он должен будет произвести для восстановления нарушенного права и стоимости поврежденного или утраченного имущества (ГК РФ, часть I, ст.15).

Проведение анализа возможного ущерба - необходимый этап при выборе методов парирования угроз ИБ.

Для информационных ресурсов и технологий результат реализации угрозы может быть охарактеризован величиной совокупного ущерба, нанесенного основным сервисам безопасности информационных активов, например, целостности, конфиденциальности, доступности, что в свою очередь приводит к различным видам ущерба для собственника информационных ресурсов.

Проявления возможного ущерба могут быть различны [28]:

- моральный и материальный ущерб деловой репутации организации;

- моральный, физический или материальный ущерб, связанный с разглашением персональных данных отдельных лиц;

- материальный (финансовый) ущерб от разглашения защищаемой (конфиденциальной) информации;

- материальный (финансовый) ущерб от необходимости восстановления нарушенных защищаемых информационных ресурсов;

- материальный ущерб (потери) от невозможности выполнения взятых на себя обязательств перед третьей стороной;

- моральный и материальный ущерб от дезорганизации деятельности организации;

- материальный и моральный ущерб от нарушения международных отношений.

Ущерб может быть причинен каким-либо субъектом и в этом случае имеется на лицо правонарушение, а также явиться следствием независящим от субъекта проявлений (например, стихийных случаев или иных воздействий, таких как проявления техногенных свойств цивилизации).

В первом случае налицо вина субъекта (ст. 24 УК РФ), которая определяет причиненный вред как состав преступления, совершенного по злому умыслу (ст. 25 УК РФ) или по неосторожности (ст. 26 УК РФ), в результате невиновного причинения вреда (ст. 28 УК РФ). Причиненный ущерб в этом случае должен квалифицироваться как состав преступления, оговоренный уголовным правом.

Во втором случае ущерб носит вероятностный характер и должен быть сопоставлен с тем риском, который оговаривается гражданским, административным или арбитражным правом как предмет рассмотрения.

В теории права под ущербом понимается невыгодные для собственника имущественные последствия, возникшие в результате правонарушения. Ущерб выражается в уменьшении имущества, либо в недополучении дохода, который был бы получен при отсутствии правонарушения (упущенная выгода).

Если в качестве субъекта, причинившего ущерб, рассматривать какую-либо личность, то категория «ущерб» справедлива только в том случае, когда можно доказать, что он причинен, то есть деяния личности необходимо квалифицировать в терминах правовых актов, как состав преступления. Поэтому, при классификации угроз безопасности информации в этом случае целесообразно учитывать требования действующего уголовного права, определяющего состав преступления.

Приведем некоторые примеры составов преступления, определяемых Уголовным Кодексом Российской Федерации.

Хищение – совершенные с корыстной целью противоправные безвозмездное изъятие и (или) обращение чужого имущества в пользу виновного или других лиц, причинившее ущерб собственнику или владельцу имущества (ст. 158 прим. УК РФ).

Копирование компьютерной информации – повторение и устойчивое запечатление информации на машинном или ином носителе (ст. 272 УК РФ).

Уничтожение – внешнее воздействие на имущество, в результате которого оно прекращает свое физическое существование либо приводится в полную непригодность для использования по целевому назначению.

Уничтоженное имущество не может быть восстановлено путем ремонта или реставрации и полностью выводится из хозяйственного оборота (ст. 167 УК РФ).

Уничтожение компьютерной информации – стирание ее в памяти ЭВМ (ст. 272 УК РФ).

Повреждение – изменение свойств имущества, при котором существенно ухудшается его состояние, утрачивается значительная часть его полезных свойств и оно становится полностью или частично непригодным для целевого использования (ст. 167 УК РФ).

Модификация компьютерной информации – внесение любых изменений, кроме связанных с адаптацией программы для ЭВМ или баз данных (ст. 272 УК РФ).

Блокирование компьютерной информации – искусственное затруднение доступа пользователей к информации, не связанное с ее уничтожением (ст. 272 УК РФ).

Несанкционированное уничтожение, блокирование модификация, копирование информации – любые не разрешенные законом, собственником или компетентным пользователем указанные действия с информацией (ст. 273 УК РФ).

Обман (отрицание подлинности, навязывание ложной информации) – умышленное искажение или сокрытие истины с целью ввести в заблуждение лицо, в ведении которого находится имущество и таким образом добиться от него добровольной передачи имущества, а также сообщение с этой целью заведомо ложных сведений [32].

Однако говорить о злом умысле личности в уничтожении информации в результате стихийных бедствий не приходится, как и о том, что стихия сможет воспользоваться конфиденциальной информацией для извлечения собственной выгоды. Хотя и в том и в другом случае собственнику информации причинен ущерб.

Здесь правомочно применение категории «причинение вреда имуществу». При этом речь уже идет не об уголовной ответственности за уничтожение или повреждение чужого имущества, а о случаях, подпадающих под гражданское право в части возмещения причиненного ущерба: риск случайной гибели имущества – то есть риск возможного нанесения убытков в связи с гибелью или порчей имущества по причинам, не зависящим от субъектов [25].

По общему правилу в этом случае убытки в связи с гибелью или порчей имущества несет собственник, однако, гражданское право предусматривает и другие варианты компенсации причиненного ущерба.

При рассмотрении в качестве субъекта, причинившего ущерб, какого-либо природного или техногенного явления, под ущербом можно понимать невыгодные для собственника имущественные последствия, вызванные этими явлениями, которые могут быть компенсированы за счет средств третьей стороны (страхование рисков наступления события) или за счет собственных средств собственника информации.

Например, страхование представляет собой отношения по защите имущественных интересов физических и юридических лиц при наступлении определенных событий (страховых случаев) за счет денежных фондов, формируемых из уплачиваемых ими страховых взносов [33, ст. 2].

Объектами страхования могут быть не противоречащие законодательству Российской Федерации имущественные интересы, связанные с возмещением страхователю причиненного ему вреда [33, ст. 4].

Таким образом, можно утверждать, что ущерб информационным ресурсам связан с реализацией следующих основных угроз безопасности информации:

С точки зрения обеспечения конфиденциальности:

- хищение (копирование) информации и средств ее обработки;
- утрата (неумышленная потеря, утечка) информации и средств ее обработки;

С точки зрения обеспечения доступности:

- блокирование информации;

- уничтожение информации и средств ее обработки;

С точки зрения обеспечения целостности:

- модификация (искажение) информации;
- отрицание подлинности информации;
- навязывание ложной информации.

## 1.8. Методы обеспечения информационной безопасности

Комплексный подход к защите информации предусматривает согласованное применение правовых, организационных и программно-технических мер, перекрывающих в совокупности все основные каналы реализации угроз. В соответствии с этим подходом в организации должен быть реализован следующий комплекс мер:

- меры по выявлению и устранению уязвимостей, на основе которых реализуются угрозы. Это позволит исключить причины возможного возникновения атак;

- меры, направленные на своевременное обнаружение и блокирование угроз информационным активам и их источников;

- меры, обеспечивающие выявление и ликвидацию последствий атак. Данный класс мер защиты направлен на минимизацию ущерба, нанесённого в результате реализации угроз.

Важно понимать, что эффективное внедрение вышеперечисленных мер на предприятии возможно только при наличии нормативно-методического, технологического и кадрового обеспечения информационной безопасности.

Нормативно-методическое обеспечение предполагает создание сбалансированной правовой базы в области защиты от угроз. Для этого в компании должен быть разработан комплекс внутренних нормативных документов и процедур, обеспечивающих процесс эксплуатации системы информационной безопасности. Состав таких документов во многом зависит от размеров самой организации, уровня сложности ИС, количества объектов защиты и т.д. Так, например, для крупных организаций основополагающим нормативным документом в области защиты информации должна быть концепция или политика безопасности.

В рамках кадрового обеспечения информационной безопасности в компании должен быть организован процесс обучения сотрудников по вопросам противодействия информационным атакам. В процессе обучения должны рассматриваться как теоретические, так и практические аспекты информационной защиты. При этом программа обучения может составляться в зависимости от должностных обязанностей сотрудника, а также от того к каким информационным ресурсам он имеет доступ.

Различные методы, позволяющие уменьшить отрицательное воздействие угроз информационной безопасности, можно разделить на пять основных групп:

- Правовые методы;
- Организационные методы;
- Инженерно-технические методы;
- Технические методы;
- Программно-аппаратные методы.

***Правовые методы защиты*** информации регламентирует правовой порядок работы с информацией, и устанавливают ответственность за его нарушение согласно национальному законодательству.

В Российской Федерации к нормативно-правовым актам в области информационной безопасности относятся [34]:

Акты федерального законодательства:

- Международные договоры РФ;
- Конституция РФ;
- Законы федерального уровня (включая федеральные конституционные законы, кодексы);
- Указы Президента РФ;
- Постановления правительства РФ;
- Нормативные правовые акты федеральных министерств и ведомств;
- Нормативные правовые акты субъектов РФ, органов местного самоуправления и т. д.

Нормативно-методические документы:

- Методические документы государственных органов России:
- Доктрина информационной безопасности РФ;

- Руководящие документы ФСТЭК (Гостехкомиссии России);
- Приказы ФСБ;
- Стандарты информационной безопасности:
- Международные стандарты;
- Государственные (национальные) стандарты РФ;
- Рекомендации по стандартизации;
- Методические указания.

*Организационные методы*, в основном, ориентированы на работу с персоналом, выбор местоположения и размещения объектов корпоративной сети, организацию систем физической и противопожарной защиты, организацию контроля выполнения принятых мер, возложение персональной ответственности за выполнение мер защиты. Эти методы применяются не только для защиты информации и, как правило, уже частично реализованы на объектах корпоративной сети. Однако, их применение дает значительный эффект и сокращает общее число угроз.

Организационная защита информации является «ядром» в общей системе защиты информации. От полноты и качества решения организационных задач зависит эффективность функционирования системы защиты информации в целом.

Организационная защита призвана посредством выбора конкретных сил и средств реализовать на практике спланированные меры по защите информации. Она является составной частью системы защиты информации, определяющей и вырабатывающей порядок и правила функционирования объектов защиты и деятельности должностных лиц в целях обеспечения ИБ.

Организационная защита информации включает в себя:
- организацию работы с персоналом;
- организацию внутри объектового и пропускного режимов и охраны;
- организацию работы с носителями сведений;
- комплексное планирование мероприятий по защите информации;
- организацию аналитической работы и контроля.

***Инженерно-технические методы*** связаны с построением оптимальных сетей инженерных коммуникаций с учетом требований безопасности информации. Это довольно дорогостоящие методы, но они, как правило, реализуются еще на этапе строительства или реконструкции объекта, способствуют повышению его общей живучести и дают высокий эффект при устранении целого ряда угроз безопасности информации. Некоторые источники угроз, например обусловленные стихийными бедствиями или техногенными факторами, вообще не устранимы другими методами.

***Технические методы*** основаны на применении специальных технических средств защиты информации и контроля обстановки (радиосканеры, детекторы, передатчики, сетевые фильтры, генераторы помех, системы видеонаблюдения и др.) и дают значительный эффект при устранении угроз, связанных с действиями криминогенных элементов по добыванию информации незаконными техническими средствами. Кроме того, некоторые методы, например, резервирование средств и каналов связи, оказываются эффективными для снижения влияния некоторых негативных техногенных факторов.

***Программно-аппаратные методы,*** в основном, нацелены на устранение угроз, непосредственно связанных с процессом обработки и передачи информации. Без этих методов невозможно построение комплексной системы информационной безопасности. Программные методы защиты информации использует специальные компьютерные программы, назначение которых варьируется от цифрового кодирования данных, логического контроля направлений информационных потоков, управления системой защиты до идентификации личности пользователя и проверки паролей.

На сегодняшний день можно выделить следующие основные виды программно-аппаратных методов защиты:

- средства криптографической защиты информации;
- средства разграничения доступа пользователей к информационным ресурсам;

- средства межсетевого экранирования;
- средства анализа защищённости информационной системы;
- средства обнаружения атак;
- средства антивирусной защиты;
- средства контентного анализа;
- средства защиты от спама.

*Средства криптографической защиты* информации представляют собой средства вычислительной техники, осуществляющее криптографическое преобразование информации для обеспечения ее конфиденциальности и контроля целостности. Защита информации может осуществляться в процессе её передачи по каналам связи или в процессе хранения и обработки информации на узлах ИС. Для решения этих задач используются различные типы средств криптографической защиты информации (СКЗИ): Крипто-Про, Верба и др.

*Средства разграничения доступа* предназначены для защиты от несанкционированного доступа к информационным ресурсам системы. Разграничение доступа реализуется средствами защиты на основе процедур идентификации, аутентификации и авторизации пользователей, претендующих на получение доступа к информационным ресурсам ИС.

На этапе идентификации пользователь предоставляет свой идентификатор, в качестве которого, как правило, используется регистрационное имя учётной записи пользователя ИС.

После представления идентификатора, проводится проверка того, что этот идентификатор действительно принадлежит пользователю, претендующему на получение доступа к информации ИС. Для этого выполняется процедура аутентификации, в процессе которой пользователь должен предоставить аутентификационный параметр, при помощи которого подтверждается принадлежность идентификатора пользователю.

В качестве параметров аутентификации могут использоваться сетевые адреса, пароли, симметричные секретные ключи, цифровые сертификаты,

биометрические данные (отпечатки пальцев, голосовая информация) и т.д. Необходимо отметить, что процедура идентификации и аутентификации пользователей в большинстве случаев проводится одновременно, т.е. пользователь сразу предъявляет идентификационные и аутентификационные параметры доступа.

В случае успешного завершения процедур идентификации и аутентификации проводится авторизация пользователя, в процессе которой определяется множество информационных ресурсов, с которыми может работать пользователь, а также множество операций, которые могут быть выполнены с этими информационными ресурсами. Присвоение пользователям идентификационных и аутентификационных параметров, а также определение их прав доступа осуществляется на этапе регистрации пользователей в ИС.

*Межсетевые экраны* (МЭ) реализуют методы контроля за информацией, поступающей и/или выходящей из ИС, и обеспечивают защиту посредством фильтрации информации на основе критериев, заданных администратором. Процедура фильтрации включает в себя анализ заголовков каждого пакета, проходящего через МЭ, и передачу его дальше по маршруту следования только в случае, если он удовлетворяет заданным правилам фильтрации. При помощи фильтрования МЭ позволяют обеспечить защиту от сетевых атак путём удаления из информационного потока тех пакетов данных, которые представляют потенциальную опасность для ИС.

*Средства анализа защищённости* выделены в классификации в обособленную группу, поскольку предназначены для выявления уязвимостей в программно-аппаратном обеспечении ИС. Системы анализа защищённости являются превентивным средством защиты, которое позволяет выявлять уязвимости при помощи анализа исходных текстов ПО, анализа исполняемого кода или анализа настроек программно-аппаратного обеспечения ИС.

*Средства антивирусной защиты* предназначены для обнаружения и удаления вредоносного ПО, присутствующего в ИС. К таким вредоносным программам относятся компьютерные вирусы, а также ПО типа "троянский конь", "spyware", "adware" и др.

*Средства защиты от спама* обеспечивают выявление и фильтрацию не запрошенных почтовых сообщений рекламного характера. В ряде случаев для рассылки спама используется вредоносное программное обеспечение, внедряемое на хосты ИС и использующее адресные книги, которые хранятся в почтовых клиентах пользователей. Наличие спама в ИС может привести к одному из следующих негативных последствий:

- нарушение работоспособности почтовой системы вследствие большого потока входящих сообщений. При этом может быть нарушена доступность как всего почтового сервера, так и отдельных почтовых ящиков вследствие их переполнения. В результате пользователи ИС не смогут отправлять или получать сообщения при помощи почтовой системы;

- реализация "phishing"-атак, в результате которых пользователю присылается почтовое сообщение от чужого имени с просьбой выполнить определённые действия. В таком сообщении пользователя могут попросить запустить определённую программу, ввести своё регистрационное имя и пароль или выполнить какие-либо другие действия, которые могу помочь злоумышленнику успешно провести атаку на информационные ресурсы ИС. Примером атаки этого типа является посылка пользователю сообщения от имени известного банка, в котором содержится запрос о необходимости смены пароля доступа к ресурсам Web-сайта банка. В случае если пользователь обратится по Интернет-адресу, указанному в таком почтовом сообщении, то он будет перенаправлен на ложный Web-сайт злоумышленника, представляющий собой копию реального сайта банка. В результате такой атаки вся парольная информация, введённая пользователем на ложном сайте, будет автоматически передана нарушителю;

- снижение производительности труда персонала вследствие необходимости ежедневного просмотра и ручного удаления спам сообщений из почтовых ящиков.

*Средства контентного анализа* предназначены для мониторинга сетевого трафика с целью выявления нарушений политики безопасности. В настоящее время можно выделить два основных вида средств контентного анализа - системы аудита почтовых сообщений и системы мониторинга Интернет-трафика.

Системы аудита почтовых сообщений предполагают сбор информации о SMTP-сообщениях, циркулирующих в ИС, и её последующий анализ с целью выявления несанкционированных почтовых сообщений, нарушающих требования безопасности, заданные администратором. Так, например, системы этого типа позволяют выявлять и блокировать возможные каналы утечки конфиденциальной информации через почтовую систему.

Системы мониторинга Интернет-трафика предназначены для контроля доступа пользователей к ресурсам сети Интернет. Средства защиты данного типа позволяют блокировать доступ пользователей к запрещённым Интернет-ресурсам, а также выявить попытку передачи конфиденциальной информации по протоколу HTTP. Системы мониторинга устанавливаются таким образом, чтобы через них проходил весь сетевой трафик, передаваемый в сеть Интернет.

*Системы обнаружения атак* представляют собой специализированные программные или программно-аппаратные комплексы, предназначенные для выявления информационных атак на ресурсы АС посредством сбора и анализа данных о событиях, регистрируемых в системе. Система обнаружения атак включает в себя следующие компоненты:

- модули-датчики, предназначенные для сбора необходимой информации о функционировании АС. Иногда датчики также называют сенсорами;
- модуль выявления атак, выполняющий анализ данных, собранных датчиками, с целью обнаружения информационных атак;
- модуль реагирования на обнаруженные атаки;

- модуль хранения данных, в котором содержится вся конфигурационная информация, а также результаты работы средств обнаружения атак;
- модуль управления компонентами средств обнаружения атак.

Таким образом, система комплексного обеспечения ИБ (СКИБ) является «большой системой». Составляющие ее средства могут быть разнесены на значительные расстояния. Каждое из них представляет собой сложную подсистему, имеющую собственную структуру и выполняющую свои функции. Все действия СКИБ должны быть автоматизированы и роль человека сведена к минимуму. Но при этом он может вмешаться, если того потребуют обстоятельства. При выходе из строя отдельных подсистем СКИБ в целом должна иметь возможность продолжать выполнять задачи ценой некоторого – поддающегося оценке – снижения эффективности. Для этого может предусматриваться как дублирование наиболее ответственных узлов системы, так и перекрытие зон ответственности различных подсистем.

Сопоставление угроз безопасности информации и групп методов их парирования позволяет определить, какими же методами какие угрозы наиболее целесообразно парировать, и определить соотношение в распределении средств, выделенных на обеспечение безопасности информации между группами методов.

Все группы методов парирования угрозам безопасности информации имеют примерно равную долю в организации комплексной защиты информации. Однако необходимо учесть, что некоторые методы могут быть использованы только для решения ограниченного круга задач защиты. Это особенно характерно для устранения угроз техногенного и стихийного характера.

Наибольший эффект достигается при применении всей совокупности методов парирования.

Гипотетическое средство защиты, прежде всего, должно обеспечивать разграничение доступа субъектов к объектам (мандатный и дискреционный принципы), управлять внешними потоками информации (фильтрация, ограничение, исключение). Кроме того, оно, как минимум, должно обеспечивать управление внутренними потоками информации с одновременным контролем целостности программного обеспечения, конфигурации сети с учетом возможности атак и разрушающих воздействий [35].

## 1.9. Модель реализации угроз информационной безопасности

На основании анализа изложенного выше материала, модель реализации угроз информационной безопасности можно представить в виде схемы, приведенной на рис.1.9.

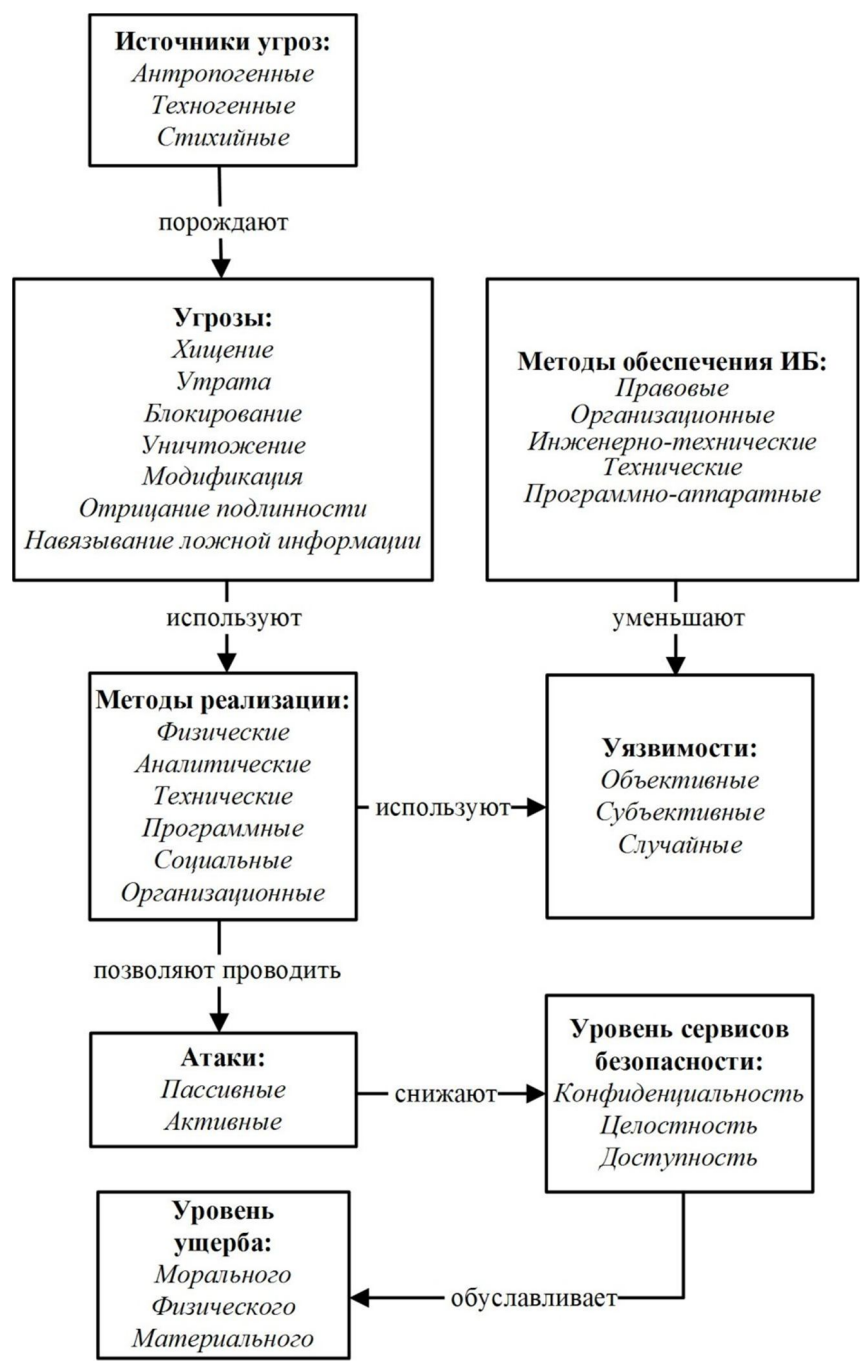

Рис.1.9. Модель реализации угроз информационной безопасности.

При этом под *атакой* на информационную систему понимается действие или последовательность связанных между собой действий, которые приводят к реализации угрозы различными методами с использованием уязвимостей данной информационной системы. Таким образом, как видно из схемы, если бы можно было устранить уязвимости ИС, то, тем самым, мы устранили бы и возможность реализации атак.

На сегодняшний день неизвестно, сколько существует методов атак. Связано это в первую очередь с тем, что до сих пор отсутствуют какие-либо серьезные математические исследования в этой области. Из близких к этой области исследований можно привести работу, написанную в 1989 году Фредом Коэном, в которой описывались математические основы вирусной технологии [36]. В работе было приведено доказательство бесконечности числа вирусов. Эти же результаты можно перенести и на область атак, поскольку вирусы - это одно из подмножеств этой области.

**Модели атак**

Традиционная модель преднамеренной атаки строится по принципу "один к одному" или "один ко многим", т.е. атака исходит из одного источника.

В модели распределенной или скоординированной атаки (distributed или coordinated attack) используются иные принципы. В отличие от традиционной модели, использующей отношения "один к одному" и "один ко многим", в распределенной модели используются отношения "много к одному" и "много ко многим"

Распределенные атаки основаны на "классических" атаках типа "отказ в обслуживании". Смысл данных атак заключается в посылке большого количества пакетов на заданный узел или сегмент сети, что может привести к выведению этого узла или сегмента из строя, поскольку он "захлебнется" в лавине посылаемых пакетов и не сможет обрабатывать запросы авторизованных пользователей.

По такому принципу работают атаки SYN-Flood, Smurf, UDP Flood, Targa3 и т.д. Однако в том случае, если пропускная способность канала до цели атаки превышает пропускную способность атакующего или целевой узел некорректно сконфигурирован, то к "успеху" такая атака не приведет. Например, с помощью этих атак бесполезно пытаться нарушить работоспособность своего провайдера.

В случае же распределенной атаки ситуация коренным образом меняется. Атака происходит уже не из одной точки, а сразу из нескольких, что приводит к резкому возрастанию трафика и выведению атакуемого узла из строя.

**Этапы реализации атак**

Можно выделить следующие этапы реализации преднамеренной атаки: предварительные действия перед атакой или "сбор информации", "реализация атаки" и завершение атаки ("заметание следов") (рис.1.10.). Обычно, когда говорят об атаке, то подразумевают именно второй этап, забывая о первом и последнем. Сбор информации и завершение атаки в свою очередь также могут являться атакой и иметь свои этапы реализации.

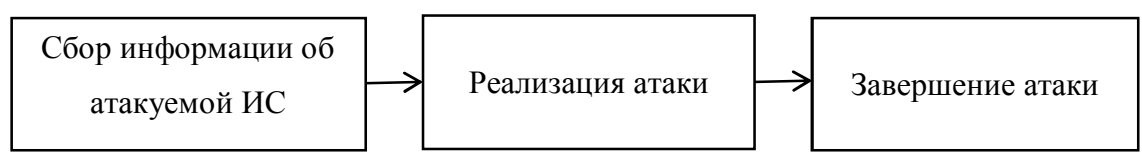

Рис. 1.10. Этапы реализации атаки

Основной этап - это сбор информации. Именно эффективность работы злоумышленника на данном этапе является залогом "успешности" атаки. В первую очередь выбирается цель и собирается информация о ней (тип и версия операционной системы, открытые порты и запущенные сетевые сервисы, установленное системное и прикладное программное обеспечение и его конфигурация и т.д.).

Затем идентифицируются наиболее уязвимые места атакуемой системы, воздействие на которые приводит к нужному злоумышленнику результату.

На первом этапе пытаются выявить все каналы взаимодействия цели атаки с другими узлами. Это позволит не только выбрать тип реализуемой атаки, но и источник ее реализации.

Например, пусть атакуемый узел взаимодействует с двумя серверами под управлением ОС Unix и Windows NT. С одним сервером атакуемый узел имеет доверенные отношения, а с другим - нет. От того, через какой сервер злоумышленник будет реализовывать нападение, зависит, какая атака будет задействована, какое средство реализации будет выбрано и т.д.

Затем, в зависимости от полученной информации и преследуемого результата, выбираются метод реализации и уязвимость, т.е. определяется тип атаки, дающий наибольший эффект. Например, для нарушения функционирования узла можно использовать SYN Flood, Teardrop, UDP Bomb и т.д., а для проникновения на узел и кражи информации - CGI-скрипт PHF для кражи файла паролей, удаленный подбор пароля и т.п.

Затем наступает второй этап - реализация выбранного типа атаки.

Традиционные средства защиты, такие как межсетевые экраны или механизмы фильтрации в маршрутизаторах, вступают в действие на втором этапе, совершенно "забывая" о первом и третьем. Это приводит к тому, что зачастую совершаемую атаку очень трудно остановить даже при наличии мощных и дорогих средств защиты. Пример тому - распределенные атаки.

Логично было бы, чтобы средства защиты начали работать еще на первом этапе, т.е. предотвращали возможность сбора информации об атакуемой системе. Это позволило бы, если и не полностью предотвратить атаку, то существенно усложнить работу злоумышленника.

Также традиционные средства не позволяют обнаружить уже совершенные атаки и оценить ущерб после их реализации (третий этап) и, следовательно, невозможно определить меры по предотвращению таких атак впредь.

В зависимости от достигаемого результата нарушитель концентрируется на том или ином этапе.

Например, для отказа в обслуживании он в первую очередь подробно анализирует атакуемую сеть и выискивает в ней лазейки и слабые места для атаки на них и выведению узлов сети из строя.

Для хищения информации злоумышленник основное внимание уделяет незаметному проникновению на анализируемые узлы при помощи обнаруженных ранее уязвимостей.

Для понимания методов обнаружения атак рассмотрим основные механизмы их реализации.

**Сбор информации**

Первый этап реализации атак - это сбор информации об атакуемой системе или узле. Он включает такие действия как, определение сетевой топологии, типа и версии операционной системы атакуемого узла, а также доступных сетевых и иных сервисов и т.п. Эти действия реализуются различными методами.

*Изучение окружения*

На этом этапе нападающий исследует сетевое окружение вокруг предполагаемой цели атаки. К таким областям, например, относятся узлы Internet-провайдера "жертвы" или узлы удаленного офиса атакуемой компании.

Злоумышленник может пытаться определить адреса "доверенных" систем (например, сеть партнера) и узлов, которые напрямую соединены с целью атаки (например, маршрутизатор ISP) и т.д.

Такие действия достаточно трудно обнаружить, поскольку они выполняются в течение достаточно длительного периода времени из области, не контролируемой средствами защиты (межсетевыми экранами, системами обнаружения атак и т.п.).

*Идентификация топологии сети*

Существует два метода определения топологии сети (network topology detection), используемых злоумышленниками: "изменение TTL" ("TTL modulation") и "запись маршрута" ("record route").

Программы traceroute для Unix и tracert для Windows используют первый способ определения топологии сети. Они используют для этого поле Time to Live ("время жизни") в заголовке IP-пакета, которое изменяется в зависимости от числа пройденных сетевым пакетом маршрутизаторов.

Утилита ping может быть использована для записи маршрута ICMP-пакета. Зачастую сетевую топологию можно выяснить при помощи протокола SNMP, установленного на многих сетевых устройствах, защита которых неверно сконфигурирована.

При помощи протокола RIP можно попытаться получить информацию о таблице маршрутизации в сети и т.д.

### *Идентификация узлов*

Идентификация узла (host detection), как правило, осуществляется путем посылки при помощи утилиты ping команды ECHO_REQUEST протокола ICMP. Ответное сообщение ECHO_REPLY говорит о том, что узел доступен.

Существуют свободно распространяемые программы, которые автоматизируют и ускоряют процесс параллельной идентификации большого числа узлов, например, fping или nmap. Опасность данного метода в том, что стандартными средствами узла запросы ECHO_REQUEST не фиксируются. Для этого необходимо применять средства анализа трафика, межсетевые экраны или системы обнаружения атак.

Еще один метод определения узлов сети - использование "смешанного" ("promiscuous") режима работы сетевой карты, который позволяет определить различные узлы в сегменте сети. Но он не применим в тех случаях, в которых трафик сегмента сети недоступен нападающему со своего узла, т.е. этот метод применим только в локальных сетях.

Другим способом идентификации узлов сети является так называемая разведка DNS, которая позволяет идентифицировать узлы корпоративной сети при помощи обращения к серверу службы имен.

### *Идентификация сервисов или сканирование портов*

Идентификация сервисов (service detection), как правило, осуществляется путем обнаружения открытых портов (port scanning). Такие порты очень часто связаны с сервисами, основанными на протоколах TCP или UDP.

Например, открытый 80-й порт подразумевает наличие Web-сервера, 25-й порт - почтового SMTP-сервера, 31337-й - серверной части троянского коня

BackOrifice, 12345 или 12346 - серверной части троянского коня NetBus и т.д. Для идентификации сервисов и сканирования портов могут быть использованы различные программы, в том числе и свободно распространяемые.

*Идентификация операционной системы*

Основной механизм удаленного определения ОС (OS detection) - анализ ответов на запросы, учитывающие различные реализации TCP/IP-стека в различных операционных системах. В каждой ОС по-своему реализован стек протоколов TCP/IP, что позволяет при помощи специально запросов и ответов на них определить, какая ОС установлена на удаленном узле.

Другой, менее эффективный и крайне ограниченный, способ идентификации ОС узлов - анализ сетевых сервисов, обнаруженных на предыдущем этапе. Например, открытый 139 порт позволяет сделать вывод, что удаленный узел вероятнее всего работает под управлением ОС семейства Windows.

*Определение роли узла*

Предпоследним шагом на этапе сбора информации об атакуемом узле является определение его роли, например, выполнении функций межсетевого экрана или Web-сервера. Выполняется этот шаг на основе уже собранной информации об активных сервисах, именах узлов, топологии сети и т.п.

Например, открытый 80-й порт может указывать на наличие Web-сервера, блокировка ICMP-пакета указывает на потенциальное наличие межсетевого экрана, а DNS-имя узла proxy.domain.ru или fw.domain.ru говорит само за себя.

*Определение уязвимостей узла*

Последний шаг - поиск уязвимостей. На этом шаге злоумышленник при помощи различных автоматизированных средств или вручную определяет уязвимости, которые могут быть использованы для реализации атаки.

**Реализация атаки**

С этого момента начинается попытка доступа к атакуемому узлу. При этом доступ может быть как непосредственный, т.е. проникновение на узел, так и опосредованный, например, при реализации атаки типа "отказ в обслуживании". Реализация атак в случае непосредственного доступа также может быть разделена на два этапа: проникновение и установление контроля.

*Проникновение*

Проникновение подразумевает под собой преодоление средств защиты периметра (например, межсетевого экрана). Реализовываться это может быть различными путями.

Например, использование уязвимости сервиса компьютера, "смотрящего" наружу или путем передачи враждебного содержания по электронной почте (макровирусы) или через апплеты Java. Такое содержание может использовать так называемые "туннели" в межсетевом экране, через которые затем и проникает злоумышленник. К этому же этапу можно отнести подбор пароля администратора или иного пользователя при помощи специализированной утилиты (например, L0phtCrack или Crack).

*Установление контроля*

После проникновения злоумышленник устанавливает контроль над атакуемым узлом. Это может быть осуществлено путем внедрения программы типа "троянский конь" (например, NetBus или BackOrifice).

После установки контроля над нужным узлом и "заметания" следов, злоумышленник может осуществлять все необходимые несанкционированные действия дистанционно, без ведома владельца атакованного компьютера. При этом установление контроля над узлом корпоративной сети сохраняется и после перезагрузки операционной системы, путем замены одного из загрузочных файлов или вставка ссылки на враждебный код в файлы автозагрузки или системный реестр.

Известен случай, когда злоумышленник смог перепрограммировать EEPROM сетевой карты и даже после переустановки ОС смог повторно реализовать несанкционированные действия.

Более простой модификацией является внедрение необходимого кода или фрагмента в сценарий сетевой загрузки (например, для ОС Novell Netware).

Необходимо отметить, что злоумышленник на втором этапе может преследовать две цели. Во-первых, получение несанкционированного доступа к самому узлу и содержащейся на нем информации. Во-вторых, получение несанкционированного доступа к узлу для осуществления дальнейших атак на другие узлы.

Первая цель, как правило, осуществляется только после реализации второй. То есть, сначала злоумышленник создает себе базу для дальнейших атак и только после этого проникает на другие узлы. Это необходимо для того, чтобы скрыть или существенно затруднить нахождение источника атаки.

**Завершение атаки**

Этапом завершения атаки является "заметание следов" со стороны злоумышленника. Обычно это реализуется путем удаления соответствующих записей из журналов регистрации узла и других действий, возвращающих атакованную систему в исходное, "предатакованное" состояние.

## 1.10. Выводы по 1 главе

Обобщая все вышеизложенное, можно констатировать, что задача повышения безопасности информационных систем и технологий в современных условиях характеризуется сложностью, неопределённостью, наличием большого количества взаимосвязанных внутренних и внешних факторов, влияющих на информационную безопасность.

При этом под информационной безопасностью понимаются все аспекты, связанные с определением, достижением и поддержанием конфиденциальности, целостности, доступности, отказоустойчивости, подотчетности, аутентичности и достоверности информации или средств ее обработки [37]. Решение задачи выявления факторов, воздействующих на безопасность информационных систем и технологий, является основой для планирования и применения эффективных мероприятий и технологий, направленных на повышение безопасности активов ИС.

К активам ИС относятся: материальные ресурсы; информационные ресурсы (аналитическая, служебная, управляющая информация на всех этапах своего жизненного цикла: создание, обработка, хранение, передача, уничтожение); информационные технологические процессы жизненного цикла автоматизированных систем; предоставляемые информационные услуги. Объектами информационной системы (ОИС) также являются средства обеспечения: территории, здания, помещения, технические средства и т.п.

При решении задачи обеспечения информационной безопасности, прежде всего, нужно выполнить идентификацию активов и установить начальный уровень безопасности, которому отвечает ИС. В процессе идентификации следует рассмотреть основные характеристики активов: информационную ценность, чувствительность активов к угрозам, наличие защитных мер. При этом необходимо учесть, что в числе факторов, влияющих на безопасность, особое место занимают субъективные факторы, которые потенциально являются наиболее опасными.

Обеспечение информационной безопасности активов ИС требует выполнения защитных мероприятий: действий, процедур и механизмов, способных обеспечить безопасность при возникновении проблем и угроз, уменьшить уязвимость и ограничить воздействие на ИС, облегчить восстановление активов.

Эффективность информационной безопасности повышается при использовании комбинации различных защитных мероприятий. Конкретное мероприятие по обеспечению ИБ может выполнять много функций безопасности, однако может оказаться, что обеспечение одного сервиса (функции) безопасности потребует нескольких защитных мер.

Защитные мероприятия по ИБ могут выполнять одну или несколько следующих функций: предотвращение, сдерживание, обнаружение, ограничение угроз, исправление ошибок, восстановление активов системы, мониторинг состояния; уведомление о событиях в системе.

Структурную схему решения задачи обеспечения комплексной безопасности информационных систем можно описать следующим образом:

- проблемная область задачи обеспечения комплексной безопасности информационных систем (понятия, определяющие суть ИБ, и связи между ними);
- угрозы, влияющие на безопасность ИС (внешние объективные факторы, внутренние объективные факторы, внешние субъективные факторы, внутренние субъективные факторы);
- мероприятия и технологии обеспечения комплексной безопасности информационных систем;
- методология обеспечения комплексной безопасности ИС (эвристические знания о состоянии безопасности ИС, стратегии обеспечения комплексной безопасности)
- решаемые задачи (идентификация активов ИС, определение критерия и показателей эффективности обеспечения безопасности; разработка

процедур оценок этих критериев и показателей; разработка модели комплексного обеспечения ИБ и др.)

- принципы комплексного обеспечения безопасности ИС (системности, адаптивности, транспарентности и конфиденциальности, непрерывности, обучаемости и накопление опыта и т.п).
- экономические, организационно-технологические, информационные решения.

Таким образом, проблемная область решения задачи обеспечения комплексной безопасности (совокупность основных понятий, определяющих суть исследования, и связи между ними), включает в себя: предметную область, цель исследования, решаемые задачи, возможные тактики и стратегии, используемые для достижения поставленной цели.

Применительно к рассматриваемым задачам предметную область можно определить как информационную систему со всем комплексом понятий и знаний о ее функционировании. При исследовании проблемной области необходимы знания о задачах, решаемых в информационной системе, и целях ее создания. Нужно также определить возможные стратегии управления на основе эвристических знаний о процессах обеспечения комплексной безопасности ИС.

Проведенный в главе 1 анализ позволил построить онтологическую схему предметной области, провести классификацию источников угроз и уязвимостей безопасности ИС, выявить особенности процесса построения модели нарушителя, выделить основные типы атак на информационные ресурсы, выявить основные виды ущерба, наносимого информационным активам, провести классификацию методов защиты информации и построить онтологическую модель реализации угроз информационной безопасности.

Выстраивание информационной безопасности должно базироваться на *системном* подходе, так как только он позволяет сформировать комплекс мероприятий по парированию угроз безопасности - начиная с анализа состояния объекта и заканчивая выдачей гарантии качества принятых мер.

# ГЛАВА 2.

# МЕТОДОЛОГИЧЕСКИЕ ОСНОВЫ КОМПЛЕКСНОГО ОБЕСПЕЧЕНИЯ ИНФОРМАЦИОННОЙ БЕЗОПАСНОСТИ

## 2.1. Понятие системности и комплексности в задачах защиты информации

Поиск новых путей защиты информации заключается не просто в создании соответствующих механизмов, но и требует комплексного использования всех имеющихся средств защиты на каждом из этапов жизненного цикла ИС. При этом все средства, методы и мероприятия, используемые для ЗИ, должны быть наиболее рациональным образом объединены в единый целостный механизм, противостоящий не только действиям злоумышленников, но и ошибкам некомпетентных или недостаточно подготовленных пользователей и персонала, а также нештатным ситуациям технического и стихийного характера.

Проблема обеспечения желаемого уровня защиты информации весьма сложная, требующая для своего решения создания целостной системы организационно-технологических мероприятий и применения комплекса специальных средств и методов ЗИ.

На основе теоретических исследований и практических работ в области ЗИ в работе [5] был сформулирован системно-концептуальный подход к защите информации.

При этом под системностью как основной частью системно-концептуального подхода понимается:

а) системность целевая, т.е. защищённость информации рассматривается как часть общего понятия качества информации;

б) системность пространственная, предлагающая взаимоувязанное решение всех вопросов защиты на всех компонентах ИС;

в) системность временная, означающая непрерывность работ по ЗИ, осуществляемых в соответствии с планом;

г) системность организационная, означающая единство организации всех работ по ЗИ и управления ими.

Концептуальность подхода предполагает разработку единой концепции как полной совокупности научно обоснованных взглядов, положений и решений, необходимых и достаточных для оптимальной организации и обеспечения надёжности защиты информации, а также целенаправленной интеграции всех работ по ЗИ.

Главным условием создания системы ЗИ является ее надежность. Но надежность может быть обеспечена лишь в том случае, если защита является комплексной и системной. Исходя из сказанного, можно дать такое определение:

«*Система защиты информации (СЗИ)* - это организованная совокупность органов и объектов (компонентов) защиты информации, использование методов и средств защиты, а также осуществление защитных мероприятий».

Средства защиты информации с одной стороны являются составной частью системы, с другой стороны - они сами организуют систему, осуществляя защитные мероприятия.

Поскольку система определяется как совокупность взаимосвязанных элементов, то *назначение системы защиты информации* состоит в том, чтобы объединить все составляющие элементы защиты в единое целое, в котором каждый компонент, выполняя свою прямую функцию, одновременно обеспечивает выполнение функций другими компонентами и связан с ними логически и технологически.

При отсутствии отдельных компонентов системы или их несогласованности между собой неизбежно возникновение уязвимостей в технологии защиты информации. Следовательно, основным условием при разработке систем защиты информации должна быть системность.

Системный подход к задачам обеспечения ИБ включает в себя, прежде всего, оценку угроз безопасности объекта; анализ средств, которые могут использоваться при построении системы; оценку экономической целесообразности СЗИ и возможности увеличения её эффективности.

К организации системы защиты информации с позиции системного подхода выдвигается ряд требований, определяющих ее целостность, адекватность и эффективность.

Система защиты информации должна быть:

- централизованной — обеспечивающей эффективное управление со стороны руководителя и должностных лиц, отвечающих за различные направления деятельности предприятия;
- плановой — объединяющей усилия различных должностных лиц и структурных подразделений для выполнения стоящих перед предприятием задач в области защиты информации;
- конкретной и целенаправленной — рассчитанной на защиту конкретных информационных ресурсов, представляющих ценность для организации;
- активной — обеспечивающей защиту информации с достаточной степенью настойчивости и возможностью концентрации усилий на наиболее важных направлениях деятельности предприятия;
- надежной и универсальной — охватывающей всю деятельность предприятия, связанную с созданием и обменом информацией.

Поскольку регулярных (а тем более формальных) методов решения задач, направленных на удовлетворение этих требований, на сегодняшний день не существует, приходится использовать методы неформальные, эмпирические.

Применение системного анализа для решения задач на всех этапах жизненного цикла СЗИ позволяет обеспечить полноту и эффективность реализаций их функций, а также оптимальные ресурсные, финансовые и временные параметры для достижения поставленных целей.

Системный подход – это принцип рассмотрения проекта, при котором анализируется система в целом, а не её отдельные части. Его задачей является оптимизация всей системы в совокупности, а не улучшение эффективности отдельных частей.

Это объясняется тем, что, как показывает практика, улучшение одних параметров часто приводит к ухудшению других, поэтому необходимо стараться обеспечить баланс противоречивых требований и характеристик [38].

Для того чтобы обеспечить надежность защиты информации нужна не только системность защиты, но и ее комплексность.

*Комплексность* – один из основополагающих принципов защиты информации. Ее назначение состоит в объединении в одно целое локальных средств ЗИ. При этом они должны функционировать в единой «связке». В качестве локальных СЗИ могут выступать, например, правовые, организационные, инженерно-технические, программно-аппаратные механизмы защиты информации.

Кроме того, комплексность должна обеспечивать безопасность всей совокупной информации, подлежащей защите, при любых обстоятельствах. Это означает, что должны защищаться все носители информации во всех компонентах ее сбора, хранения, передачи и использования, в любое время и при всех режимах функционирования системы обработки информации.

В то же время комплексность не исключает, а наоборот, предполагает дифференцированный подход к защите информации. Дифференцированность зависит от состава ее носителей, видов тайны, к которым отнесена информация, степени ее конфиденциальности, средств хранения и обработки, форм и условий проявления ее уязвимости, каналов и методов несанкционированного доступа к информации.

Исходя из этого, можно сформулировать следующее определение СКЗИ:

«*Система комплексной защиты информации* – это система, полно и всесторонне охватывающая все предметы, процессы и факторы, которые обеспечивают безопасность всей защищаемой информации».

Из вышесказанного следует, что только комплексность и системность построения СЗИ могут гарантировать достижение максимальной эффективности защиты информации, поскольку системность обеспечивает необходимые составляющие защиты и устанавливает между ними логическую и технологическую связь, а комплексность, требует полноты этих составляющих, всеохватности защиты, обеспечивает ее надежность.

## 2.2. Принципы обеспечения комплексной безопасности информационных систем

В основе концепции безопасности должен быть контроль над информационными активами системы с целью нейтрализации воздействия негативных факторов (угроз).

Анализ научных и практических работ в данной области позволил сформулировать общеметодологические принципы (общие положения) построения и функционирования КСЗИ.

Создание и развитие эффективной системы обеспечения информационной безопасности должно основываться на следующих основных принципах [9]:

1. *Принцип системности.*

Активы КСЗИ представляют собой совокупность консолидированных и взаимосвязанных элементов, служащих для обеспечения эффективного функционирования ИС. Такие системы выступают как единое и сложное целое. При их оптимальной консолидации наблюдается, так называемый, синергетический эффект: результат функционирования элементов в системе выше суммы результатов функционирования каждого элемента в отдельности.

Кроме этого принцип системного подхода к построению системы защиты позволяет заложить комплекс мероприятий по парированию угроз безопасности информации уже на стадии проектирования ИС, обеспечив оптимальное сочетание организационных и инженерно-технических мер защиты информации. Важность реализации этого принципа основана на том, что дооборудование действующей незащищенной ИС средствами защиты информации сложнее и дороже, чем изначальное проектирование и построение ее в защищенном варианте.

## 2. *Принцип адаптивности.*

Система защиты информации должна строиться с учетом возможного изменения конфигурации ИС, числа пользователей и степени конфиденциальности и ценности информации. При этом, введение каждого нового элемента ИС или изменение действующих условий не должно снижать достигнутый уровень защищенности в целом.

## 3. *Принцип транспарентности и конфиденциальности.*

Принцип с одной стороны предполагает полную доступность активов и процессов ИС для легальных пользователей, с другой - требует недоступности и закрытости информационных активов для неавторизованного пользователя.

Парирование угрозам безопасности информации всегда носит недружественный характер по отношению к пользователям и обслуживающему персоналу ИС. Это происходит из-за того, что любая система защиты, по определению, всегда налагает на работу ограничения организационного и технического характера.

Поэтому одним из основных принципов создания системы комплексной защиты информации должен стать ***принцип максимальной дружественности***. То есть не надо вводить запреты там, где без них можно обойтись ("на всякий случай"), а если уж и вводить ограничения, то сделать это нужно с минимальными неудобствами для пользователя.

При этом следует учесть совместимость создаваемой системы комплексной защиты с используемой операционной и программно-аппаратной структурой ИС, и со сложившимися традициями фирмы.

Вплотную к этой проблеме стоит ***принцип прозрачности***. Информационной системой пользуются не только высококлассные программисты. Кроме того, основное ее назначение – это обеспечение производственных потребностей пользователей. Поэтому система защиты информации должна работать в "фоновом" режиме, быть "незаметной" и не мешать пользователям в основной работе, но при этом выполнять все возложенные на нее функции.

### 4. *Принцип непрерывности, обучаемости и накопление опыта.*

КСЗИ должна обеспечить непрерывность реализации безопасного функционирования всех элементов и объектов ИС. Это, в свою очередь, предполагает необходимость накопления, обобщения и использования всего имеющегося в сфере ИБ опыта.

### 5. *Принцип прогнозируемости и функциональной взаимосвязанности.*

Безопасность неотделима от общих проблем функционирования активов ИС, в том числе в области: информационно-технологических процессов, процессов потребления ресурсов, финансово-экономических, социальных, экологических и т.д.

Принцип предполагает выявление причинно-следственных связей возникновения возможных проблем и угроз, и построение на этой основе наиболее точного прогноза развития ситуации.

### 6. *Принцип своевременности, оперативного реагирования и адекватности.*

Принцип предполагает своевременность выявления проблем и угроз, потенциально способных повлиять на безопасность активов системы и дальнейшее оперативное реагирование с целью их блокирования и устранения последствий.

При этом устранение последствий проявления угроз безопасности информации всегда требует гораздо больших финансовых, временных и материальных затрат, чем затраты на создание системы комплексной защиты информации.

Кроме того, выбор необходимых защитных мер обеспечения безопасности должен быть адекватен возникающим проблемам и угрозам, с учетом затрат на реализацию этих мер и объема возможных потерь от реализации угроз. Принимаемые решения должны быть дифференцированы в зависимости от важности, частоты и вероятности возникновения угроз безопасности информации, степени конфиденциальности самой информации и ее коммерческой стоимости.

7. *Принцип доказательности.*

При создании системы защиты информации необходимо соблюдение организационных мер внутри корпоративной сети, включая привязку логического и физического рабочих мест друг к другу, и применения специальных аппаратно-программных средств идентификации, аутентификации и подтверждения подлинности информации. Реализация данного принципа позволяет сократить расходы на усложнение системы, например, применять цифровую электронную подпись только при работе с удаленными и внешними рабочими местами и терминалами, связанными с корпоративной сетью по открытым каналам связи.

8. *Принцип контролируемости.*

Для обеспечения безопасности необходимо применять только те защитные меры и технологии, правильность работы которых может быть проверена. При этом необходимо регулярно оценивать адекватность защитных мер и эффективность применяемых технологий.

Соблюдение данных принципов способствует повышению эффективности защиты ИС.

Описание проблемной области решения задачи обеспечения комплексной безопасности информационных систем должно включать в себя основные понятия, определяющие суть ИБ, решаемые задачи, цели исследования, возможные тактики и стратегии, используемые для достижения поставленных целей, связи между ними. Необходимо выявить факторы, влияющие на безопасность ИС (внешние объективные и субъективные факторы, внутренние объективные и субъективные факторы), изучить мероприятия и технологии обеспечения комплексной безопасности информационных систем, выбрать методологию обеспечения ИБ, используя эвристические знания о состоянии безопасности ИС.

При этом решаются следующие задачи: идентификация активов ИС, определение критериев и показателей эффективности обеспечения ИБ;

разработка процедур оценок этих критериев и показателей; разработка модели комплексного обеспечения ИБ и др.

Отметим, что в каждый момент времени состояние информационной системы характеризуется некоторым уровнем безопасности. Уровень безопасности понимается как интегральная оценка, основанная на наборе показателей и критериев, характеризующих состояние системы в плане защищенности ее активов [37].

В случае прогнозирования возникновения, либо реального возникновения угроз безопасности активов ИС должны применяться меры, ослабляющие воздействие этих угроз (вплоть до полного их блокирования) или устраняющие последствия реализации угроз, если таковые все же возникли. При этом после реализации соответствующих защитных мероприятий, уровень безопасности повышается, что позволяет сохранить активы информационной системы и обеспечить ее функциональность.

## 2.3. Методология защиты информации как теоретический базис комплексного обеспечения информационной безопасности

*Теория защиты информации* определяется как система основных идей, относящихся к защите информации, дающая целостное представление о сущности проблемы защиты, закономерностях ее развития и существенных связях с другими отраслями знания, формирующаяся и развивающаяся на основе опыта практического решения задач защиты и определяющая основные ориентиры в направлении совершенствования практики защиты информации [5].

Теория защиты должна:
- предоставлять полные и адекватные сведения о происхождении, сущности и развитии проблем защиты;
- полно и адекватно отображать структуру и содержание взаимосвязей с родственными и смежными областями знаний;

- аккумулировать опыт предшествующих исследований, разработок и практического решения задач защиты информации;
- ориентировать исследователей в направлении наиболее эффективного решения основных задач защиты и предоставлять необходимые для этого научно-методологические и инструментальные средства (какие конкретно методы, способы и средства используются для эффективной защиты информации);
- формировать научно обоснованные перспективные направления развития теории и практики защиты информации.

Сформулированные назначения теории защиты предопределяют ее состав, общее содержание и свидетельствуют о ее многоаспектности.

Составляющими частями ее, очевидно, должны быть:
- полные и систематизированные сведения о происхождении, сущности и содержании проблемы защиты;
- систематизированные результаты анализа развития теоретических исследований и разработок, а также результаты опыта практического решения задач защиты (метод экспериментальных оценок, метод мозгового штурма, метод психоинтеллектуальной генерации);
- научно обоснованная постановка задач защиты информации в современных системах ее обработки, полно и адекватно учитывающая текущие и перспективные концепции построения систем и технологий обработки, потребности в защите информации и объективные предпосылки их удовлетворения;
- общие стратегические установки на организацию защиты информации, учитывающие все многообразие потенциально возможных условий защиты;
- методы, необходимые для адекватного и наиболее эффективного решения всех задач защиты и содержащие как общеметодологические подходы к решению, так и конкретные прикладные методы решения;

- методологическая и инструментальная база, содержащая необходимые методы и инструментальные средства для решения любой совокупности задач защиты в рамках любой выбранной стратегической установки;
- научно обоснованные предложения по организации и обеспечению работ по защите информации;
- научно обоснованный прогноз перспективных направлений развития теории и практики защиты информации.

Приведенный перечень составных частей свидетельствует о большом объеме и многоаспектности теории защиты, что, естественно, порождает значительные трудности ее формирования. Эти трудности усугубляются еще тем, что по мере развития исследований, разработок и практической их реализации появляются новые направления исследований, формирующиеся на стыке с другими науками.

Например, для оценки влияния на безопасность информации основного ее носителя, человека, надо владеть навыками психологического анализа личности.

Отсюда следует, что защита информации становится все более системной и комплексной и все более масштабной проблемой. Кроме того, существенным фактором является повышенное влияние на процессы защиты случайных трудно предсказуемых (слабоструктурированных) событий, особенно тех из них, которые связаны со злоумышленными действиями людей или возникновением чрезвычайных ситуаций.

Постановка задачи защиты информации на современном этапе приобретает целый ряд особенностей:
- необходимо обеспечить комплексную защиту информации;
- защита информации становится актуальной для всё большего количества объектов (больших и малых) различной принадлежности (государственной и негосударственной);

- резко расширяется разнообразие подлежащей защите информации (государственная, промышленная, коммерческая, персональная и т.п.).

Осуществление мероприятий по защите информации носит массовый характер, занимается этой проблемой большое количество специалистов различного профиля. Но успешное осуществление указанных мероприятий при такой их масштабности возможно только при наличии хорошего инструментария в виде методов и средств решения соответствующих задач. Разработка такого инструментария требует наличия развитых научно-методологических основ защиты информации.

Поэтому возникает настоятельная необходимость выбора и обоснования методологических принципов формирования самой теории защиты.

Общеметодологические принципы формирования теории, методы решения задач и методологический базис в совокупности составляют научно-методологическую основу теории защиты информации [39].

Под ***научно-методологическими основами*** комплексной защиты информации (как и решения любой другой проблемы) понимается совокупность принципов, подходов и методов (научно-технических направлений), необходимых и достаточных для анализа (изучения, исследования) проблемы комплексной защиты, построения оптимальных механизмов защиты и управления механизмами защиты в процессе их функционирования.

Основными компонентами научно-методологических основ являются ***принципы, подходы и методы***. При этом под принципами понимается основное исходное положение какой-либо теории, учения, науки, мировоззрения; под подходом - совокупность приемов, способов изучения и разработки какой-либо проблемы; под методом - способ достижения какой-либо цели, решения конкретной задачи. Например, при реализации принципа разграничения доступа в качестве подхода можно выбрать моделирование, а в качестве метода реализации – построение матрицы доступа.

Методологический базис это методологические принципы формирования самой теории защиты.

СЗИ относится к системам организационно-технологического (социотехнического) типа, т.к. общую организацию защиты и решение значительной части задач осуществляют люди (организационная составляющая), а защита информации осуществляется параллельно с технологическим процессами ее обработки (технологическая составляющая).

Главная цель создания СЗИ - это достижение максимальной эффективности защиты за счет одновременного использования всех необходимых ресурсов, методов и средств, исключающих несанкционированный доступ к защищаемой информации и обеспечивающих физическую сохранность ее носителей.

Исходя из этого, общее назначение методологического базиса заключается:

- в формировании обобщенного взгляда на организацию и управление КСЗИ, отражающего наиболее существенные аспекты проблемы;
- в формировании полной системы принципов, следование которым обеспечивает наиболее полное решение основных задач;
- в формировании совокупности методов, необходимых и достаточных для решения всей совокупности задач управления.

При этом, поскольку речь идет об организации и построении комплексной *системы* ЗИ, то в качестве общеметодологической основы должны выступать основные положения теории систем. Т.к. речь идет об управлении, то в качестве научно-методической основы выступают общие законы кибернетики (как науки об управлении в системах любой природы). Процессы управления связаны с решением большого количества разноплановых задач, поэтому необходимо опираться на принципы и методы моделирования больших систем и процессов их функционирования.

## 2.4. Основные понятия теории систем

Система, это совокупность связанных между собой элементов, объединенных единством цели или назначения и функциональной целостностью.

Можно выделить некоторые признаки, характеризующие наличие системы. К основным признакам можно отнести следующие:

1. Целостность системы.

2. Делимость системы.

3. Наличие существенных, определяющих систему устойчивых связей.

4. Наличие интегральных свойств, присущих системе в целом.

5. Организация системы, т.е. число значимых элементов и связей.

Таким образом, системный подход делает акцент на анализе целостных интегральных свойств объекта, выявлении его структуры и функций. Следует иметь в виду, что свойства системы как целого определяются не только свойствами его элементов, но и свойствами структуры системы. При этом под структурой системы понимают совокупность элементов, связей и отношений между ними [40].

Под элементами понимаются: люди, подразделения, компоненты изделия или изделие в целом, являющееся в свою очередь составной частью более крупного изделия и т.п. [38].

Любая система образуется в результате взаимодействия составляющих ее элементов, причем это взаимодействие придает системе новые свойства, отсутствовавшие у отдельно взятых элементов. Проявляется, так называемый синергетический эффект. Так, например, отдельные элементы (резистор, индуктивность, конденсатор) и узлы (умножитель частоты, усилитель, частотомер и др.), даже очень хорошие, не являются радиоприемником в целом, хотя они являются совокупностью связанных узлов этого приемника.

Функциональная целостность системы характеризует завершенность ее внутреннего строения. Именно система выступает как нечто целое относительно окружающей среды: при возмущающем воздействии внешней среды проявляются внутренние связи между ее элементами.

Отсюда следует, что только совокупность взаимосвязанных элементов образует систему, и это взаимодействие между элементами порождает новое, особое качество целостности, называемое системным [38].

Стратегия изучения сложных систем и прежде всего систем со слабо формализуемыми проблемами, с которыми часто приходится иметь дело на практике, называется *системным анализом*.

*Системный анализ* представляет собой совокупность методов и средств, позволяющих исследовать свойства, структуру и функции объектов, явлений или процессов в целом, представив их в качестве систем со всеми сложными межэлементными взаимосвязями, взаимовлиянием элементов на систему и на окружающую среду, а также влиянием самой системы на ее структурные элементы.

Такими элементами в СЗИ являются люди (руководство и сотрудники организации, прежде всего, сотрудники службы безопасности), инженерные конструкции, технические и программно-аппаратные средства, обеспечивающие защиту информации. В рамках СЗИ они представляют собой не просто набор взаимосвязанных элементов, а объединены общей целью и решаемыми задачами.

В основе стратегии системного анализа лежат следующие общие положения:

- четкая формулировка цели исследования;
- постановка задачи по реализации этой цели и определение критерия эффективности решения задачи;
- разработка развернутого плана исследования с указанием основных этапов и направлений решения задачи;
- последовательное продвижение по всему комплексу взаимосвязанных этапов и возможных направлений;

- организация последовательных приближений и повторных циклов исследований на отдельных этапах, т.е. постепенное увеличение масштаба и увеличение части исследования объекта;
- принцип нисходящей иерархии анализа и восходящей иерархии синтеза в решении составных задач и т.п.

Системный анализ позволяет организовать знания об объекте таким образом, чтобы помочь выбрать нужную стратегию, либо предсказать результаты одной или нескольких стратегий, представляющихся целесообразными для принятия правильных решений.

Для этого в качестве одного из методов исследования систем используется математическое моделирование. При этом используется теория массового обслуживания, теория игр, теория нечетких множеств и т.д.

Как уже отмечалось, свойства системы как целого во многом определяются свойствами структуры системы. Структура системы наполняется различным содержанием, которое зависит от степени формализации процессов, происходящих внутри элементов. В настоящее время принято различать три уровня формализации [41]:

1. Вербальное описание, когда для характеристики функционирования элемента используют лишь содержательное словесное или графическое представление.

2. Класс "мягких" моделей, в котором описание функционирования элемента производится упрощенно; математическая модель представляет искусственно сформированную конструкцию, которая отражает самое существенное свойство системы (как правило, одно, максимум два). Несмотря на кажущуюся упрощенность такого подхода, "мягкие" модели демонстрируют свою работоспособность и находят широкое распространение в различных областях науки (экологии, биологии, экономике).

3. Класс "жестких" моделей образован строгими математическими структурами, полученными при декомпозиции процессов на основе анализа

причинно-следственных связей и установления количественных зависимостей между входными и выходными воздействиями.

При решении задач обеспечения информационной безопасности редко удается сформулировать ее строгую математическую модель. Примеры строго математической постановки и решения некоторых задач из области ИБ приведены в Приложениях 1 и 2.

Для описания слабоструктурированных, слабо формализованных и многосвязных систем большой размерности наилучшими являются "мягкие" модели, поскольку только при полной формализации отдельных процессов можно получить строго формальное представление системы.

Любая "мягкая" модель в математической постановке задачи является набором черных ящиков с заданными входами и определенными выходами. Таким образом, традиционная модель управления в случае больших слабоструктурированных и слабо формализованных систем сводится к анализу функционирования совокупности черных ящиков, функционально связанных между собой законом прохождения импульса.

Главная проблема применения методов исследования состоит в том, чтобы правильно подобрать типовую или разработать новую математическую модель. При этом надо собрать необходимые исходные данные и убедиться путем анализа исходных предпосылок и результатов предварительного математического расчета, что эта модель отражает существо решаемой задачи.

Однако при любом методе математического моделирования основным принципом является декомпозиция сложной системы на более простые подсистемы (принцип иерархии системы). В этом случае общая математическая модель строится по блочному принципу: общая модель подразделяется на блоки, которым можно дать сравнительно простые математические описания. Но каждый отдельный блок не должен вступать в противоречие с целями общей сложной системы. Кроме того, при математическом моделировании необходимо выяснять близость полученного результата к желаемому.

С позиций системного анализа решаются задачи не только моделирования, но и управления и оптимального проектирования систем.

Особый вклад (важность) системного анализа в решении различных проблем заключается в том, что он позволяет выявить факторы и взаимосвязи между элементами или блоками, дает возможность спланировать методику наблюдений и построить эксперимент так, чтобы эти факторы были включены в рассмотрение и освещали слабые места гипотез и допущений. Как научный подход системный анализ создает инструментарий познания физического мира и объединяет его в систему гибкого исследования сложных явлений.

*Системный подход* - это направление методологии научного познания и социальной практики, в основе которого лежит рассмотрение сложных объектов как систем, как объединения элементов, связанных комплексом отношений друг с другом и выступающих единым целым по отношению к внешней среде.

Он ориентирует исследование на раскрытие целостности объекта, на выявление различных типов связей в нем и сведения в единую теоретическую картину.

Системный подход основан на представлении о системе как о чем-то целостном, обладающем новыми свойствами (качествами) по сравнению со свойствами составляющих ее элементов. Новые свойства при этом понимаются очень широко. Они могут выражаться, в частности, в достижении новых целей или в способности решать новые проблемы. Для этого требуется определить границы системы, выделив ее из окружающего мира, и затем соответствующим образом изменить (преобразовать), или, говоря математическим языком, перевести систему в желаемое состояние. В системном подходе можно выделить следующие этапы:

- постановка задачи (проблемы): определение объекта исследования, постановка целей, задание критериев для изучения объекта и управления им;

- очерчивание границ изучаемой системы и ее первичная структуризация. На этом этапе вся совокупность объектов и процессов, имеющих отношение к поставленной цели, разбивается на два класса – собственно изучаемая система и внешняя среда;
- составление математической модели изучаемой системы: параметризации системы, задание области определения параметров, установление зависимостей между введенными параметрами;
- исследование построенной модели: прогноз развития изучаемой системы на основе ее модели, анализ результатов моделирования;
- выбор оптимального управления. Этот шаг как раз и позволяет перевести систему в желаемое (целевое) состояние и тем самым решить проблему.

Сложные системы характеризуются выполняемыми процессами (функциями), структурой и поведением во времени. Для адекватного моделирования этих аспектов в информационных системах применяют функциональные, информационные и поведенческие модели.

***Функциональная модель*** системы описывает совокупность выполняемых системой для достижения определенной цели функций, характеризует морфологию системы (ее строение), состав функциональных подсистем, их взаимосвязи.

***Информационная модель*** отражает отношение между элементами системы в виде структур данных, их состава и взаимосвязи.

***Поведенческая (событийная) модель*** описывает информационные процессы (динамику функционирования). В ней фигурируют такие категории, как состояние системы, событие, переход их одного состояния в другое, условия перехода, последовательность событий. Поведенческая модель предполагает понимание системного объекта как совокупности процессов, характеризуемых последовательностью состояний во времени. Основным понятием здесь является понятие периода жизни, т.е. временного интервала, в течение которого функционирует данный процесс.

Функциональное и информационное представления отражают статические аспекты системы. Поведенческая модель рассматривает систему как динамический процесс. При этом по отношению к состоянию целей система может находиться в двух режимах: функционирования и развития [38].

В первом случае считается, что система полностью удовлетворяет потребностям, и процесс перехода ее отдельных элементов из состояния в состояние происходит при постоянстве заданных целей.

Во втором случае считается, что система в некоторый момент времени перестает удовлетворять потребностям и требуется корректировка прежних целевых установок.

## 2.5. Функциональная модель процесса комплексного обеспечения информационной безопасности

Разработка, внедрение и эксплуатация СКЗИ представляет собой многофункциональную динамическую систему, для которой характерны признаки, присущие сложным системам.

Сложная система при функциональной декомпозиции расчленяется на более простые подсистемы, каждая из которых, в свою очередь, представляет определенную совокупность еще менее сложных структурных единиц, выполняющих самостоятельные функции.

Такая процедура базируется на учете следующих основных закономерностей:

- каждая более сложная подсистема декомпозируется на менее сложные до тех пор, пока они не окажутся простыми либо квазипростыми;

- каждый уровень структурных единиц системы должен содержать не только необходимые, но и достаточное количество самостоятельных структурных единиц, которые в совокупности удовлетворяют вышестоящим структурным единицам системы с точки зрения необходимости и достаточности для их формализации и оценки;

- каждая вышестоящая структурная единица системы должна включать нижестоящие структурные единицы по принципу «И», а не «ИЛИ»;

- структурные единицы должны быть тем более конкретными, точнее и детальнее описанными, чем ниже они расположены по иерархии.

Таким образом, функциональная декомпозиция процесса разработки внедрения и эксплуатации СКЗИ предполагает формирование целей и задач, стоящих перед структурными элементами СКЗИ, которые достигаются в процессе решения частных задач, каждая из которых, в свою очередь, достигается за счет решения задач, стоящих перед структурными единицами, расположенными ниже по иерархии.

Для описания элементов декомпозированной системы может быть использовано ее представление как стохастической системы с неполной информацией о состояниях. При этом, поскольку задачи, стоящие перед СКЗИ, являются сложно формализуемыми, то, как показывает анализ, наиболее предпочтительными методами описания являются такие, как экспертные оценки, имитационное моделирование, методы теории нечетких множеств, лингвистических переменных, мягких измерений и т.п.

Построение функциональной модели сложных систем, таких как система КОИБ, целесообразно проводить в соответствии со стандартами IDEF (IDEF0, IDEF3, IDEF5, DFD и др.). Разработка моделей в данных стандартах позволяет наглядно и эффективно отобразить весь механизм создания, внедрения и эксплуатации системы комплексной защиты информации в нужном разрезе.

Стандарт IDEF0 – это технология описания системы в целом как множества взаимозависимых действий или функций. Важно отметить функциональную направленность IDEF0 – функции системы исследуются независимо от объектов, которые обеспечивают их выполнение. «Функциональная» точка зрения позволяет четко отделить аспекты назначения системы от аспектов ее физической реализации.

Результаты IDEF0 анализа могут применяться при проведении исследований с использованием моделей IDEF3 и диаграмм потоков данных DFD.

Отличительной особенностью языка IDEF0 является использование в качестве основы естественного языка экспертов, который структурируется с помощью графических средств. Это дает возможность эксперту свободно описывать функционирование системы, пользуясь знакомой и удобной терминологией, а затем легко перенести описание на естественном языке в графическое представление языка IDEF0. Кроме того, IDEF0-модели могут быть также использованы для функционально-стоимостного анализа бизнес-процессов.

В стандарте IDEF0 описание системы организовано в виде иерархически упорядоченных и взаимосвязанных диаграмм.

Вершина этой древовидной структуры представляет собой самое общее описание системы и ее взаимодействия с внешней средой. В основании структуры находятся наиболее детализированные описания выполняемых системой функций. Диаграммы содержат функциональные блоки, соединенные дугами. Дуги отображают взаимодействия и взаимосвязи между блоками.

Функциональный блок на диаграммах изображается прямоугольником и представляет собой функцию или активную часть системы. Названиями блоков служат глаголы или глагольные обороты.

Каждая сторона блока имеет особое, вполне определенное назначение. К левой стороне блока подходят дуги входов, к верхней – дуги управления, к нижней – дуги механизмов реализации выполняемой функции, а из правой - направлены дуги выходов.

Такое соглашение предполагает, что используя управляющую информацию об условиях и ограничениях, а также реализующий ее механизм, функция блока преобразует свои входы в соответствующие выходы.

На диаграмме блоки упорядочиваются, начиная с левого верхнего угла и заканчивая нижним правым углом. Для обеспечения наглядности и лучшего понимания моделируемых процессов рекомендуется использовать от 3 до 6 блоков на одной диаграмме. Такое представление модели устраняет неоднозначность, присущую естественному языку, и благодаря этому достигаются необходимые для понимания и анализа лаконичность и точность описания, без потери деталей и качества [42].

Общим управляющим воздействием для процессов КОИБ являются положения соответствующих законов, ГОСТы, руководящие документы (РД).

Реализация различных процессов, отображаемых в моделях, осуществляется сотрудниками отдела информационной безопасности, субъектами информационных отношений, экспертами в области ИБ.

Поскольку, в соответствии со стандартом IDEF0, система представляется как множество вложенных функций, образующих иерархию, в первую очередь должна быть определена функция, описывающая систему в целом, т.е. построена так называемая контекстная диаграмма.

Затем иерархическая декомпозиция каждого блока может быть продолжена до необходимого уровня детализации.

Общая структурная схема процесса функциональной декомпозиции задачи комплексного обеспечения информационной безопасности представлена на рис.2.1. Далее приведены соответствующие диаграммы декомпозиции.

Планирование ЗИ как функция управления представляет собой процесс последовательного снятия неопределенности относительно структуры и состава средств защиты на объекте управления.

Разработанная в рамках стандарта IDEF0 функциональная модель КОИБ дает возможность установить, какие функции выполняются системой, в какой последовательности, кто является ответственным за проведение конкретных работ, что является результатом реализации той или иной процедуры.

Таким образом, инструментарий IDEF0 позволяет эффективно моделировать процесс комплексного обеспечения информационной безопасности с целью оптимизации управления защитой информационных активов.

Рис.2.1. Состав функциональных блоков модели КСЗИ

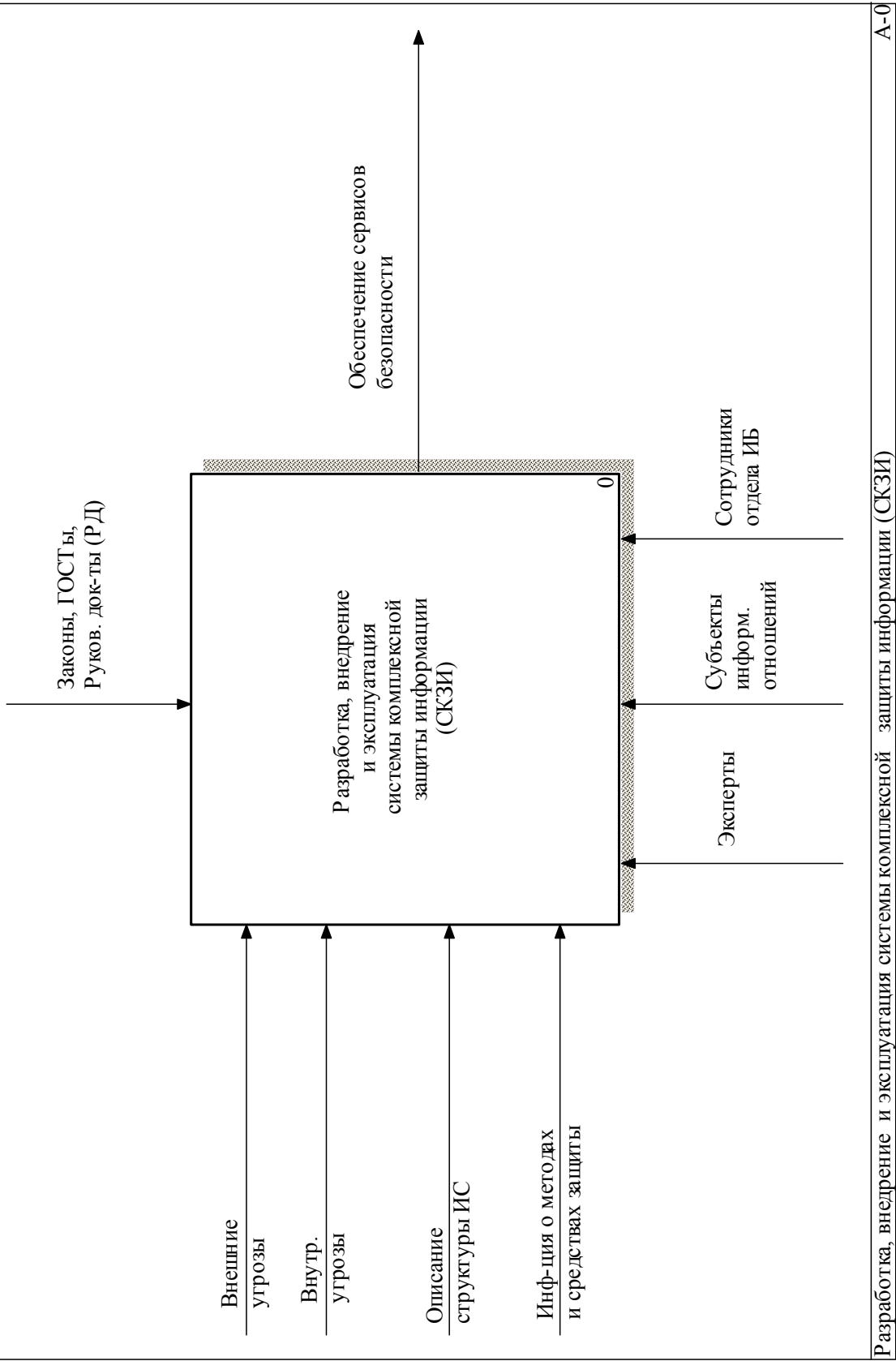

Контекстная диаграмма модели КОИБ

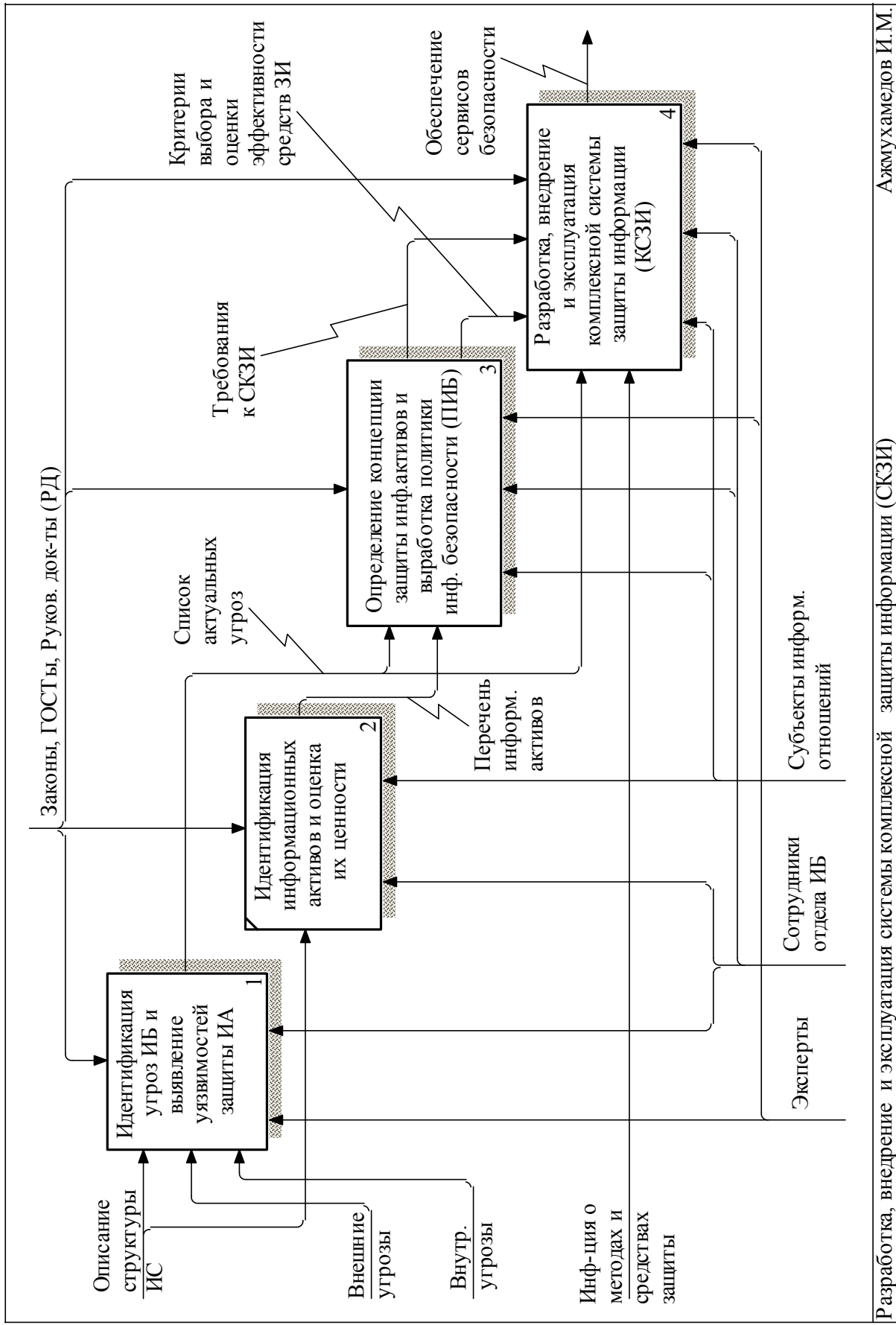

Разработка, внедрение и эксплуатация системы комплексной защиты информации (СКЗИ)    Ажмухамедов И.М.

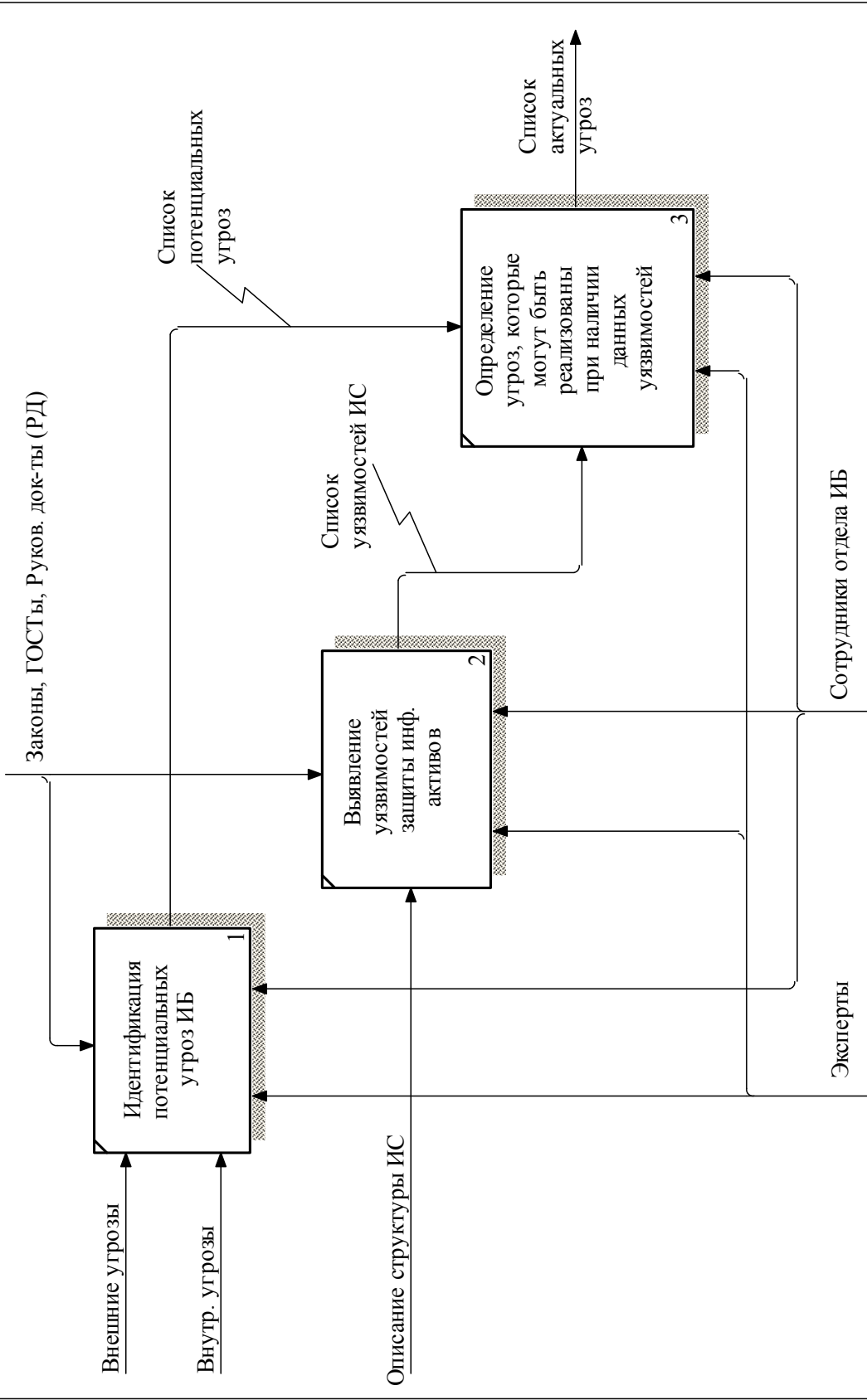

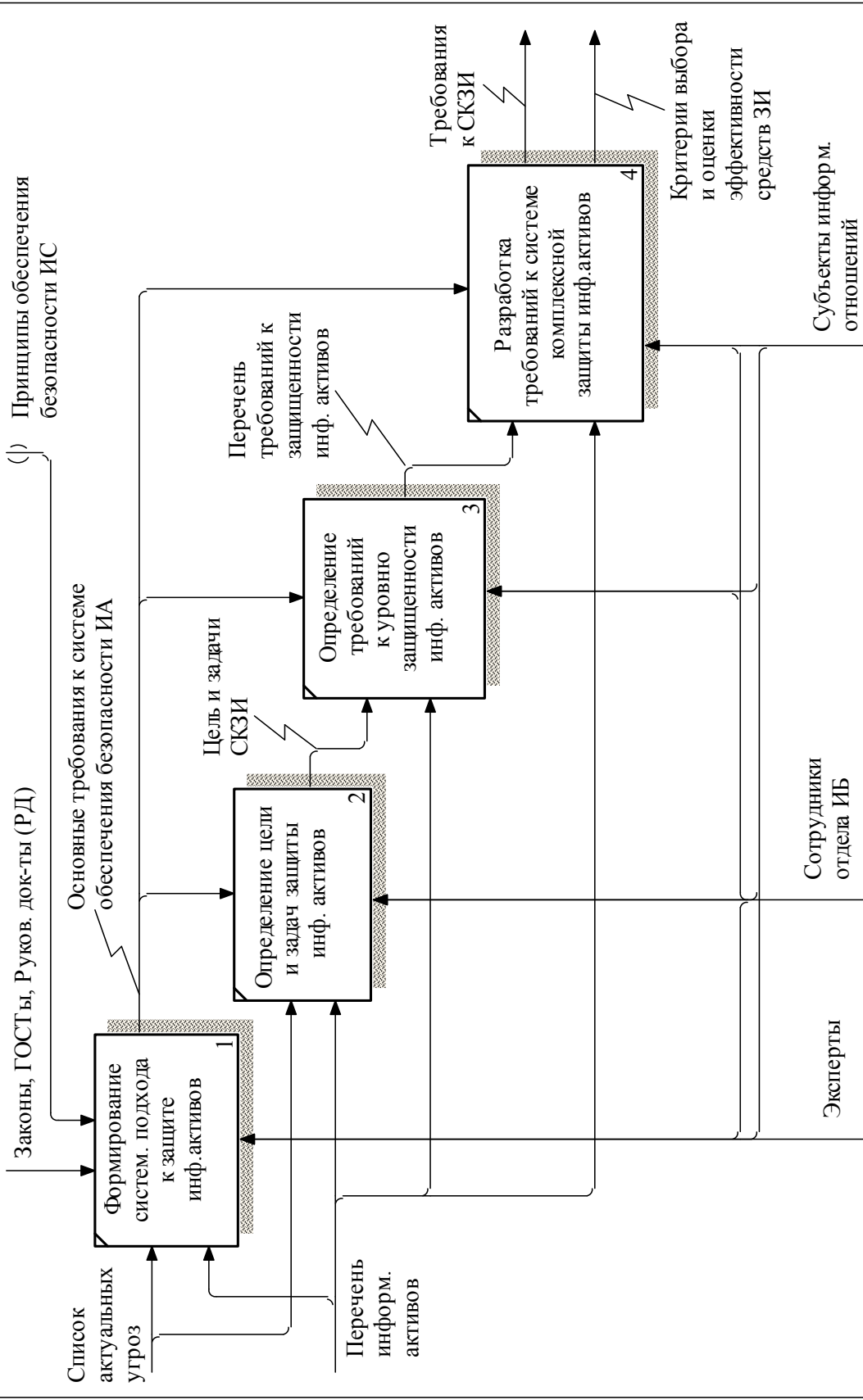

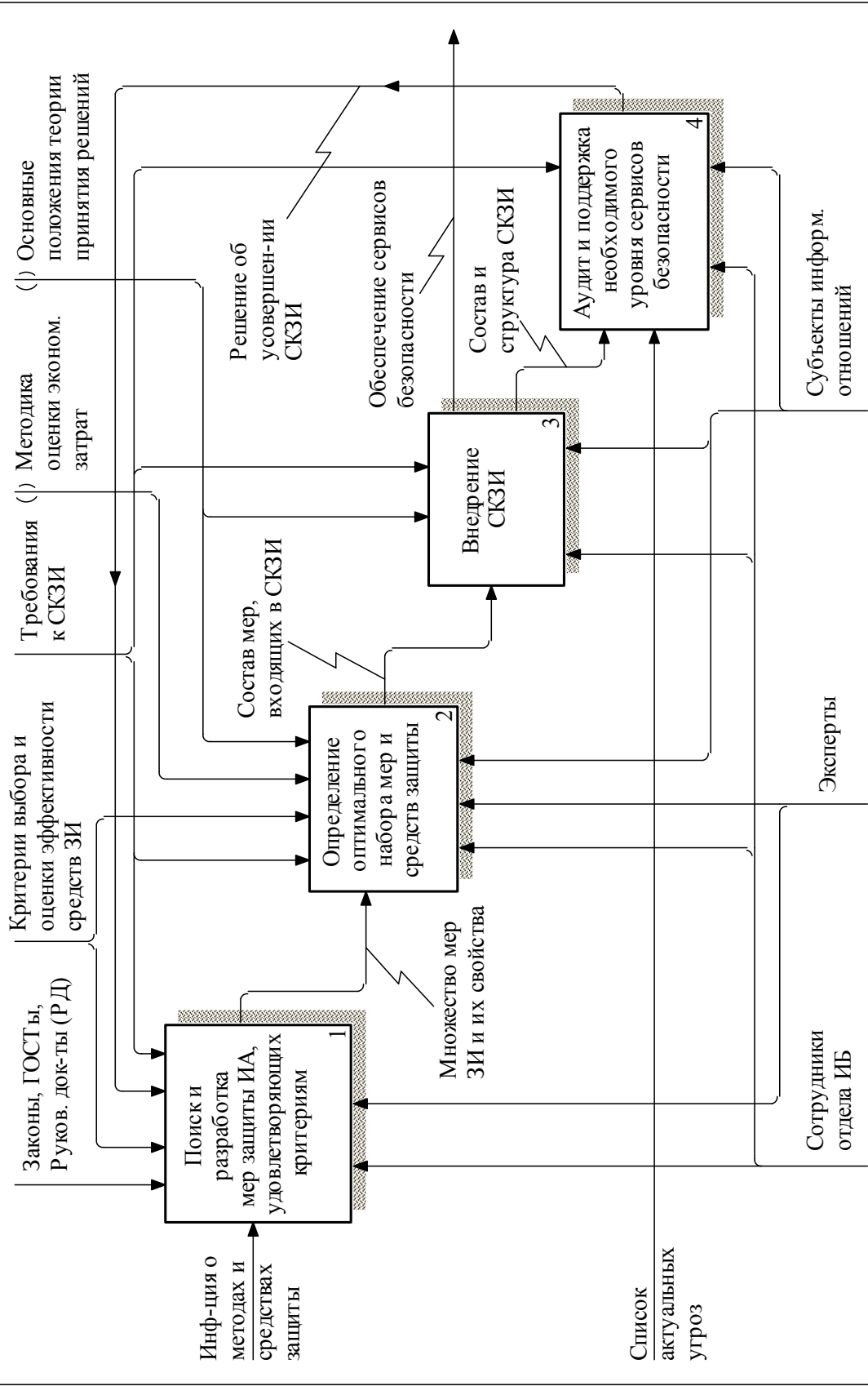
125

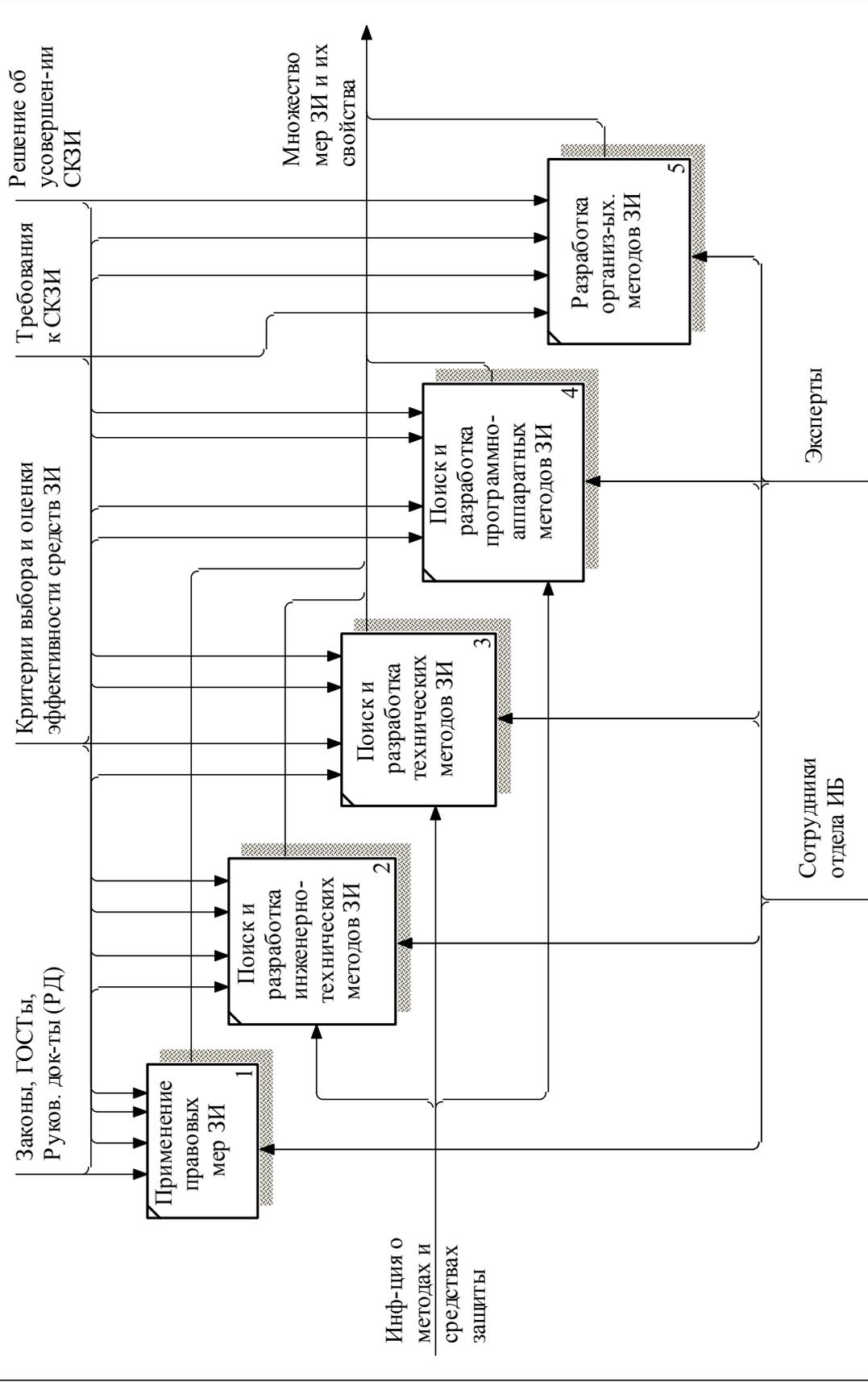

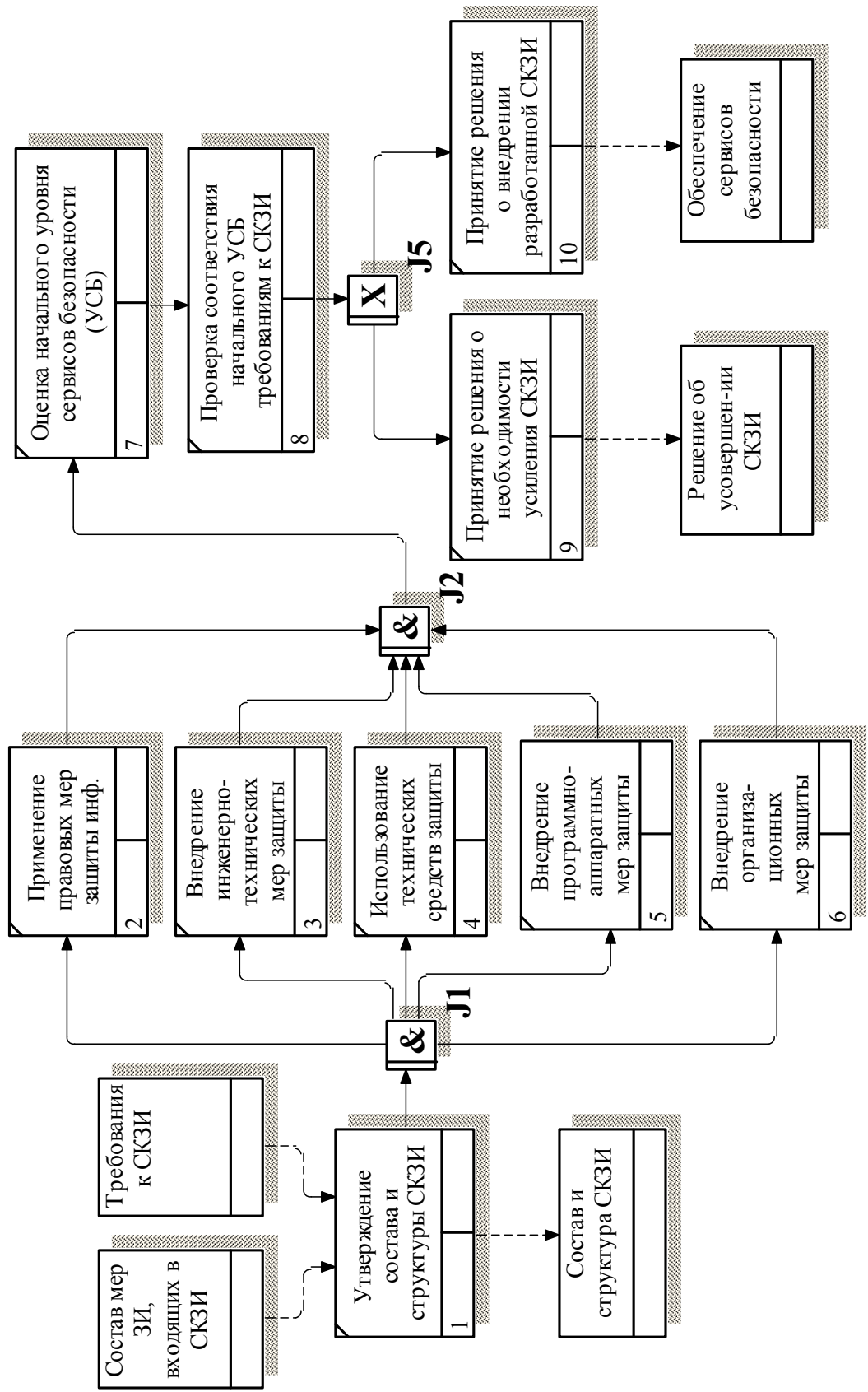

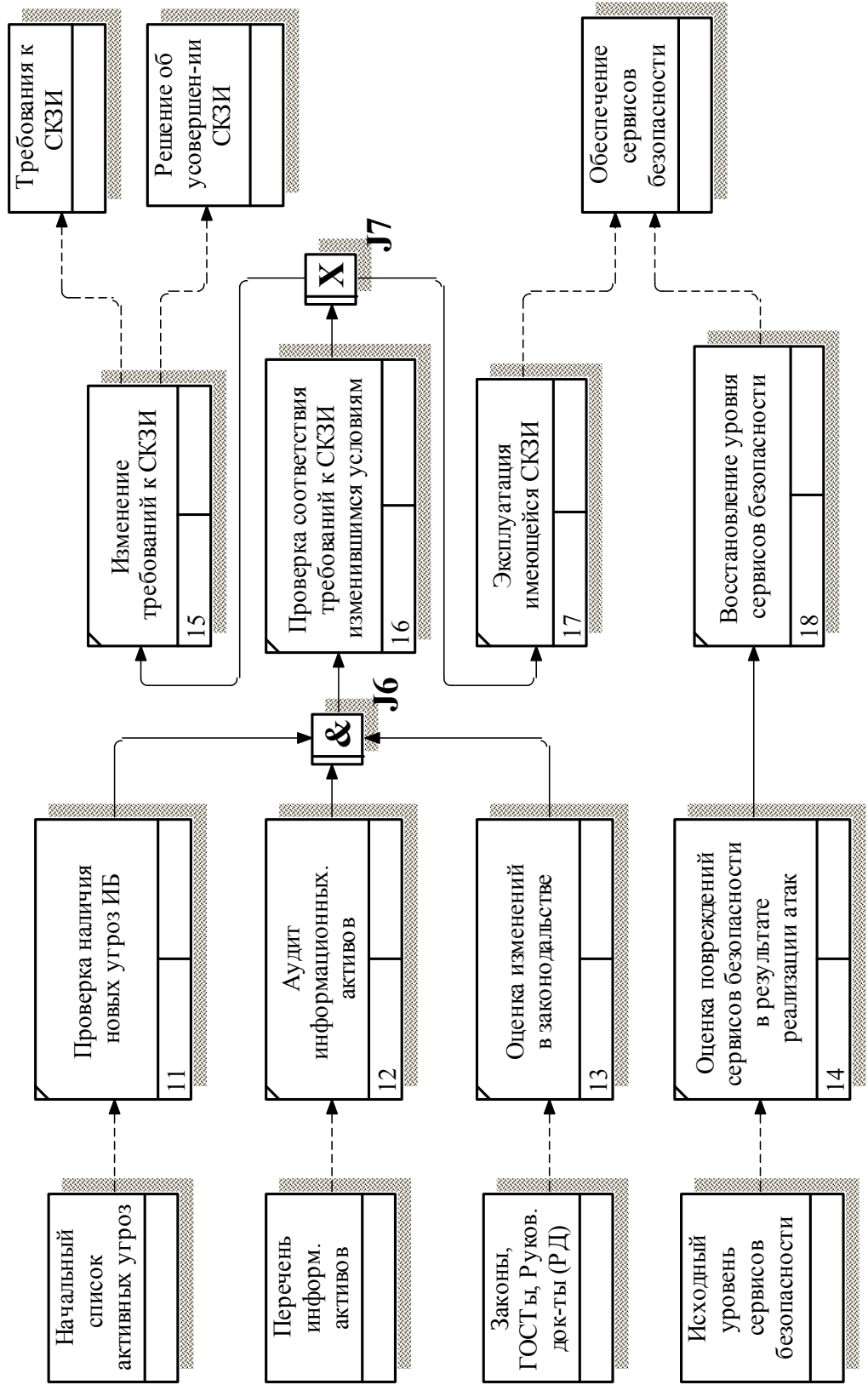

Ажмухамедов И.М.

Аудит и поддержка необходимого уровня сервисов безопасности

A4.4.1

## 2.6. Выводы по главе 2

Обобщая рассмотренный в главе 2 материал можно сделать следующие основные выводы:

- Постановка задачи защиты информации на современном этапе приобретает ряд специфических особенностей. Проблема обеспечения желаемого уровня защиты информации требует для своего решения создания целостной системы организационно-технологических мероприятий и применения комплекса специальных средств и методов ЗИ. Поиск новых путей защиты информации заключается не просто в создании соответствующих механизмов, но и требует комплексного использования всех имеющихся средств защиты на каждом из этапов жизненного цикла ИС.

- Работы в области ИБ должны основываться на системно-концептуальном подходе к защите информации. Концептуальность подхода предполагает разработку единой концепции как полной совокупности научно обоснованных взглядов, положений и решений, необходимых и достаточных для оптимальной организации и обеспечения надёжности защиты информации, а также целенаправленной организации всех работ по защите информационных активов.

- Безопасность – понятие комплексное и не может рассматриваться как простая сумма составляющих ее частей. Каждая часть критически значима, эти части взаимосвязаны и взаимозависимы, поэтому при решении задач КОИБ ярко проявляется синергетический эффект: средства защиты информации с одной стороны являются составной частью системы, с другой стороны - они сами организуют систему, осуществляя защитные мероприятия. При отсутствии отдельных компонентов системы или их несогласованности между собой неизбежно возникновение уязвимостей в технологии защиты информации. Как следствие, одним из основных принципов при создании и развитии эффективной системы обеспечения информационной безопасности должен быть принцип системности [43].

- К организации системы защиты информации с позиции системного подхода выдвигается ряд требований, определяющих ее целостность, адекватность и эффективность. Поскольку регулярных (а тем более формальных) методов решения задач, направленных на удовлетворение этих требований, на сегодняшний день не существует, приходится использовать методы неформальные, эмпирические.

- Применение системного анализа для решения задач на всех этапах жизненного цикла СКЗИ позволяет обеспечить полноту и эффективность реализаций их функций, а также оптимальные ресурсные, финансовые и временные параметры для достижения поставленных целей.

- Анализ научных и практических работ в области ИБ позволил сформулировать общеметодологические принципы (общие положения) построения и функционирования СКЗИ.

- Разработка, внедрение и эксплуатация СКЗИ представляет собой многофункциональную динамическую систему, для которой характерны признаки, присущие сложным системам. Построение функциональной модели сложных систем, таких как система КОИБ, целесообразно проводить в соответствии со стандартами IDEF (IDEF0, IDEF3, IDEF5, DFD и др.). Разработка моделей в данных стандартах позволяет наглядно и эффективно отобразить весь механизм создания, внедрения и эксплуатации системы комплексной защиты информации в нужном разрезе.

- Разработанная в рамках стандартов IDEF0 и IDEF3 функциональная модель КОИБ дает возможность установить, какие функции выполняются системой, в какой последовательности, кто является ответственным за проведение конкретных работ, что является результатом реализации той или иной процедуры. Таким образом, инструментарий стандартов IDEF позволяет эффективно моделировать процесс комплексного обеспечения информационной безопасности с целью оптимизации управления защитой информационных активов.

# ГЛАВА 3.

# ПРИНЦИПЫ ПОСТРОЕНИЯ НЕЧЕТКИХ КОГНИТИВНЫХ МОДЕЛЕЙ

## 3.1. Основные понятия когнитивного моделирования

Как неоднократно отмечалось в предыдущих главах, управление системой комплексного обеспечения информационной безопасности, которая является системой организационно-технологического (социотехнического) типа, является сложной, слабо структурируемой и трудно формализуемой задачей.

Современные научные подходы к исследованию подобных систем не возможны без применения системного подхода, позволяющего связать в единое целое множество процессов, протекающих в информационной системе.

Наиболее удобным математическим аппаратом для описания и исследования подобных систем является когнитивное моделирование, как одно из направлений современной теории поддержки принятия решений при управлении слабоструктурированными системами и ситуациями [7; 44-55.]. Оно является, по мнению академика И.В.Прангишвили, одним из оправдавших себя на практике научных методов повышения эффективности управления сложными системами [40; 56].

Когнитивный анализ и моделирование позволяют исследовать проблему, учесть изменения внешней среды, определить реакцию системы.

Построение когнитивной модели осуществляется на основе системного подхода, который представляет собой совокупность методов и средств, позволяющих исследовать свойства, структуру и функции объектов, явлений или процессов в целом, представив их в качестве систем со всеми сложными межэлементными связями. Системный подход позволяет увидеть и оценить целостность проблемы во всем ее многообразии и выбрать наилучший способ управления сложной системой [57].

В работе [58] выделены основные классы управленческих задач, для решения которых применяется когнитивное моделирование.

*Управленческие задачи, для решения которых целесообразно применение когнитивного моделирования*

Традиционные теоретические методы управления (в рамках теории рационального выбора) концентрируют внимание на процессах поиска оптимального решения из фиксированного набора альтернативных решений для достижения четко поставленной цели. Вопросы идентификации проблем, формирования целей и множества альтернатив их достижения зачастую остается в стороне.

В реальных управленческих ситуациях очень часто возникает задача, которая состоит не в том, чтобы сделать выбор между альтернативными решениями, а в том, чтобы проанализировать ситуацию для выявления реальных проблем и причин их появления.

Понимание проблемы – обязательное предварительное условие нахождения приемлемого решения. Без знания объекта управления вообще нельзя им управлять, не говоря уже о качестве целенаправленных воздействий и их результатах. При этом для слабоструктурированных систем (СС) характерны проблемы, которые с трудом поддаются вычленению в исследуемой управленческой ситуации, что ограничивает возможности применения традиционных методов поиска оптимального (или даже удовлетворительного) управленческого решения в задачах управления такими системами.

Одной из причин является недостаток информации о состоянии СС в условиях слабо контролируемой и изменяющейся внешней среды. Отсутствие достаточных знаний о системе, относительно которой принимается решение, не является единственной неопределенностью, обусловленной субъективными причинами. Также можно выделить неопределенность целей развития СС и критериев выбора управленческого решения [59-60]. Как правило, неудовлетворенность текущим состоянием системы осознается субъектом управления, но его представления о причинах и возможных способах изменения ситуации в СС размыты, нечетки и противоречивы. Формализация нечетких представлений – одна из главных задач, которую надо решать при разработке моделей и методов принятия решений в слабоструктурированных ситуациях [59].

Важно также учесть, что субъекту управления очень часто приходиться принимать решения в постоянно изменяющихся условиях и при ограниченных временных ресурсах.

Другая трудность связана с тем, что субъекту управления приходится манипулировать качественной информацией в виде гипотез (предположений), интуитивных понятий и смысловых образов. Многочисленные исследования процессов принятия решений подтверждают, что субъекту управления несвойственно мыслить и принимать решения только в количественных характеристиках. Он мыслит, прежде всего, качественно, и для него поиск решения – это, поиск, в первую очередь, замысла решения, где количественные оценки играют вспомогательную роль [59]. Поэтому структуры знания в мышлении субъекта управления оказываются важнейшими элементами ситуации, неустранимыми из модели принятия решений.

Особенностью исследования СС является то, что процесс подготовки и принятия решений по управлению СС, как правило, является групповой деятельностью. Каждый участник этого процесса представляет проблемную ситуацию исходя из "своих" внутренних представлений и знаний о ситуации. Картина мира включает в себя набор убеждений, особенностей восприятия, ценностных и практических установок субъекта, которыми он руководствуется в своей деятельности и влияет на процесс разрешения проблемной ситуации.

Таким образом, подготовку и принятие решений в задачах управления СС, следует рассматривать как сложный интеллектуальный процесс разрешения проблем, несводимый исключительно к рациональному выбору. Для поддержки этого процесса требуются новые подходы к разработке формальных моделей и методов решения проблем и формирования целей развития СС, особенно на ранних этапах подготовки управленческих решений.

В [61] отмечается, что первый этап при применении методов принятия решений: «предварительный анализ проблемы и ее структуризация», – является наиболее сложным и трудно формализуемым. На этом этапе к работе привлекаются «опытные консультанты–аналитики», а арсенал применяемых методов, как правило, включает эвристические экспертные методы (мозговой штурм, интервьюирование и т.п.).

Когнитивный подход к моделированию и управлению СС направлен на разработку формальных моделей и методов, поддерживающих интеллектуальный процесс решения проблем благодаря учету в этих моделях и методах когнитивных возможностей (восприятие, представление, познание, понимание, объяснение) субъектов управления при решении управленческих задач.

***Основные понятия и определения когнитивного моделирования***

Когнитивное моделирование в задачах анализа и управления СС − это исследование функционирования и развития слабоструктурированных систем и ситуаций посредством построения модели СС (ситуации) на основе когнитивной карты. Модель включает когнитивную карту как обязательный элемент наряду с возможными другими параметрами. Во многих публикациях, развивающих когнитивный подход к управлению СС, когнитивную карту часто отождествляют с когнитивной моделью. Однако следует помнить, что понятие "когнитивная модель" связывается с ментальной моделью субъекта, порожденной под воздействием его познавательных возможностей.

В этой модели когнитивная карта отражает субъективные представления (индивидуальные или коллективные) исследуемой проблемы, ситуации, связанной с функционированием и развитием СС. Основными элементами когнитивной карты являются базисные факторы (или концепты) и причинно-следственные связи между ними [7; 46; 48-49; 62].

Содержательно, базисные факторы – это факторы, которые:

- определяют и ограничивают наблюдаемые явления и процессы в СС и окружающей ее среде;

- интерпретированы субъектом управления как существенные, ключевые параметры, признаки этих явлений и процессов.

Когнитивная карта представляет собой знаковый или взвешенный граф над множеством факторов, т.е. ориентированный граф, вершинам которого сопоставлены факторы [41]. В случае знакового графа ребрам сопоставляются знаки (+ или −), а в случае взвешенного - веса в той или иной шкале. Веса ищутся либо с помощью статистической обработки информации, либо экспертным путем.

Изменения факторов проводятся по шагам до определения реакции системы, после этого с помощью многокритериального выбора определяется множество благоприятных сценариев и они ранжируются [63].

Различные интерпретации вершин, ребер и весов на ребрах, а также различные функции, определяющие влияние связей на факторы, приводят к различным модификациям когнитивных карт и средствам их исследования. При этом интерпретации могут различаться как в содержательном плане, так и в математическом. Благодаря наличию множества модификаций когнитивных карт можно говорить о различных типах моделей, основу которых составляют эти карты [48].

Различают пять видов когнитивных карт по типу используемых отношений [54]:

- оценивающих фокусирование внимания, ассоциации и важность понятий (концептов);
- показывающих размерность категорий и когнитивных таксономий;
- представляющих влияние, причинность и системную динамику (каузальные когнитивные карты);
- отражающих структуру аргументов и заключений;
- иллюстрирующих фреймы и коды восприятия.

Практика применения когнитивных карт показывает, что для исследования СС наиболее целесообразно применение каузальных когнитивных карт [53].

Выбор способа структурирования слабоструктурированных систем и ситуаций в виде множества факторов и причинно-следственных связей между ними не случаен. Он обусловлен тем, что явления и процессы функционирования и развития СС включают в себя различные события, тенденции, определяемые многими факторами, причем каждый в свою очередь влияет на некоторое число других факторов. Образуются сети причинных отношений между ними [51].

В книге известного немецкого психолога Д. Дёрнера, посвященной исследованию мышления субъекта управления и анализу причин ошибок при разрешении проблемных ситуаций в функционировании и развитии сложных систем, указывается, что «сиюминутная ситуация с ее признаками – это только актуальное состояние системы и ее переменных. Следует не только понимать, что происходит, но и предвидеть, что произойдет или может произойти в будущем, а также предположить, как будет изменяться ситуация в зависимости от конкретных вмешательств. Для этого требуется структурное знание, то есть знание о том, как системные переменные взаимосвязаны и влияют друг на друга» [64].

Дёрнер отмечает, что в идеальном варианте это знание представляется в форме "математических функций", но в случае невозможности построения последних, применимы схемы причинно-следственных отношений, позволяющие реконструировать различного рода предположения (гипотезы), содержащиеся в голове субъекта управления, причем не в виде "каузальных цепей", а в виде "каузальных сетей".

Исследование взаимодействия факторов позволяет оценивать распространение влияния по когнитивной карте. Поведение (состояние) системы может быть описано на основе значений системных переменных, что делает возможным использование классических подходов из теории систем, в частности, для моделирования, анализа динамики, управления. Анализ когнитивной карты позволяет выявить структуру проблемы (системы), найти наиболее значимые факторы, влияющие на нее, оценить воздействие факторов (концептов) друг на друга. Если в когнитивной карте выделены целевые и входные концепты, на которые можно воздействовать, то круг решаемых задач включает оценку достижимости целей, разработку сценариев и стратегий управления, поиск управленческих решений.

Согласно [48], задачи анализа ситуаций на основе когнитивных карт можно разделить на два типа: статические и динамические.

Статический анализ, или анализ влияний – это анализ исследуемой ситуации посредством изучения структуры взаимовлияний когнитивной карты. Анализ влияний выделяет факторы с наиболее сильным влиянием на целевые факторы, т.е. факторы, значения которых требуется изменить. Динамический анализ лежит в основе генерации возможных сценариев развития ситуации во времени.

Таким образом, возможности решения задач анализа и управления определяются типом используемых моделей – статических или динамических.

Для проведения обоих видов анализа, как правило, используется математический аппарат двух типов: аппарат линейных динамических систем и аппарат нечеткой математики.

Сложные социотехнические системы, к которым принадлежит и система комплексного обеспечения информационной безопасности, по классификации относятся к искусственным, т.к. одним из активных элементов в них является человек [57].

Деятельность человека представляет собой набор разовых импульсных воздействий разнесенных во времени. Поэтому, одним из важнейших требований к построению системы является ее стабильность к такого рода воздействиям.

При моделировании подобные свойства процессов и явлений обеспечиваются существованием отрицательных (стабилизирующих) и положительных (стимулирующих рост) обратных связей. А также тонкой настройкой параметров обратных связей, гарантирующих выполнение указанных выше условий [65].

Рассмотрим упрощенный пример когнитивной карты для анализа проблемы обеспечения информационной безопасности при обработке данных с использованием средств вычислительной техники (рис. 3.1).

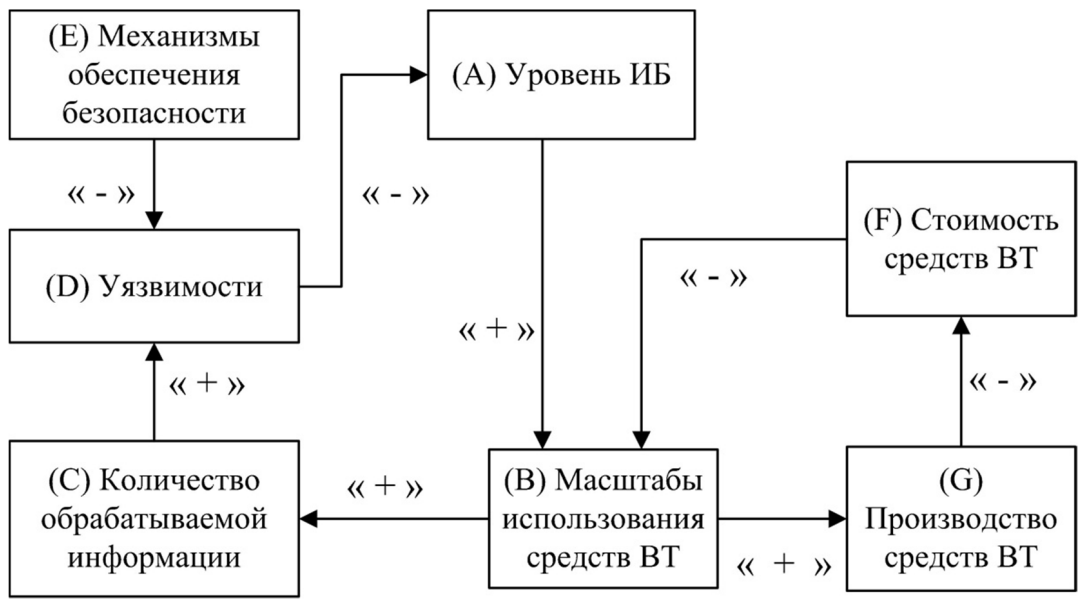

Рис.3.1. Пример когнитивной карты

Допустим, что исследуемую проблему можно описать семью факторами: {A, B, C, D, E, F, G}. Дугами на рис. 3.1 отмечены существенные причинно-следственные отношения (считается, что влиянием остальных можно пренебречь).

Знак «+» рядом с дугой означает, что при увеличении исходного фактора (концепта) зависимый фактор также увеличивается. Знак «-» означает, что эти два фактора изменяются в противоположных направлениях: увеличение исходного концепта ведет к уменьшению значения зависящего от него концепта.

Рассмотрим взаимодействие факторов в контуре {A-B-C-D-A}. Предположим, что масштабы использования средств ВТ выросли. Это приведет к увеличению количества обрабатываемой информации и, следовательно, к появлению большего количества уязвимостей, что в свою очередь повлечет снижение уровня информационной безопасности. А это приведет к уменьшению использования средств ВТ при обработке данных

Таким образом, влияние импульса в вершине B будет компенсироваться действием контура {A-B-C-D-A}, и поведение системы стабилизируется. Четыре фактора A, B, C и D образуют контур, противодействующий отклонению.

В отличие от этого, в контуре {B, F, G} увеличение (уменьшение) любой переменной будет усилено.

Контуры в когнитивной карте соответствуют контурам обратной связи. Контур, усиливающий отклонение, является контуром положительной обратной связи, а контур, противодействующий отклонению, — контуром отрицательной обратной связи. Японский ученый М. Маруяма назвал эти контуры соответственно морфогенетическими и гомеостатическими [66].

В этой же работе Маруяма доказал, что контур усиливает отклонение тогда и только тогда, когда он содержит четное число отрицательных дуг или не содержит их совсем, в противном случае это контур, противодействующий отклонению. Действительно, в случае четного числа отрицательных дуг противодействие отклонению будет само встречать противодействие. Если число отрицательных дуг нечетно, то последнее противодействие отклонению не встречает противодействия.

Данная схема анализа в основном соответствует интуитивным представлениям о причинности. Если взаимодействие двух факторов A и B подчиняется более сложным закономерностям, то в этом случае для описания исследуемого процесса следует использовать языки функциональных взаимосвязей.

Слабая формализованность задач обеспечения безопасности приводит к необходимости применения аппарата нечеткой логики

Процесс моделирования при этом целесообразно проводить на основе *нечетких когнитивных карт* (fuzzy cognitive maps) (НКК), которые были предложены Б.Коско [55] и используются для моделирования причинных взаимосвязей, выявленных между концептами некоторой области.

В отличие от *простых* когнитивных карт, *нечеткие* когнитивные карты представляют собой нечеткий ориентированный граф, узлы которого являются нечеткими множествами. Направленные ребра графа, как и в случае простой карты, не только отражают причинно-следственные связи между концептами, но и определяют степень влияния (вес) связываемых концептов. Веса ребер - это либо числа из отрезка [-1, 1], либо значения из некоторой лингвистической шкалы типа {малый, средний, большой}, которые характеризуют силу влияния соответствующей связи либо (в некоторых интерпретациях) степень уверенности в наличии этой связи. Методы анализа НКК используют операции нечеткой математики.

Существенным обобщением когнитивной карты является модель когнитивной карты, управляемой нечеткими правилами (RBFCM - Rule Based FCM) [67].

Нечеткие правила (продукции) имеют форму предложений вида ЕСЛИ-ТО, условная часть которых представляет собой выражение нечеткой логики над лингвистическими значениями факторов и отношениями между ними. Например: «ЕСЛИ $x_1$ есть А1 И $x_2$ есть А2 … ТО $y$ есть В», где $x_1$ и $x_2$ - входные переменные, $y$ - выходная переменная, А1, А2, В - нечеткие (лингвистические) значения. Пример построения такой модели для управления сетевым трафиком приведен в главе 8.

Посылка правила описывает условия его применимости, а заключение правила определяет функции принадлежности лингвистических значений выходных переменных. Ребра графа соответствуют отношениям влияния, выраженным условными частями правил. Каждому фактору сопоставляется база правил, состоящая из всех продукций, имеющих в заключении данный фактор.

Активное использование нечетких когнитивных карт в качестве средства моделирования систем обусловлено возможностью наглядного представления анализируемой системы и легкостью интерпретации причинно-следственных связей между концептами. Основные проблемы связаны с процессом построения когнитивной карты, который не поддается формализации. Кроме того, необходимо доказать, что построенная когнитивная карта адекватна реальной моделируемой системе. Для решения данных проблем разработаны алгоритмы автоматического построения когнитивных карт на основе выборки данных.

Однако когнитивная карта отображает лишь наличие влияний факторов друг на друга. В ней не отражается ни детальный характер этих влияний, ни динамика изменения влияний в зависимости от изменения ситуации, ни временные изменения самих факторов.

Учет всех этих обстоятельств требует перехода на следующий уровень структуризации информации, отображенной в когнитивной карте, т.е. к **когнитивной модели**.

На этом уровне каждая связь между факторами когнитивной карты раскрывается до соответствующего уравнения, которое может содержать как количественные (измеряемые) переменные, так и качественные (не измеряемые) переменные.

При этом количественные переменные входят естественным образом в виде их численных значений.

Каждой же качественной переменной ставится в соответствие совокупность лингвистических переменных, отображающих различные состояния этой качественной переменной (например, риски реализации атак на информационные ресурсы могут быть "слабыми", "умеренными", "высокими" и т.п.), а каждой лингвистической переменной соответствует определенный числовой эквивалент в шкале [0,1].

По мере накопления знаний о процессах, происходящих в исследуемой ситуации, становится возможным более детально раскрывать характер связей между факторами.

Формально, когнитивная модель ситуации может, как и когнитивная карта, быть представлена графом, однако каждая дуга в этом графе представляет уже некую функциональную зависимость между соответствующими базисными факторами, т.е. когнитивная модель ситуации представляется *функциональным графом*.

Опыт использования различных моделей и методов на базе когнитивного подхода (в России и за рубежом), повышающийся интерес управленцев-практиков к разработкам в данном направлении показывают целесообразность применения данного подхода в управлении.

### 3.2. Анализ влияний в когнитивных картах.

При анализе ситуаций, опирающемся на описанные выше модели когнитивных карт, решаются два типа задач: статические и динамические. Статический анализ - это анализ текущей ситуации, заключающийся в выделении и сопоставлении путей влияния одних факторов на другие через третьи (каузальных цепочек). Динамический анализ - это генерация и анализ возможных сценариев развития ситуации во времени. Математическим аппаратом анализа является теория знаковых и нечетких графов [68].

Задачи статического анализа, рассматриваемые в терминах знаковых графов, - это исследование влияний одних факторов на другие, исследование устойчивости ситуации в целом и поиск способов изменения графа для получения устойчивых структур.

Фактор $x_i$ влияет на фактор $x_j$, если существует ориентированный путь от вершины $x_i$ в вершину $x_j$. Суммарное влияние $x_i$ на $x_j$ положительно, если все пути от $x_i$ к $x_j$ положительны; отрицательно, если все пути отрицательны; неопределенно, если среди этих путей есть как положительные, так и отрицательные.

Одна из основных задач, решаемых в терминах знаковых графов - это задача об устойчивости. Анализ устойчивости графа предполагает поиск способов его изменения с целью получения устойчивой сбалансированной структуры.

В этой задаче ребра графа интерпретируются как некоторые отношения. Если отношения симметричны, то ситуация представляется неориентированным знаковым графом, вершины которого соответствуют субъектам отношений.

Неориентированный знаковый граф сбалансирован, если все его циклы положительны. Такая ситуация устойчива в том смысле, что ввиду однородности отношений нет предпосылок для ее изменения. Динамика развития ситуации при этом не рассматривается. Если ситуация устойчива, она будет таковой и в дальнейшем; если же она неустойчива, то характер связей скорее всего будет меняться.

Если отношения между факторами несимметричны, то когнитивная карта является ориентированным знаковым графом. При этом, как было сказано выше, контуры в когнитивной карте соответствуют морфогенетическими (положительным) и гомеостатическими (отрицательным) контурам обратной связи.

Положительный цикл при увеличении фактора в цикле ведет к его дальнейшему увеличению и, в конечном счете, неограниченному росту. Отрицательный цикл противодействует отклонениям от начального состояния, однако возможна неустойчивость в виде значительных колебаний, возникающих при прохождении возбуждения по циклу.

Различают случаи линейного, экспоненциального роста значений факторов, а также случай знакопеременного изменения и роста значений факторов (резонанса) [69].

Знаковые графы успешно используются для решения многих прикладных задач [63]. Однако наличие только двух видов оценок связей: +1 и -1 порождает два основных недостатка этой модели: отсутствие учета силы влияния по разным путям и отсутствие механизма разрешения неопределенностей при одновременном существовании положительных и отрицательных путей между двумя вершинами. Эти недостатки затрудняют выбор решений при управлении ситуацией.

Задача управления в когнитивных картах ставится следующим образом. Среди факторов ситуации выделяются управляющие факторы (факторы, на которые ЛПР имеет возможность непосредственно воздействовать) и целевые факторы, изменение или стабилизация которых является целью управления.

Конкретное управляющее решение (стратегия) - это выбор некоторого множества управляющих факторов.

В описанных выше интерпретациях знакового графа возможности сравнения и ранжирования решений весьма ограничены. Их можно сравнивать только по множеству целевых факторов, на которые они оказывают нужное (положительное или отрицательное) влияние. При совпадении этих множеств нет возможности сравнивать решения по силе влияния. Кроме того, довольно частое возникновение неопределенностей вообще не дает возможности оценить решение даже по знаку его влияния на целевые факторы.

В рамках знаковых графов эти трудности устраняет подход, развитый в работе [70]. Он оставляет граф знаковым, но предлагает гораздо более тонкий его анализ. Метод анализа влияний, предложенный в этой работе, основывается на следующих допущениях:

1. Сила влияния одного фактора на другой по данному пути зависит от длины этого пути (т.е. числа ребер в нем).
2. Чем больше параллельных влияний (по разным путям) существует между факторами, тем сильнее влияние между ними.

Пусть $P_{ij}^m$ и $N_{ij}^m$ - число положительных и отрицательных путей длины $m$, идущих от фактора $x_i$ к фактору $x_j$, соответственно. Тогда суммарные положительное и отрицательное влияния фактора $x_i$ на фактор $x_j$ определяются следующим образом:

– положительное влияние:

$$\bar{P}_{ij} = \sum_{m=1}^{\infty} f(m) P_{ij}^m$$

– отрицательное влияние:

$$\bar{N}_{ij} = \sum_{m=1}^{\infty} f(m) N_{ij}^m$$

где *f(m)* - монотонная неубывающая функция от длины пути *m*, определяющая степень ослабления влияния на пути от $x_i$ к $x_j$.

В качестве *f(m)* обычно выбирается монотонно неубывающая и дифференцируемая функция:

$$f(m) = \alpha m,$$

где $\alpha \in [0; 1]$ - коэффициент, определяющий степень ослабления.

С уменьшением $\alpha$ уменьшается влияние длинных путей на конечный результат; поэтому, изменяя $\alpha$, можно анализировать влияние путей разной длины.

Для сравнения различных стратегий рассматриваются различные варианты оценочной функции $V(s_{ij}, c_{ij})$, где $s_{ij}$ - суммарное влияние фактора *i* на фактор *j* и $c_{ij}$ - консонанс влияния фактора *i* на фактор *j*, которые определяются из следующих соотношений:

$$s_{ij} = \bar{P}_{ij} + \bar{N}_{ij};$$
$$c_{ij} = (\bar{P}_{ij} - \bar{N}_{ij})/(\bar{P}_{ij} + \bar{N}_{ij});$$

Консонанс $c_{ij}$ - это мера различия между положительным и отрицательным влиянием. Чем он больше, тем определеннее характер влияния.

Функция $V(s_{ij}, c_{ij})$ должна удовлетворять, в частности, следующим требованиям:

1. Пусть стратегия *i* характеризуется парой ($s_{ij}, c_{ij}$), а стратегия *i′* - парой ($s_{i′j}, c_{i′j}$). Тогда, если $V(s_{ij}, c_{ij}) > V(s_{i′j}, c_{i′j})$, то *i* предпочтительнее *i′*.

2. Если $c_{ij} = 0$, то $V(s_{ij}, c_{ij}) = 0$ при любых $s_{ij}$.

3. Если $c_{ij} > 0$, то $V(s_{ij}, c_{ij})$ монотонно возрастает по обеим переменным; если $c_{ij} < 0$, то $V(s_{ij}, c_{ij})$ монотонно убывает по обеим переменным.

При некоторых разумных допущениях целесообразно выбирать оценочную функцию в виде $V(s,c) = VS(s) \cdot VC(c)$.

Более детальные характеристики взаимодействия факторов можно выявить при использовании нечетких когнитивных карт.

Наиболее распространенный подход к вычислению нечетких влияний, предложенный в [71], заключается в следующем.

Пусть между $f_i$ и $f_j$ имеется $m$ путей и $I_r(f_i, f_j)$ обозначает влияние $f_i$ на $f_j$ по $r$-му пути, а $T(f_i, f_j)$ - суммарное влияние $f_i$ на $f_j$ по всем $m$ путям. Тогда

$$I_r(f_i, f_j) = \min_p w_{p,p+1}$$

$$T(f_i, f_j) = \max_{1 \leq r \leq m} I_r(f_i, f_j)$$

где $w_{p,p+1}$ - вес ориентированного ребра от $f_p$ к $f_{p+1}$ на $r$-м пути.

Таким образом, операция $I_r(f_i, f_j)$ выделяет наиболее слабую связь в $r$-м пути, а операция $T(f_i, f_j)$ выделяет наиболее сильную из связей $I_r(f_i, f_j)$.

Проиллюстрируем эти определения примером из [55]:

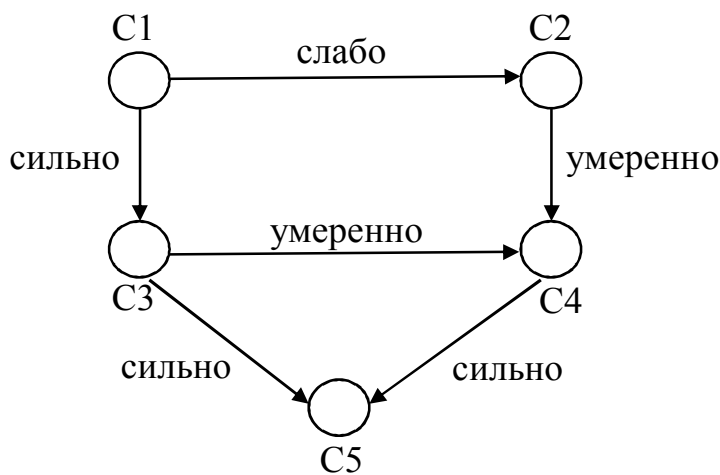

Рис.3.2. Нечеткий когнитивный граф влияний

Здесь веса ребер принимают значения из лингвистической шкалы {слабо, умеренно, сильно}. Рассмотрим влияния фактора С1 на фактор С5. В графе три каузальных пути от С1 к С5 ($r$ =1, 2, 3): $r$ =1 соответствует путь (С1, С3, С5), $r$ =2 - путь (С1, С3, С4, С5), $r$ =3 - путь (С1, С2, С4, С5). Три непрямых влияния С1 на С5 таковы:

$I_1$(С1, С5) = min{$w_{13}$, $w_{35}$} = min{сильно, сильно) = сильно,

$I_2$(С1, С5) = min{$w_{13}$, $w_{34}$, $w_{45}$} = min{сильно, умеренно, сильно) = умеренно,

$I_3$(C1, C5) = min{$w_{12}$, $w_{24}$, $w_{45}$) = min{слабо, умеренно, сильно) = слабо.

Отсюда суммарное влияние C1 на C5 равно:

$T$(C1, C5) = max{$I_1$(C1, C5), $I_2$(C1, C5), $I_3$(C1, C5)} = max{сильно, умеренно, слабо} = сильно.

В работе [71] предложена модифицированная модель влияний. В этой модели вершина может находиться в активном или пассивном состоянии; кроме того, каждой вершине приписан порог. Вершина переходит в активное состояние, только если сумма входных влияний достигает порога. Только в активном состоянии вершина передает влияние дальше.

*Задачи динамического анализа.*

В задачах динамического анализа нечеткие величины приписываются не только связям, но и факторам. При этом в отличие от весов связей, которые в процессе анализа считаются постоянными, величина, приписанная фактору $v_i$, - это значение некоторой функции $y_i(t)$, которое зависит от весов входящих ребер и значений факторов, входных для $v_i$, и меняется со временем.

Вектор $Y(t) = (y_1(t), y_2(t), …, y_n(t))$ значений всех факторов ситуации в момент $t$ образует состояние ситуации в момент $t$: $X(t)$. Совокупность весов ребер $w_{ij}$ задается матрицей смежности графа $W = \{w_{ij}\}$. Наличие у фактора величины позволяет не только оценить силу влияния на фактор, но и выразить результат суммарных влияний в виде конкретного значения.

Понятие состояния ситуации позволяет говорить о развитии ситуации во времени под действием различных внешних воздействий, выражающихся в изменении значений факторов, т.е. ставить задачу прогноза (прямая задача), а также исследовать возможности управления ситуацией, т.е. искать воздействия, приводящие к нужному (целевому) состоянию (обратная задача).

В общем случае функции, приписанные разным факторам, различны, что приводит к структуре, сходной с НКК, управляемой правилами, описанной выше. Вычислительная сложность анализа такой НКК весьма велика [68].

Рассмотрим более простой, но достаточно часто встречающийся во многих прикладных задачах случай, когда эти функции, во-первых, для всех факторов одинаковы, а во-вторых, зависят не от значений входных факторов, а от их приращений (т.е. величина приращения в любом состоянии влияет одинаково).

Будем считать, что область значений каждого фактора $v_i$ - это линейно упорядоченное множество (шкала) лингвистических значений

$$Z_i=\{z_{i1}, z_{i2}, \ldots, z_{iq}(i)\},$$

где $z_{i1}$ и $z_{iq}(i)$ - минимальный и максимальный элементы множества, соответственно, из $k < l$ следует $z_{ik} < z_{il}$.

Мощности шкал $q(i)$ для разных факторов различны. Приращение значения фактора характеризуется знаком приращения и его величиной. Для текущего значения $y_i(t) = z_{im}$ переход к элементам $Z_i$: $\{z_{i(m+1)}, z_{i(m+2)}, \ldots, z_{iq(i)}\}$ дает положительное приращение $\widetilde{P}^+$; отрицательное приращение $\widetilde{P}^-$ получается при переходе к $\{z_{i(m-1)}, z_{i(m-2)}, \ldots, z_{i1}\}$.

В общем случае при переходе от $z_{im}$ к $z_{il}$ фактор получает приращение: $\widetilde{P}^{sign(l-m)}$. Очевидно, что при $m < l$ приращение будет положительным, а при $m > l$ - отрицательным.

Для удобства вычислений при анализе ситуаций определим отображение Е: $Z_i \rightarrow [0, 1]$ дискретной лингвистической шкалы $Z_i = \{z_{i1}, z_{i2},\ldots, z_{ir}\}$ на отрезок [0, 1] следующим образом. Разобьем отрезок [0, 1] на на $r$ равных отрезков, границы которых обозначим в порядке возрастания $b_0 = 0, b_1, \ldots, b_{r-1}, b_r = 1$. Положим $(z_{ik}) = (b_{k-1} - b_k)/2$ (элемент $z_{ik}$ отображается в центр $k$-го отрезка). Такое отображение позволяет алгоритмы нечеткой когнитивной модели сделать числовыми.

Обратное отображение $E^{-1}$: $[0, 1] \rightarrow Z_i$ является гомоморфизмом: все точки, лежащие в интервале $(b_{k-1}, b_k)$, отобразятся в одну точку $z_{ik}$.

С помощью прямого отображения состояние ситуации представляется в числовом виде, и дальнейшие вычисления производятся с числовым представлением состояния ситуации $X(t)$. Обратное отображение используется только для качественных интерпретаций результатов анализа.

*Получение прогноза развития ситуации (прямая задача).*

Данная задача формулируется следующим образом [68].

Заданы:

- когнитивная карта $G(V, W)$, где $V$ - множество вершин (факторов ситуации), $W$ - матрица смежности;

- множество $\{Z_1, …, Z_n\}$ шкал всех факторов ситуации;

- начальное состояние ситуации $X(0)=(x_1(0), …, x_n(0))$;

- начальный вектор приращений факторов ситуации $\boldsymbol{P}(0)=(p_1(0),…, p_n(0))$.

Необходимо найти состояния ситуации $X(1), …, X(n)$ и векторы приращений $\boldsymbol{P}(1), …, \boldsymbol{P}(n)$ в последовательные дискретные моменты времени $\{1, …, n\}$, где $n=\|V\|$ для того, чтобы влияние исходного возмущения могло достичь всех вершин.

Прогноз развития ситуации определяется с помощью матричного соотношения:

$$\boldsymbol{P}(t+1)=\boldsymbol{P}(t) \circ W$$

где $\circ$ - правило max-product:

$$p_i(t+1) = \max_{j}{}_{mod}(p_j(t)\cdot w_{ji})$$

Таким образом, приращение $p_i(t+1)$ - это максимальная по модулю величина из $p_j(t)\cdot w_{ji}$, где максимум берется по всем факторам, входным для фактора $v_i$ (для остальных факторов $w_{ji}=0$).

Схема работы с моделью выглядит следующим образом. Эксперт задает в лингвистических значениях начальное состояние ситуации $X(0)$ и следующее состояние $X(1)$, возникающее после воздействия на концепты когнитивной карты. По этим данным после представления состояния ситуации в числовом виде вычисляется числовое значение начального приращения:

$$P(0) = X(1) - X(0) = (Y(1)) - (Y(0)).$$

Последующие вычисления являются числовыми: с помощью операции max-product вычисляются приращения в последовательные моменты $t = 1,…, n$, а состояние ситуации определяется из соотношения:

$$X(t+1) = X(t)+P(t).$$

Для интерпретации прогнозов и выдачи результатов ЛПР производится обратное отображение Е$^{-1}$ числовых величин в лингвистические.

При получении прогноза наряду с вычислением вектора $P(t+1)$ вычисляется вектор $C = \{c_1(t+1), …, c_n(t+1)\}$. Величина $c_i(t+1)$ - это консонанс фактора $v_i$. Он имеет тот же смысл, что был определен выше, но меняется со временем и определяется следующим образом.

Обозначим через $P_i^+(t+1)$ максимум положительных приращений, поступающих на вход фактора $v_i$: $P_i^+(t+1) = \{(p_j(t) \cdot w_{ji}) \mid p_j(t) \cdot w_{ji} > 0\}$.

Аналогично, $P_i^-(t+1)$ - максимум абсолютных величин отрицательных приращений, поступающих на вход фактора $v_i$: $P_i^-(t+1) = \{(p_j(t) \cdot w_{ji}) \mid p_j(t) \cdot w_{ji} < 0\}$.

Тогда $c_i(t+1) = |P_i^+(t+1) - P_i^-(t+1)| / (P_i^+(t+1) + P_i^-(t+1))$

Консонанс $c_i(t+1)$ характеризует степень уверенности в прогнозе на момент $t+1$. При $c_i(t+1) \approx 1$, т.е. $P^+(t+1) >> |P^-(t+1)|$ или $|P^-(t+1)| >> P^+(t+1)$ уверенность субъекта в значении приращения $i$-го фактора $P_i(t+1)$ максимальна, а при $c_i(t+1) \approx 0$, т.е. при $P^+(t+1) \approx |P^-(t+1)|$, минимальна. Интервалы значений консонанса могут иметь лингвистическую интерпретацию типа «Невозможно», «Возможно», «Достоверно» и т.д.

Таким образом, правдоподобный прогноз развития ситуации к моменту $t+1$ определяется парой: $X(t+1)$, $C(t+1)$, где $X(t+1)$ - вектор значений факторов ситуации в момент $t+1$, $C(t+1)$ - вектор консонанса в момент $t+1$.

Решение прямой задачи должно учитывать два существенных момента:
1. При анализе нечетких ситуаций нечеткими являются и интервалы времени, поскольку время реализации влияния одних факторов на другие точно неизвестно и оценивается довольно грубо. Моменты времени ($t+i$) должны пониматься не как точки на абсолютной временной шкале, а как линейно упорядоченные во времени промежуточные шаги прогноза. Целевое состояние $X(t+n)$ не вычисляется итеративно по формуле

$$X(t+i) = X(t+i-1) + P(t+i), i = 1, …, n,$$

а является результатом обобщенной качественной оценки всего прогнозируемого развития ситуации от *t* до *t+n*. В алгоритмах используются нечеткие матричные операции, называемые в нечеткой математике композициями [72-73]: max-product (роль сложения играет взятие максимума, умножение - обычное) или max-min (роль умножения играет взятие минимума).

2. При вычислении приращений и состояний ситуации в последовательные моменты времени {*t*, *t*+1,…, *t+n*} приходится вычислять не только следующее значение приращения, но и степень уверенности его выбора (консонанс). Поэтому при выборе положительного (или отрицательного) приращения необходимо сохранять и отвергнутое отрицательное (или положительное) приращение.

*Задача нахождения управляющих воздействий (обратная задача).* Требуется найти управляющие воздействия, которые дают требуемое приращение значений факторов ситуации. В отличие от прямой, в формулировке обратной задачи моменты времени не участвуют. Это объясняется тем, что неважно, на каком шаге требуемое приращение будет достигнуто.

При поиске решения просматриваются пути распространения влияний, имеющие различную длину. Для этого используется нечеткое транзитивное замыкание матрицы смежности $W$: $W' = \{w'_{ij} = (w_{ij}, (w_{ij})^2, …, (w_{ij})^n)\}$, где элемент $(w_{ij})^k$ матрицы $W'$ ($k = 1, 2, …, n$) определяется из соотношения:

$$(w_{ij})^k = (w_{il} \cdot (w_{lj})^{k-1}).$$

Очевидный смысл транзитивного замыкания состоит в описании влияния факторов друг на друга не только непосредственно, но и через промежуточные факторы. Например, фактор $v_j$ может влиять на фактор $v_i$ не только напрямую, но и посредством влияния на фактор $v_l$.

*Постановка обратной задачи:*

Заданы причинно-следственные связи между факторами в виде матрицы транзитивного замыкания $W'$ и целевой вектор требуемых приращений значений факторов ситуации $\bar{\bar{P}} = (\bar{\bar{P}}_1, \bar{\bar{P}}_2, ..., \bar{\bar{P}}_n)$. Найти множества векторов входных воздействий $\{U\}$, такие, что для всех $U$ выполняется равенство

$$U \cdot W' = \bar{\bar{P}}.$$

Методы решения данного нечеткого реляционного уравнения описаны в работах [74-75].

В [76-77] показано, что обратная задача для нечетких реляционных уравнений типа max-product эквивалентна задаче покрытия и поэтому является NP-трудной задачей. Там же приведены приемы, позволяющие уменьшать число переменных в уравнении без потери множества решений, что позволяет понизить трудоемкость решения конкретных задач.

Специфика когнитивного моделирования заключается в том, что формальные математические методы анализа применяются к моделям, описывающим субъективное видение ситуации. На каждом этапе формирования модели приходится принимать решения, от совокупности которых, в конечном счете, зависит адекватность построенной модели.

Предлагаемый в литературе набор моделей и методов анализа влияний в слабоструктурированных ситуациях сам по себе не гарантирует построения адекватной модели. Адекватность окончательно выясняется только в процессе реальной работы с моделью. Отсюда, в частности, следует, что информационные технологии поддержки принятия решений, основанные на аппарате когнитивных карт, должны быть максимально открытыми для модификаций.

Нужно отметить, что уже сам процесс построения модели оказывается весьма полезным для аналитиков еще до начала расчетов, поскольку он заставляет структурировать проблемную область. При формальном выделении факторов и связей между ними неизбежно выявляются ранее неучтенные аспекты ситуации, связи, казавшиеся несущественными, и формируется система понятий, в терминах которой даже неформальное обсуждение проблемы становится более четким и обоснованным.

Главным достоинством аппарата когнитивных карт является возможность систематического качественного (в смысле - неколичественного) учета отдаленных последствий принимаемых решений и выявления побочных эффектов, которые могут помешать реализации, казалось бы, очевидных решений и которые трудно оценить интуитивно при большом числе факторов и многообразии многочисленных путей взаимодействия между ними.

Можно утверждать, что данный подход является математической основой для интеллектуальных информационных технологий поддержки принятия решений в слабоструктурированных предметных областях [68].

В последние годы появляется активный спрос на информационные технологии, использующие этот подход, со стороны управляющих органов различного уровня. Поэтому работы в данном направлении являются перспективными как в теоретическом, так и в прикладном плане.

### 3.3. Основы теории нечетких множеств

Задачи, стоящие перед человеком в различных областях знаний являются по своей природе слишком сложными и многогранными для того, чтобы использовать для их решения только точные, строго определенные модели и алгоритмы.

Многие понятия, вследствие человеческого мышления, приближенного характера умозаключений, лингвистического их описания являются нечеткими по своей природе.

Существующие методы решения задач управления в условиях неопределенности, как правило, учитывают только достаточно малые изменения коэффициентов целевой функции и системы ограничений модели, и практически не позволяют учесть вариации структуры модели.

Получение оценок, необходимых при решении задач управления, с помощью только теоретико-вероятностного подхода возможно лишь при наличии информации, достаточной для построения необходимых частотных распределений, что в задачах обеспечения безопасности встречается крайне редко.

В таких ситуациях используются экспертные оценки и другая информация, для которой характерны неопределенности субъективной природы. Источником субъективной неопределенности служит также многокритериальность, внутренне присущая оценкам ИБ. В этих условиях решение задач оценки и управления безопасностью требует использования соответствующего математического аппарата.

Сегодня одним из наиболее перспективных направлений научных исследований в области анализа, прогнозирования и моделирования слабоструктурированных и плохо формализуемых явлений и процессов является нечеткая логика (fuzzy logic), сформулированная и получившая развитие в работах Л. Заде.

Она значительно расширяет возможности учета неопределенностей различной природы, неизбежно сопутствующих математическому описанию реальности. Такой подход позволяет решать задачи совершенствования функционирования различных систем в условиях неполноты и неточности информации о протекающих процессах, недостаточности и недостоверности знаний, при наличии субъективности оценок.

В отличие от традиционной математики, требующей на каждом шаге моделирования точных и однозначных формулировок закономерностей, нечеткая логика предлагает совершенно иной уровень мышления, благодаря которому творческий процесс моделирования происходит на наивысшем уровне, при котором постулируется лишь минимальный набор закономерностей.

В области ИБ эта проблема весьма актуальна. С одной стороны, точность и оптимальность принятия решений – это залог успешной стратегии обеспечения безопасности, которая позволяет добиваться ее наибольшей эффективности.

С другой стороны, важной особенностью оценки и управления ИБ являются имеющиеся факторы случайности, неточности. Поэтому, математическая модель должна строиться не только с точки зрения наиболее адекватного отражения сущности моделируемых процессов и явлений, но и с учетом условий неопределенности.

В традиционных подходах к управлению все неопределенности естественных процессов трактуются в вероятностном смысле, однако на практике это не всегда соответствует природе неопределенностей, часто представляющих собой следствия субъективных оценок. Кроме того, частотные распределения событий в задачах обеспечения безопасности, как правило, не известны.

Сильными сторонами применения математического подхода, основанного на нечетких множествах и нечетких логиках, являются: описание условий и метода решения задачи на языке, близком к естественному, универсальность и эффективность.

Вместе с тем, имеются характерные недостатки: исходный набор постулируемых нечетких правил формируется экспертом и может оказаться неполным или противоречивым; вид и параметры функции принадлежности, описывающие входные и выходные переменные системы, выбираются субъективно, и могут оказаться недостаточно адекватно отражающими реальную действительность.

Важно отметить, что термин фаззи (fuzzy) (и особенно в укоренившемся русском переводе "нечеткий") вызывает настороженность у лиц, принимающих решения при выборе методов реализации того или иного проекта. Понятие "нечеткость" в этом случае воспринимается как неоднозначность или даже ненадежность функционирования будущей системы.

На самом деле компьютерные модели на основе нечеткой математики абсолютно точны и однозначны по отношению к конкретной ситуации на входе модели. Их замечательным свойством является способность обрабатывать разнородную по качеству входную информацию, в целом повышая достоверность описания поведения объекта. Иными словами нечеткие системы отражают на выходе суммарную степень размытости, неполноты и неточности входных данных, тем не менее, предлагая единственное для данной конкретной ситуации решение.

Поскольку аппарат нечеткой математики является одним из основных методов анализа при моделировании слабоструктурированных и плохо формализуемых процессов (к которым относятся и задачи обеспечения информационной безопасности), рассмотрим основные положения и понятия теории нечетких множеств.

Математическая теория нечетких множеств (fuzzy sets) и нечеткая логика (fuzzy logic) являются обобщениями классической теории множеств и классической формальной логики. Данные понятия были впервые предложены американским ученым Лотфи Заде (Lotfi Zadeh) в 1965г [78].

Основной причиной появления новой теории стало наличие нечетких и приближенных рассуждений при описании человеком процессов, систем, объектов.

***Нечеткая и лингвистическая переменные.***

Характеристикой нечеткого множества выступает функция принадлежности (Membership Function). Обозначим через $MF_c(x)$ – степень принадлежности к нечеткому множеству $C$. $MF_c(x)$ представляет собой обобщение понятия характеристической функции обычного множества. Тогда нечетким множеством $C$ называется множество упорядоченных пар вида $C=\{MF_c(x)/x\}$, $MF_c(x) \in [0,1]$. Значение $MF_c(x)=0$ означает отсутствие принадлежности к множеству, 1 – полную принадлежность.

Проиллюстрируем это на простом примере. Формализуем неточное определение "горячий чай". В качестве $x$ (область рассуждений) будет выступать шкала температуры в градусах Цельсия. Очевидно, что она изменяется от 0 до 100 градусов. Нечеткое множество для понятия "горячий чай" может выглядеть следующим образом:

C={0/0; 0/10; 0/20; 0,15/30; 0,30/40; 0,60/50; 0,80/60; 0,90/70; 1/80; 1/90; 1/100}.

Так, чай с температурой 60°C принадлежит к множеству "Горячий" со степенью принадлежности 0,80. Для одного человека чай при температуре 60°C может оказаться горячим, для другого – не слишком горячим. Именно в этом и проявляется нечеткость задания соответствующего множества.

Для нечетких множеств, как и для обычных, определены основные логические операции. Самыми основными, необходимыми для расчетов, являются пересечение и объединение.

Пересечение двух нечетких множеств (нечеткое "И") $A \cap B$:

$$MF_{AB}(x) = min(MF_A(x), MF_B(x)).$$

Объединение двух нечетких множеств (нечеткое "ИЛИ") $A \cup B$:

$$MF_{AB}(x) = max(MF_A(x), MF_B(x)).$$

В теории нечетких множеств разработан общий подход к выполнению операторов пересечения, объединения и дополнения, реализованный в так называемых треугольных нормах и конормах. Приведенные выше реализации операций пересечения и объединения – наиболее распространенные случаи $t$-нормы и $t$-конормы.

Для описания нечетких множеств вводятся понятия нечеткой и лингвистической переменных.

Нечеткая переменная описывается набором ($N, X, A$), где $N$ – это название переменной, $X$ – универсальное множество (область рассуждений), $A$ – нечеткое множество на $X$.

Значениями лингвистической переменной могут быть нечеткие переменные, т.е. лингвистическая переменная является более общим понятием, чем нечеткая переменная. Каждая лингвистическая переменная состоит из:

- названия;

- множества своих значений, которое также называется базовым терм-множеством *T*. Элементы базового терм-множества представляют собой названия нечетких переменных. Количество термов в лингвистической переменной редко превышает 7;
- универсального множества *X*;
- синтаксического правила *G*, по которому генерируются новые термы с применением слов естественного или формального языка;
- семантического правила *P*, которое каждому значению лингвистической переменной ставит в соответствие нечеткое подмножество множества *X*.

Рассмотрим такое нечеткое понятие как "Состояние безопасности информационной системы". Это и есть название лингвистической переменной. Сформируем для нее базовое терм-множество, которое может, например, состоять из трех нечетких переменных: "Низкое", "Среднее", "Высокое" и зададим область рассуждений в виде Х=[0;100] (процент защищенных информационных активов от их общего количества). Последнее, что необходимо сделать – это построить функции принадлежности для каждого лингвистического терма из базового терм-множества *T*.

Существует свыше десятка типовых форм кривых для задания функций принадлежности. Наибольшее распространение получили: треугольная, трапецеидальная и гауссова функции принадлежности.

Треугольная функция принадлежности определяется тройкой чисел (*a,b,c*), и ее значение в точке *x* вычисляется согласно выражению:

$$MF(x) = \begin{cases} 1 - \dfrac{b-x}{b-a}, a \leq x \leq b \\ 1 - \dfrac{x-b}{c-b}, b \leq x \leq c \\ 0, \text{в остальных случаях} \end{cases}$$

При (*b-a*)=(*c-b*) имеем случай симметричной треугольной функции принадлежности, которая может быть однозначно задана двумя параметрами из тройки (*a,b,c*).

Аналогично для задания трапецеидальной функции принадлежности необходима четверка чисел (*a,b,c,d*):

$$MF(x) = \begin{cases} 1 - \dfrac{b-x}{b-a}, a \leq x \leq b \\ 1, a \leq x \leq b \\ 1 - \dfrac{x-c}{d-c}, c \leq x \leq d \\ 0, \text{в остальных случаях} \end{cases}$$

При (*b-a*)=(*d-c*) трапецеидальная функция принадлежности принимает симметричный вид (рис.3.3.).

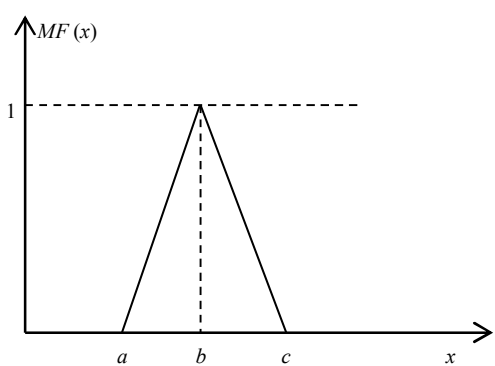 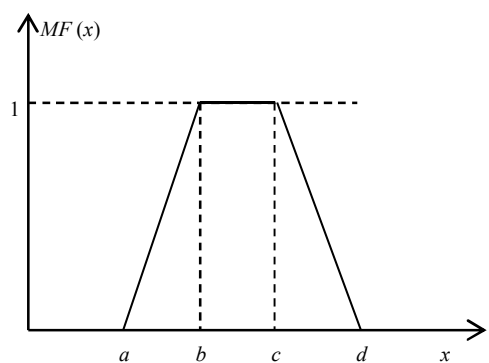

3.3а. 3.3б.

Рис.3.3. Типовые кусочно-линейные функции принадлежности (3.3а – треугольная; 3.3б – трапецеидальная).

Функция принадлежности гауссова типа описывается формулой

$$MF(x) = exp\left[-\left(\dfrac{x-c}{\sigma}\right)^2\right]$$

и оперирует двумя параметрами. Параметр *c* обозначает центр нечеткого множества, а параметр $\sigma$ отвечает за крутизну функции (рис.3.4).

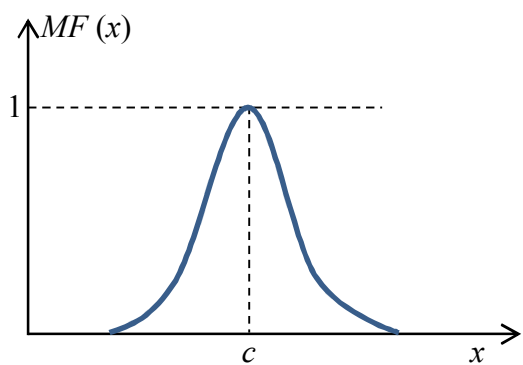

Рис.3.4. Гауссова функция принадлежности.

Совокупность функций принадлежности для каждого терма из базового терм-множества *T* обычно изображаются вместе на одном графике. На рис.3.5. приведен пример представления терм-множества для описанной выше лингвистической переменной "Состояние безопасности информационной системы".

Из рисунка видно, что если, например, защищены около 30% информационных активов, то степень принадлежности лингвистической переменной к множеству "Низкая" будет равна примерно 0.2, "Средняя" – 0,6, "Высокая" – 0.

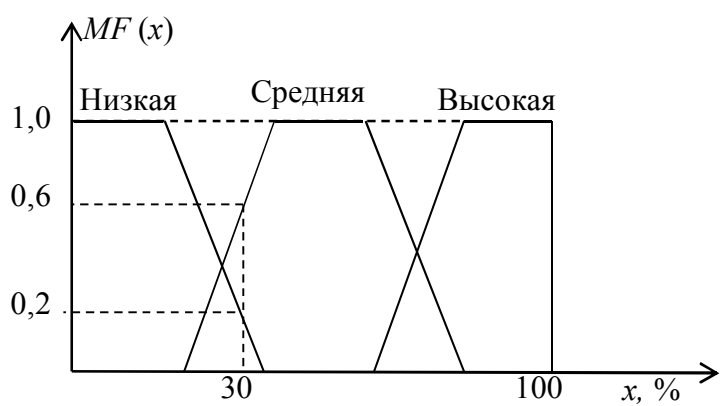

Рис.3.5. Описание лингвистической переменной

"Состояние безопасности информационной системы".

*Нечеткий классификатор*

Определим в качестве носителя лингвистической переменной отрезок вещественной оси [0,1]. Любые конечномерные отрезки вещественной оси могут быть сведены к отрезку [0,1] путем простого линейного преобразования, поэтому выделенный отрезок единичной длины носит универсальный характер и называется 01- носителем [79].

Введем лингвистическую переменную «Уровень показателя» с терм-множеством значений «Низкий (Н), Ниже среднего (НС), Средний(С), Выше среднего (ВС), Высокий (В)». Для описания подмножеств терм-множества введем систему из пяти соответствующих функций принадлежности трапецеидального вида (через $\mu_i(x)$ обозначены соответствующие функции принадлежности $MF_i(x)$):

$$\text{H: } \mu_1(x) = \begin{cases} 1, & 0 \leq x < 0.15 \\ 10(0.25 - x), & 0.15 \leq x < 0.25 \\ 0, & 0.25 \leq x \leq 1 \end{cases} \quad (3.1)$$

$$\text{HC: } \mu_2(x) = \begin{cases} 0, & 0 \leq x < 0.15 \\ 10(x - 0.15), & 0.15 \leq x < 0.25 \\ 1, & 0.25 \leq x < 0.35 \\ 10(0.45 - x), & 0.35 \leq x < 0.45 \\ 0, & 0.45 \leq x \leq 1 \end{cases} \quad (3.2)$$

$$\text{C: } \mu_3(x) = \begin{cases} 0, & 0 \leq x < 0.35 \\ 10(x - 0.35), & 0.35 \leq x < 0.45 \\ 1, & 0.45 \leq x < 0.55 \\ 10(0.65 - x), & 0.55 \leq x < 0.65 \\ 0, & 0.65 \leq x \leq 1 \end{cases} \quad (3.3)$$

$$\text{BC: } \mu_4(x) = \begin{cases} 0, & 0 \leq x < 0.55 \\ 10(x - 0.55), & 0.55 \leq x < 0.65 \\ 1, & 0.65 \leq x < 0.75 \\ 10(0.85 - x), & 0.75 \leq x < 0.85 \\ 0, & 0.85 \leq x \leq 1 \end{cases} \quad (3.4)$$

$$\text{B: } \mu_5(x) = \begin{cases} 0, & 0 \leq x < 0.75 \\ 10(x - 0.75), & 0.75 \leq x < 0.85 \\ 1, & 0.85 \leq x \leq 1 \end{cases} \quad (3.5)$$

Везде в (3.*) $x$ – это 01–носитель (отрезок [0,1] вещественной оси). Построенные функции принадлежности приведены на рис. 3.6.

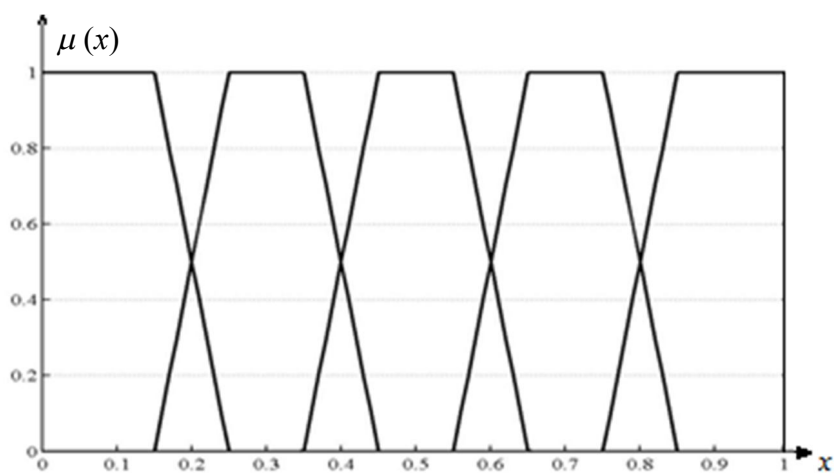

Рис. 3.6. Система трапецеидальных функций принадлежности

на 01-носителе

Введем также набор так называемых узловых точек $a_j$ = (0.1; 0.3; 0.5; 0.7; 0.9), которые являются, с одной стороны, абсциссами максимумов соответствующих функций принадлежности на 01-носителе, а, с другой стороны, равномерно отстоят друг от друга на 01-носителе и симметричны относительно узла 0.5.

Введенную лингвистическую переменную «Уровень фактора», определенную на 01-носителе, в совокупности с набором узловых точек называют **стандартным пятиуровневым нечетким 01-классификатором** [79].

Суть сконструированного нечеткого классификатора в том, что если о факторе неизвестно ничего, кроме того, что он может принимать любые значения в пределах 01-носителя (принцип равнопредпочтительности), а надо провести ассоциацию между качественной и количественной оценками фактора, то предложенный классификатор делает это с максимальной достоверностью. При этом сумма всех функций принадлежности для любого $x$ равна единице, что указывает на непротиворечивость данного классификатора.

Таким образом, стандартный классификатор осуществляет проекцию нечеткого лингвистического описания на 01-носитель, при этом делает это непротиворечивым способом, симметрично располагая узлы классификации (0.1, 0.3, 0.5, 0.7, 0.9).

В этих узлах значение соответствующей функции принадлежности равно единице, а всех остальных функций – нулю. Неуверенность эксперта в классификации убывает (возрастает) линейно с удалением от узла (с приближением к узлу, соответственно). При этом сумма функций принадлежности во всех точках носителя равна единице.

Построенный классификатор есть разновидность так называемой «серой» шкалы Поспелова [80], представляющей собой полярную (оппозиционную) шкалу, в которой переход от свойства $A^+$ к свойству $A^-$ (например, от свойства «большой дом» к свойству «дом среднего размера» лингвистической переменной «Размер дома») происходит плавно, постепенно. Подобные шкалы удовлетворяют условиям: а) взаимной компенсации между свойствами $A^+$ и $A^-$ (чем в большей степени проявляется $A^+$, тем в меньшей степени проявляется $A^-$, и наоборот); б) наличия нейтральной точки $A^0$, интерпретируемой как точка наибольшего противоречия, в которой оба свойства присутствуют в равной степени (например, когда дом кажется одновременно и большим, и средним по размерам). В случае нашего нечеткого классификатора абсциссы нейтральных точек: (0.2, 0.4, 0.6, 0.8).

Таким образом, мы переходим от качественного описания уровня параметра к стандартному количественному виду соответствующей функции принадлежности (нечеткое трапециевидное число).

### 3.4. Вычисления с нечеткими числами

Возможность успешного применения подходов, основанных на нечеткости, во многом определяется гибким математическим аппаратом, используемым при анализе и обработке данных, способным адекватно отразить не только не подлежащие строгой формализации зависимости и взаимосвязи, но и учесть неточные, субъективные оценки специалистов, лежащие в их основе.

Среди очевидных преимуществ нечетких систем можно выделить следующие [81]:

- возможность оперировать нечеткими входными данными: например, непрерывно изменяющиеся во времени значения (динамические задачи), значения, которые невозможно задать однозначно (результаты статистических опросов, рекламные компании и т.д.);
- возможность нечеткой формализации критериев оценки и сравнения: оперирование критериями "большинство", "возможно", "преимущественно";
- возможность проведения качественных оценок, как входных данных, так и выходных результатов, т.е. возможность оперировать не только значениями данных, но и степенью их достоверности и ее распределением;
- возможность проведения быстрого моделирования сложных динамических систем и их сравнительный анализ с заданной степенью точности.

В связи с возрастающей потребностью применения теории нечетких множеств для решения различных прикладных задач возникла необходимость автоматизации процесса работы с основным элементом данной теории – нечеткими числами.

Определение *нечеткого числа* вводится на основе введенного выше понятия *нечеткого множества*. Нечетким числом называется нечеткое подмножество (совокупность пар вида $A = \{x, \mu_A(x)\}$, где $x \in X$, а $\mu_A(x)$ - функция принадлежности, ставящая в соответствие множеству $X$ отрезок [0,1]) универсального множества действительных чисел $R$, функция принадлежности $\mu$ которого удовлетворяет условиям:

- непрерывности;
- нормальности: $sup\{\mu(x)\} = 1, x \in R$;
- выпуклости: $\mu(x_j) \geq min\{\mu(x_i), \mu(x_k)\}$, $x_i \leq x_j \leq x_k$.

Как было сказано выше, в зависимости от вида функции принадлежности принято различать треугольные, трапециевидные и колоколообразные нечеткие числа (рис.3.3 и 3.4).

Часто при решении слабо формализуемых задач в качестве семейства функций принадлежности выступает стандартный 01-классификатор. Наиболее часто используются 3-х, 5-ти и 7-ми уровневые классификаторы.

Примеры 3-х и 5-ти уровневого классификаторов на основе трапециевидных чисел были приведены на рис.3.5-3.6.

Рассмотрим алгоритм выполнения вычислений с нечеткими числами на примере трапециевидных чисел.

Трапециевидным нечетким числом с интервалом толерантности [$a$, $b$], левой границей $\gamma > 0$ и правой границей $\delta > 0$ называется нечеткое множество, функция принадлежности которого имеет вид:

$$\mu(x) = \begin{cases} 1 - \dfrac{a-x}{\gamma}, & a - \gamma \leq x \leq a \\ 1, & a \leq x \leq b \\ 1 - \dfrac{x-b}{\delta}, & b \leq x \leq b + \delta \\ 0, & \text{иначе} \end{cases}$$

Будем обозначать трапециевидные нечеткие числа как ($\gamma$, $a$, $b$, $\delta$).

Существует два основных способа введения операций над нечеткими числами: с использованием $\alpha$ – сечений или при помощи расширения понятия арифметических операций на нечеткие числа, предложенного Заде [78]. Первый из них является более наглядным и простым для использования.

$\alpha$-сечением ($\alpha$-уровнем, срезом) $[A]^\alpha$ нечеткого числа $A$ называется множество

$$[A]^\alpha = \begin{cases} \{t \in R | \mu(t) \geq \alpha\}, & \alpha > 0 \\ \overline{\sup(A)}, & \alpha = 0 \end{cases},$$

где $\overline{sup(A)}$ - замыкание носителя нечеткого подмножества, т.е. при $\overline{sup(A)}$ = ($a$, $b$) нулевым уровнем будет [$a$, $b$].

Понятие $\alpha$-уровня соответствует понятию интервала достоверности. Количество $\alpha$-уровней выбирают так, чтобы обеспечить необходимую точность вычислений.

Операции над нечеткими числами осуществляются последовательно уровень за уровнем, аналогично выполнению операций над этими интервалами:

- Сложение $[a, b] (+) [c, d] = [a + c, b + d]$;
- Умножение $[a, b] (*) [c, d] = [a * c, b * d]$;
- Вычитание $[a, b] (-) [c, d] = [a - c, b - d]$;
- Деление $[a, b] (/) [c, d] = [a / c, b / d]$;
- Умножение на коэффициент $[a, b] (*) k = [a * k, b * k]$.

Возведение нечеткого числа $\tilde{A}$ в десятичную степень $p$ производится следующим образом:

$$\tilde{A} = \mu^p(x) = \begin{cases} 0, 0 \leq x < a - \gamma \\ 10(x^{1/p} - (a - \gamma)), a - \gamma \leq x < a \\ 1, a \leq x < b \\ 10((b + \delta) - x^{1/p}), b \leq x < b + \delta \\ 0, b + \delta \leq x \leq 1 \end{cases}$$

Здесь $a, b, c$ и $d$ зависят от $\alpha$: $a$ и $c$ – это левые точки пересечения прямой $\mu(x) = \alpha$ и соответствующих графиков функций принадлежности $\mu(x)$ каждого из чисел, участвующих в операции; $b$, и $d$ – правые точки пересечения прямой и графиков принадлежности.

Пример.

Пусть заданы два трапециевидных числа $\tilde{A}(1, 2, 3, 4)$ и $\tilde{B}(2, 3, 4, 8)$. Необходимо найти нечеткое число $\tilde{C} = \tilde{A} + \tilde{B}$. Пусть $\alpha$-уровни заданы следующим образом: $\alpha_1 = 0$; $\alpha_2 = 0,5$ и $\alpha_3 = 1$. Результат нахождения $\tilde{C}$ показан на рис.3.7.

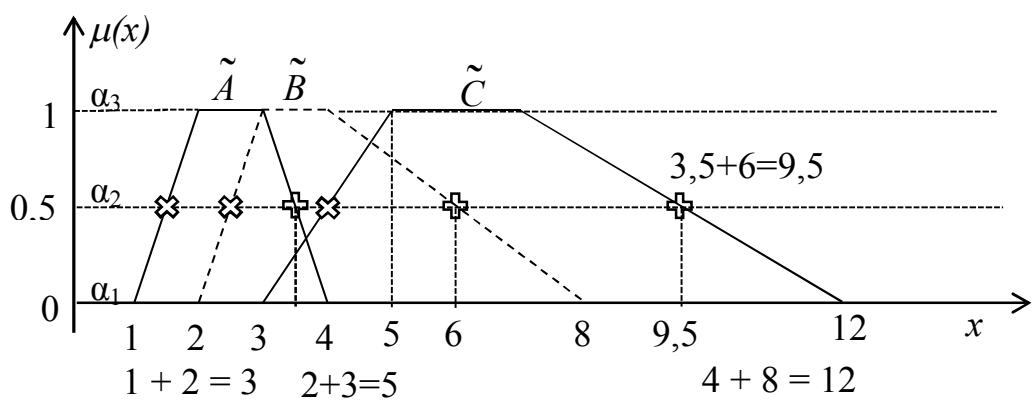

Рис.3.7. Пример нахождения суммы двух нечетких трапециевидных чисел

Изложенные выше теоретические сведения были положены в основу разработки в среде Visual Studio 2008 на языке C# программного продукта «Вычисления с нечеткими числами» [82-83], главное окно которого представлено на рис.3.8.

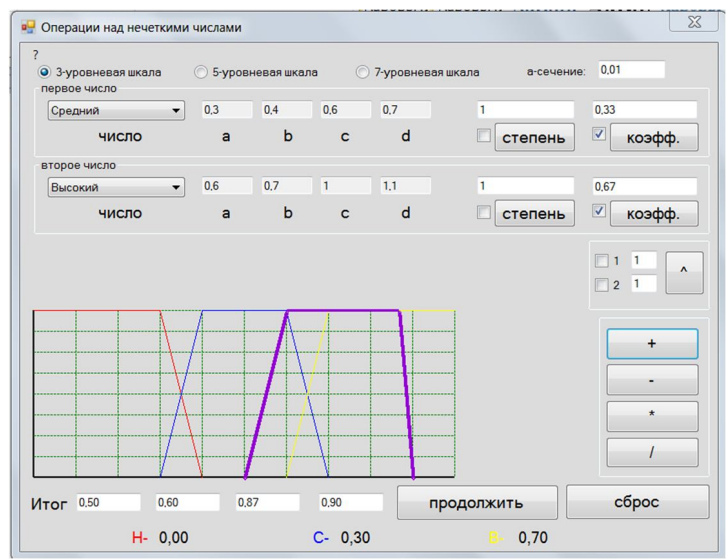

Рис.3.8. Главное окно программы «Вычисления с нечеткими числами»

В качестве примера вычислено значение нечеткого числа, представляющего собой сумму двух нечетких чисел соответствующих лингвистическим переменным «средний» и «высокий» в трехуровневой шкале оценок. Суммирование произведено с учетом весов: фактор, имеющий оценку «средний» имеет вес 1/3, фактор, лингвистическое значение которого оценено как «высокий» имеет вес 2/3. Полученное значение распознано как «высокий» с индексом схожести 0,7 (более подробно вопрос о порядке определения индекса схожести будет рассмотрен в следующей главе).

Разработанный программный продукт позволяет производить все основные арифметические действия (сложение, умножение, вычитание, деление, возведение в степень, умножение на коэффициент, растяжение и сжатие) над нормированными нечеткими числами и определять индекс схожести полученного результата с «эталонными» значениями выбранного классификатора.

## 3.5. Нечеткий логический вывод

Основой для проведения операции нечеткого логического вывода является база правил, содержащая нечеткие высказывания в форме "Если-то" и функции принадлежности для соответствующих лингвистических термов. При этом должны соблюдаться следующие условия (в противном случае база нечетких правил будет неполной):

1. Существует хотя бы одно правило для каждого лингвистического терма выходной переменной.
2. Для любого терма входной переменной имеется хотя бы одно правило, в котором этот терм используется в качестве предпосылки (в левой части правила).

Пусть в базе правил имеется $m$ правил вида:

$$R_1: \text{ЕСЛИ } x_1 \text{ это } A_{11} \ldots \text{ И } \ldots x_n \text{ это } A_{1n}, \text{ ТО } y \text{ это } B_1$$
$$\ldots$$
$$R_i: \text{ЕСЛИ } x_1 \text{ это } A_{i1} \ldots \text{ И } \ldots x_n \text{ это } A_{in}, \text{ ТО } y \text{ это } B_i$$
$$\ldots$$
$$R_m: \text{ЕСЛИ } x_1 \text{ это } A_{i1} \ldots \text{ И } \ldots x_n \text{ это } A_{mn}, \text{ ТО } y \text{ это } B_m,$$

где $x_k$ ($k=1..n$) – входные переменные; $y$ – выходная переменная; $A_{ik}$ – заданные нечеткие множества с функциями принадлежности.

Результатом нечеткого вывода является четкое значение переменной $y^*$ на основе заданных четких значений $x_k$ ($k=1..n$).

В общем случае механизм логического вывода включает четыре этапа: введение нечеткости (фазификация), нечеткий вывод, композиция и приведение к четкости (дефазификация) (рис 3.9).

Алгоритмы нечеткого вывода различаются главным образом видом используемых правил, логических операций и разновидностью метода дефазификации. Разработаны модели нечеткого вывода Мамдани, Сугено, Ларсена, Цукамото [84].

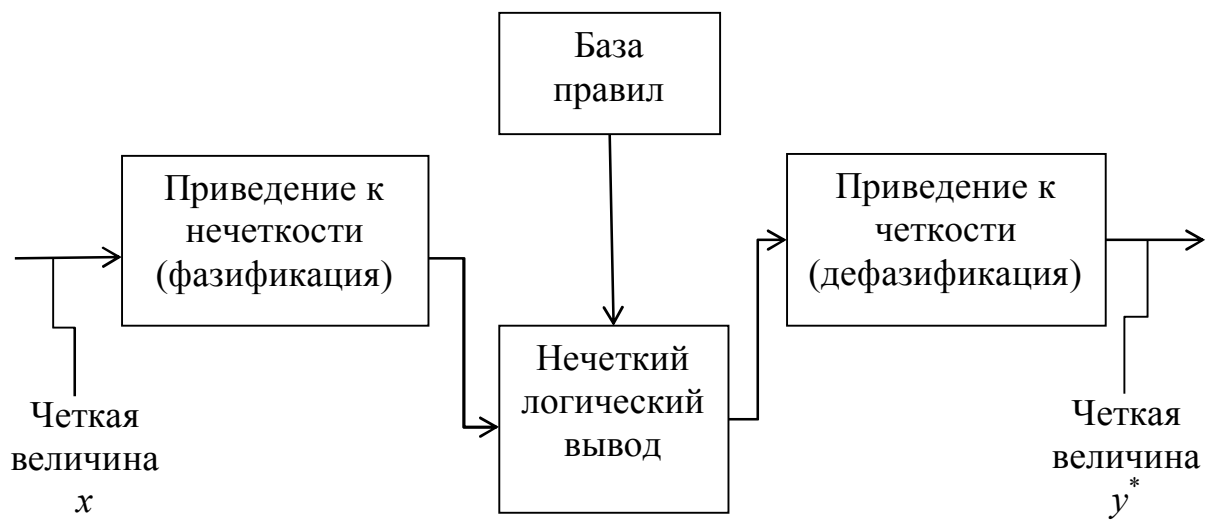

Рис.3.9. Система нечеткого логического вывода.

Рассмотрим подробнее нечеткий вывод на примере механизма Мамдани (Mamdani). Это наиболее распространенный способ логического вывода в нечетких системах. В нем используется минимаксная композиция нечетких множеств. Данный механизм включает в себя следующую последовательность действий.

1. Процедура фазификации: определяются степени истинности, т.е. значения функций принадлежности для левых частей каждого правила (предпосылок). Для базы правил с *m* правилами обозначим степени истинности как $A_{ik}(x_k)$, $i=1..m$, $k=1..n$.

2. Нечеткий вывод. Сначала определяются уровни "отсечения" для левой части каждого из правил:
$$\alpha_i = \min_i (A_{ik}(x_k))$$

3. Далее находятся "усеченные" функции принадлежности:
$$B_i^*(y) = \min_i(\alpha_i, B_i(y))$$

4. Композиция или объединение полученных усеченных функций, для чего используется максимальная композиция нечетких множеств:
$$MF(y) = \max_i B_i^*(y)$$

где *MF(y)* – функция принадлежности итогового нечеткого множества.

5. Дефазификация, или приведение к четкости. Существует несколько методов дефазификации. Например, метод среднего центра, или центроидный метод. Геометрический смысл такого значения – центр тяжести для кривой $MF(y)$.

Рисунок 3.10 графически показывает процесс нечеткого вывода по Мамдани для двух входных переменных и двух нечетких правил $R_1$ и $R_2$.

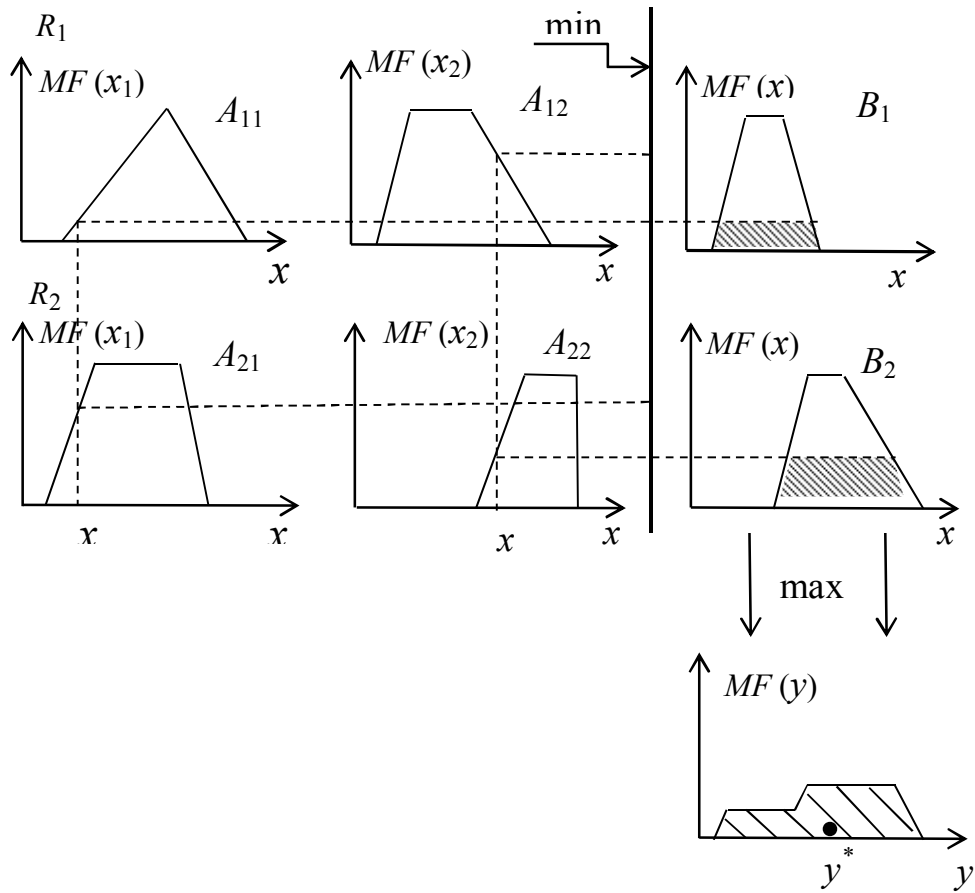

Рис. 3.9. Схема нечеткого вывода по Мамдани.

Проиллюстрируем процедуру нечеткого вывода на примере решения одной из подзадач обеспечения экономической безопасности: регулирования Центробанком курса национальной валюты по отношению к какой-либо мировой (базовой) валюте и поддержания его в рамках заданного коридора.

В качестве управляющего воздействия Центробанк использует валютные интервенции: скупает или продает базовую мировую валюту, относительно которой определяется курс национальной денежной единицы.

При решении задачи мы должны, прежде всего, определить нечеткие понятия, связанные со степенью вмешательства ЦБ в деятельность валютного рынка. Для этой цели воспользуемся пятиуровневым классификатором, отражающим следующие действия ЦБ: «интенсивная» покупка валюты (ИПк); «слабая» покупка (СПк); бездействие ЦБ (БД); «слабая» продажа валюты (СПр); «интенсивная» продажа валюты (ИПр).

Каждому уровню соответствует нечеткое треугольное число в диапазоне от минус 1 до плюс 1. Аналогично определяются понятия высокая (слабая) скорость падения (роста) курса валюты, а также высокая (слабая) степень положительного (отрицательного) отклонения курса от заданного ЦБ «равновесного» (оптимального) значения.

Определим множество правил управления курсом валюты на рынке.

Если, например, курс не выходит за рамки заданного ЦБ «коридора» и скорость изменения курса близка к нулю, то очевидно, что это и есть искомое положение, и ничего предпринимать не нужно (ЦБ бездействует).

Рассмотрим другой случай: курс базовой валюты близок к «оптимальному» значению, определенному ЦБ, но слабо растет (скорость роста низкая). Необходимо компенсировать это отклонение. ЦБ в этом случае должен начать «слабую» продажу базовой валюты (продажу со слабой интенсивностью). Аналогично получаются правила управления курсом для других возможных случаев изменения ситуации на валютном рынке. Их можно свести в следующую таблицу:

Таблица 3.1.
Действия ЦБ

|  |  | Отклонение курса валюты от оптимального значения | | | | |
|---|---|---|---|---|---|---|
|  |  | ВО | СО | Н | СП | ВП |
| Скорость изменения курса базовой валюты | ВП | БД | СПр | ИПр | ИПр | ИПр |
| | СП | СПк | БД | СПр | ИПр | ИПр |
| | Н | ИПк | СПк | БД | СПр | ИПр |
| | СО | ИПк | ИПк | СПк | БД | СПр |
| | ВО | ИПк | ИПк | ИПк | СПк | БД |

Для расчетного примера зададим численные значения отклонения курса базовой валюты и скорости изменения ее курса. Допустим, что реальное значение отклонения курса валюты принадлежит нечеткому множеству «ноль» (Н) со степенью 0.75 и нечеткому множеству «слабое положительное» (СП) со степенью принадлежности 0.25, а реальное значение скорости изменения курса принадлежит нечеткому множеству "ноль" со степенью 0.4 и нечеткому множеству «слабое отрицательное» (СО) со степенью принадлежности 0.6.

Приведенным значениям отклонения курса валюты и скорости ее изменения соответствуют только четыре правила из таблицы, определяющей действия ЦБ (ДЦБ). Объединим их в одно решение. При этом, учитывая, что условия в правилах объединены конъюнкцией, необходимо взять минимальное из степеней принадлежности условий к соответствующему нечеткому множеству и уменьшить принадлежность переменной "ЦБ" к этому нечеткому множеству до данного минимального значения [78].

Результатом применения правила:

«Если отклонения курса валюты нулевое (Н) И скорость изменения курса валюты нулевая (Н), то действия ЦБ – «бездействие» (БД)» является ДЦБ=БД со степенью принадлежности 0.4.

Результатом применения правила:

«Если отклонения курса валюты слабое положительное (СП) И скорость изменения курса нулевая (Н), то действия ЦБ – «слабая продажа» (СПр)» является ДЦБ = СПр со степенью принадлежности 0.25.

Результатом применения правила:

«Если отклонения курса валюты нулевое (Н) И скорость изменения курса слабая отрицательная (СО), то действия ЦБ – «слабая покупка» (СПк)» является ДЦБ=СПр со степенью принадлежности 0.6.

Результатом применения правила:

«Если отклонения курса валюты слабое положительное (СП) И скорость изменения курса слабая отрицательная (СО), то действия ЦБ - «бездействие» (БД)» является ДЦБ=БД со степенью принадлежности 0.25.

Таким образом, результатом применения первого правила является усеченное на уровне 0.4 нечеткое число, которое соответствовало значению лингвистической переменной ДЦБ=«БД» («бездействие»). Результатом применения второго и четвертого правил является усеченное на уровне 0.25 нечеткое число, соответствовавшее значению лингвистической переменной ДЦБ=«СПр» («слабая продажа»). Результатом применения третьего правила является усеченное на уровне 0.6 нечеткое число, которое соответствовало значению лингвистической переменной ДЦБ=«СПк» («слабая покупка»).

Совокупное применение данных правил дает общее решение, представляющее собой нечеткое множество, состоящее из объединения этих чисел. Графически данное множество представляет собой ступенчатую фигуру из наложенных друг на друга трапеций.

Далее необходимо на основе одного из эвристических методов осуществить дефаззификацию: перейти от нечеткого описания к четкому числовому значению, выбранному из полученного множества.

Например, можно выбрать в качестве конечного значения абсциссу центра тяжести нечеткого множества. В нашем случае это значение примерно равно 0.2. Таким образом, ЦБ должен осуществить покупку валюты в размере 0.2 условных единицы.

## 3.6. Выводы по главе 3

Обобщая рассмотренный в главе 3 материал можно сделать следующие основные выводы:

- наиболее удобным математическим аппаратом для описания и исследования систем социотехнического типа, к которым относится система комплексного обеспечения информационной безопасности, является когнитивное моделирование, как одно из направлений современной теории поддержки принятия решений при управлении слабоструктурированными и трудно формализуемыми системами и ситуациями.

- когнитивный подход к моделированию и управлению слабоструктурированными системами направлен на разработку формальных моделей и методов, поддерживающих интеллектуальный процесс решения проблем благодаря учету в этих моделях и методах когнитивных возможностей (восприятие, представление, познание, понимание, объяснение) субъектов управления при решении управленческих задач.

- явления и процессы функционирования и развития СС включают в себя различные события, тенденции, определяемые многими факторами, причем каждый в свою очередь влияет на некоторое число других факторов. Таким образом, образуются сети причинных отношений между ними. Поэтому для исследования СС наиболее целесообразно применение каузальных когнитивных карт.

- *когнитивная карта* отображает лишь наличие влияний факторов друг на друга. В ней не отражается ни детальный характер этих влияний, ни динамика изменения влияний в зависимости от изменения ситуации, ни временные изменения самих факторов. Учет всех этих обстоятельств требует перехода на следующий уровень структуризации информации, отображенной в когнитивной карте, т.е. к *когнитивной модели*. На этом уровне каждая связь между факторами когнитивной карты раскрывается до соответствующего уравнения, которое может содержать как количественные (измеряемые) переменные, так и качественные (не измеряемые) переменные. Таким образом, когнитивная модель ситуации представляется ***функциональным графом***.

- адекватность построенной на основе когнитивных карт модели окончательно выясняется только в процессе реальной работы с моделью. Отсюда, в частности, следует, что информационные технологии поддержки принятия решений, основанные на аппарате когнитивных карт, должны быть максимально открытыми для модификаций.

- при анализе ситуаций, опирающемся на когнитивные модели, решаются два типа задач: статические и динамические. Статический анализ - это анализ текущей ситуации, заключающийся в выделении и сопоставлении путей влияния одних факторов на другие через третьи (каузальных цепочек). Динамический анализ - это генерация и анализ возможных сценариев развития ситуации во времени. Математическим аппаратом анализа является теория знаковых и нечетких графов.

- одним из наиболее перспективных направлений научных исследований в области анализа, прогнозирования и моделирования слабоструктурированных и плохо формализуемых явлений и процессов является нечеткая логика (fuzzy logic), сформулированная и получившая развитие в работах Л. Заде. Она значительно расширяет возможности учета неопределенностей различной природы, неизбежно сопутствующих математическому описанию реальности. Такой подход позволяет решать задачи совершенствования функционирования различных систем в условиях неполноты и неточности информации о протекающих процессах, недостаточности и недостоверности знаний, при наличии субъективности оценок.

# ГЛАВА 4.

# АНАЛИЗ И УПРАВЛЕНИЕ КОМПЛЕКСНОЙ БЕЗОПАСНОСТЬЮ ИНФОРМАЦИОННЫХ СИСТЕМ НА ОСНОВЕ КОГНИТИВНОГО МОДЕЛИРОВАНИЯ

## 4.1. Концептуальный анализ понятия комплексной безопасности

Концептуальный анализ различных аспектов комплексной безопасности требует выявления генезиса данной проблемы.

В современном понятийно-категориальном аппарате под безопасностью понимается состояние и тенденции развития защищенности жизненно важных элементов системы от внешних и внутренних негативных факторов [85].

При этом необходимо учесть, что безопасность - динамичная категория, имеющая множественную предметность в зависимости от конкретной области применения.

Деятельность по обеспечению безопасности возникает в ходе разрешения противоречия между опасностью и потребностью управлять безопасностью: предвидеть, предотвращать, локализовать и устранять ущерб от воздействия опасности [86].

Оценка уровня безопасности всегда относительна. Попытки напрямую приписать этой оценке численное значение в большинстве случаев бесперспективны в плане дальнейшей интерпретации результатов.

Безопасность – понятие комплексное и не может рассматриваться как простая сумма составляющих ее частей. Эти части взаимосвязаны и взаимозависимы. Кроме того, каждая часть критически значима. Следовательно, никакие методы предусматривающие осреднение частных критериев безопасности (пусть и неявное) при комплексной оценке безопасности неприемлемы.

Комплексная оценка уровня безопасности (КОУБ) не может быть больше минимальной оценки, полученной для различных аспектов системы.

Безопасность не существует сама по себе, в отрыве от человека. Она обеспечивается для человека и им же оценивается. Поэтому, понятие безопасности имеет не только объективную, но и субъективную сторону, поскольку оценка ее уровня проводится в конечном итоге человеком.

Это весьма важный аспект, который приводит, с одной стороны, к необходимости оперирования при оценке уровня безопасности лингвистическими переменными (основными структурными единицами в языке людей) и, как следствие, к необходимости применению аппарата нечеткой логики.

С другой стороны, наличие субъективности при оценке уровня безопасности приводит к тому, что в шкале оценок может появляться диапазон «условной приемлемости» [43].

Поясним этот тезис. Допустим, что уровень комплексной безопасности оценивается от нуля до единицы. При этом нулю соответствует абсолютно неприемлемый уровень безопасности (низшая оценка), а единице – самая высокая оценка.

При оценке отдельных сторон безопасности системы существует некоторый пороговый уровень, выше которого система считается безопасной, а ниже – нет. Уровень остаточных рисков при этом лишь сдвигает пороговое значение в ту или иную сторону, не меняя картину в целом.

При комплексной оценке ситуация качественно меняется. Какие-то риски неприемлемые при одних обстоятельствах, при других оцениваются как допустимые. Причем в роли таких обстоятельств, влияющих на оценку безопасности, может выступать, в том числе, как это ни парадоксально, и наличие или отсутствие других рисков.

Например, угроза высокой безработицы и связанный с ней риск потерять работу и остаться без средств к существованию, могут перевести оценку экологической безопасности, используемую при принятии решения о работе в том или ином регионе из категории «неприемлемый уровень» в категорию

«условно приемлемый уровень». Данная ситуация соответствует философскому определению свободы как «осознанной необходимости».

Совершенно очевидно, что при изменении условий категория уровня безопасности будет пересмотрена.

Исходя из этого, данный диапазон оценок безопасности, находящийся между значениями «нижний пороговый уровень» и «верхний пороговый уровень», можно назвать условно приемлемым.

Любые неконтролируемые внешние или внутренние процессы потенциально могут привести к возникновению угроз. Реализация этих угрозы в свою очередь оказывает негативное влияние на состояние безопасности системы, что вызывает различные деструктивные процессы. Нарушается нормальное функционирование системы, что находит свое отражение в значениях различных критериев и показателей, используемых для оценки безопасности [87].

Исследованию в этой области посвящен ряд работ, в которых предлагаются различные подходы к оценке уровня КОИБ.

Так, например, в [88] изложен системный подход к построению комплексной защиты информационной системы предприятия и описана методика построения такой системы с применением отечественных технических и криптографических средств защиты. В работе [89] рассмотрены принципы и методы аудита информационной безопасности (ИБ) на основе процессорного подхода, приведены некоторые методы оценивания ИБ.

Наиболее яркое выражение системный подход к решению задач безопасности нашел в работах В. В. Домарева. Им предложена трехмерная модель, включающая в себя основные этапы, направления и методы обеспечения безопасности различных систем [3]. Подчеркнуто, что специфическими особенностями задачи создания систем защиты являются:

- неполнота и неопределенность исходной информации о составе и характере угроз;

- многокритериальность задачи, связанная с необходимостью учета большого числа частных показателей;
- наличие как количественных, так и качественных показателей, которые необходимо учитывать при решении задач разработки и внедрения систем защиты;
- невозможность применения классических методов оптимизации.

Поэтому разработка модели, позволяющей унифицировать подходы к управлению комплексной безопасностью системы, является весьма актуальной задачей.

## 4.2. Методы решения многокритериальных задач

Критерии и показатели, используемые для оценки безопасности систем, часто носят противоречивый характер. Например, при комплексной оценке уровня безопасности региона усиление экономической безопасности нередко приводит к резкому обострению экологических проблем и т.п.

В подобных случаях критерий выбора в ситуации принятия решения (СПР) представляет собой совокупность отдельных критериев, и соответствующая задача становится многокритериальной.

При решении многокритериальных задач часто используются различные методы свертки критериев в один обобщенный (интегральный) критерий.

Наиболее простой метод построения интегрального критерия заключается в выделении одного из критериев в качестве основного, а все остальные критерии добавляются к ограничениям, в которых задается область допустимых значений вектора независимых переменных. Таким образом, задача с векторным критерием сводится к задаче принятия решения со скалярным аргументом.

Основной недостаток такого подхода заключается в том, что фактически поиск ведется лишь по одному критерию. Значения остальных критериев, если они удовлетворяют ограничениям, по существу не влияют на результаты поиска.

Безопасность – понятие комплексное и попытка оценить ее уровень по одному какому-то параметру (например, имеющему стандартное количественное выражение) обречена на неудачу. Поэтому данный способ получения свертки при решении задач, связанных с безопасностью неприемлем.

Другими методами построения комплексного критерия являются аддитивная и мультипликативная свертка.

Аддитивный критерий, являясь наиболее простым, в тоже время из-за возможности неограниченной компенсации значений одних критериев за счет других, нечувствителен к крайним значениям отдельных критериев.

Кроме того, комплексная безопасность не может рассматриваться как простая сумма составляющих ее частей. Эти части, как отмечалось выше, взаимосвязаны и взаимозависимы, каждая часть критично значима. Следовательно, модели, в основе которых лежит предположение о линейном поведении системы (аддитивная свертка предполагает именно такую модель), при оценке уровня комплексной безопасности обычно не могут адекватно отражать ситуацию. Поэтому аддитивная свертка для комплексной оценки уровня безопасности, в большинстве случаев, также не подходит.

Значение же мультипликативного критерия, в отличие от аддитивного, резко уменьшается при малых значениях отдельных критериев, что позволяет исключить нежелательные варианты при принятии решения.

Таким образом, для задач, связанных с обеспечением комплексной безопасности, в большинстве случаев, наиболее целесообразным представляется применение мультипликативной свертки векторного критерия:

$$K = \prod_i K_i^{\alpha_i}$$

где $K_i$ – частные критерии, $\alpha_i$ – некоторым образом определенные веса, приписываемые каждому частному критерию $K_i$.

При построении свертки с целью унификации разнородных критериев используют переход от абсолютных значений критериев к относительным величинам. Для этого фиксируется шкала возможных значений для критериев и возможные границы изменения для каждого из них. Например, если в качестве шкалы принять [0; 1], а границы изменения критерия $K_i$ лежат между $K_i^{min}$ и $K_i^{max}$, то в качестве относительного значения критерия будет выступать величина:

$$\overline{K} = \frac{K_i - K_i^{min}}{K_i^{max} - K_i^{min}}$$

Однако, часто получение от лица, принимающего решение (ЛПР), надежной количественной информации для построения $K$, бывает затруднительным. В таких случаях стремятся получить от ЛПР, в основном, только качественную информацию. Например, о том, какой из критериев наиболее или наименее значим, какой из критериев может быть ухудшен, а для каких ухудшение крайне нежелательно и т.д.

В качестве такой процедуры получения информации может быть использован алгоритм Беленсона-Капура [90].

Многие аспекты, касающиеся безопасности системы, могут вообще не подлежать количественному измерению. Тогда при их оценивании прибегают к искусственным приемам. Например, каждому фактору сопоставляется количественная балльная шкала [79.].

При этом необходимо предложить эксперту методику, по которой он должен назначить баллы. Однако проблема заключается в том, что многие понятия, связанные с безопасностью являются сугубо качественными и, как отмечалось выше, предлагать измерять их количественно в большинстве случаев бесперспективно.

Другое дело, если сразу применять нечетко выраженные степени, например «Низкая, Ниже Среднего, Средняя, Выше Среднего, Высокая». Тогда

от эксперта не требуется количественной точности, а требуется как раз субъективная оценка на естественном языке. Затем лингвистическое описание может быть сопоставлено с количественной (например, балльной) шкалой носителя с помощью методов теории нечеткого гранулирования [91].

## 4.3. Динамическая модель влияния угроз на безопасность системы

Сформулируем математическую модель, описывающую динамику изменения уровня комплексной безопасностью различных систем [92-93].

Назовем «*Уровнем комплексной безопасности системы*» (УКБС) оценку, основанную на наборе показателей и критериев, характеризующих состояние системы в плане защищенности критичных для неё элементов.

Уровень комплексной безопасности системы можно охарактеризовать следующей матрицей

$$B = \begin{pmatrix} K_1 & F_1 & V_1 & T_1 & S_1 \\ K_2 & F_2 & V_2 & T_2 & S_2 \\ K_3 & F_3 & V_3 & T_3 & S_3 \\ K_4 & F_4 & V_4 & T_4 & S_4 \\ \underline{\phantom{K}} & \underline{\phantom{F}} & \underline{\phantom{V}} & \underline{\phantom{T}} & \underline{\phantom{S}} \\ K_n & F_n & V_n & T_n & S_n \end{pmatrix},$$

где $K_i$ – показатель уровня безопасности по *i*-му критерию; $F_i$ – тенденция изменения *i*-го критерия (возрастает, убывает, нейтрален); $V_i$ - скорость изменения *i*-го критерия (например: низкая, ниже среднего, средняя, выше среднего, высокая); $T_i$ – характерное для *i*-го критерия время, которое, в частности, позволяет правильно интерпретировать значения параметра $V_i$ ; $S_i$ - степень критичности негативных последствий при реализации рисков, ухудшающих значение *i*-го критерия.

Матрицу вида *B* будем называть в дальнейшем *матрицей безопасности* (МБ).

Первый и четвертый столбец МБ представляют собой вектор частных критериев безопасности и их весов и характеризуют текущее состояние комплексной безопасности, позволяя оценить сложившуюся на текущий момент времени ситуацию. Остальные столбцы матрицы отражают динамику развития процессов и позволяют строить прогноз развития на будущее.

В этом случае, мультипликативная свертка интегрального критерия комплексной безопасности представляет собой величину:

$$K = \prod_{i=1} K_i^{S_i}$$

Оценки $S_i$ могут быть получены экспертным путем. Однако для эксперта в большинстве случаев затруднительно дать непосредственные численные оценки этим коэффициентам. Поэтому предпочтительнее могут оказаться различные ранговые методы, при реализации которых требуется лишь упорядочить критерии.

Может быть использован, например, метод нестрогого ранжирования. В соответствии с этим методом экспертом производится нумерация всех критериев по убыванию степени приемлемости негативных последствий, связанных с данным критерием безопасности. Причем допускается, что эксперту не удастся различить между собой некоторые критерии. В этом случае при ранжировании он помещает их рядом в произвольном порядке. Затем проранжированные критерии последовательно нумеруются. Оценка (ранг) критерия определяется его номером.

Если на одном месте находятся несколько неразличимых между собой критериев, то, обычно, оценка каждого из них принимается равной среднему арифметическому их новых номеров [94]. Однако, представляется целесообразным несколько модифицировать такой метод оценки, приняв за ранг каждого из неразличимых критериев номер всей группы как целого объекта в упорядочении [95].

Таким способом могут быть оценены как степени влияния каждого параметра на частные критерии безопасности $K_i$, так и степени приемлемости последствий реализации угроз $S_i$.

Например, пусть эксперт упорядочил критерии следующим образом:

$$K_5, (K_3, K_7, K_2), K_1, (K_6, K_8), K_9, K_4.$$

Критерии, не различимые между собой, объединены в круглые скобки. Тогда оценки для каждого из критериев, вычисленные в соответствии с описанной выше процедурой, равны:

$S_5=1; S_3=S_7=S_2=2, S_1=3; S_6=S_8=4; S_9=5; S_4=6.$

Применим нормирование по величине, равной сумме всех оценок:

$$R = \sum_i S_i$$

В нашем случае $R = 29$. Таким образом, после линейного преобразования в шкалу [0;1] по норме $R$ получим:

$S_5=1/29; S_3=S_7=S_2=2/29, S_1=3/29; S_6=S_8=4/29; S_9=5/29; S_4=6/29.$

Найденные предложенным способом оценки представляют собой обобщение системы весов Фишберна [96] для случая смешанного распределения предпочтений, когда наряду с предпочтениями в систему входят и отношения безразличия.

Критерии в матрице безопасности можно сгруппировать по соответствующим направлениям обеспечения безопасности, например: экономические, экологические, социальные, технические и т.п.

Таким образом, каждый кортеж $(K_i, F_i, V_i, S_i, T_i)$ характеризует состояние безопасности по $i$ – му критерию.

Частичные матрицы, состоящие из строк, представляющих определенное направление обеспечения безопасности, описывают состояние в соответствующей области.

Показатели уровня безопасности $K_i$ тесно связаны с последствиями от возможной реализации имеющихся в системе угроз, мерами предотвращения таких последствий и мерами, направленными на локализацию и устранение последствий, если таковые все же возникают.

Следует особо отметить, что угрозы можно разделить на первичные и вторичные. Первичные угрозы существуют вне зависимости от состояния системы и априорно имеют определённую безусловную вероятность появления.

Вероятность появления вторичных угроз является условной и зависит от состояния системы и состояния внешней среды.

В частности, некоторые состояния системы могут спровоцировать возникновение угроз, появление которых в иных условиях было бы невозможным.

Введем следующие обозначения:

$\overline{UG_i}$ и $\widetilde{UG_j}(i,j = 1,2,3,...)$ - совокупность первичных и вторичных угроз, возникающих с вероятностями $\overline{PUG_i}$ и $\widetilde{PUG_j}$, соответственно, и оказывающих влияние $\bar{n}_{km}$ и $\tilde{n}_{km}$ на элемент ($k,m$) матрицы безопасности $B$ ($k = 1,2,3,...$; $m = 1,2,3,4,5$).

Влияние каждой из первичных или вторичных угроз можно описать *матрицей влияния* (МВ) имеющей вид:

$$N_i = \begin{pmatrix} n_{11} & n_{12} & n_{13} & n_{14} & n_{15} \\ n_{21} & n_{22} & n_{23} & n_{24} & n_{25} \\ n_{n1} & n_{n2} & n_{n3} & n_{n4} & n_{n5} \end{pmatrix}$$

Фактически МВ представляет собой матрицу весов влияния *i*-го негативного фактора на элементы МБ.

Необходимо отметить, что влияние $\bar{n}_{km}$ и $\tilde{n}_{km}$ на некоторые элементы матрицы безопасности $B$ может быть как отрицательным, так и положительным.

Элементы матрицы влияния, оказывающие отрицательное воздействие назовем *негативными относительно элементов МБ*; элементы, оказывающие положительное воздействие – *позитивными относительно элементов МБ*; элементы, не оказывающие никакого воздействия, назовем *нейтральными*.

Кортеж $\bar{R}_i = \{\bar{N}_i; \overline{PUG_i}\}$ назовем *риском реализации i-й первичной угрозы*.

Данный кортеж отражает появление с вероятностью $\overline{PUG_i}$ последствий, которые изменяют состояние системы через соответствующие матрицы влияния $\bar{N}_i$.

Вероятности возникновения первичных угроз $\overline{PUG_i}$ от нас не зависят. Однако совокупность превентивных мер защиты позволяет ослабить влияние первичных угроз на степень безопасности системы.

Этот факт может быть описан с помощью матриц превентивных мер (МПМ):

$$Z_j = \begin{pmatrix} z_{11} & z_{12} & z_{13} & z_{14} & z_{15} \\ z_{21} & z_{22} & z_{23} & z_{24} & z_{25} \\ \underline{\phantom{z}} & \underline{\phantom{z}} & \underline{\phantom{z}} & \underline{\phantom{z}} & \underline{\phantom{z}} \\ z_{n1} & z_{n2} & z_{n3} & z_{n4} & z_{n5} \end{pmatrix},$$

где $j = \overline{1, M}$, M - общее количество превентивных мер.

Элементы матрицы $Z_i$ назовем *демпфирующими коэффициентами*.

Тогда под *остаточным влиянием* будем подразумевать матрицу $\widehat{N}_i$ (назовем ее матрицей остаточного влияния – МОВ) элементы которой находятся из выражения:

$$\widehat{n}_{mn} = n_{mn} \otimes \max_{k=1\ldots M} z_{mn}^k$$

где $z_{mn}^k$ – элемент $(m, n)$ матрицы превентивных мер $Z_k$. Символом «$\otimes$» обозначена некоторым образом определенная для двух матриц операция.

В случае числовых значений элементов матриц, это может быть, например, операция простого *поэлементного* умножения или сложения. В случае лингвистических значений данная операция (называемая в нечеткой математике композицией) определяется с помощью принципа расширения обычных (четких) математических функций на нечеткие числа, предложенного Л.Заде [78].

Под *остаточным риском* будем понимать кортеж:

$$\bar{R} = \{\widehat{N}_i;\ \overline{PUG_i}\},$$

Если все же, несмотря на превентивные меры защиты, реализация определенного множества первичных угроз привела к возникновению последствий, то необходимо предпринять меры для их локализации и устранения.

Прежде всего, необходимо оценить отклонение текущего состояния системы $\hat{B}$ от безопасного состояния $B_S$.

Введем понятие разности между двумя матрицами, определив результат применения операции «#» к двум элементам матриц аналогично тому, как это было сделано для операции «⊗»: в случае числовых значений элементов матриц - это операция поэлементного вычитания, в случае лингвистических значений - операция определяется с помощью принципа расширения Л.Заде.

Тогда матрицу $Q = B_S \# \hat{B}$ назовем *матрицей потерь безопасности* (МПБ) на данном этапе.

Матрица *потерь безопасности* $Q$ представляет собой входные данные для блока ликвидации последствий (БЛП).

Реализация мероприятий этого блока может быть формализована с помощью матрицы ликвидации последствий (МЛП):

$$L = \begin{pmatrix} l_{11} & l_{12} & l_{13} & l_{14} & l_{15} \\ \underline{l_{21}} & \underline{l_{22}} & \underline{l_{23}} & \underline{l_{24}} & \underline{l_{25}} \\ l_{n1} & l_{n2} & l_{n3} & l_{n4} & l_{n5} \end{pmatrix}.$$

Результат применения БЛП может быть записан следующим образом:

$$\hat{Q} = Q \otimes L = \begin{pmatrix} \hat{q}_{11} & \hat{q}_{12} & \hat{q}_{13} & \hat{q}_{14} & \hat{q}_{15} \\ \underline{\hat{q}_{21}} & \underline{\hat{q}_{22}} & \underline{\hat{q}l_{23}} & \underline{\hat{q}_{24}} & \underline{\hat{q}_{25}} \\ \hat{q}_{n1} & \hat{q}_{n2} & \hat{q}_{n3} & \hat{q}_{n4} & \hat{q}_{n5} \end{pmatrix}$$

Матрицу $\hat{Q}$ назовем *матрицей остаточных потерь безопасности* (МОПБ).

Если $\hat{Q} \neq B_S$, то подобное состояние системы может инициировать появление вторичных угроз с вероятностями $\widehat{PUG_i}$.

Таким образом, кроме первичных угроз, в зависимости от текущего состояния системы и ее окружения возможно возникновение вторичных угроз, вероятность появления которых равна $\widehat{PUG_i}$.

Кортеж $\tilde{R}_i = \{\tilde{N}_i ; \widehat{PUG_i}\}$ назовем *риском реализации i-й вторичной угрозы*.

Заметим, что вероятности появления вторичных угроз не являются безусловными, как для первичных угроз. Они зависят от текущего состояния системы. С первичными угрозами мы начинаем бороться еще до их наступления, т.е. фактически пытаемся свести к минимуму их последствия, не имея возможности повлиять на сам факт их появления. В случае с вторичными угрозами мы должны пытаться вообще не допустить их, т.е. должны бороться с вызывающими их причинами. Это принципиальное различие в блоках мероприятий, воздействие которых формализовано множеством матриц $Z_j$ и матрицей $L$.

### 4.4. Когнитивная модель оценки уровня комплексной безопасности

Для решения широкого круга задач, связанных с моделированием плохо формализованных процессов, их прогнозированием и поддержкой принятия решений, часто используются нечеткие когнитивные модели. Как отмечалось выше, неоспоримыми их достоинствами по сравнению с другими методами являются возможность формализации численно неизмеримых факторов, использования неполной, нечеткой и даже противоречивой информации [97].

При построении нечеткой когнитивной модели (НКМ) объект исследования обычно представляют в виде знакового ориентированного графа. В качестве такой модели при оценке комплексной безопасности системы ($KBS$) может быть принят кортеж [98]:

$$KBS = <G, QL, E> \quad (4.1).$$

Здесь $G$ – ориентированный граф, имеющий одну корневую вершину и не содержащий петель и горизонтальных ребер в пределах одного уровня иерархии:

$$G = <\{GF_i\}; \{GD_{ij}\}> \quad (4.2),$$

где $\{GF_i\}$ – множество вершин графа (факторов или концептов в терминологии НКМ); $\{GD_{ij}\}$ – множество дуг, соединяющих $i$-ую и $j$-ую вершины (множество причинно-следственных связей между концептами; при этом дуги расположены так, что началу дуги соответствует вершина нижнего уровня иерархии (ранга), а концу дуги – вершина ранга, на единицу меньшего); $GF_0 = K$ – корневая вершина, отвечающая уровню комплексной безопасности в целом (интегральному критерию безопасности – целевому концепту); $QL$ – набор качественных оценок уровней каждого фактора в иерархии; $E$ – система отношений предпочтения одних факторов другим по степени их влияния на заданный элемент следующего уровня иерархии:

$$E = \{GF_i(е)GF_j \mid е \in (\succ; \approx)\} \quad (4.3),$$

где $GF_i$ и $GF_j$ – факторы одного уровня иерархии; $\succ$ – отношение предпочтения; $\approx$ – отношение безразличия. Такая система, как было показано выше, позволяет определить обобщенные на случай предпочтения/безразличия факторов по отношению друг к другу веса Фишберна для каждой дуги $GD_{ij}$ (веса связей) [95].

Веса Фишберна отражают тот факт, что системе убывающего предпочтения $N$ альтернатив наилучшим образом отвечает система снижающихся по правилу арифметической прогрессии весов [99-100].

Поэтому эти веса представляют собой рациональные дроби, в знаменателе которых стоит сумма $N$ первых членов натурального ряда (арифметической прогрессии с шагом 1), а в числителе – убывающие на единицу элементы натурального ряда, от $N$ до 1 (например, 3/6, 2/6, 1/6). Таким образом, предпочтение по Фишберну выражается в убывании на единицу числителя рациональной дроби весового коэффициента более слабой альтернативы.

Пример наложения системы отношений предпочтения типа (4.3) $E = \{U_1 \succ U_2; U_2 \succ U_3 \approx U_4; U_4 \approx U_5\}$ на фрагмент графа изображен на рис.4.1.

Связь между любыми двумя вершинами (концептами) при необходимости можно также представить в виде нечеткой когнитивной модели более низкого уровня. При этом на верхний уровень будет передаваться максимальное значение связи, выявленное в ходе анализа НКМ нижнего уровня. Такой иерархический способ позволяет упростить построение НКМ для систем высокой степени сложности.

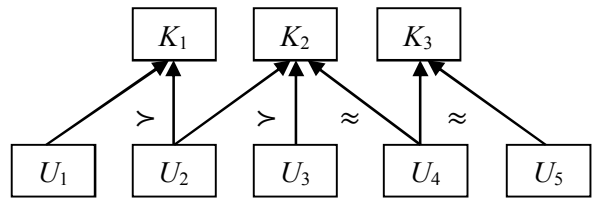

Рис. 4.1. Пример системы отношений предпочтения

на одном из уровней иерархии

Как было показано выше, состояние системы с точки зрения безопасности можно охарактеризовать матрицей $B$, строки которой состоят из элементов ($K_i$, $F_i$, $V_i$, $T_i$, $S_i$), где $K_i$ – показатель уровня безопасности по $i$-му критерию.

В этом случае текущее значение $K_i$ в произвольный момент времени $t$ может быть найдено по формуле:

$$K_i(t) = K_i(0) + F_i * V_i(t/T_i) \qquad (4.4)$$

Показатели же степени критичности негативных последствий $S_i$ фактически представляют собой веса, с которыми частные критерии безопасности $K_i$ влияют на комплексный показатель безопасности системы в целом, который, как было показано выше, может быть найден в результате мультипликативной свертки частных критериев $K_i$.

Обобщенный пример графа для комплексной оценки безопасности информационной системы представлен на рис.4.2. Через $Z_{\{1,2,3,...\}}$ обозначены превентивные меры защиты (механизмы обеспечения безопасности), призванные уменьшить уязвимости ИС - $UZ_{\{1,2,3,...\}}$, $UG_{\{1,2,3,...\}}$ – угрозы безопасности системы, $A_{\{1,2,3,...\}}$ – атаки на ИС, $K_{\{1,2,3,...\}}$ – частные показатели уровня безопасности по соответствующему критерию, $K$ – комплексный (интегральный) показатель безопасности ИС.

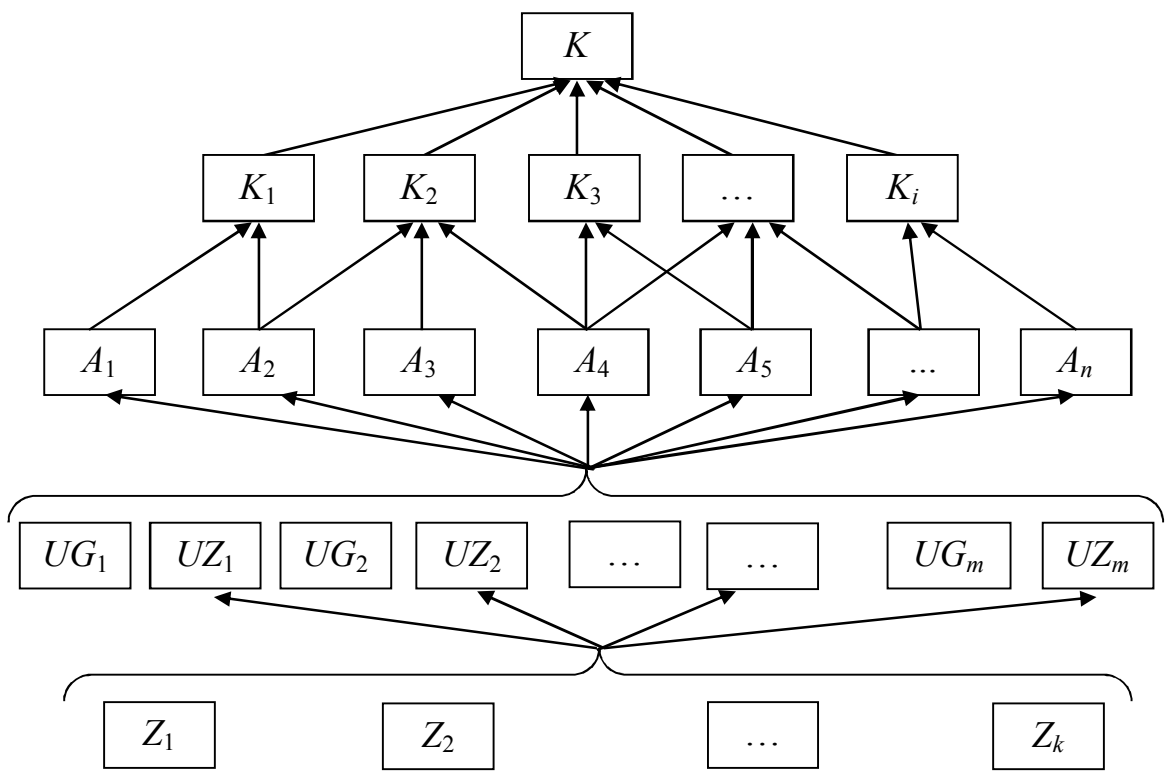

Рис.4.2. Влияние факторов на уровень комплексной безопасности.

Следует заметить, что данный связный граф не является деревом, поскольку не выполняется требование отсутствия простых циклов. Это обусловлено тем, что факторы, находящиеся на нижнем уровне иерархии, могут одновременно оказывать влияние на несколько факторов более высокого уровня.

Например, применение превентивных мер защиты от одной уязвимости может одновременно устранить и какую-либо другую или привести к появлению новой уязвимости. Некоторые атаки могут вызвать изменение сразу нескольких частных критериев безопасности (иногда в противоположном направлении).

Для дальнейшего построения когнитивной модели оценки уровня комплексной безопасности системы на полученный граф необходимо наложить систему отношений предпочтения $E$ одних факторов другим (аналогично тому, как показано на рис. 4.1).

Введем в рассмотрение набор качественных оценок $QL$ уровней каждого фактора в иерархии с терм-множеством:

$$QL = \{\text{Низкий (Н), Ниже среднего (НС), Средний (С),}$$
$$\text{Выше среднего (ВС), Высокий (В)}\}, \qquad (4.5)$$

В качестве семейства функций принадлежности лингвистической переменной «Уровень фактора» будет выступать стандартный пятиуровневый 01-классификатор, введенный в гл.3 (см. п.3.3).

Построим показатель уровня комплексной безопасности на базе агрегирования значений со всех уровней иерархии факторов на основе качественных данных об уровнях факторов и их отношениях порядка на одном уровне иерархии. При этом агрегирование совершается по направлению дуг графа иерархии.

Агрегированию должно подлежать не отдельное значение выбранной функции принадлежности в структуре лингвистической переменной «Уровень фактора», а вся функция принадлежности целиком.

Пусть для каждого показателя на выбранном подуровне иерархии $G$ известна его лингвистическая оценка $QL$, определяемая функцией принадлежности на 01-носителе $x$, а также определена система весов Фишберна на основе системы предпочтений $E$ вида (4.3).

В подобных случаях для агрегирования обычно применяется OWA-оператор Ягера [101], причем весами в свертке выступают упомянутые выше коэффициенты Фишберна. (OWA - Ordered Weighted Averaging – осреднение с упорядоченными весами).

Однако, как было показано выше, аддитивная свертка и осреднение для комплексной оценки уровня безопасности системы неприемлемы и необходимо использовать мультипликативную свертку для нахождения интегрального критерия:

$$K = \prod_{i=1} K_i^{S_i}$$

Поскольку в качестве семейства функций принадлежности лингвистической переменной «Уровень фактора $K_i$» выступает стандартный пятиуровневый 01-классификатор, введенный в гл.3 (см. п.3.3), то для нахождения функции принадлежности $K$ можно записать:

$$\mu_K(x) = \prod_{i=1}^{n} \mu_{K_i}^{S_i}(x) \quad (4.6)$$

где $S_i$ – соответствующие веса Фишберна;

$$\mu_{K_i}(x) = \begin{cases} (3.1), \text{если } QL_{K_i} = \text{"низкий"} \\ (3.2), \text{если } QL_{K_i} = \text{"ниже среднего"} \\ (3.3), \text{если } QL_{K_i} = \text{"средний"} \\ (3.4), \text{если } QL_{K_i} = \text{"выше среднего"} \\ (3.5), \text{если } QL_{K_i} = \text{"высокий"} \end{cases} \quad (4.7)$$

Полученную функцию вида (4.6) необходимо лингвистически распознать, чтобы выработать суждение о качественном уровне показателя $K$. Для этого необходимо соотнести полученную функцию $\mu_K(x)$ и функции $\mu_i(x)$ вида (3.1)-(3.5).

Если

$$(\forall x \in [0,1]) \sup \min (\mu_K(x), \mu_i(x)) = 0, \quad (4.8)$$

то уровень показателя $K$ однозначно не распознается как уровень, которому отвечает $i$-ая «эталонная» функция принадлежности. Стопроцентное распознавание наступает, если выполняется

$$(\forall x \in [0,1]) \min (\mu_K(x), \mu_i(x)) = \mu_i(x). \quad (4.9)$$

Во всех промежуточных случаях необходимо задаться мерой уровня распознавания, т.е. ввести, так называемый индекс схожести [81].

Индекс схожести $\Omega$ может быть найден на основе Хемингового:

$$\rho(A, B) = \int_0^1 |\mu_A(x) - \mu_B(x)| \, dx \quad (4.10)$$

или квадратичного (Евклидового):

$$\rho(A, B) = \int_0^1 \sqrt{\left(\mu_A(x) - \mu_B(x)\right)^2} \, dx \quad (4.11)$$

расстояния между нечетким числом, характеризующим результат, и числами эталонного терм-множества (формулы (3.1)-(3.5)) следующим образом [93]:

$$\Omega = \frac{(1+\widetilde{\rho})}{2} \qquad (4.12)$$

$$\widetilde{\rho} = \frac{\rho_{in} - \rho_{out}}{\rho_{in} + \rho_{out}} \qquad (4.13)$$

где $\rho_{out}$ представляет собой площадь нечеткого трапециевидного числа, характеризующего результат, лежащую вне эталонного числа, а $\rho_{in}$ - площадь, лежащую внутри этого же эталонного числа.

Определенный таким образом индекс схожести, изменяясь в диапазоне от 0 до 1, будет характеризовать близость найденной свертки к тому или иному нечеткому числу, которое, в свою очередь, соответствует элементу эталонного терм-множества. При этом обеспечивается семантическое соответствие: чем больше индекс схожести, тем выше степень соответствия вычисленного значения одному из элементов терм-множества $QL$ (4.5).

Аналогично находятся свертки на более низких уровнях иерархии $G$. При нахождении некоторых из них может использоваться max (min) значение влияния факторов:

$$\overline{GF} = \max_{k(n,m)} S_{k(n,m)} \overline{\overline{GF}}_{k(n,m)} \ (\min_{k(n,m)} S_{k(n,m)} \overline{\overline{GF}}_{k(n,m)}), \qquad (4.14)$$

или аддитивная свертка вида:

$$\overline{GF} = \sum_{k(n,m)} S_{k(n,m)} \overline{\overline{GF}}_{k(n,m)}, \qquad (4.15)$$

где $\overline{\overline{GF}}_{k(n,m)}$ – факторы, влияющие на фактор $\overline{GF}$ следующего уровня иерархии, $S_{k(n,m)}$ – веса влияния.

Следует заметить, что при нахождении сверток значения некоторых показателей необходимо предварительно инвертировать, т.к. влияние данных концептов на факторы более высокого уровня отрицательно.

Например, при переходе от уровня превентивных мер защиты $Z_m$ на уровень угроз безопасности $UZ_k$ перед нахождением свертки необходимо инвертировать значения показателя $Z_m$, а при переходе с уровня атак $A_n$ к уровню частных критериев безопасности $K_i$ (рис.4.2), инвертировать значения $A_n$ согласно табл.1.

Таблица 1.

Инверсия лингвистических переменных

| Уровень показателя $F$ | Инвертированное значение $F$ |
|---|---|
| Н | В |
| НС | ВС |
| С | С |
| ВС | НС |
| В | Н |

В общем случае инверсия произвольного нечеткого числа $F$ может быть определена согласно формуле:

$$Inv(F) = 1 - \mu_F \qquad (4.16)$$

Таким образом, пройдя последовательно снизу вверх по всем уровням иерархии $G$ и применяя соотношения (4.5) – (4.15), мы не только можем путем комплексного агрегирования данных выработать суждение о качественном уровне показателя на каждой ступени иерархии (вплоть до $GF_0 = K$), но и оценить степень обоснованности данного суждения с помощью индекса соответствия $\Omega$.

Если кроме качественных значений показателей имеются и количественные данные, то простейшим способом для их совместного учета при комплексной оценке является загрубление полученных количественных оценок до качественного их описания, и последующий переход к изложенной выше модели оценки.

## 4.5. Оценка повреждений сервисов безопасности информационной системы на основе нечетко-когнитивного подхода

Как отмечалось выше, нарушение безопасного режима функционирования системы может наступить в результате целого ряда взаимосвязанных между собой причин. Реализация потенциальных угроз может вызвать деструктивные процессы и оказать негативное влияние на безопасность. При этом нарушается нормальное функционирование системы, повреждаются основные сервисы безопасности (конфиденциальность, целостность, доступность).

Для выработки обоснованных рекомендаций по применению мер, направленных на ликвидацию последствий снижения безопасности («блок ликвидации последствий» в терминах модели управления комплексной безопасностью, изложенной в работе [43]), лицо принимающее решение (ЛПР), опираясь на множество признаков, должно определить (классифицировать) степень повреждения безопасности ИС.

В подобных случаях критерий выбора в ситуации принятия решения представляет собой совокупность отдельных критериев, и соответствующая задача также становится многокритериальной.

Для ее решения сформируем лингвистическую переменную «Уровень повреждения» с терм-множеством значений вида (4.5): $QL$ = {Низкий (Н), Ниже среднего (НС), Средний (С), Выше среднего (ВС), Высокий (В)}.

В качестве семейства функций принадлежности будет выступать стандартный пятиуровневый 01-классификатор, где функции принадлежности – трапециевидные нечеткие числа (3.1)-(3.5).

Полученная при этом классификация является нестрогой. Однако предполагается, что каждой категории (степени) повреждений безопасности поставлена в соответствие рекомендация по повышению ее уровня. Например, если уровень повреждения отнесен к категории «высокий», то рекомендация может состоять в том, чтобы полностью перестроить ИС, развернув все сервисы и средства безопасности на качественно более высоком уровне.

Очевидно, что каждой категории повреждений соответствует свой объем восстановительных работ. Цель заключается в том, чтобы степень повреждения отнести к одной из категорий, заданных вербальным выражением.

Данную задачу можно сформулировать следующим образом: разработать рациональный способ для доказательства истинности (ложности) гипотезы о том, что безопасности ИС нанесено повреждение определенной степени тяжести, или же доказательства того, что такая гипотеза более разумна, чем другие.

Рассматриваемая проблема относится к методам принятия решений на основе базы знаний. Будучи сформулированной как задача о проверке гипотезы на основе данных о наблюдениях, она может рассматриваться как задача распознавания образов в условиях неопределенности.

При этом трудно построить простой классификатор, который многомерное пространство наблюдаемых признаков отображал бы в определенный набор категорий повреждений. Поэтому необходимо максимально эффективно использовать знания экспертов.

Информационная система должна подвергаться как экспериментальному, так и аналитическому изучению при проведении периодического аудита безопасности и всякий раз, когда появляются признаки нарушения нормальной работы системы.

Поэтому на нижнем уровне иерархии НКМ необходимо расположить узлы, характеризующие состояние элементов ИС. Например:

- степень аппаратных повреждений сервера (АПС)
- степень снижения физической защищенности сервера (СФБС)
- степень повреждения безопасности программного обеспечения (ПО) сервера (СПБС)
- аппаратные повреждения компьютеров (рабочих станций) сети (АПК)
- степень снижения физической защищенности рабочих станций (СФБК)

• степень повреждения безопасности ПО на отдельных компьютерах сети (СПБК)

• нарушение физической безопасности сетевого оборудования (СФБСО), которая в свою очередь включает в себя нарушения безопасности линий связи и снижение безопасности функционирования коммутационного оборудования.

Состояние каждого из узлов зависит от количества и характера наблюдаемых повреждений, а также от степени влияния этих повреждений на оценку состояния узла. Кроме того, повреждения могут носить кумулятивный характер. В этом случае степень их опасности возрастает. Количество и характер повреждений оцениваются с помощью лингвистических переменных, способ введения которых описан выше.

Возникшие повреждения можно объединить в две группы более высокого уровня: общесистемные (СП) и локальные (ЛП). Категория повреждений по группам оценивается исходя из данных, полученных на нижнем уровне иерархии. На нулевом, самом высоком, уровне иерархии находится комплексная оценка степени повреждений системы в целом (КУП), представляющая собой обобщенный (интегральный) критерий.

Состояние каждого узла НКМ оценивается в процессе вывода, использующего множество правил, составляющих в совокупности Базу Знаний данной предметной области. В правилах может использоваться и различная справочная информация. Например, сведения о топологии и свойствах сети, используемых транспортных протоколах, нагрузке сети и ее отдельных компонентов, качестве используемого программного обеспечения (ПО), о наличии и надежности антивирусного ПО, квалификации персонала и степени его лояльности к политике безопасности и др.

В качестве примера ниже приведены несколько правил вывода для узла СП («Общесистемные повреждения»):

Если «СФБС – высокое» или «АПС - высокие», то точно (1.0), что «СП – высокие»;

Если «АПС – средние» и «АПС носят кумулятивный характер», то почти точно (0.8), что «СП – высокие»;

Если «АПК – высокие» и («АПК – много» или «АПК носят кумулятивный характер»), то весьма возможно (0.7), что «СП – высокие».

Величина в скобках отражает степень уверенности эксперта в выводе и имеет следующие вербальные интерпретации: 0 – не возможно; 0,1-0,3 – маловероятно; 0,4-0,5 – возможно; 0,6-0,7 – весьма возможно; 0,8-0,9 – почти точно; 1,0 – точно. Эта величина приводит к пересчету степени принадлежности повреждения, указанного в выводе, к той или иной категории.

Например, если в выводе степень повреждения указана «высокая» (соответствующая функция принадлежности $\mu_B=(0,75; 0,85; 1; 1)$), и степень уверенности эксперта в данном выводе вербально сформулирована как «весьма возможно» (0,7), то функция принадлежности вывода будет равна $0,7*\mu_B$.

Таким образом, приведенные выше правила определяют условия, при которых мы с той или иной степенью уверенности можем отнести СП к категории «высокий уровень повреждений».

Аналогичные правила формулируются и для других категорий повреждения безопасности в каждом из узлов НКМ. При принятии окончательного решения о состоянии безопасности конкретного узла на рассматриваемом уровне иерархии мы должны руководствоваться максимальной из степеней уверенности для максимально возможного повреждения. Подобная схема отражает общепринятый при рассмотрении вопросов безопасности принцип «самого слабого звена».

Каждый элемент из множества возможных повреждений, в свою очередь, с определенным «весом» влияет на снижение эффективности одного или нескольких сервисов безопасности, таких как конфиденциальность, целостность, доступность и др. Оценка этого снижения позволяет принять обоснованные решения по восстановлению их уровня до приемлемого значения.

Поскольку оценка степени повреждений средств и сервисов информационной безопасности относится к слабоструктурированным и трудноформализуемым задачам, построение нечеткой когнитивной модели, учитывающей мнения экспертов в данной предметной области, является часто единственно возможным способом ее решения.

### 4.5. Алгоритм управления комплексной безопасностью

На основании вышеизложенного общий алгоритм анализа и управления комплексной безопасностью на основе нечеткого когнитивного моделирования можно представить в следующем виде:

1. Сбор информации об объекте защиты: идентификация активов и установление начального уровня безопасности, которому отвечает ИС. В процессе идентификации следует рассмотреть основные характеристики активов: информационную ценность, чувствительность активов к угрозам, наличие защитных мер. При этом необходимо учесть, что в числе факторов, влияющих на безопасность, особое место занимают субъективные факторы, которые являются наименее прогнозируемыми [9].

2. Выбор критериев, характеризующих состояние различных сторон обеспечения безопасности, определение их приемлемого уровня (возможно в виде интервальных оценок или лингвистических термов).

3. Построение когнитивной модели в виде знакового ориентированного графа с наложенной системой отношений предпочтения типа (4.3).

4. Вычисление весов Фишберна на основании модифицированного метода нестрого ранжирования.

5. Расчет и анализ уровня комплексной безопасности системы (УКБС).

6. Если УКБС не находится в приемлемом диапазоне значений, то производятся изменения в составе концептов, участвующих в построении когнитивной модели, в составе связей между концептами, изменяются их веса посредством введения защитных мероприятий, влияния которых отражаются

МПМ и МЛП. Данные изменения соответствуют различным стратегиям управления безопасностью, приведенным в главе 1: уменьшение рисков, уклонение от рисков, принятие рисков [102].

Таким образом, процесс обеспечения безопасности системы подразумевает решение двух взаимосвязанных задач: прямой (анализ состояния системы) и обратной задачи управления (воздействие на систему).

При решении первой задачи требуется определить значения критериев безопасности $K_i$ и интегрального критерия $K$ при заданных значениях всех влияющих на них концептов.

Если полученные значения находятся вне диапазона приемлемости, то при решении обратной задачи необходимо подобрать такие управляющие воздействия $Z_i$ и $L$, которые обеспечат возвращение целевых критериев в безопасный диапазон.

Если существует не единственный набор необходимых управляющих воздействий, то на этом этапе может возникнуть задача оптимизации, состоящая в нахождении такой комбинации $Z_i$ и $L$, которая обеспечивает максимальное воздействие на негативные факторы при заданных или минимальных затратах на реализацию способов и средств защиты.

### 4.7. Программа для построения нечетких когнитивных моделей

Приведенный выше алгоритм был положен в основу разработки в среде Visual Studio 2008 на языке C# программного продукта для построения нечетких когнитивных моделей [103]. Основной интерфейс программы представлен на рис.4.3.

В главном окне программы имеется несколько вкладок, которые позволяют задать структуру НКМ: количество уровней, количество элементов на каждом уровне, их названия, связи между концептами, а также веса влияния одних факторов на другие и вид свертки, которая используется для вычисления текущего уровня фактора.

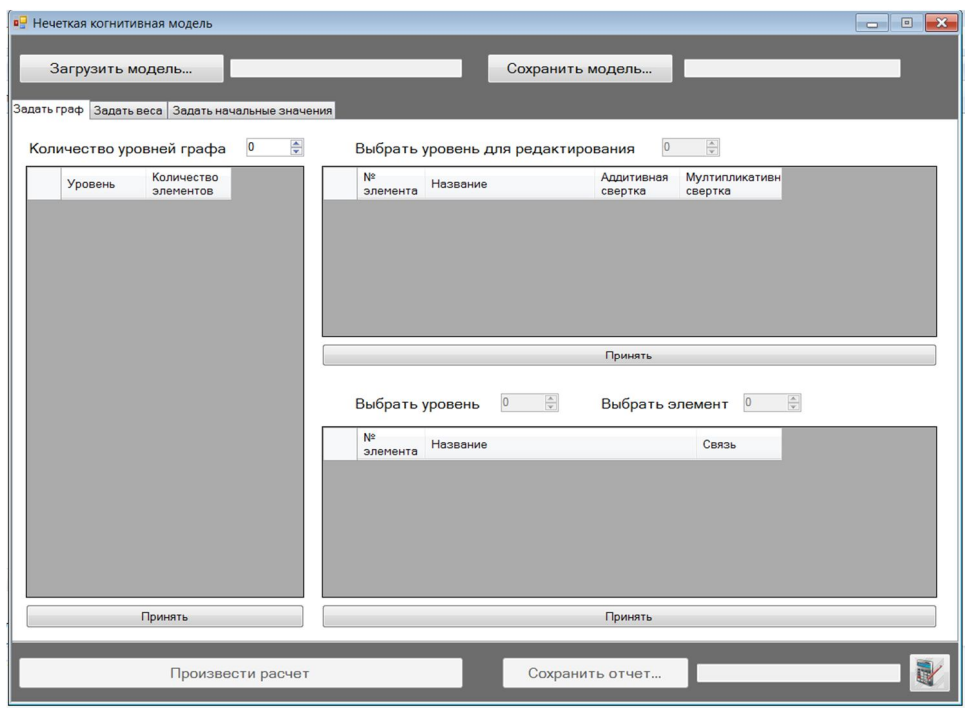

Рис.4.3. Главное окно программы для построения нечетких когнитивных моделей

Перед произведением расчетов на соответствующей вкладке задаются начальные значения концептов нижнего уровня иерархии. При этом можно воспользоваться 3-х, 5-ти или 7-ми уровневым классификатором и выбрать лингвистические значения из выпадающего списка (рис 4.4).

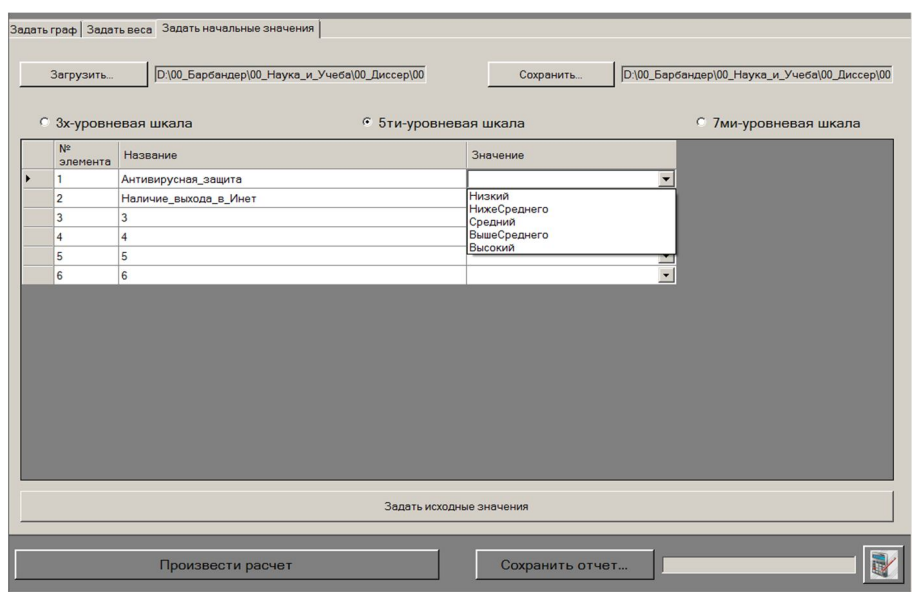

Рис. 4.4. Выбор лингвистических значений концептов нижнего уровня иерархии

Построенную НКМ и набор начальных значений факторов можно сохранить для дальнейшей работы. Можно также вывести в файл подробный отчет с результатами, полученными для заданной структуры НКМ при заданном наборе начальных условий.

При нажатии кнопки в правом нижнем углу главного окна программы запускается калькулятор для расчетов с нечеткими числами, описание которого было дано в главе 3.

Разработанный программный продукт, опираясь на изложенный в предыдущих параграфах алгоритм, позволяет оценивать лингвистические значения целевых концептов в нечеткой когнитивной модели при заданных значениях концептов нижнего уровня иерархии, используя различные виды сверток для комплексных критериев более высокого уровня.

При этом на каждом этапе выполнения расчета определяется индекс схожести полученного результата с «эталонными» значениями выбранного классификатора. Однако на более высокий уровень передается «точное» значение нечеткого числа, не подверженное «огрублению» за счет распознавания. Это позволяет снизить погрешность вычислений и точнее определить лингвистическое значение интегрального критерия.

## 4.8. Выводы по главе 4

Обобщая рассмотренный в главе 4 материал можно сделать следующие основные выводы:

- Безопасность – понятие комплексное и не может рассматриваться как простая сумма составляющих ее частей. Эти части взаимосвязаны и взаимозависимы, каждая часть критически значима.

- Оценка уровня безопасности всегда относительна. Попытки напрямую приписать этой оценке численное значение в большинстве случаев бесперспективны в плане дальнейшей интерпретации результатов.

- Специфическими особенностями задачи создания систем обеспечения безопасности являются:

    – неполнота и неопределенность исходной информации о составе и характере угроз;

    – многокритериальность задачи, связанная с необходимостью учета большого числа частных показателей;

    – наличие как количественных, так и качественных показателей, которые необходимо учитывать при решении задач разработки и внедрения систем защиты;

    – невозможность применения классических методов оптимизации.

- Получение надежной количественной информации для построения комплексной оценки уровня безопасности бывает затруднительным. В таких случаях стремятся получить от ЛПР, в основном, только качественную информацию, которую затем целесообразно формализовать с помощью введения лингвистических переменных и использовать аппарат нечетких множеств.

- Математическая модель оценки уровня комплексной безопасности системы может быть представлена в виде кортежа, включающего в себя ориентированный граф, наложенную на этот граф систему предпочтений одних факторов другим и набор качественных оценок уровней каждого фактора в иерархии.

- Построенная таким образом концептуальная модель позволяет унифицировать подходы к управлению комплексной безопасностью и приступить к разработке соответствующих вычислительных процедур и модулей, которые могут быть в дальнейшем использованы при построении систем поддержки принятия решений.

# ГЛАВА 5.

# ДИНАМИЧЕСКАЯ НЕЧЕТКАЯ КОГНИТИВНАЯ МОДЕЛЬ ОЦЕНКИ БЕЗОПАСНОСТИ ИНФОРМАЦИОННЫХ АКТИВОВ ВУЗА

## 5.1. Специфика образовательных учреждений

Проиллюстрируем общие положения построения НКМ для анализа и управления комплексной безопасностью систем на примере оценки информационной безопасности высшего учебного заведения [104].

В Доктрине информационной безопасности Российской Федерации [1] указывается, что обеспечение *информационной* безопасности играет ключевую роль в обеспечении национальной безопасности РФ. При этом одним из приоритетных направлений государственной политики в данной области является совершенствование подготовки кадров, развитие образования в области информационной безопасности. Особую роль в решении этих задач играют ВУЗы.

Российская высшая школа переживает период адаптации не только к объективным процессам информационного общества, но и к новым социально-политическим условиям с разноплановыми проявлениями конкурентной борьбы.

Создание эффективных механизмов управления информационными ресурсами системы высшего образования в современных условиях невозможно без научного обоснования и практической реализации сбалансированной политики информационной безопасности ВУЗа, которая может быть сформирована на основе решения следующих задач [105]:

- анализ процессов информационного взаимодействия (ИВ) во всех сферах основной деятельности российского ВУЗа (информационных потоков, их масштаба и качества, противоречий, конкурентной борьбы с выявлением собственников и соперников);
- разработка качественного и простого количественного (математического) описания ИВ;

- введение количественных индикаторов и критериев открытости, безопасности и справедливости информационного обмена;
- разработка сценариев необходимости и значимости баланса в информационной открытости и конфиденциальности;
- определение роли и места политики информационной безопасности в управлении информационными ресурсами ВУЗа и выработка согласующихся принципов и подходов;
- формулировка основных составляющих политики: целей, задач, принципов и ключевых направлений обеспечения информационной безопасности
- разработка базовых методик управления процессом обеспечения политики информационной безопасности;
- подготовка проектов нормативно-правовых документов.

В современном ВУЗе хранится и обрабатывается огромное количество различных данных, связанных не только с обеспечением учебного процесса, но и с научно-исследовательскими и проектно-конструкторскими разработками, персональные данные студентов и сотрудников, служебная, коммерческая и иная конфиденциальная информация [106].

Рост преступлений в сфере высоких технологий диктует свои требования к защите ресурсов вычислительных сетей учебных заведений и ставит задачу построения собственной интегрированной системы безопасности. Ее решение предполагает наличие нормативно-правовой базы, формирование концепции безопасности, разработку мероприятий, планов и процедур по безопасной работе, проектирование, реализацию и сопровождение технических средств защиты информации (СЗИ) в рамках образовательного учреждения. Эти составляющие определяют единую политику обеспечения безопасности информации в ВУЗе.

Специфика защиты информации в образовательной системе заключается в том, что ВУЗ - публичное заведение с непостоянной аудиторией, а также место повышенной активности "начинающих кибер-преступников".

Основную группу потенциальных нарушителей здесь составляют студенты, некоторые из которых имеют достаточно высокий уровень подготовки. Возраст - от 18 до 23 лет - и юношеский максимализм побуждает таких людей блеснуть знаниями перед сокурсниками: устроить вирусную эпидемию, получить административный доступ и «наказать» преподавателя, заблокировать выход в Интернет и т.д. Достаточно вспомнить, что первые компьютерные правонарушения родились именно в ВУЗе (червь Морриса) [107].

Особенности ВУЗа как объекта информатизации связаны также с многопрофильным характером деятельности, обилием форм и методов учебной работы, пространственной распределенностью инфраструктуры (филиалы, представительства). Сюда же можно отнести и многообразие источников финансирования, наличие развитой структуры вспомогательных подразделений и служб (строительная, производственная, хозяйственная деятельность), необходимость адаптации к меняющемуся рынку образовательных услуг, потребность в анализе рынка труда, отсутствие общепринятой формализации деловых процессов, необходимость электронного взаимодействия с вышестоящими организациями, частое изменение статуса сотрудников и обучаемых [108].

При этом несколько облегчает проблему то, что ВУЗ представляет собой стабильную, иерархическую по функциям управления систему, обладающую всеми необходимыми условиями жизнедеятельности и действующую на принципах централизованного управления (последнее означает, что в управлении задачами информатизации может активно использоваться административный ресурс).

Указанные выше особенности обуславливают необходимость соблюдения следующих требований:

- комплексной проработки задач информационной безопасности, начиная с концепции ИБ и заканчивая сопровождением программно-технических решений;
- привлечения большого числа специалистов, владеющих содержательной частью деловых процессов;
- использования модульной структуры корпоративных приложений, когда каждый модуль покрывает взаимосвязанную группу деловых процедур или информационных сервисов при обеспечении единых требований к безопасности;
- применения обоснованной последовательности этапов в решении задач информационной безопасности;
- документирования разработок на базе разумного применения стандартов, что гарантирует создание успешной системы;
- использования надежных и масштабируемых аппаратно-программных платформ и технологий различного назначения, обеспечивающих необходимый уровень безопасности.

С точки зрения архитектуры в корпоративной информационной среде можно выделить три уровня, для обеспечения безопасного функционирования которых необходимо применять различные подходы:
- оборудование вычислительной сети, каналов и линий передачи данных, рабочих мест пользователей, системы хранения данных;
- операционные системы, сетевые службы и сервисы по управлению доступом к ресурсам, программное обеспечение среднего слоя;
- прикладное программное обеспечение, информационные сервисы и среды, ориентированные на пользователей.

При создании комплексной информационной сети (КИС) необходимо обеспечить межуровневое согласование требований по безопасности к выбираемым решениям или технологиям. Так, на втором уровне архитектура КИС многих ВУЗов представляет собой разрозненные и слабо связанные подсистемы с разными операционными средами, согласованные друг с другом только на уровне закрепления IP-адресов или обмена сообщениями. Причинами плохой системной организации КИС является отсутствие утвержденной архитектуры КИС, наличие нескольких центров ответственности за развитие технологий, которые действуют не согласованно. Проблемы начинаются с нежелания управлять выбором операционных сред в подразделениях, когда ключевые технологические решения полностью децентрализованы, что резко снижает уровень безопасности системы.

ВУЗы, имеющие четкую стратегию развития ИТ, единые требования к информационной инфраструктуре, политику информационной безопасности и утвержденные регламенты на основные компоненты КИС, отличаются, как правило, сильным административным ядром в управлении и высоким авторитетом руководителя ИТ-службы [109].

В таких ВУЗах могут, конечно, использоваться различные операционные среды или системы среднего слоя, но это обусловлено организационно-техническими или экономическими причинами и не препятствует развертыванию КИС ВУЗа и внедрению унифицированных принципов безопасного доступа к информационным ресурсам

Состояние развития в ВУЗах третьего уровня архитектуры КИС можно охарактеризовать следующим образом: в основном завершен переход от локальных программных приложений, автоматизирующих отдельный деловой процесс и опирающихся на локальный набор данных, к корпоративным клиент-серверным информационным системам, обеспечивающим доступ пользователей к оперативным базам данных ВУЗа. В том или ином виде решена задача интеграции данных, порожденных различными информационными системами, что позволяет усовершенствовать бизнес-процессы, повысить качество управления и принятия решений.

Если в начале 90-х годов был высокий спрос на бухгалтерское программное обеспечение и программное обеспечение управленческого учета (учет кадров, отчетность и т. д.), то сейчас этот спрос в большей части удовлетворен. В настоящее время стоит задача обеспечить достоверными данными о деятельности образовательного учреждения не только управленческий персонал, но и каждого преподавателя и студента. То есть стоит задача эффективного управления данными, циркулирующими в КИС, что в свою очередь делает задачу обеспечения информационной безопасности в таких сетях еще более актуальной.

### 5.2. Особенности корпоративных сетей ВУЗов

Активное внедрение ИНТЕРНЕТ и новых информационных технологий в образовательный процесс и систему управления ВУЗом создали предпосылки к появлению корпоративных сетей.

Корпоративной сетью ВУЗа - это информационная система, включающая в себя компьютеры, сервера, сетевое оборудование, средства связи и телекоммуникации, систему программного обеспечения предназначенная для решения задач управления ВУЗом и ведения образовательной деятельности.

Корпоративная сеть обычно объединяет не только структурные подразделения ВУЗа, но и их региональные представительства. Ранее недоступные для ВУЗа, сегодня эти сети стали активно внедряться в образовательные структуры в связи с массовым распространением ИНТЕРНЕТ и его доступностью [110].

Комплексная информационная безопасность ВУЗа - система сохранения, ограничения и авторизованного доступа к информации, содержащейся на серверах в корпоративных сетях ВУЗов, а также передаваемая по телекоммуникационным каналам связи в системах дистанционного обучения.

В более широком смысле термин комплексная информационная безопасность ВУЗа включает два аспекта: систему защиты интеллектуальной информационной собственности ВУЗа от внешних и внутренних агрессивных воздействий и систему управления доступом к информации и защиты от агрессивных информационных пространств. В последнее время, в связи с неконтролируемым массовым развитием ИНТЕРНЕТ, последний аспект безопасности становится особенно актуальным.

Под термином информационное пространство понимается информация, содержащаяся на серверах в корпоративных сетях учебных заведений, учреждений, библиотек и в глобальной сети ИНТЕРНЕТ, на электронных носителях информации, а также передаваемая по телевизионным каналам связи или по телевидению.

Агрессивное информационное пространство – это информационное пространство, содержание которого может вызвать проявления агрессии у пользователя как сразу же после информационного воздействия, так и через некоторое время (отдаленный эффект).

Термин основан на гипотезе, что информация в определенных формах и содержании может вызвать определенные эффекты с проявлением агрессии и враждебности [110].

Проблемы комплексной информационной безопасности корпоративных сетей ВУЗов гораздо шире, разнообразнее и острее, чем в других системах. Это связано со следующими особенностями:

• корпоративная сеть ВУЗа строится обычно на концепции «скудного финансирования» (оборудование, кадры, нелицензионное программное обеспечение);

• как правило, корпоративные сети не имеют стратегических целей развития. Это значит, что топология сетей, их техническое и программное обеспечение рассматриваются с позиций текущих задач;

• в одной корпоративной сети ВУЗа решаются две основные задачи: обеспечение образовательной и научной деятельности и решение задачи управления образовательным и научным процессами. Это означает, что одновременно в этой сети работает несколько автоматизированных систем или подсистем в рамках одной системы управления (АСУ «Студент», АСУ «Кадры», АСУ «Учебный процесс», АСУ «Библиотека», АСУ «НИР» АСУ «Бухгалтерия» и т.д.);

- корпоративные сети гетерогенны как по оборудованию, так и по программному обеспечению в связи с тем, создавались в течение длительного периода времени для разных задач;
- планы комплексной информационной безопасности, как правило, либо отсутствуют, либо не соответствуют современным требованиям.

В такой сети возможны как внутренние, так и внешние угрозы безопасности информации:
- попытки несанкционированного администрирования баз данных;
- исследование сетей, несанкционированный запуск программ по аудиту сетей;
- удаление информации, в том числе библиотек;
- запуск игровых программ;
- установка вирусных программ и троянских коней;
- попытки взлома АСУ «ВУЗ»;
- сканирование сетей, в том числе других организаций через ИНТЕРНЕТ;
- несанкционированная откачка из ИНТЕРНЕТА нелицензионного софта и установка его на рабочие станции;
- попытки проникновения в системы бухгалтерского учета;
- поиск «дыр» в ОС, firewall, Proxy – серверах;
- попытки несанкционированного удаленного администрирования ОС;
- сканирование портов и т.п.

## 5.3. Угрозы информационной безопасности ВУЗа и анализ рисков

Для анализа информационных рисков необходимо:
- провести классификацию объектов, подлежащих защите, и ранжировать их по степени важности;
- определить привлекательность объектов защиты для злоумышленников;
- определить возможные угрозы информационной безопасности;
- рассмотреть вероятные пути их реализации (уязвимости);
- оценить эффективность существующих мер обеспечения безопасности;
- провести оценку ущерба от возможных атак на информационные ресурсы.

Основными объектами ВУЗа, нуждающимися в защите, являются:
- бухгалтерские ЛВС, данные планово-финансового отдела, а также статистические и архивные данные;
- серверы баз данных;
- консоль управления учетными записями;
- www/ftp сервера;
- ЛВС и серверы исследовательских проектов.

Согласно приведенной в главе 1 классификации с учетом рассмотренных в п.5.2. особенностей можно выделить следующие угрозы информационным активам ВУЗа:

*Непреднамеренные субъективные угрозы ИБ:*
- $UG_1$ - Угроза неумышленного повреждения оборудования;
- $UG_2$ - Угроза неправомерного отключения оборудования;
- $UG_3$ - Угроза неумышленного удаления файлов с важной информацией;
- $UG_4$ - Угроза неумышленного искажения файлов с важной информацией;
- $UG_5$ - Угроза неумышленного удаления программ;
- $UG_6$ - Угроза неумышленного внесения изменений в программы;
- $UG_7$ - Угроза неправомерного изменения режимов работы устройств;
- $UG_8$ - Угроза неправомерного изменения режимов работы программ;

- $UG_9$ - Угроза неумышленной порчи носителей информации;
- $UG_{10}$ - Угроза некомпетентного запуска технологических программ;
- $UG_{11}$ - Угроза использования нелегального программного обеспечения;
- $UG_{12}$ - Угроза заражения компьютера вирусами;
- $UG_{13}$ - Угроза неосторожных действий, приводящих к нарушению конфиденциальности;
- $UG_{14}$ - Угроза разглашения, передачи или утраты атрибутов разграничения доступа;
- $UG_{15}$ - Угроза игнорирования организационных ограничений при работе в ИС;
- $UG_{16}$ - Угроза входа в систему в обход средств защиты;
- $UG_{17}$ - Угроза совершения непреднамеренной ошибки при проектировании и обслуживании ИС;
- $UG_{18}$ - Угроза некомпетентного использования и настройки средств ЗИ;
- $UG_{19}$ - Угроза неправомерного отключения средств ЗИ персоналом СБ;
- $UG_{20}$ - Угроза пересылки данных по ошибочному адресу абонента;
- $UG_{21}$ - Угроза ввода ошибочных данных.

*Преднамеренные субъективные угрозы ИБ:*
- $UG_{22}$ - Угроза преднамеренного физического разрушения системы;
- $UG_{23}$ - Угроза вывода из строя наиболее важных компонентов ИС;
- $UG_{24}$ - Угроза отключения подсистем обеспечения ИС;
- $UG_{25}$ - Угроза вывода из строя подсистем обеспечения ИС;
- $UG_{26}$ - Угроза преднамеренного нарушения режимов эксплуатации устройств или режимов использования ПО;
- $UG_{27}$ - Угроза саботажа со стороны персонала;
- $UG_{28}$ - Угроза внедрения агентов в персонал ИС;
- $UG_{29}$ - Угроза вербовки персонала;
- $UG_{30}$ - Угроза применения технических средств разведки;

- $UG_{31}$ - Угроза перехвата ПЭМИН;
- $UG_{32}$ - Угроза перехвата информации в каналах связи и их анализа;
- $UG_{33}$ - Угроза осуществления «маскарада»;
- $UG_{34}$ - Угроза хищения носителей информации;
- $UG_{35}$ - Угроза несанкционированного копирования информации;
- $UG_{36}$ - Угроза хищения производственных отходов;
- $UG_{37}$ - Угроза чтения остаточной информации из ОЗУ;
- $UG_{38}$ - Угроза чтения остаточной информации с внешних ЗУ;
- $UG_{39}$ - Угроза незаконного получение реквизитов разграничения доступа;
- $UG_{40}$ - Угроза вскрытия шифров криптозащиты информации;
- $UG_{41}$ - Угроза внедрения аппаратных "спецвложений";
- $UG_{42}$ - Угроза внедрения программных "закладок" и "вирусов";
- $UG_{43}$ - Угроза работы "между строк";
- $UG_{44}$ - Угроза навязывания ложных сообщений (фальсификация);
- $UG_{45}$ - Угроза модификации потока данных;

*Техногенные угрозы:*

- $UG_{46}$ - Угроза сбоя технических средств обработки информации (ТСОИ);
- $UG_{47}$ - Угроза отказа ТСОИ;
- $UG_{48}$ - Угроза сбоя вспомогательных технических средств;
- $UG_{49}$ - Угроза отказа вспомогательных технических средств;
- $UG_{50}$ - Угроза сбоя системы электроснабжения;
- $UG_{51}$ - Угроза сбоя системы климат-контроля;
- $UG_{52}$ - Угроза сбоя внешних информационных каналов коммуникации;
- $UG_{53}$ - Угроза отказа внешних информационных каналов коммуникации;
- $UG_{54}$ - Угроза отказа систем водоснабжения и канализации на объекте информатизации (ОИ);
- $UG_{55}$ - Угроза ошибочного срабатывания систем пожаротушения на ОИ;
- $UG_{56}$ - Угроза ошибочного срабатывания охранной сигнализации на ОИ.

*Стихийные угрозы:*

- $UG_{57}$ - Угроза пожара;
- $UG_{58}$ - Угроза наводнения;
- $UG_{59}$ - Угроза землетрясения;
- $UG_{60}$ - Угроза поражения молнией;
- $UG_{61}$ - Угроза урагана;
- $UG_{62}$ - Угроза пыльной бури;
- $UG_{63}$ - Угроза экстремально высокой температуры на ОИ;
- $UG_{64}$ - Угроза экстремально низкой температуры на ОИ.

Каждая из перечисленных угроз $UG_i$ имеет априорную вероятность возникновения, зависящую от привлекательности информационного актива для злоумышленника, уровня его квалификации, состояния внешней инфраструктуры, окружающей объект информатизации, климатических условий, месторасположения объекта и т.д. Любая из угроз множества $UG = \{UG_j\}$ может реализоваться в виде атаки на информационные активы при наличии соответствующей уязвимости. Рассмотрим основные уязвимости, характерные для ИС ВУЗа.

## 5.4. Уязвимости информационных систем ВУЗа

К основным уязвимостям информационных систем образовательных учреждений можно отнести:

- $UZ_1$ - Отсутствие утверждённой концепции ИБ
- $UZ_2$ - Наличие ПЭМИН
- $UZ_3$ - Ошибки в ПО
- $UZ_4$ - Несоблюдение режима охраны на ОИ
- $UZ_5$ – Нештатные режимы эксплуатации ТС
- $UZ_6$ - Нештатные режимы использования ПО
- $UZ_7$ - Несоблюдение мер обеспечения ИБ
- $UZ_8$ - Низкая надежность ТСОИ
- $UZ_9$ - Низкая надежность вспомогательных ТС
- $UZ_{10}$ - Старение и размагничивание носителей информации
- $UZ_{11}$ - Низкая надежность ПО
- $UZ_{12}$ - Низкое качество внутренней системы электроснабжения
- $UZ_{13}$ - Отсутствие резервных источников электропитания
- $UZ_{14}$ - Низкое качество внешних сетей электроснабжения
- $UZ_{15}$ - Низкая надежность внутренней системы водоснабжения и канализации
- $UZ_{16}$ - Низкая надежность внешних сетей водоснабжения и канализации
- $UZ_{17}$ - Отсутствие на ОИ систем климат-контроля
- $UZ_{18}$ - Плохое качество систем климат-контроля
- $UZ_{19}$ - Плохое качество внешних коммуникационных каналов
- $UZ_{20}$ - Низкая надежность внешних коммуникационных каналов
- $UZ_{21}$ - Отсутствие противопожарной системы на ОИ
- $UZ_{22}$ - Несоблюдение мер противопожарной безопасности
- $UZ_{23}$ - Плохое качество системы обнаружения пожара на ОИ
- $UZ_{24}$ - Плохое качество системы охранной сигнализации на ОИ

- $UZ_{25}$ - Несоблюдение мер защиты ОИ от затопления
- $UZ_{26}$ - Отсутствие мер обеспечения сейсмоустойчивости ОИ
- $UZ_{27}$ - Отсутствие регламента действий при чрезвычайных ситуациях
- $UZ_{28}$ - Отсутствие мер защиты от молний
- $UZ_{29}$ - Отсутствие мер защиты от ураганов
- $UZ_{30}$ - Отсутствие мер по очистке воздуха
- $UZ_{31}$ - Наличие повреждений жизнеобеспечивающих коммуникаций
- $UZ_{32}$ - Наличие повреждений ограждающих ОИ конструкций
- $UZ_{33}$ - Наличие «слабых» паролей доступа к ресурсам ИС
- $UZ_{34}$ - Наличие незаблокированных встроенных учётных записей
- $UZ_{35}$ - Неправильно установлены права доступа к информ ресурсам
- $UZ_{36}$ - Наличие неиспользуемых потенциально опасных служб и ПП
- $UZ_{37}$ - Неправильная конфигурация средств защиты
- $UZ_{38}$ - Низкий уровень квалификации обслуживающего ИС персонала
- $UZ_{39}$ - Низкий уровень квалификации пользователей
- $UZ_{40}$ - Неправильно организован доступ к оборудованию ИС
- $UZ_{41}$ - Неправильно реализовано разграничение доступа к ПО
- $UZ_{42}$ - Неправильно определены права пользователей
- $UZ_{43}$ - Неправильно организовано хранение носителей информации
- $UZ_{44}$ - Неправильно организован учет носителей информации
- $UZ_{45}$ - Отсутствует или неправильно организована система антивирусной защиты (АВЗ)
- $UZ_{46}$ - Слабая подготовка персонала и пользователей в вопросах ЗИ
- $UZ_{47}$ - Неправильно организованы выдача и учет атрибутов разграничения доступа
- $UZ_{48}$ - Слабый контроль за соблюдением политики безопасности на ОИ
- $UZ_{49}$ - Невнимательность персонала и пользователей ИС
- $UZ_{50}$ - Слабая система охраны на ОИ
- $UZ_{51}$ - Отсутствие контроля за подсистемами жизнеобеспечения на ОИ

- $UZ_{52}$ - Низкий уровень лояльности персонала
- $UZ_{53}$ - Отсутствие контроля за помещениями на ОИ
- $UZ_{54}$ - Слабая кадровая политика в организации
- $UZ_{55}$ - Отсутствие средств противодействия ТСР
- $UZ_{56}$ - Неправильно выполненное заземление оборудования
- $UZ_{57}$ - Недостаточная защита каналов связи
- $UZ_{58}$ - Недостаточная эффективность механизмов идентификации и аутентификации пользователей ИС
- $UZ_{59}$ - Отсутствие инструкции по уничтожению документов
- $UZ_{60}$ - Несоблюдение инструкции по уничтожению документов
- $UZ_{61}$ - Неправильное удаление информации с запоминающих устройств
- $UZ_{62}$ - Наличие доступа к ОЗУ во время обработки информации
- $UZ_{63}$ - Недостаточная стойкость систем криптозащиты
- $UZ_{64}$ - Отсутствие или низкое качество аудита безопасности

Таким образом, определено множество уязвимостей для информационных активов ВУЗа $UZ = \{UZ_k\}$.

## 5.5. Рубежи защиты информационных систем ВУЗа и меры по обеспечению информационной безопасности

Первый рубеж обороны от атак извне (Интернет) - роутер (маршрутизатор). Он применяется для связи участков сети друг с другом, а также для более эффективного разделения трафика и использования альтернативных путей между узлами сети. От его настроек зависит функционирование подсетей и связь с глобальными сетями (WAN). Его главная задача в плане безопасности - защита от распределенных атак в отказе обслуживания (DDOS).

Вторым рубежом может служить межсетевой экран (МСЭ): например, аппаратно-программный комплекс Cisco PIX Firewall.

Затем следует демилитаризованная зона (DMZ). В этой зоне необходимо расположить главный прокси-сервер, DNS-сервер, www/ftp, mail сервера. Прокси-сервер обрабатывает запросы от рабочих станций учебного персонала, серверов, не подключенных напрямую к роутеру, и фильтрует трафик. Политика безопасности на этом уровне должна определяться блокированием нежелательного трафика и его экономией (фильтрация мультимедиаконтента, ISO-образов, блокировка страниц нежелательного/нецензурного содержания по ключевым словам). Чтобы не происходило скачивания зараженной вирусами информации, на этом сервере оправдано размещение антивирусных средств.

Информация от прокси-сервера должна параллельно отсылаться на сервер статистики, где можно посмотреть и проанализировать деятельность пользователей в Интернете. На почтовом сервере обязательно должен присутствовать почтовый антивирус, к примеру, Kaspersky AntiVirus for Mail servers.

Так как эти серверы связаны непосредственно с глобальной сетью, аудит программного обеспечения, установленного на них, - первоочередная задача инженера по информационной безопасности ВУЗа. Для экономии средств и гибкости настраивания желательно применять opensource ОС и программное обеспечение.

Одни из самых распространенных ОС - FreeBSD и GNU Linux. Но ничто не мешает использовать и более консервативную Open BSD или даже сверхстабильную ОС реального времени QNX.

Для централизованного управления антивирусной деятельностью необходим продукт с клиент-серверной архитектурой, такой как Dr.Web Enterprise Suite. Он позволяет централизованно управлять настройками и обновлением антивирусных баз с помощью графической консоли и предоставлять удобочитаемую статистику о вирусной деятельности, если таковая присутствует.

Для большего удобства работников ВУЗа можно организовать доступ к внутренней сети университета с помощью технологии VPN.

Некоторые университеты имеют свой пул дозвона для выхода в Интернет и используют каналы связи учреждения. Во избежание использования этого доступа посторонними лицами в незаконных целях работники учебного заведения не должны разглашать телефон пула, логин, пароль.

Степень защищенности сетей и серверов большинства ВУЗов России оставляет желать лучшего. Причин тому много, но одна из главных - плохая организация мер по разработке и обеспечению политики ИБ и недооценка важности этих мероприятий. Вторая проблема заключается в недостаточном финансировании закупок оборудования и внедрения новых технологий в сфере ИБ.

Система комплексной информационной безопасности должна включать в себя выработку следующих политик.

Прежде всего, это финансовая политика развертывания, развития и поддержания в актуальном состоянии корпоративной сети ВУЗа. Она является доминирующей и ее можно разделить на три направления: скудного финансирования, финансирование с разумной достаточностью и приоритетное финансирование.

Вторая политика определяется уровнем организации развертывания и сопровождения корпоративной сети ВУЗа.

Третья политика относится к кадровому составу информационного центра. Для ВУЗа она особенно актуальна в связи с повышенной востребованностью опытных сисадминов.

Политика программного обеспечения сегодня один из затратных факторов развития корпоративной сети. Рациональные подходы к его решению в условия монопольного рынка ОС и программных продуктов MicroSoft – это отдельный вопрос, требующий внимательного рассмотрения.

Политика технического обеспечения – может быть и не вполне актуальна в условиях достаточного финансирования. Но всегда существует проблема обновления устаревшего оборудования.

Наконец, последняя политика связана с формирование морально-этических норм толерантного поведения в информационных системах и разумного ограничения от посещений агрессивных информационных пространств. Недооценка этих направлений будет компенсироваться повышенными финансовыми затратами на сопровождение корпоративных сетей ВУЗов.

Таким образом, к основным мерам обеспечения ИБ в ВУЗе можно отнести:

- $Z_1$ - Организацию процедуры хранения документов
- $Z_2$ - Разработку процедуры оперативного реагирования на инциденты
- $Z_3$ - Административные и технические средства контроля за работой пользователей
- $Z_4$ - Использование сертифицированного лицензионного ПО
- $Z_5$ - Разграничение доступа к ПО
- $Z_6$ - Техническую поддержку аппаратных ресурсов
- $Z_7$ - Резервное копирование
- $Z_8$ - Обучение сотрудников основам ИБ
- $Z_9$ - Формирование корпоративной культуры
- $Z_{10}$ - Мероприятия по предотвращению конфликтов в коллективе
- $Z_{11}$ - Разработку внутренней нормативной документации по ИБ
- $Z_{12}$ - Организацию системы физической защиты
- $Z_{13}$ - Организацию противопожарной защиты
- $Z_{14}$ - Организацию контроля выполнения мер по ИБ
- $Z_{15}$ - Организацию пропускного режима и охраны
- $Z_{16}$ - Комплексное планирование мероприятий по ЗИ
- $Z_{17}$ - Организацию аналитической работы и контроля
- $Z_{18}$ - Построение оптимальных с точки зрения ИБ инженерных сетей
- $Z_{19}$ - Применение специальных технических средств ЗИ и контроля обстановки
- $Z_{20}$ - Резервирование средств и каналов связи
- $Z_{21}$ - Применение средств криптографической ЗИ

- $Z_{22}$ - Применение средств разграничения доступа к информ ресурсам
- $Z_{23}$ - Применение средств межсетевого экранирования
- $Z_{24}$ - Применение средств анализа защищённости ИС
- $Z_{25}$ - Применение средств обнаружения атак
- $Z_{26}$ - Разработку и внедрение концепции антивирусной защиты
- $Z_{27}$ - Разработку процедуры восстановления после атак на ИС
- $Z_{28}$ - Средства контентного анализа
- $Z_{29}$ - Средства защиты от спама
- $Z_{30}$ – Правильное заземление оборудования
- $Z_{31}$ – Наличие системы резервного электропитания

Перечисленные средства и меры защиты информации образуют множество мер противодействия $Z = \{Z_i\}$.

## 5.6. Динамическая нечеткая когнитивная модель оценки уровня информационной безопасности ВУЗа

Для построения НКМ оценки уровня ИБ ВУЗа согласно методике, изложенной в главе 4 (см. п.4.4), необходимо связать приведенные выше данные по составу угроз, уязвимостей, средств защиты между собой и рассмотреть влияние потенциально возможных атак на основные сервисы информационной безопасности.

При этом на нижнем уровне иерархии $G$ располагаются средства и механизмы защиты $Z$, действия которых уменьшают вероятность возникновения угроз $UG$ и ослабляют степень уязвимостей $UZ$, расположенных на уровень выше. Угрозы и уязвимости, в свою очередь предопределяют вероятность возникновения атак $A$, которые негативно влияют на сервисы безопасности $SRV$.

В качестве наиболее значимых для ВУЗа сервисов безопасности выберем конфиденциальность, целостность и доступность. Эти сервисы в совокупности определяют интегральный показатель комплексной информационной безопасности учебного заведения *K* и влияют на рейтинговые показатели ВУЗа: репутацию, материально-техническое состояние, финансовую устойчивость, качество образовательного процесса и т.п.

В каждой из информационных систем ВУЗа (АСУ «Деканат», АСУ «Учебный процесс», АСУ «Бухгалтерия», АСУ «Абитуриент» и т.д.) необходимо выделить и проранжировать по степени значимости информационные активы, которые могут быть подвержены атакам.

Из множеств *UG* и *UZ* нужно отобрать угрозы и уязвимости, характерные для каждого из активов. Затем необходимо составить перечень имеющихся в ВУЗе средств обеспечения ИБ и проанализировать влияние этих средств на угрозы и уязвимости, выявленные на предыдущем этапе.

При этом все оценки задаются с помощью лингвистической переменной «Уровень фактора» с терм-множеством *QL*={Низкий(Н), Ниже среднего (НС), Средний (С), Выше среднего (ВС), Высокий(В)} и соответствующим данному терм-множеству пятиуровневым классификатором (3.1-3.5).

Вычисления значений факторов производится по следующим формулам:

$$\overline{UZ}_j^{mn} = UZ_j^{mn} \cdot \prod_i Inv(Z_i^{mn})^{r_i^{mn}} ; \quad (5.1)$$

$$\overline{UZ}^{mn} = Inv\left[\prod_j Inv(\overline{UZ}_j^{mn})^{s_j^{mn}}\right] ; \quad (5.2)$$

$$\overline{UG}_n^m = UG_n^m \cdot \prod_i Inv(Z_i^{mn})^{v_i^{mn}} , \quad (5.3)$$

где *Inv*(*F*) = 1 - μ$_F$ – инверсия фактора *F* (см. п.4.4); $\overline{UZ}_j^{mn}$ - остаточный (после применения средств защиты) уровень *j*-й уязвимости *m*-го актива относительно *n*-й угрозы; $UZ_j^{mn}$ – исходный (до применения средств защиты) уровень *j*-й уязвимости *m*-го актива относительно *n*-й угрозы; $Z_i^{mn}$ – уровень *i*-й защитной меры по отношению к *n*-й угрозе *m*-му активу; $r_i^{mn}$ – весовой коэффициент, отражающий «вклад» *i*-й защитной меры в снижение уровня *n*-й угрозы *m*-у активу; $\overline{UZ}^{mn}$ - интегральный уровень уязвимости *m*-го актива по отношению к *n*-й угрозе; $s_j^{mn}$ – весовой коэффициент, отражающий «вклад» $\overline{UZ}_j^{mn}$ в $\overline{UZ}^{mn}$; $\overline{UG}_n^m$ – остаточная вероятность существования угрозы $UG_n$ для *m*-го актива после применения совокупности средств защиты $Z_i^{mn}$; $UG_n^m$ – исходная вероятность существования угрозы $UG_n$ *m*-му активу; $v_i^{mn}$ – весовой коэффициент, отражающий «вклад» элемента защиты $Z_i^{mn}$ в уменьшение угрозы $UG_n^m$.

Необходимо отметить, что все упомянутые выше параметры в общем случае могут являться функциями времени *t*, и при проведении динамических расчетов необходимо задать их начальные значения при *t* = 0 и выбрать шаг по времени $\Delta t$.

Реализация *n*-й угрозы *m*-му активу $\overline{UG}_n^m$ с использованием оставшихся уязвимостей $\overline{UZ}^{mn}$ порождает атаку $A_n^m$, вероятность возникновения которой может быть оценена по формуле:

$$A_n^m = \overline{UG}_n^m \cdot \overline{UZ}^{mn} \quad (5.4)$$

Если данная величина отлична от нуля, т.е. несмотря на предпринятые защитные мероприятия, в некоторый момент времени $t^{mn}$ атака $A_n^m$ все же возникла, необходимо задействовать меры $Z_j^{mn}$ по снижению уровня таких нарушений безопасности (инцидентов).

Данные меры могут быть активизированы не сразу, а спустя некоторое время $(t_j^{mn})_{\text{нач}}$, необходимое для идентификации атаки и принятия решения о реагировании на нее. При достижении времени $(t_j^{mn})_{\text{кон}}$ действие этих мер прекращается либо в связи с окончанием инцидента, либо в связи с исчерпанностью ресурсов, обеспечивающих сдерживание атаки.

Во время действия данных мер остаточный уровень *n*-й атаки на *m*-й ресурс $\bar{A}_n^m$ может быть найден по формуле:

$$\bar{A}_n^m = A_n^m \cdot \prod_j Inv(Z_j^{mn})^{\gamma_j^{mn}}; \quad (5.5)$$

где $\gamma_j^{mn}$ – весовой коэффициент, отражающий вклад меры $Z_j^{mn}$ в снижение уровня *n*-го инцидента (атаки) по отношению к *m*-му активу.

Инциденты безопасности при условии, что их уровень выше некоторого критического порога $(\bar{A}_n^m)_{\text{крит}}$ и продолжительность больше некоторого критического интервала времени $(t^{mn})_{\text{крит}}$, могут порождать новые или усиливать уже имеющиеся уязвимости системы, что должно найти отражение при оценке их уровня на следующем шаге по времени:

$$UZ_j^{mn}(t + \Delta t) = UZ_j^{mn}(t) \cdot Inv\left[\prod_n Inv(\bar{A}_n^m)^{\delta_j^{mn}}\right]; \quad (5.6)$$

Совокупность атак на *m*-й актив, в свою очередь, предопределяет уровень обеспеченности сервисов безопасности данного актива $SRV_k^m$:

$$SRV_k^m = \prod_n Inv(\bar{A}_n^m)^{w_n^{mk}} \quad (5.7)$$

Если уровень какого-либо сервиса падает ниже критического значения необходимо предпринять меры по его восстановлению, т.е. реализовать мероприятия блока ликвидации последствий (см. п.4.3). Действия на этом шаге можно формализовать с помощью следующей формулы:

$$\overline{SRV}_k^m = SRV_k^m \cdot Inv\left[\prod_j Inv(Z_j^{mk})^{\theta_j^{mk}}\right]; \quad (5.8)$$

где $\overline{SRV}_k^m$ – уровень *k*-го сервиса *m*-го актива после реализации мер ликвидации последствий $Z_j^{mk}$; $\theta_j^{mk}$ - весовой коэффициент, отражающий вклад меры $Z_j^{mk}$ в повышение уровня *k*-го сервиса безопасности *m*-го актива.

Так же как и в случае с мерами по снижению уровня инцидентов, до активизации данных мер проходит некоторое время $(t_j^{mk})_{\text{нач}}$, необходимое для идентификации уровня сервисов безопасности и принятия решения о необходимости их повышения. Действие этих мер прекращается в некоторый момент времени $(t_j^{mk})_{\text{кон}}$ либо в связи с достижением нужного уровня сервиса безопасности, либо в связи с исчерпанностью ресурсов для его повышения.

В качестве примера, иллюстрирующего приведенные выше рассуждения можно рассмотреть следующую ситуацию. Предположим, что, несмотря на предпринятые превентивные меры, на объекте информатизации возник пожар. В этом случае для уменьшения уровня инцидента включается система пожаротушения. Возникновение пожара резко увеличит уровень других уязвимостей системы и создаст уязвимости, отсутствовавшие при штатной работе ОИ. Система пожаротушения отработает положенное время и отключится. Если за это время пожар не будет локализован и потушен, с высокой долей вероятности доступность информационных ресурсов будет нарушена. Возникнет необходимость ее восстановления, например, путем подключения к резервному (удаленному) серверу. Однако ресурсов резервного сервера и каналов связи с ним может оказаться недостаточно, чтобы длительное время поддерживать доступность ИС на нужном уровне. Следовательно, необходимо предусмотреть варианты полного восстановления сервисов за время, не превышающее время безопасной работы по резервному каналу.

Основываясь на значениях сервисов безопасности $SRV_k^m$ для $m$-го актива можно определить интегральный уровень безопасности $K^m$:

$$K^m = \prod_k (\overline{SRV_k^m})^{\alpha_k^m} \ , \qquad (5.9)$$

где $\alpha_k^m$ - весовой коэффициент, отражающий «вклад» $k$-го сервиса в интегральную оценку уровня безопасности $m$-го актива.

Интегральный показатель $K$ комплексной безопасности всего учебного заведения может быть найден по формуле:

$$K = \prod_m (K^m)^{\beta_m} \ , \qquad (5.10)$$

где $\beta_m$ - весовой коэффициент, отражающий значимость $m$-го информационного актива в интегральной оценке комплексного уровня безопасности образовательного учреждения.

Можно продолжить построение когнитивной модели с целью оценки рейтинговых показателей деятельности ВУЗа $R_i$ в зависимости от степени безопасности основных информационных активов $K^m$. Однако необходимо учесть, что значения $R_i$ зависят от ряда других показателей, не имеющих прямого отношения к безопасности информационных систем ВУЗа. Поэтому решение этой задачи выходит за рамки данного исследования.

Таким образом, построенная динамическая нечеткая когнитивная модель дает возможность, последовательно пройдя все уровни ее иерархии и применяя для свертки параметров формулы (5.1)-(5.10), моделировать поведение системы во времени, оценивать уровень безопасности информационных активов ВУЗа и выработать рекомендации по его повышению.

## 5.7. Выводы по главе 5

1. Управление информационными ресурсами высшего учебного заведения в современных условиях невозможно без научного обоснования и практической реализации сбалансированной политики информационной безопасности.

2. Высшие учебные заведения обладают рядом особенностей, которые необходимо учесть при построении системы информационной безопасности. Специфика защиты информации в образовательном учреждении заключается в том, что это - публичные заведения с непостоянной аудиторией, а также место повышенной активности "начинающих кибер-преступников". Кроме того, в современном ВУЗе хранится и обрабатывается огромное количество различных данных, связанных не только с обеспечением учебного процесса, но и с научно-исследовательскими и проектно-конструкторскими разработками, персональные данные студентов и сотрудников, служебная, коммерческая и иная конфиденциальная информация. Особенности ВУЗа как объекта информатизации связаны также с многопрофильным характером деятельности, обилием форм и методов учебной работы, пространственной распределенностью инфраструктуры (филиалы, представительства) и т.д..

3. Для оценки уровня информационной безопасности ВУЗа необходимо связать данные по составу угроз, уязвимостей, средств защиты между собой и рассмотреть влияние потенциально возможных атак на основные сервисы безопасности информационных активов учебного заведения.

4. Построенная динамическая нечеткая когнитивная модель дает возможность, последовательно пройдя все уровни ее иерархии и применяя для свертки параметров приведенные в главе 5 формулы, оценить уровень безопасности информационных активов ВУЗа и выработать рекомендации по его повышению.

# ГЛАВА 6.

# ИСПОЛЬЗОВАНИЕ НКМ ПРИ РЕШЕНИИ ЗАДАЧ ПОДБОРА И ПОДГОТОВКИ КАДРОВ В СФЕРЕ ИНФОРМАЦИОННОЙ БЕЗОПАСНОСТИ

## 6.1. Определение наиболее востребованных компетенций специалиста по информационной безопасности

Неотъемлемой составляющей государственной политики России, направленной на защиту информационных ресурсов государства и защиту информации с ограниченным доступом, является подготовка специалистов в сфере защиты информации (ЗИ) и информационной безопасности (ИБ) [1].

Разработка Федеральных государственных образовательных стандартов (ФГОС) третьего поколения основывается на компетентностном подходе, который предполагает наличие в таких стандартах формулировок, отражающих способности выпускников высшего учебного заведения применять знания, умения и личностные качества для успешной деятельности в определенной области.

Для их формирования, прежде всего, необходимо определить основные характеристики образовательной области, соответствующей конкретному направлению подготовки, рассмотреть структуру предметной области, и определить исходные данные, необходимые для формулирования перечня профессиональных компетенций.

Обсуждение и решение проблемы построения эффективного учебного процесса предполагает поиск ответов на целый ряд вопросов, главный среди которых: что должен знать и уметь специалист, какими компетенциями он должен обладать?

Изучению данного вопроса в сфере информационной безопасности посвящен ряд работ [111-113], в которых обсуждаются проблемы перехода к компетентностному подходу в образовании и конкретные механизмы формулирования множества профессиональных компетенций специалистов укрупненного направления 090000 «Информационная безопасность».

В [112] предложен порядок формулирования профессиональных компетенций отдельно для каждого направления и специальности ВПО и СПО. При этом рекомендован следующий алгоритм действий:

1. Выбирается направление или специальность подготовки специалистов ВПО или СПО.
2. Определяется область профессиональной деятельности.
3. Определяются объекты профессиональной деятельности.
4. Уточняются виды профессиональной деятельности.
5. Формулируются задачи профессиональной деятельности на основе уточнения обобщенных задач по обеспечению информационной безопасности.
6. Для каждой задачи профессиональной деятельности формулируется одна или несколько профессиональных компетенций.
7. Формулируются квалификационные характеристики (уровни знаний, умений и владений предметом), необходимые для овладения определенными профессиональными компетенциями.

Данный алгоритм проиллюстрирован в работе [113] на примере специальности 090105 «Комплексное обеспечение информационной безопасности автоматизированных систем». При этом 32 компетенции объединены в пять групп в соответствии с задачами профессиональной деятельности: эксплуатационная, организационно-управленческая, проектная, экспериментально-исследовательская, научно-исследовательская.

Необходимо подчеркнуть, что в области информационной безопасности «открытые» исследования сопряжены с гораздо более серьезными политическими и этическими соображениями, чем в большинстве других наук. Поэтому специалисты по комплексному обеспечению информационной безопасности, прежде всего, должны обладать высокими моральными качествами. Кроме того, компьютерные преступники активно работают над повышением своей квалификации, вовлекают в свою среду подрастающее поколение и активно его обучают через Интернет. Все это подчеркивает важность разработки эффективных методов проведения воспитательной работы среди молодежи с целью активного противодействия вовлечению ее в преступную среду.

Таким образом, насущной задачей современного образования становится разработка таких методов учебно-воспитательной работы, где бы гармонично сочеталось обучение современным информационным технологиям с формированием высоких нравственных качеств для выработки иммунитета к совершению компьютерных преступлений.

Поэтому проект ФГОС третьего поколения предусматривает, что выпускник кроме профессиональных компетенций (*ПК*) должен обладать и рядом общекультурных компетенций (*ОК*).

При этом остаются открытыми ряд вопросов:

1. Какие компетенции являются наиболее значимыми?
2. Совпадают ли представления о степени значимости компетенций у преподавательского состава, который готовит молодого специалиста, и у потенциальных работодателей?
3. Как оценивать уровень компетенций?

Для ответа на первые два вопроса в течение 2009-2011 года было проведено исследование представлений экспертов о степени значимости компетенций специалистов по информационной безопасности [114].

Объектами исследования стали потенциальные работодатели: руководители и специалисты правоохранительных и контролирующих органов государственной власти (МВД, ФСБ, ФСТЭК) – группа №1, представители банков и финансово-экономических подразделений различных учреждений – группа №2, работники организаций, работающих в сфере оказания услуг по обеспечению информационной безопасности – группа №3, а также преподаватели профильных кафедр Астраханского государственного технического университета (АГТУ), в стенах которого с 2001 года ведется подготовка специалистов по специальности 090105 «Комплексное обеспечение информационной безопасности автоматизированных систем» – группа №4.

Предметом исследования являлось представление экспертов о профессиональных и общекультурных компетенциях специалиста по комплексному обеспечению информационной безопасности автоматизированных систем.

Цель исследования состояла в изучении представлений экспертов о значимости компетенций, которыми должен обладать специалист, для успешной работы в области защиты информации.

Методом сбора данных было анкетирование. Эксперту предлагалось ранжировать по степени возрастания значимости 32 профессиональные и 12 общекультурных компетенции специалиста. *ПК* были, согласно [113], объединенные в следующие категории:

- эксплуатационная (*Э*);
- организационно-управленческая (*ОУ*);
- проектная (*П*);
- экспериментально-исследовательская (*ЭИ*);
- научно-исследовательская (*НИ*).

Кроме того, было необходимо сравнить между собой значимость выделенных выше категорий компетенций, а также *ПК* и *ОК* в целом. При этом использовался описанный в главе 4 метод нестрогого ранжирования.

Представителями всех групп экспертов было заполнено 54 анкеты (15 экспертами из группы №1, 14 - из группы №2, 9 - из группы №3, 16 –из группы №4).

Результаты обработки данных показали, что расхождение мнений экспертов внутри каждой группы весьма незначительно: лишь в 4 случаях номера позиций при ранжировании различных компетенций разошлись более чем на две и только в одном случае – на три позиции. В остальных случаях расхождение не превышало одной позиции шкалы ранжирования. Это может свидетельствовать, с одной стороны, о высокой квалификации экспертов, с

другой – о минимальном уровне субъективной составляющей оценки значимости компетенций для решения практических задач обеспечения информационной безопасности.

Полученные данные были усреднены внутри каждой группы экспертов, что в итоге дало возможность сформировать проранжированные по возрастанию значимости множества компетенций для каждой из групп (табл.6.1).

Таблица 6.1.

| Усл. зн. | Содержание компетенции | Ранг гр.1 | Ранг гр.2 | Ранг гр.3 | Ранг гр.4 |
|---|---|---|---|---|---|
| Категория: *эксплуатационная* | | | | | |
| Э1 | Способность обеспечить эффективное функционирование информационных технологий, используемых в АС, с учетом требований по обеспечению ИБ | 2 | 1 | 1 | 2 |
| Э2 | Способность обеспечить эффективное функционирование средств ЗИ, используемых в АС | 2 | 1 | 1 | 2 |
| Э3 | Способность проводить мониторинг защищенности АС | 1 | 2 | 2 | 1 |
| Э4 | Способность администрировать подсистему ИБ АС | 2 | 2 | 1 | 3 |
| Э5 | Способность обеспечить восстановление работоспособности АС в режиме нештатных ситуаций | 3 | 3 | 1 | 4 |
| Категория: *организационно-управленческая* | | | | | |
| ОУ1 | Способность выявлять особенности функционирования АС с целью определения ее информационной инфраструктуры и активов, подлежащих защите | 1 | 3 | 2 | 4 |
| ОУ2 | Способность формировать политику ИБ АС на основе определения для АС активов, подлежащих защите, разработки моделей угроз и нарушителя, проведения анализа информационных рисков и формулирования правил, процедур, практических приемов и руководящих принципов по обеспечению ИБ АС | 3 | 3 | 2 | 1 |
| ОУ3 | Способность осуществлять контроль эффективности реализации политики ИБ АС | 2 | 1 | 1 | 2 |
| ОУ4 | Способность осуществлять менеджмент ИБ АС | 4 | 1 | 3 | 2 |
| ОУ5 | Способность проводить оценку эффективности функционирования системы менеджмента ИБ АС на основе использования методов мониторинга и аудита | 2 | 3 | 1 | 3 |
| ОУ6 | Способность разрабатывать предложения по совершенствованию системы менеджмента ИБ АС | 4 | 2 | 1 | 3 |
| ОУ7 | Способность планировать и организовать работу малых коллективов исполнителей | 5 | 4 | 3 | 5 |
| Категория: *проектная* | | | | | |
| П1 | Способность проводить анализ технических заданий на проектирование защищенных АС. Способность выполнять техническое и рабочее проектирование подсистем ИБ АС, с учетом действующих нормативных и методических документов | 1 | 1 | 2 | 1 |
| П2 | Способность разработать систему обеспечения ИБ АС | 4 | 1 | 1 | 1 |
| П3 | Способность разрабатывать и реализовывать методы и средства защиты информации для использования в АС | 4 | 4 | 3 | 4 |
| П4 | Способность участвовать в разработке системы менеджмента ИБ АС | 2 | 2 | 2 | 3 |
| П5 | Способность разрабатывать методы и средства оценки эффективности средств защиты для АС как основы проведения их сертификации | 1 | 2 | 1 | 3 |
| П6 | Способность разрабатывать методы и средства мониторинга ИБ АС | 3 | 2 | 3 | 2 |
| П7 | Способность разрабатывать методы и средства аудита ИБ АС | 2 | 2 | 3 | 4 |
| П8 | Способность разрабатывать проекты нормативных и методических материалов, регламентирующих работу по обеспечению ИБ АС, а также положений, инструкций и других организационно-распорядительных документов | 1 | 3 | 2 | 2 |
| Категория: *экспериментально-исследовательская* | | | | | |
| ЭИ1 | Способность участвовать в экспериментально-исследовательских работах при сертификации средств | 1 | 3 | 2 | 3 |
| | защиты АС | | | | |
| ЭИ2 | Способность участвовать в экспериментально-исследовательских работах при аттестации АС с учетом требований к обеспечению ИБ | 1 | 2 | 1 | 2 |
| ЭИ3 | Способность участвовать в проведении аудита (в том числе инструментального) защищенности АС | 1 | 1 | 1 | 2 |
| ЭИ4 | Способность участвовать в экспериментальных исследованиях АС с целью выявления уязвимостей активов в условиях существования угроз в информационной сфере | 1 | 1 | 3 | 1 |
| ЭИ5 | Способность участвовать в экспериментальных исследованиях информационных технологий, используемых в АС, с целью выявления их уязвимостей | 2 | 3 | 3 | 3 |
| Категория: *научно-исследовательская* | | | | | |
| НИ1 | Способность изучать и обобщать опыт работы других учреждений, организаций и предприятий по способам обеспечения ИБ АС с целью повышения эффективности и совершенствования работ по ЗИ на конкретной АС | 2 | 3 | 1 | 3 |
| НИ2 | Способность осуществлять подбор, изучение и обобщение научно-технической литературы, нормативных и методических материалов по методам обеспечения ИБ АС | 2 | 2 | 1 | 2 |
| НИ3 | Способность составлять аналитические обзоры по вопросам обеспечения ИБ АС | 1 | 3 | 2 | 5 |
| НИ4 | Способность проводить анализ ИБ современных информационных технологий, используемых в АС | 1 | 2 | 3 | 1 |
| НИ5 | Способность участвовать в исследовании и разработке методов ЗИ и систем, обеспечивающих ИБ АС | 4 | 4 | 1 | 4 |
| НИ6 | Способность участвовать в исследовании и разработке математических моделей средств защиты информации и систем, обеспечивающих ИБ АС | 5 | 5 | 3 | 4 |
| НИ7 | Способность обосновывать и выбирать оптимальные решения по уровню обеспечения ИБ АС как компромисса между различными требованиями | 3 | 1 | 1 | 1 |

| | Категория: *Общекультурные компетенции* | | | | |
|---|---|---|---|---|---|
| ОК1 | Способен действовать в соответствии с Конституцией РФ исполнять свой гражданский и профессиональный долг, руководствуясь принципами законности и патриотизма | 1 | 2 | 3 | 2 |
| ОК2 | Способен осуществлять свою деятельность в различных сферах общественной жизни с учетом принятых в обществе морально-нравственных и правовых норм, соблюдать принципы профессиональной этики | 2 | 1 | 3 | 1 |
| ОК3 | Способен анализировать социально значимые явления и процессы, в том числе политического и экономического характера, мировоззренческие и философские проблемы, применять основные положения и методы гуманитарных, социальных и экономических наук при решении социальных и профессиональных задач | 2 | 3 | 3 | 2 |
| ОК4 | ☐Способен понимать движущие силы и закономерности исторического процесса, роль личности в истории, политической организации общества, способен уважительно и бережно относиться к историческому наследию и культурным традициям, толерантно воспринимать социальные и культурные различия | 3 | 5 | 6 | 4 |
| ОК5 | Способен понимать социальную значимость своей будущей профессии, цели и смысл государственной службы, обладать высокой мотивацией к выполнению профессиональной деятельности в области обеспечения информационной безопасности, защиты интересов личности, общества и государства, готов и способен к активной состязательной деятельности в условиях информационного противоборства | 1 | 2 | 2 | 1 |
| ОК6 | Способен к работе в многонациональном коллективе, к кооперации с коллегами, в том числе и над междисциплинарными, инновационными проектами, способен в качестве руководителя подразделения, лидера группы сотрудников формировать цели команды, принимать организационно- | 2 | 2 | 1 | 3 |
| | управленческие решения в ситуациях риска и нести за них ответственность, применять методы конструктивного разрешения конфликтных ситуаций | | | | |
| ОК7 | Способен логически верно, аргументировано и ясно строить устную и письменную речь на русском языке, создавать и редактировать тексты профессионального назначения, публично представлять собственные и известные научные результаты, вести дискуссии | 2 | 4 | 2 | 2 |
| ОК8 | Способен к письменной и устной деловой коммуникации, к чтению и переводу текстов по профессиональной тематике на одном из иностранных языков | 2 | 4 | 4 | 2 |
| ОК9 | Способен к логическому мышлению, обобщению, анализу, критическому осмыслению, систематизации, прогнозированию, постановке исследовательских задач и выбору путей их достижения | 3 | 6 | 5 | 1 |
| ОК10 | Способен самостоятельно применять методы и средства познания, обучения и самоконтроля для приобретения новых знаний и умений, в том числе в новых областях, непосредственно не связанных со сферой деятельности, для развития социальных и профессиональных компетенций, для изменения вида и характера своей профессиональной деятельности | 3 | 2 | 2 | 2 |
| ОК11 | Способен к осуществлению воспитательной и образовательной деятельности в сферах публичной и частной жизни | 4 | 7 | 4 | 5 |
| ОК12 | способен самостоятельно применять методы физического воспитания для повышения адаптационных резервов организма и укрепления здоровья, готов к достижению должного уровня физ. подготовленности для обеспечения полноценной социальной и проф. деятельности | 2 | 3 | 5 | 6 |
| *Сравнение категорий профессиональных компетенций* | | | | | |
| Э | Эксплуатационная категория | 2 | 2 | 2 | 1 |
| ОУ | Организационно-управленческая категория | 1 | 1 | 3 | 2 |
| П | Проектная категория | 3 | 3 | 1 | 2 |
| ЭИ | Экспериментально-исследовательская категория | 2 | 2 | 2 | 1 |
| НИ | Научно-исследовательская категория | 4 | 4 | 4 | 3 |
| Сравнение обобщенных показателей *ПК* и *ОК* | | | | | |
| *ПК* | Профессиональные компетенции | 1 | 1 | 1 | 1 |
| *ОК* | Общекультурные компетенции | 1 | 1 | 2 | 1 |

Данные, приведенные в таблице, позволяют сделать следующие выводы:

1. Ранг обобщенного показателя *ПК* по мнению большинства групп экспертов (всех, кроме третьей) оказался равным рангу обобщенного показателя *ОК*. Это свидетельствует о том, что важной частью компетенций специалиста по защите информации являются личностные качества, и *ОК* являются необходимым базисом, на который должны быть наложены профессиональные компетенции.

2. Оценки степени значимости категорий компетенций различными группами экспертов отличаются: эксперты 1 и 2 группы в качестве самых важных указали организационно-управленческие компетенции, группа 3 отдала предпочтение проектным, а представители ВУЗа - экспериментально-исследовательским и эксплуатационным компетенциям. Имеют место расхождения во мнениях и внутри каждой категории. Данные расхождения связаны, по-видимому, со спецификой предметной области, с которой чаще сталкиваются эксперты той или иной группы и отражают тот факт, что значимость профессиональных компетенций в большой степени зависит от задач обеспечения информационной безопасности, с решением которых предстоит иметь дело специалисту.

3. Большинство групп экспертов (все кроме четвертой) на последнее место по значимости поставили компетенции, связанные с научно-исследовательской деятельностью. Подобное отношение работодателей (группы 1-3), вполне объяснимо: развитие и финансирование исследовательских проектов, направленных на среднесрочную, и тем более на долгосрочную перспективу, для них либо не актуально, либо экономически нецелесообразно.

Переход к приему в ВУЗы по результатам ЕГЭ не позволяет проводить диагностику личностных качеств и оценивать профессиональную пригодность желающих обучаться по специальностям, связанным с информационной безопасностью.

Молодые люди, которые не обладают необходимыми личностными качествами, как правило, не могут в дальнейшем эффективно работать в сфере обеспечения безопасности. Поэтому нужна особым образом организованная профориентационная деятельность среди потенциальных абитуриентов по информированию их о специфике будущей профессиональной деятельности.

Таким образом, требования современного рынка труда, специфика компетенций, необходимых для успешного освоения специальности ведут к необходимости создания своеобразного механизма отбора кадров для обучения и дальнейшей работы в сфере обеспечения информационной безопасности.

Центрами организации научно-исследовательской деятельности и работодателями для выпускников, имеющих высокий уровень соответствующих компетенций категории *НИ*, на сегодняшний день из рассмотренных групп организаций могут стать только ВУЗы (возможно путем создания под своей эгидой малых инновационных предприятий) [115].

Полученные результаты применения предложенной методики определения наиболее востребованных компетенций специалистов в области ИБ могут быть использованы для решения ряда практических задач:

- Уточнения методологии подготовки специалистов при переходе на двухуровневое образование в рамках новых ФГОС.
- Для постановки образовательных задач, предполагающих большую ориентацию на практическую деятельность, формирование умений и способностей решать актуальные проблемы в сфере защиты информации.
- Для более четкого понимания ожиданий работодателей.

- Для решения образовательных задач, связанных с профессиональной переподготовкой и повышением квалификации работников, занятых в сфере обеспечения информационной безопасности.
- Полученные оценки значимости различных компетенций могут быть использованы в качестве весовых коэффициентов в методике определения комплексного показателя компетентности обучаемого, т.е. при решении поставленного выше вопроса об оценке компетенций.

Рассмотрим данный вопрос в следующем параграфе более подробно в рамках постановки и решения задачи формирования команды для реализации различных проектов, в том числе в сфере информационной безопасности.

### 6.2. Методика формирования команды для реализации IT-проектов на основе НКМ оценки компетенций специалистов

Как было подчеркнуто выше, определение уровня компетентности специалиста является на сегодняшний день одной из важных задач при подготовке кадров.

Построение соответствующих ФГОС учебных программ должно обеспечивать достижение максимального уровня *ПК* и *ОК*. При этом если рассматривать задачу управления процессом обучения на основе компетентностного подхода как задачу управления сложным объектом, то необходимо разработать методику оценки текущего уровня компетентности обучаемого, который в данном случае является объектом управления [116;; 118].

Не менее важным является оценка уровня компетентности специалиста при приеме на работу или назначении на определенную должность, поскольку оптимальный подбор кадров способствует высокой отдаче каждого сотрудника при исполнении им своих служебных обязанностей [119].

Также важным фактором эффективного управления персоналом при реализации различных проектов является оптимальный подбор сотрудников для решения отдельных задач, что в конечном итоге предопределяет успешность выполнения проекта в целом.

На начальной стадии такого подбора для каждой задачи проекта $Z_l$ ($l=1,…,N$) необходимо сформировать перечень необходимых для ее успешного выполнения компетенций $KZ_l=\{KZ_{lm}\}$ ($m=1,…,M_l$; число компетенций $M_l$ для каждой из задач $Z_l$ может быть разным). Данный перечень может быть определен, например, с помощью предложенного в [120] метода функциональных моделей IDEF0.

Использование функциональной модели IDEF0 для анализа всего спектра работ позволяет выявить полный перечень компетенций, необходимых для их успешного выполнения. Применение же аппарата нечётких множеств для описания квалификационных характеристик обеспечивает адекватный учёт неоднозначности их формулировок. При этом необходимо разработать методику формирования команды для реализации проекта, предусматривающего решение совокупности $N$ отдельных задач.

Проблему можно формализовать следующим образом: из множества $S$ кандидатов подобрать команду исполнителей и распределить их по задачам проекта таким образом, чтобы наиболее полно обеспечить соответствие множеству $KZ=\{KZ_l\}$, т.е. суммарное по всем значениям $l$ и $m$ различие между значениями необходимых для выполнения $l$-й задачи компетенций $\{KZ_{lm}\}$ и значениями компетенций $\{KP_{jm}\}$ $j$-го исполнителя $l$-й задачи проекта должно быть минимальным [121].

Отклонение нежелательно как в отрицательную, так и в положительную сторону. В первом случае оно приводит к падению качества выполняемых работ, во втором – к неэффективности использования работника, т.к. более высокая квалификация исполнителя, как правило, требует более высокого уровня оплаты.

Решение задачи после получения множества необходимых компетенций $KZ$ может быть осуществлено в два этапа. На первом этапе необходимо оценить уровень компетенций каждого из претендентов. На втором – отобрать на основе анализа полученных на первом этапе данных наиболее подходящих исполнителей для реализации каждой из задач проекта.

*Этап 1. Оценка уровня компетенций специалистов.*

Задача первого этапа может быть эффективно решена путем использования технологий тестового контроля, когда испытуемому предлагается выполнить заранее подготовленные задания различной степени сложности, после чего результаты выполнения оцениваются группой экспертов.

Однако при этом процедура оценки обладает специфическими особенностями, основными из которых являются:

1. Неполнота, субъективность и неопределенность информации. Субъективность связана с личными предпочтениями экспертов, а неопределённость проявляется в неточных формулировках требований типа «уверенное владение», «умение анализировать», «обладание навыками», допускающими трактовку в очень широких пределах.

2. Многокритериальность задачи, связанная с необходимостью учета большого числа частных показателей.

3. Наличие как количественных, так и качественных показателей, которые необходимо учитывать при решении задачи оценки уровня компетенции.

4. Невозможность применения классических методов оптимизации.

Таким образом, несмотря на ряд достоинств, метод тестирования не лишен недостатков: процедура оценки результатов является слабо формализованной.

Поэтому построим методику тестовой комплексной оценки уровня компетентности, основанную на применении когнитивного моделирования, теории нечетких множеств и отношениях предпочтения между различными критериями.

Как отмечалось в главе 4, при решении многокритериальных задач часто используются различные методы свертки критериев в один обобщенный (интегральный) критерий. Одними из таких методов построения комплексного критерия являются аддитивная и мультипликативная свертка.

Для задач, где все оцениваемые параметры критично значимы, взаимосвязаны и взаимозависимы, наиболее целесообразным представляется применение мультипликативной свертки векторного критерия:

$$K = \prod_i K_i^{s_i} \qquad (6.1)$$

где $K_i$ – частные критерии, $S_i$ – некоторым образом определенные веса, приписываемые каждому частному критерию $K_i$.

В случае, когда значения каждого отдельного параметра не является критично значимым и допускается компенсация влияния параметров друг на друга, может быть применена аддитивная свертка:

$$K = \sum_i S_i K_i \qquad (6.2)$$

Оценки $S_i$ могут быть получены экспертным путем. В частности, при решении задач, связанных с ИБ, могут быть использованы результаты применения методики определения наиболее востребованных компетенций специалиста по информационной безопасности, изложенной в предыдущем параграфе.

Влияние оценок различных факторов на комплексный (интегральный) показатель компетентности может быть представлено в виде ориентированного графа $G$, аналогично тому, как это было сделано для оценки уровня безопасности информационных активов в главе 4 [122-123]:

$$G = <\{GF_i\};\{GD_{ij}\}>,$$

где $\{GF_i\}$ – множество вершин графа (факторов или концептов в терминологии нечеткой когнитивной модели (НКМ)); $\{GD_{ij}\}$ – множество дуг, соединяющих $i$-ю и $j$-ю вершины; $GF_0 = K_0$ – корневая вершина, отвечающая комплексному критерию компетентности.

При этом дуги расположены так, что началу дуги соответствует вершина нижнего уровня иерархии (ранга), а концу дуги – вершина ранга, на единицу меньшего. Значения факторов могут быть заданы лингвистическими значениями, взятыми из множества $QL$, заданного формулой (4.5). При этом в качестве семейства функций принадлежности выступает стандартный пятиуровневый 01-классификатор (3.1)-(3.5) (см.п.3.3).

Для дальнейшего построения методики ОУК на данный граф необходимо наложить полученную изложенным выше методом нестрогого ранжирования систему отношений предпочтения $E$ одних параметров над другими по степени их влияния на заданный элемент следующего уровня иерархии:

$$E = \{GF_i(e)GF_j \mid e \in (\succ; \approx)\}$$

При построении НКМ оценки компетентности на нижнем уровне необходимо расположить критерии оценки сложности тестовых заданий и критерии оценки результатов выполнения тестов.

На уровень выше - концепты, отражающие оценку сложности тестов и результатов их выполнения, рассчитанные на основе значений критериев нижнего уровня с учетом их весов влияния согласно формулам (6.1) и (6.2).

С целью повышения корректности оценки необходимо предложить испытуемому тесты различного уровня сложности (среднего, выше среднего, высокого).

На следующем уровне иерархии располагаются вершины, отражающие общую оценку выполнения конкретного теста, рассчитанную как «произведение» сложности теста на результат его выполнения. Операция «произведение» в случае лингвистических значений определяется с помощью принципа расширения обычных (четких) математических функций на нечеткие числа, предложенного Л.Заде [78] (см.п.3.4).

Следующий уровень содержит оценки отдельных компетенций испытуемого, найденные как максимум соответствующих общих оценок, полученных на предыдущем уровне.

И, наконец, на верхнем уровне может быть при необходимости получена комплексная оценка компетентности испытуемого на основе аддитивной или мультипликативной свертки, согласно формулам (6.1) или (6.2).

Чтобы рассчитать комплексную оценку компетентности, необходимо произвести агрегирование данных, собранных в рамках иерархии $G$ по направлению дуг графа.

Таким образом, в качестве математической модели оценки компетентности испытуемого может быть принят кортеж $<G, QL, E>$, в котором при переходе с одного уровня на другой применяются различные виды сверток векторного критерия [115].

Пройдя последовательно снизу вверх по всем уровням иерархии $G$ и применяя описанные выше соотношения, мы не только можем путем комплексного агрегирования данных выработать суждение о качественном уровне показателя на каждой ступени иерархии, но и оценить степень обоснованности данного суждения. Для этого необходимо:

1. Поставить в соответствие лингвистическим значениям концептов иерархии $G$ нечеткие числа стандартного пятиуровневого 01-классификатора (3.1)-(3.5).

2. Методом нестрогого ранжирования каждой дуге графа $D_{ij}$ сопоставить некоторый вес $S_{ij}$, отражающий влияние $i$-го концепта на $j$-й. При этом $S_{ij}$ представляют собой модифицированные веса Фишберна. В некоторых случаях веса могут быть заданы экспертом непосредственно.

3. Применить аддитивную и мультипликативную свертки согласно формулам (6.1) и (6.2). При этом операции над нечеткими числами могут быть проведены с помощью программного продукта, описание которого приведено в п.3.4. [82-83].

4. Лингвистически распознать полученный результат. Индекс схожести $\Omega$ может быть найден на основе Хемингова или Евклидова расстояний между нечетким числом, характеризующим результат, и числами эталонного терм-множества (3.1)-(3.5), например, по формулам (4.12 - 4.13) (см.п.4.4).

*Расчетный пример определения уровня компетентности.*

Пусть $T_i$ - тест, позволяющий проверить уровень владения некоторой компетенцией $K_j$. Предположим, что общая сложность теста зависит от сложности трех составляющих, входящих в тест: $k_1$, $k_2$ и $k_3$ [116].

Например, если $T_i$ – тест, предназначенный для выявления уровня компетенции $K_j$ = «Владение компьютером», то в качестве $k_1$, $k_2$ и $k_3$ могут выступать «Умение работать с офисными программами», «Знание основ функционирования аппаратной части компьютера» и «Соблюдение правил информационной безопасности». Причем вклад этих составляющих в оценку уровня сложности теста $T_i$ может быть различен. Это различие учитывается с помощью заданных экспертами весов, соответствующих каждому из $k$.

Оценка сложности теста $D$ получается после применения аддитивной свертки согласно формуле (6.2).

Результат выполнения теста $R$ определяется с помощью мультипликативной свертки (6.1) частных критериев успешности выполнения, взятых с соответствующими априорно заданными весами. В качестве частных критериев могут, например, выступать: $r_1$ - правильность выполнения заданий теста, $r_2$ - полнота выполнения заданий, $r_3$ - скорость выполнения.

Веса, используемые при оценке сложности теста и определении результата его выполнения, могут быть либо непосредственно заданы экспертами, либо могут быть получены как веса Фишберна в результате нестрогого ранжирования согласно описанной выше методике.

Общая оценка выполнения теста $QT_i$ вычисляется как произведение сложности теста на результат его выполнения.

После прохождения испытуемым всего множества тестов, предназначенных для оценки уровня данной компетенции $K_j$, оценка ее уровня $UK_j$ находится как максимальное значение общих оценок выполнения тестов множества.

Далее, если это необходимо, компетенции могут быть объединены с помощью мультипликативной свертки (6.1) в комплексный (интегральный) показатель компетентности.

При этом значения параметров соответствуют набору качественных оценок $QL$ с семейством функций принадлежности (3.1) - (3.5).

В табл.6.2 приведены исходные данные и результаты расчетов для определения уровня некоторой компетенции $K_1$.

Таблица 6.2.

Исходные данные и результаты расчетов для компетенции $K_1$

| $K_j$ | $T_i$ | $k_1$ / (вес) | $k_2$ / (вес) | $k_3$ / (вес) | $D\ (\Omega)$ | $r_1$ / (вес) | $r_2$ / (вес) | $r_3$ / (вес) | $R\ (\Omega)$ | $QT_i$ | $UK_j$ |
|---|---|---|---|---|---|---|---|---|---|---|---|
| $K_1$ | $T_1$ | НС / (1/5) | С / (2/5) | С / (2/5) | **С (0.81)** | В / (0.7) | ВС / (0.2) | В / (0.1) | **В (0.79)** | **С (0.72)** | **ВС (0.83)** |
| | $T_2$ | НС / (1/5) | НС / (2/5) | С / (2/5) | **НС (0.61)** | В / (0.7) | В / (0.2) | В / (0.1) | **В (0.79)** | **НС (0.84)** | |
| | $T_3$ | ВС / (1/5) | В / (2/5) | В / (2/5) | **В (0.80)** | ВС / (0.7) | ВС / (0.2) | В / (0.1) | **ВС (0.91)** | **ВС (0.83)** | |

Таким образом, в результате тестового контроля для компетенции $K_1$ получена оценка «Выше среднего» со степенью принадлежности 0,83.

Предположим, что для $K_2$ и $K_3$ аналогичным образом получены оценки «Выше среднего» и «Средняя». Причем эксперты при ранжировании расположили $K_j$ в следующей последовательности: ($K_1$, $K_2$); $K_3$. Тогда веса Фишберна для $K_1$ и $K_2$ равны по 1/4, а для $K_3$ вес составляет 2/4. После применения мультипликативной свертки (6.1) комплексный (интегральный) уровень компетентности испытуемого можно оценить как «Средний» со степенью принадлежности 0,56 (степень принадлежности к оценке «Выше среднего» составляет 0,44).

Несмотря на то, что две компетенции из трех имеют уровень «Выше среднего», комплексный уровень компетентности лишь «Средний», поскольку влияние третьей компетенции на интегральную оценку выше.

***Этап 2. Распределение задач между исполнителями.***

На втором этапе, прежде всего, необходимо вычислить интегральный индекс соответствия $\delta_l^j$ (ИИС) каждого $j$-го претендента каждой из $l$ задач проекта (как бы «примерить» каждую задачу на каждого из претендентов). Для этого по формулам (4.12-4.13) нужно найти индексы схожести между нечеткими числами $KZ_{lm}$, отражающими компетенции, требуемые для выполнения $l$-й задачи, и нечеткими числами $KP_{jm}$ отражающими компетенции $j$-го претендента.

После выполнения этого шага, задавшись для каждой из компетенций некоторым значением предельно допустимого отклонения индекса схожести от требуемого $\Omega_{jl}^{крит.}$, можно исключить из дальнейшего рассмотрения варианты распределения задач, содержащие индексы схожести меньшие критического (значение интегрального критерия соответствия данного претендента для этой задачи принимается равным нулю).

Для учета значимости различных компетенций при определении ИИС необходимо ввести в рассмотрение веса $\alpha_{lm}$ для каждой $m$-й компетенции $l$-й задачи. Сделать это эксперты могут либо непосредственно, либо описанным выше способом нестрогого ранжирования.

Таким образом, для расчета ИИС получаем формулу:

$$\delta_l^j = \begin{cases} 0,\ если\ \exists m : \Omega_{lm}^j < \Omega_{l\ крит.}^j (m=1,...,M_l) \\ \sum_{m=1}^{M_l} \alpha_{lm} \Omega_{lm}^j,\ если\ \forall m : \Omega_{lm}^j \geq \Omega_{l\ крит.}^j \end{cases}, (6.3)$$

где $\Omega_{lm}^j = \Omega(KP_{jm}, KZ_{lm})$ - индекс схожести между нечетким числом $KP_{jm}$, отражающим $m$ компетенцию $j$-го претендента, и нечетким числом $KZ_{lm}$, соответствующим уровню необходимой для выполнения задачи «$l$» компетенции «$m$».

Обозначим $V$ – множество $\{V_k\}$ различных вариантов распределения $S$ претендентов для решения $N$ задач. Мощность множества $V$ может быть найдена по формуле расчета числа размещений из $S$ по $N$:

$$\|V\| = \frac{S!}{(S-N)!} \qquad (6.4)$$

Элементами множества $V$ являются наборы упорядоченных пар (номер задачи; номер исполнителя):

$$V_k = (NZ_l; NP_j), (l \in [1, N]; j \in [1, S]).$$

Например, если проект предусматривает решение 5 задач и на место в команде претендуют 9 человек, то один из наборов может иметь вид: {(1;2), (2;5), (3;7), (4,4), (5;9)}.

Далее необходимо найти обобщенную оценку эффективности $\theta_k$ каждого варианта $V_k$ распределений претендентов по задачам, т.е. определить эффективность различных вариантов для проекта в целом. Для этого необходимо просуммировать частные индексы соответствия $\delta_l^j$ по всем парам $(l; j)$, входящим в $V_k$:

$$\theta_k = \begin{cases} 0, \textit{если } \exists (j, l) \in V_k : \delta_l^j = 0 \ (k = 1, \ldots, \|V\|) \\ \sum_{m=1}^{M_l} \alpha_{lm} \Omega_{lm}^j, \textit{если } \forall (j, l) \in V_k : \delta_l^j \neq 0 \end{cases} \qquad (6.5)$$

Оптимальным для выполнения проекта следует признать тот вариант распределения претендентов, оценка которого максимальна:

$$opt \ V_k \ (k: \max_k (V_k)) \qquad (6.6)$$

***Расчетный пример оценки эффективности вариантов назначения.***

Пусть успешное выполнение проекта предусматривает решение $N=3$ задач, которые потенциально могут быть распределены между $S=4$ претендентами, которые обозначены номерами 1, 2, 3 и 4. Уровень компетенций претендентов будем считать установленным на первом этапе. Критическое значение $\Omega_{jl}^{крит.}$ примем равным 0,8. Исходные данные для расчетного примера, а также результаты определения $\delta_l^j$ приведены в табл.6.3.

Таблица 6.3.

Исходные данные для расчетного примера.

| Задачи $Z_i$ | $Z_1$ | | | $Z_2$ | | | $Z_3$ | |
|---|---|---|---|---|---|---|---|---|
| Коды компетенций $KZ_{lm}$ | $K_1$ | $K_2$ | $K_3$ | $K_1$ | $K_4$ | $K_5$ | $K_2$ | $K_6$ |
| Веса $KZ_{lm}$ в ИИС ($\alpha_{lm}$) | 0,2 | 0,4 | 0,4 | 0,33 | 0,33 | 0,33 | 0,66 | 0,34 |
| Необходимый уровень $KZ_{lm}$ | С | ВС | ВС | С | С | В | ВС | ВС |
| $KP_{1m}$ | 0,86 | 0,91 | 1,00 | 0,86 | 1,00 | 0,88 | 0,91 | 0,74 |
| ИИС ($\delta_l^1$) | 0,2·0,86+0,4·0,91+0,4·1,00=**0,94** | | | **0,91** | | | **0** | |
| $KP_{2m}$ | 0,87 | 0,93 | 0,82 | 0,87 | 0,78 | 1 | 0,82 | 0,88 |
| ИИС ($\delta_l^2$) | **0,87** | | | **0** | | | **0,84** | |
| $KP_{3m}$ | 0,8 | 0,89 | 0,68 | 0,8 | 0,94 | 0,72 | 0,89 | 0,91 |
| ИИС ($\delta_l^3$) | **0** | | | **0** | | | **0,90** | |
| $KP_{4m}$ | 0,85 | 0,97 | 0,79 | 0,85 | 0,90 | 0,92 | 0,97 | 1,00 |
| ИИС ($\delta_l^4$) | **0** | | | **0,89** | | | **0,98** | |

После исключения из таблицы размещений, изначально содержавшей 24 строки, способов распределения задач, в которых хотя бы один ИИС равен нулю, остается 5 вариантов. Эти варианты, а также соответствующие им обобщенные оценки эффективности приведены в таблице 6.4.

Таблица 6.4.

Варианты распределения задач и их оценки.

| Задачи \ Варианты | $Z_1$ | $Z_2$ | $Z_3$ | Обобщенная оценка эффективности $\theta_k$ |
|---|---|---|---|---|
| 3 | 2 | 1 | 3 | **2,68**=0.87+0,91+0,90 |
| 9 | 2 | 4 | 3 | **2,66** |
| 13 | 1 | 4 | 3 | **2,73** |
| 20 | 1 | 4 | 2 | **2,67** |
| 21 | 2 | 1 | 4 | **2,76** |

Как показывают результаты расчетов, наиболее оптимальным является 21 вариант распределения задач, имеющий максимальную оценку эффективности равную 2,76. Он предусматривает выполнение первой задачи 2 претендентом, второй задачи – 1 претендентом и выполнение третьей задачи - 4 претендентом.

Предложенная методика оценки уровня компетенций испытуемого на базе нечеткой когнитивной модели была положена в основу работы компьютерной программы, позволяющей рассчитывать комплексный (интегральный) показатель компетентности [124]. При этом лицо, принимающее решение, может варьировать веса дуг, связывающих концепты НКМ с целью подготовки или подбора специалиста под конкретные задачи.

Методика оценки компетенций и формирования команды для наиболее эффективного выполнения проектов может быть использована не только при управлении образовательным процессом, но и может применяться кадровыми службами для более обоснованного и целенаправленного подбора персонала.

### 6.3. Формирование оценки эффективности образовательной деятельности на основе нечеткой когнитивной модели

Современное состояние проблем эффективности российского образования отражает усиливающееся противоречие между возрастающими требованиями общества к нравственности и интеллекту человека и фактическим уровнем образования и развития выпускников вузов. В этих условиях принципиальное значение приобретает поиск новых подходов к повышению эффективности организации и управления высшим образованием с ориентацией на его качественные аспекты.

В конце 80–х и в течение 90–х годов XX века произошло определенное снижение качества и конкурентоспособности образовательного и исследовательского процессов в высшей школе России, и как следствие, неполное соответствие уровня подготовки специалистов – выпускников вузов. Это приводит к потере авторитета российской образовательной системы, переизбытку на рынке труда невостребованных дипломированных специалистов. Появление и стремительное развитие высоких технологий, рост уровня технической оснащенности производств, обеспечение высоких темпов развития науки и техники, обусловленные необходимостью достижения конкурентоспособности отечественного производства и сферы услуг, требуют наличия квалифицированных специалистов и соответствующей системы их подготовки.

В 2003 г. Россия присоединилась к Болонским соглашениям, официально закрепив свое участие подписанием Берлинского коммюнике Конференции Министров образования [125]. В этой связи российское образование, как и образовательные системы других стран, претерпевает серьезные изменения: меняются приоритеты, структура и содержание образования, вводятся новые стандарты, формируется независимая система оценки результатов обучения и качества образования в целом.

В последнее время именно качество обучения и воспитания все более определяет уровень развития стран, становится стратегической областью, обеспечивающей их безопасность и потенциал за счет подготовки подрастающего поколения.

При этом качество образования рассматривается как комплексный показатель, синтезирующий все этапы становления личности, условия и результаты учебно-воспитательного процесса. Оно также является критерием эффективности деятельности образовательного учреждения, отражает соответствие реально достигнутых результатов нормативным требованиям, социальным и личностным ожиданиям. В то же время для профессионального образования все более значимой становится ориентация на запросы работодателя. Оценка качества образования в этом случае представляет собой не только самостоятельный интерес, но и рассматривается как ключ к решению назревших практических проблем в каждом отдельном образовательном учреждении и в экономике страны в целом [126].

На международном уровне волна внимания к данной проблеме обусловлена в первую очередь тем, что, согласно положениям Болонской декларации, европейские страны начали связывать взаимное признание документов об образовании с наличием систем независимой оценки его качества.

Для продвижения в этом направлении национальные системы образования всех европейских стран, в том числе и России, в ближайшее время должны решить ряд общих проблем: распределить ответственность по обеспечению качества образования между государством и образовательными учреждениями; разработать национальные системы оценки качества образования, включая внутреннюю оценку в вузах (внутри вузовские системы менеджмента качества) с привлечением студентов и публикацией результатов в печати; обеспечить разработку системы сопоставимых контрольно-оценочных материалов и процедур, их сертификацию, аккредитацию программ и образовательных учреждений, проведение сравнительных международных обследований в области оценки качества образования.

Наиболее важными направлениями решения этих проблем для российского образования являются разработка набора критериев оценки эффективности образовательной деятельности и обоснование принципов оценивания качества образования на всех его уровнях.

Поскольку в данном случае, как и при рассмотрении предыдущих задач, мы имеем дело с совокупностью отдельных критериев, то соответствующая задача в ее математической постановке также является многокритериальной.

Кроме того, анализ основных принципов обеспечения качества образовательных услуг позволяет сделать вывод о том, что задача оценки эффективности образовательной деятельности является также слабоструктурированной и плохо формализуемой [127].

Таким образом, для решения данной задачи представляется целесообразным использование изложенной выше методики, основанной на применении НКМ [128- 130].

Влияние различных факторов $GF_i$, определяющих оценку качества образования, на уровень комплексной оценки качества образовательных услуг может быть представлено в виде ориентированного графа $G$:

$$G = <\{GF_i\};\{GD_{ij}\}>,$$

с наложенной на него системой отношений предпочтения $E$ одних параметров над другими по степени их влияния на заданный элемент следующего уровня иерархии:

$$E = \{GF_i(e)GF_j \mid e \in (\succ; \approx)\}$$

где $F_i$ и $F_j$ – факторы одного уровня иерархии, $\succ$ - отношение предпочтения, $\approx$ - отношение безразличия. Такая система может быть получена, например, изложенным выше способом нестрогого ранжирования.

После введения в рассмотрение набора качественных оценок уровней каждого фактора в иерархии в виде лингвистических значений, взятых из множества $QL$, заданного формулой (4.5), в качестве математической модели комплексной оценки качества образования («*ОКО*») может быть принят кортеж:

$$OKO = <G, QL, E>$$

Опираясь на данную модель, можно построить показатель комплексной оценки качества образовательных услуг на базе агрегирования значений со всех уровней иерархии факторов на основе качественных данных об уровнях факторов и их отношениях порядка на одном уровне иерархии, используя подход, аналогичный приведенному в главе 4 для оценки уровня безопасности информационных активов.

*Расчетный пример применения метода*

Дадим комплексную оценку качества образования по критериям «Качество специальных знаний выпускника», «Качество естественнонаучных и гуманитарных знаний выпускника» и «Качество воспитательной работы ВУЗа».

Исходные данные для расчетов приведены в табл.6.5.

Таблица 6.5.

Факторы и их уровни

| Шифр фактора | Наименование фактора | Уров. факт. | На какие факторы влияет |
|---|---|---|---|
| $K_0$ | Комплексная оценка качества образования | * | ---- |
| $K_1$ | Качество специальных знаний выпускника | * | $K_0$ |
| $K_2$ | Качество естественнонаучных и гуманитарных знаний выпускника | * | $K_0$ |
| $K_3$ | Качество воспитательной работы ВУЗа | * | $K_0$ |
| $U_1$ | Уровень выполнения лицензионных требований и нормативов | * | $K_1\ K_2, K_3$ |
| $U_2$ | Уровень выполнения требований ГОС | * | $K_1\ K_2, K_3$ |
| $U_3$ | Уровень выполнения требований по государств. аккредитации | * | $K_1\ K_2, K_3$ |
| $N_1$ | Уровень материально-технической базы ВУЗа | Сред. | $U_1$ |
| $N_2$ | Уровень востребованности выпускников | Низк. | $U_2$ |
| $N_3$ | Уровень довузовской подготовки и отбора абитуриентов | Выс. | $U_2$ |
| $N_4$ | Уровень воспитательной деятельности ВУЗа | Сред. | $U_3$ |
| $N_5$ | Уровень социально-бытового обеспечения | Сред. | $U_1$ |
| $N_6$ | Уровень внутривузовской системы контроля качества подготовки специалистов | Низк. | $U_2$ |
| $N_7$ | Уровень информатизации ВУЗа | Сред. | $U_3$ |
| $N_8$ | Качественный состав ППК | Выс. | $U_1, U_3$ |

Знаком «*» обозначены факторы, уровень которых предстоит определить.

Пусть при этом существует следующая система отношения предпочтений факторов:

для $K_0 : K_1 \succ K_2 \approx K_3 \to (2/4;\ 1/4;\ 1/4)$.

для $K_1 : U_1 \approx U_2 \succ U_3 \to (2/5;\ 2/5;\ 1/5)$.

для $K_2 : U_1 \succ U_2 \succ U_3 \to (3/6;\ 2/6;\ 1/6)$.

для $K_3 : U_1 \approx U_3 \succ U_2 \to (2/5; 2/5; 1/5)$.

для $U_1 : N_1 \approx N_5 \succ N_8 \to (2/5; 2/5; 1/5)$.

для $U_2 : N_3 \approx N_6 \succ N_2 \to (2/5; 2/5; 1/5)$.

для $U_3 : N_4 \succ N_7 \approx N_8 \to (2/4; 1/4; 1/4)$.

В скобках указаны соответствующие системе предпочтений веса Фишберна, найденные описанным выше способом нестрогого ранжирования.

Необходимо оценить уровень качества образования.

Результаты расчетов приведены в табл.6.6. (в скобках рядом с уровнем фактора указано значение индикатора схожести с эталонной функцией принадлежности).

Таблица 6.6.

Результаты расчетов

| Показатель | Наименование фактора | Уровень фактора |
|---|---|---|
| $K_0$ | Комплексная оценка качества образования | низкий (0.95) |
| $K_1$ | Качество специальных знаний выпускника | низкий (0,82) / средний (0,78) |
| $K_2$ | Качество естественнонаучных и гуманитарных знаний выпускника | средний (0,90) / низкий (0,80) |
| $K_3$ | Качество воспитательной работы ВУЗа | средний (0,94) / низкий (0,79) |
| $U_1$ | Уровень выполнения лицензионных требований и нормативов | средний (0,96) |
| $U_2$ | Уровень выполнения требований ГОС | высокий (0,65) / средний (0,55) |
| $U_3$ | Уровень выполнения требований по государственной аккредитации | средний (0,95) |

Видно, что, несмотря на близость показателей $K_2$ и $K_3$ к уровню «средний», показатель качества образования $K_0$ имеет значение «низкий». Это обусловлено тем, что показатель $K_1$, имеющий значение «низкий», оказывает по оценкам экспертов большее влияние на оценку качества образования, чем $K_2$ и $K_3$.

## 6.4. Выводы по главе 6

Приведенные в главе 6 сведения позволяют сделать следующие основные выводы:

1. Грамотная политика подготовки национальных кадров в сфере информационной безопасности может оказать существенное противодействие росту преступлений в сфере информационных технологий. При этом важным элементом в процессе формировании специалиста по защите информации, наряду с профессиональными навыками, являются его личностные качества и общекультурные компетенции.

2. При решении различных задач обеспечения информационной безопасности меняется степень значимости компетенций, которыми должен обладать специалист по защите информации. Требования современного рынка труда, специфика компетенций, необходимых для успешного освоения специальности обуславливают актуальность создания механизмов отбора кадров для обучения и дальнейшей работы в сфере обеспечения информационной безопасности.

3. Методика оценки компетенций и формирования команды для наиболее эффективного выполнения проектов на базе нечеткой когнитивной модели, изложенная в данной главе, не только позволяет использовать ее при управлении образовательным процессом и эффективно решать задачу комплексной оценки качества образовательных услуг, но и может применяться кадровыми службами для более обоснованного и целенаправленного подбора персонала.

# ГЛАВА 7.

# ОЦЕНКА ЭФФЕКТИВНОСТИ МЕР ПО ОБЕСПЕЧЕНИЮ ИНФОРМАЦИОННОЙ БЕЗОПАСНОСТИ

## 7.1. Особенности оценки экономической эффективности мер по обеспечению информационной безопасности

Быстрое развитие информационных технологий требует постоянного совершенствования систем защиты информации и анализа экономической эффективности их внедрения. Оценка экономической целесообразности применения мер по обеспечению информационной безопасности является важной задачей, решение которой должно осуществляться на основе комплексного подхода, учитывающего специфику конкретной организации [131].

Формальная постановка задачи выбора оптимального с точки зрения затрат набора мероприятий, обеспечивающих необходимую эффективность защиты активов информационной системы была приведена в работах [9; 132]. Она опиралась на следующие рассуждения.

Поддержание информационной безопасности активов ИС требует выполнения защитных мероприятий: действий, процедур и механизмов, способных обеспечить безопасность при возникновении проблем и угроз, уменьшить уязвимость и ограничить воздействие на ИС, облегчить восстановление активов. Защитные мероприятия по ИБ могут выполнять одну или несколько следующих функций: предотвращение, сдерживание, обнаружение, ограничение угроз, исправление ошибок, восстановление активов системы, мониторинг состояния; уведомление о событиях в системе и т.д. [37].

При этом конкретное мероприятие может быть направлено на обеспечение нескольких сервисов безопасности, или же может оказаться, что для обеспечения какого-либо сервиса безопасности потребуется несколько защитных мер. В некоторых случаях эффективность защиты может повыситься при использовании комбинации различных мер обеспечения ИБ.

При воздействии *i*–ой угрозы ожидаемый потенциальный ущерб *j*-му активу ИС можно представить в виде [133]:

$$D_{ij} = \overline{PU_i} * F(U_i \to A_j) * r_j = D_{ij}(PU_i, U_i, B_j, r_j), \qquad (7.1)$$

где $\overline{PU_i}$ - вероятность возникновения *i*-ой угрозы, $F(U_i \to A_j)$ - величина воздействия угрозы $U_i$ на актив $A_j$, безопасность которого описывается блочной матрицей $B_j$ (см. п.4.3.), $r_j$ - ценность *j*-го актива. Таким образом, $D_j$ измеряется в единицах ценности актива $r_j$.

Общий ожидаемый ущерб равен:

$$D = \sum_i \sum_j D_{ij}$$

Как отмечалось в главе 4 (п.4.3), угрозы можно разделить на первичные и вторичные. Первичные угрозы существуют вне зависимости от состояния системы и имеют априорно заданную безусловную вероятность появления. Вероятность появления вторичных угроз является условной и зависит от состояния системы и параметров внешней среды. В частности, некоторые состояния системы могут спровоцировать возникновение угроз, появление которых в иных условиях было бы невозможным.

Несмотря на то, что вероятности возникновения первичных угроз $\overline{PU_i}$ от нас не зависят, совокупность превентивных мер защиты позволяет устранить некоторые из имеющихся уязвимостей ИС и тем самым ослабить влияние первичных угроз на степень безопасности системы.

Если все же, несмотря на эти меры, реализация определенного множества первичных угроз привела к возникновению последствий, то необходимо предпринять меры для их локализации и устранения, т.е. минимизировать отклонение текущего состояния системы от безопасного [43].

Совокупную стоимость всех мероприятий по ликвидации последствий обозначим $PR_L$. Аналогичным образом определенная «цена» превентивных мер защиты может быть найдена по формуле:

$$PR_Z = \sum_{j=1}^{M} PR_j$$

где $PR_j$ – «стоимость» $j$-ой превентивной меры защиты, M - общее количество превентивных мер.

Таким образом, суммарные затраты на реализацию мер по обеспечению безопасности составят:

$$PR = PR_Z + PR_L$$

Из соображений экономической целесообразности следует, что должно выполняться условие:

$$PR < D.$$

Перевод системы на более высокий уровень безопасности можно реализовать различными способами, используя совокупность защитных мер и технологий. Поэтому возможна постановка задачи выбора оптимального с точки зрения затрат набора мероприятий, обеспечивающих необходимую эффективность защиты активов системы.

Формальная постановка задачи в общем случае может выглядеть следующим образом. Необходимо обеспечить минимизацию целевой функции $PR$ ($PR_Z$, $PR_L$) при ограничениях, наложенных на значения элементов матрицы безопасности $B$.

Если же величина финансовых средств, выделенных на осуществление мероприятий по защите информационных ресурсов, не может превышать определенный уровень, то в качестве целевой функции в этом случае может выступать аддитивная или мультипликативная «свертка» разницы между матрицей текущего состояния системы и матрицей безопасного состояния.

Следует отметить, что в некоторых случаях в условиях автоматизированного управления и при использовании экспертной информации в процессе принятия решения (даже в случае формализованного правила выбора) приходится говорить не об оптимальном, а всего лишь о рациональном решении.

Таким образом, эффективное обеспечение информационной безопасности при возникновении угроз требует адекватного применения мер по защите активов системы. Выбор соответствующих мероприятий и технологий следует проводить с учетом структурной схемы решения задачи обеспечения комплексной безопасности, руководствуясь системными принципами.

Однако прежде чем приступить к решению поставленной задачи, необходимо оценить «экономическую» составляющую мероприятий по защите информации и прежде всего ответить на вопрос, что же понимать под термином «стоимость» того или иного мероприятия.

Оценка экономической эффективности внедрения инвестиционных проектов в такой сфере деятельности как информационная безопасность вызывает особые сложности [134]. Это обусловлено рядом факторов. В частности:

1. Информационная безопасность является относительно новой сферой деятельности, и в связи с этим недостаточно проработаны методы оценки информационных рисков и методики оценки информационных ресурсов.

2. Большинство методов оценки экономической эффективности внедрения мер по обеспечению ИБ опираются на использование статистики успешной реализации угроз. Однако такую статистику получить крайне сложно. Это обусловлено нежеланием организаций разглашать информацию, которая может повлечь за собой неблагоприятные последствия (ухудшение репутации компании, потерю постоянных и потенциальных клиентов и т.п.).

3. Убытки (прямые и косвенные), наносимые организации вследствие реализации угрозы, довольно сложно оценить, что связано со сложностью прогнозирования всех возможных последствий реализации угрозы и оценки их значимости для организации.

4. Неизвестно как проводить оценку при рассмотрении редких событий, которые происходят один раз в 10, 20 или более лет. Сбор статистических данных в этом случае практически невозможен, поскольку каждое такое событие является по-своему уникальным, а размер ущерба, наносимого организации в результате реализации такого события, как правило, превышает размер ущерба, нанесенного в результате идентичного предыдущего события.

Кроме того, как отмечалось выше, многие аспекты, касающиеся безопасности, могут вообще не подлежать количественному измерению. Они являются сугубо качественными, и предлагать измерять их количественно в большинстве случаев бесперспективно. К тому же, получение от лица, принимающего решение (ЛПР), надежной количественной информации часто бывает затруднительным. В таких случаях целесообразно сразу применять лингвистическое описание. Тогда от эксперта не требуется количественной точности, а требуется лишь субъективная оценка на естественном языке.

Исходя из этого в [134-135] была предложена методика выбора оптимального комплекса мер по обеспечению ИБ при заданных ограничениях, основанная на применении аппарата нечетких множеств и принципах решения многокритериальных задач.

## 7.2. Интегральный показатель эффективности комплекса мер по обеспечению ИБ

Пусть $S$ — число значимых для предприятия сервисов безопасности; $N$ — число мер обеспечения безопасности, которые потенциально могут быть реализованы для обеспечения ИБ; $a_1, a_2, \ldots a_S$ — степени значимости каждого сервиса безопасности, определяемые, например, с помощью модифицированного метода нестрогого ранжирования, предложенного в [95] и описанного в главе 4.

Данный метод позволяет определить обобщенные на случай предпочтения/безразличия факторов по отношению друг к другу веса Фишберна, и обеспечивает выполнение условий:

$$\text{для } \forall j \in \{1, 2, \ldots S\}: \ a_j \in [0,1] \text{ и } \sum_{j=1}^{S} a_j = 1;$$

Введем определение технической эффектности $E_{ij}^t$, подразумевая под этим термином величину, отражающую, насколько эффективно *i*-й набор мер обеспечивает необходимый уровень *j*-го сервиса безопасности (*i* =1,…, *N*; *j* = 1,…, *S*).

Оценки $E_{ij}^t$ могут быть получены в результате вычислительных экспериментов, проведенных с использованием построенной в главе 4 НКМ (см. рис.4.2). Согласно этой модели наборы защитных мер ослабляют или полностью ликвидируют уязвимости ИС (или уничтожают источники угроз), это приводит к уменьшению вероятности возникновения атак (реализации угроз), что в свою очередь благоприятно отражается на состоянии сервисов безопасности информационной системы.

После нахождения $E_{ij}^t$ при решении задачи выбора оптимального НМОИБ представляется целесообразным воспользоваться аддитивной сверткой оценок технической эффективности, которая для *i*-го набора мер может быть записана в виде:

$$E_i^t = \sum_{j=1}^{S} a_j E_{ij}^t$$

Такой выбор интегрального критерия технической эффективности будет корректным, при условии, что соблюдены требования по обеспечению минимального уровня каждого из значимых сервисов безопасности. Совокупность ограничений, отражающих эти требования, фактически определяет подмножество приемлемых вариантов НМОИБ, среди которых необходимо найти лучшее (в каком-то заданном смысле, например, с точки зрения минимальной «стоимости»).

Рассчитав показатели технической эффективности $E_i^t$ для всех возможных наборов мер обеспечения информационной безопасности, необходимо определить критерии выбора наиболее оптимального комплекса с точки зрения экономической эффективности.

*Экономическая эффективность* $E_{ij}^e$ характеризуется соотношением экономического эффекта, полученного в течение одного года, к затратам общественного труда. *Сравнительная экономическая эффективность* характеризует экономические преимущества одного варианта по сравнению с другими в наиболее рациональном использовании затрат и ресурсов [136].

При решении задачи обеспечения ИБ примем в качестве интегрального показателя эффективности $Q_i$ отношение технической эффективности определённого набора мер к экономической составляющей, соответствующей данному набору:

$$Q_i = E_i^t / E_i^e$$

Однако необходимо решить, какую величину принять в качестве экономической составляющей? Рассмотрим этот вопрос более детально.

### 7.3. Методы оценки экономической эффективности

Для оценки экономической составляющей эффективности ИТ-проектов в мировой практике наиболее часто используются следующие основные показатели: *коэффициент рентабельности инвестиций (ROI), чистая приведенная стоимость (NPV), внутренняя норма рентабельности (IRR), период окупаемости (Payback Period), экономическая привлекательность (EVA), сбалансированная система показателей (BSC), совокупная стоимость владения (TCO)* и *первичное публичное предложение (IPO)* [137].

Выделим основные преимущества и недостатки этих показателей с точки зрения ИБ. Коэффициент рентабельности инвестиций *ROI*, представляет собой отношение чистой прибыли к общим затратам. Формула, на первый взгляд, довольно простая. Однако при попытке оценить чистую прибыль от реализации проекта по информационной безопасности возникают существенные затруднения.

Дело в том, что целью внедрения таких проектов не является прямое увеличение прибыли организации. Напротив, введение дополнительных мер по обеспечению информационной безопасности приводит к снижению производительности информационных систем и сотрудников организации.

Однако полностью исключить затраты на информационную безопасность нежелательно, поскольку несанкционированный доступ или модификация, а также временная недоступность критичной для организации информации может привести к самым неблагоприятным для неё последствиям (вплоть до прекращения существования вследствие банкротства).

Отдельные статьи расходов на ИБ вообще невозможно исключить, т.к., например, Федеральный закон Российской Федерации от 27 июля 2006 г. N 152-ФЗ «О персональных данных», полностью вступивший в силу 1 января 2011 года, обязывает организации защищать персональные данные своих сотрудников и клиентов [138]. По закону каждой информационной системе, в которой хранятся и обрабатываются персональные данные, необходимо присвоить класс, в соответствии с которым должна обеспечиваться защита этих данных. Очевидно, что расходы, которые несет организация вследствие вступления в силу этого и аналогичных законов, неизбежны, и ни о какой прибыли не может идти и речи.

Что касается других мер по обеспечению информационной безопасности, внедрение которых не является обязательным, подсчитать доход от их реализации довольно сложно. Информационная безопасность имеет целый ряд особенностей, делающих распространенные способы расчета *ROI* неэффективными [139].

Аналогичные трудности возникают и при попытке оценить другие показатели: чистую приведенную стоимость (*NPV*), внутреннюю норму рентабельности (*IRR*), период окупаемости (*Payback Period*), экономическую привлекательность (*EVA*) и т.п.

Учитывая специфические особенности сферы применения этих показателей, их использование для оценки эффективности проектов по информационной безопасности нецелесообразно в связи с потенциальной убыточностью таких проектов.

Другая методика оценки затрат компании на информационную безопасность связана с принципами *BCP* (*Business Continuity Management* – планирование непрерывности бизнеса) [140].

*BCP* представляет собой целый комплекс различных мероприятий, направленных на снижение рисков прерывания бизнеса и их негативных последствий.

Это более широкий вопрос, чем обеспечение информационной безопасности. Система защиты коммерческой информации является частью *BCP*. Использование этой методики в такой сфере деятельности как информационная безопасность вызывает большие сложности по двум основным причинам:

1. Оценка эффективности затрат, предполагаемая принципами *BCP*, основывается на статистических данных. При этом учитываются вероятность возникновения опасной ситуации и потери, которые понесет компания в этом случае. К сожалению, в большинстве случаев подобной точной статистики не существует или она не доступна.

2. Введение в действие принципов BCP требует значительных затрат, которые могут себе позволить только крупные компании.

*Сбалансированная система показателей (BSC)* это управленческая и стратегическо-измерительная система, которая переводит миссию и стратегию организации в сбалансированный комплекс интегрированных рабочих показателей. Данная система позволяет выявить существующие взаимосвязи между важнейшими параметрами развития предприятия.

В своем исходном варианте *BSC* не очень подходит для применения в информационной безопасности, хотя в работе А. В. Лукацкого [141], была предпринята весьма успешная попытка адаптации данной системы к целям оценки деятельности служб безопасности. Однако для решения задачи определения НМОИБ методика *BSC* не подходит, поскольку она более ориентирована на управление активами и ресурсами, а не на оценку их финансирования [142].

Чаще всего в области информационной безопасности для определения затрат на создание системы защиты и оценки ее эффективности используется показатель *TCO* (*Total Cost Of Ownership* - совокупная стоимость владения).

Наиболее общим определением *TCO* является следующее: полный комплекс затрат, связанных с приобретением, внедрением и использованием системы, и воспринимаемый как единые затраты на информационную систему в процессе её создания и эксплуатации [140].

Данная методика была разработана известной аналитической компанией Gartner Group в конце 80-х годов (1986-1987). Первоначально она рассматривалась как способ оценки затрат на внедрение в управление компанией компьютерных технологий. Однако сегодня данная методика получила наибольшее распространение именно в области информационной безопасности [143].

Под показателем TCO понимается сумма прямых и косвенных затрат на организацию (реорганизацию), эксплуатацию и сопровождение корпоративной системы защиты информации в течение года.

При этом прямые затраты включают как капитальные компоненты затрат (ассоциируемые с фиксированными активами или «собственностью»), так и трудозатраты, которые учитываются в категориях операций и административного управления. Сюда же относят затраты на услуги удаленных пользователей и другие затраты, связанные с информационной поддержкой деятельности организации.

В свою очередь косвенные затраты отражают влияние корпоративной информационной системы (КИС) и подсистемы защиты информации на сотрудников компании посредством таких измеримых показателей как простои и «зависания» корпоративной системы защиты информации и КИС в целом, затраты на операции и поддержку (не относящиеся к прямым затратам). Очень часто косвенные затраты играют значительную роль, так как они обычно изначально не отражаются в бюджете на ИБ, а выявляются явно при анализе затрат в последствии, что в конечном счете приводит к росту «скрытых» затрат компании на ИБ.

Совокупная стоимость владения для системы ИБ в общем случае складывается из стоимости: проектных работ; закупки и настройки программно-технических средств защиты, включающих следующие основные группы: межсетевые экраны, средства криптографии, антивирусы и AAA (средства аутентификации, авторизации и администрирования); затрат на обеспечение физической безопасности; обучения персонала; управления и поддержки системы (администрирование безопасности); аудита ИБ; периодической модернизации системы ИБ [144].

При этом затраты на приобретение и ввод в действие программно-технических средств могут быть получены из анализа накладных, записей в складской документации и т. п.

Величина выплат персоналу может быть взята из ведомостей. Объемы выплат заработной платы должны учитывать реально затраченное время на проведение работ по обеспечению информационной безопасности.

При применении методики *TCO* для каждого варианта системы безопасности вычисляется совокупная стоимость владения. Затем выбирается система с минимальным параметром *TCO*. Это означает, что при выборе этого варианта компания понесет минимальные затраты (в которые входят и убытки вследствие возникновения проблем с ИБ) [145].

Основной недостаток показателя совокупной стоимости владения состоит в том, что он характеризует только расходную часть внедрения информационной системы. Круг применения этого метода обычно ограничивается выбором одного из альтернативных проектов с предполагаемым одинаковым эффектом использования.

## 7.4. Определение оптимального комплекса мер по обеспечению информационной безопасности

При решении задачи нахождения оптимального набора мер по обеспечению информационной безопасности представляется целесообразным объединить методику расчета показателей технической эффективности и методику *TCO* для определения комплексной эффективности намеченных к внедрению НБОИБ.

После расчёта *TCO* и показателей технической эффективности $E_i^t$ каждому набору мер по обеспечению ИБ можно поставить в соответствие величину $Q_i = E_i^t / TCO_i$, которая фактически отражает соотношение «цена/качество» для каждого НМОИБ.

Далее можно составить таблицу, по которой будет осуществляться выбор комплекса мер с учетом заданных ограничений.

Таблица 7.1.

Характеристики НМОИБ

| Номер набора $i$ | Номера мер, входящих в набор | | | | | | Показатель техн.эффект. набора, $E_i^t$ | Показатель экон.эффект. набора, $TCO_i$ | Комплексный критерий, $Q_i = E_i^t / TCO_i$ |
|---|---|---|---|---|---|---|---|---|---|
| | 1 | 2 | 3 | .. | .. | $N$ | | | |
| 1 | 1 | 0 | 0 | 0 | 0 | 0 | $E_1^t$ | $TCO_1$ | $Q_1$ |
| 2 | 0 | 1 | 0 | 0 | 0 | 0 | $E_2^t$ | $TCO_2$ | $Q_2$ |
| 3 | 0 | 0 | 1 | 0 | 0 | 0 | $E_3^t$ | $TCO_3$ | $Q_3$ |
| ... | ... | | | | | | ... | ... | ... |
| $2^N-1$ | 1 | 1 | 1 | 1 | 1 | 0 | $E_{2^N-1}^t$ | $TCO_{2^N-1}$ | $Q_{2^N-1}$ |
| $2^N$ | 1 | 1 | 1 | 1 | 1 | 1 | $E_{2^N}^t$ | $TCO_{2^N}$ | $Q_{2^N}$ |

В таблице символ «1» в *i*-й строке означает присутствие соответствующей меры в *i*-м наборе, символ «0» - мера в наборе отсутствует.

Максимальное число строк в таблице равно $2^N$. Однако не всегда все наборы могут быть реализованы. Например, не может быть реализован набор, в котором одновременно присутствует несколько антивирусных программ в связи с тем, что обычно они конфликтуют за ресурсы системы и приводят к ее

«зависанию». Какие-то наборы могут быть исключены исходя из других соображений.

Если размер финансовых средств, выделяемых на обеспечение ИБ, не ограничен, то целесообразно выбрать комплекс мер, которому соответствует наибольшее значение $E_i^t$.

Если выделяемый на обеспечение ИБ бюджет ограничен, то целесообразно рассмотреть только такие совокупности средств, *TCO* которых не превышает максимально допустимую величину, а из этого множества в свою очередь выбрать такой НМОИБ, которому соответствует наибольшее значения комплексного критерия $Q_i$.

## 7.5. Выводы по главе 7

Материал, приведенный в главе 7, позволяет сделать следующие основные выводы:

1. Задача обеспечения безопасности активов информационной системы может быть решена различными способами, с использованием совокупности защитных мер и технологий. Поэтому возможна постановка задачи выбора оптимального с точки зрения затрат набора мероприятий, обеспечивающих необходимую эффективность защиты.

2. В качестве интегрального показателя эффективности при решении задачи обеспечения ИБ можно принять отношение технической эффективности $E_{ij}^t$ определённого набора мер к экономической эффективности $E_{ij}^e$, соответствующей данному набору.

3. Оценки $E_{ij}^t$ могут быть получены в результате вычислительных экспериментов, проведенных с использованием построенной в главе 4 НКМ. После нахождения $E_{ij}^t$ при решении задачи выбора оптимального НМОИБ представляется целесообразным воспользоваться аддитивной сверткой оценок технической эффективности. Такой выбор интегрального критерия технической эффективности будет корректным, при условии, что соблюдены требования по обеспечению минимального уровня каждого из значимых сервисов безопасности.

4. Оценка экономической эффективности внедрения инвестиционных проектов в сфере информационной безопасности вызывает особые сложности, связанные с тем, что:

   - информационная безопасность является относительно новой сферой деятельности;
   - большинство методов оценки экономической эффективности внедрения мер по обеспечению ИБ опираются на использование статистики успешной реализации угроз, которую крайне сложно получить;

- сложно оценить прямые и косвенные убытки, наносимые организации вследствие реализации угрозы;
- неизвестно как проводить оценку при рассмотрении редких событий, которые происходят, например, один раз в 10, 20 или более лет.

5. Наиболее целесообразным в области информационной безопасности для определения затрат на создание системы защиты и оценки ее эффективности представляется использование показателя *TCO* (Total Cost Of Ownership - совокупная стоимость владения).

6. При решении задачи нахождения оптимального набора мер по обеспечению информационной безопасности необходимо объединить методику расчета показателей технической эффективности $E_{ij}^t$, основанную на применении НКМ, и методику расчета *TCO* для определения экономической эффективности $E_{ij}^e$ намеченных к внедрению НБОИБ.

# ГЛАВА 8.

# УПРАВЛЕНИЕ СЕТЕВЫМ ТРАФИКОМ НА ОСНОВЕ КОГНИТИВНОЙ МОДЕЛИ, ОСНОВАННОЙ НА НЕЧЕТКИХ ПРАВИЛАХ

## 8.1. Структура модели управления сетевым трафиком

Модель когнитивной карты, основанной на нечетких правилах, общее описание которой приведено в главе 3, нашла применение при разработке системы поддержки принятия решений в области управления трафиком в компьютерных сетях общего пользования (КСОП) [146 - 149].

Актуальность данной задачи связана с тем, что одной из главных причин, влияющих на эффективность работы таких сетей, являются аномалии в объеме сетевого трафика, которые могут быть вызваны случайными или преднамеренными действиями со стороны легитимных пользователей, неверной работой приложений, действиями злоумышленников и т.д.

Поэтому, должны быть приняты меры по своевременному выявлению аномалий, поиску их источников и принятию мер, обеспечивающих надежное функционирование КСОП (включение дополнительных каналов передачи данных, фильтрация аномального трафика и т.п.).

Таким образом, для обеспечения надежной передачи данных в КСОП важное значение приобретает разработка методов обнаружения аномальных пакетов и управления трафиком.

Одним из перспективных направлений в данной области является использование фрактальных свойств сетевого трафика. В качестве одного из возможных примеров такого использования можно привести методику, описанную в [146].

Согласно этой методики производится перехват всего входящего и исходящего трафика. В перехваченном трафике осуществляется поиск заголовков IP-пакетов, из которых извлекается все необходимые данные: объем

IP-пакета; IP-адрес источника; IP-адрес назначения; дата получения IP-пакета; время получения IP-пакета.

Полученная информация сохраняется в базе данных сетевой статистики. На основе накопленных данных проводится прогнозирование объема сетевого трафика на основе циклического анализа. Полученный прогноз сравнивается с реальными данными, поступающими с сетевого устройства. В случае их значительного расхождения делается вывод об обнаружении аномалии и принимается решение о необходимости применения управляющих воздействий.

Моделирование системы управления сетевым трафиком было проведено в нотациях стандартов IDEF0 и IDEF3, описанных в главе 2 (см. п.2.5). Соответствующие диаграммы приведены на рис.8.1-8.6.

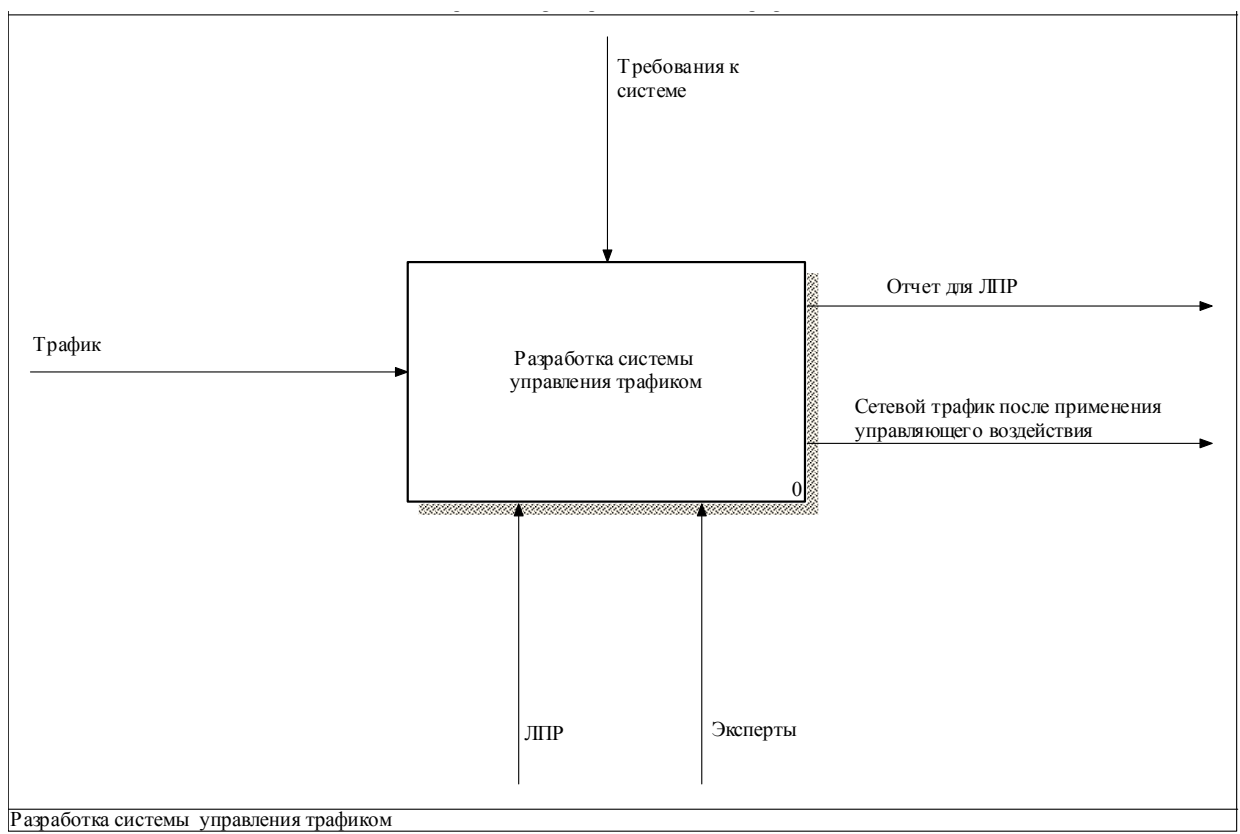

Рис.8.1. Контекстная диаграмма системы управления сетевым трафиком

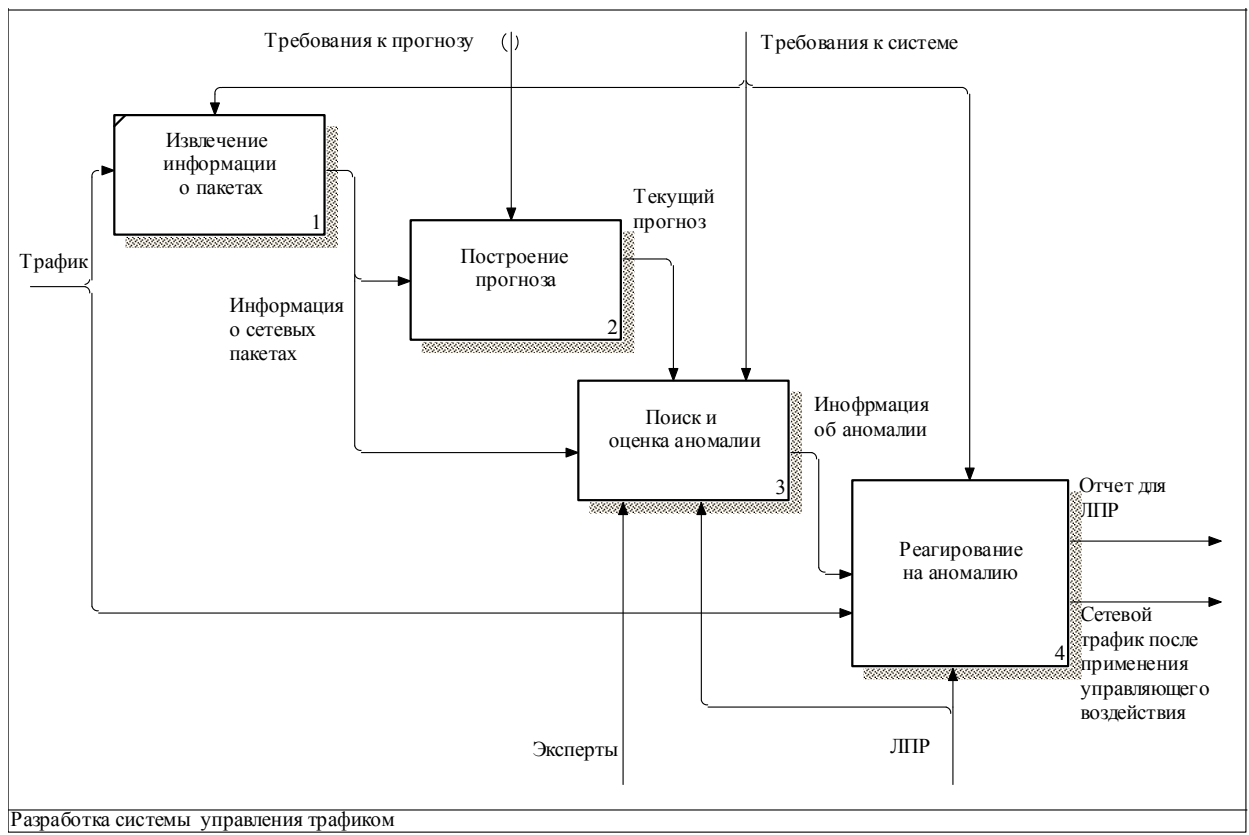

Рис.8.2. Декомпозиция контекстной диаграммы

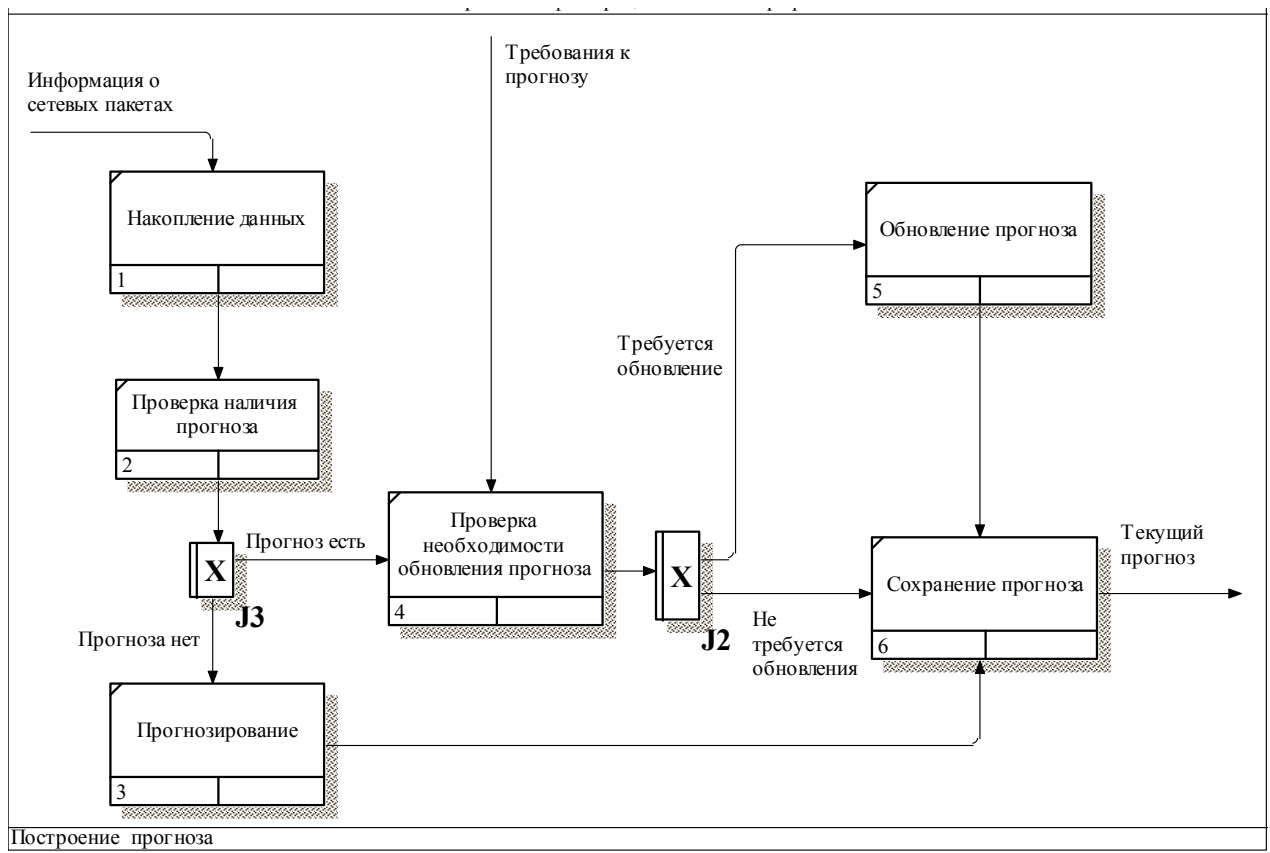

Рис.8.3. Декомпозиция блока «Построение прогноза» в нотации IDEF3

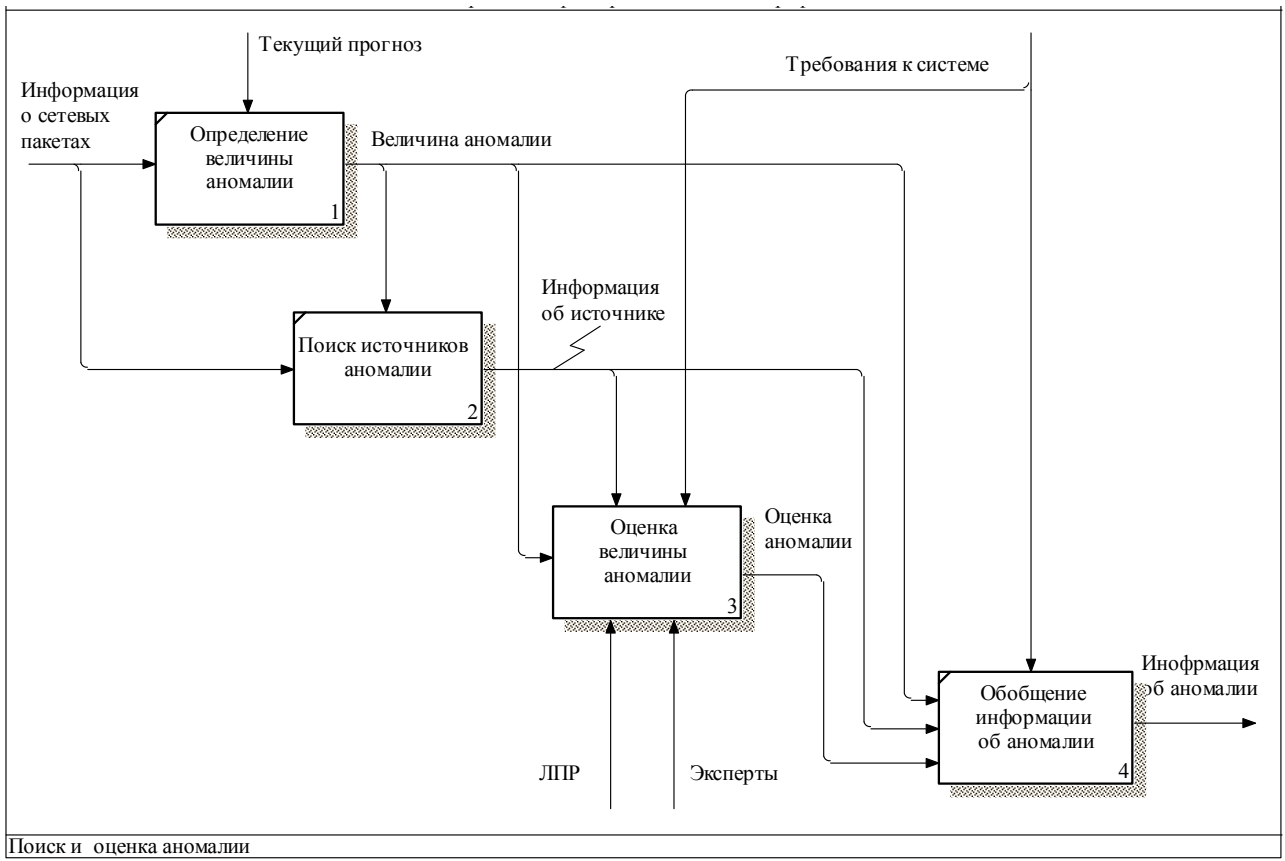

Рис.8.4. Декомпозиция блока «Поиск и оценка аномалии»

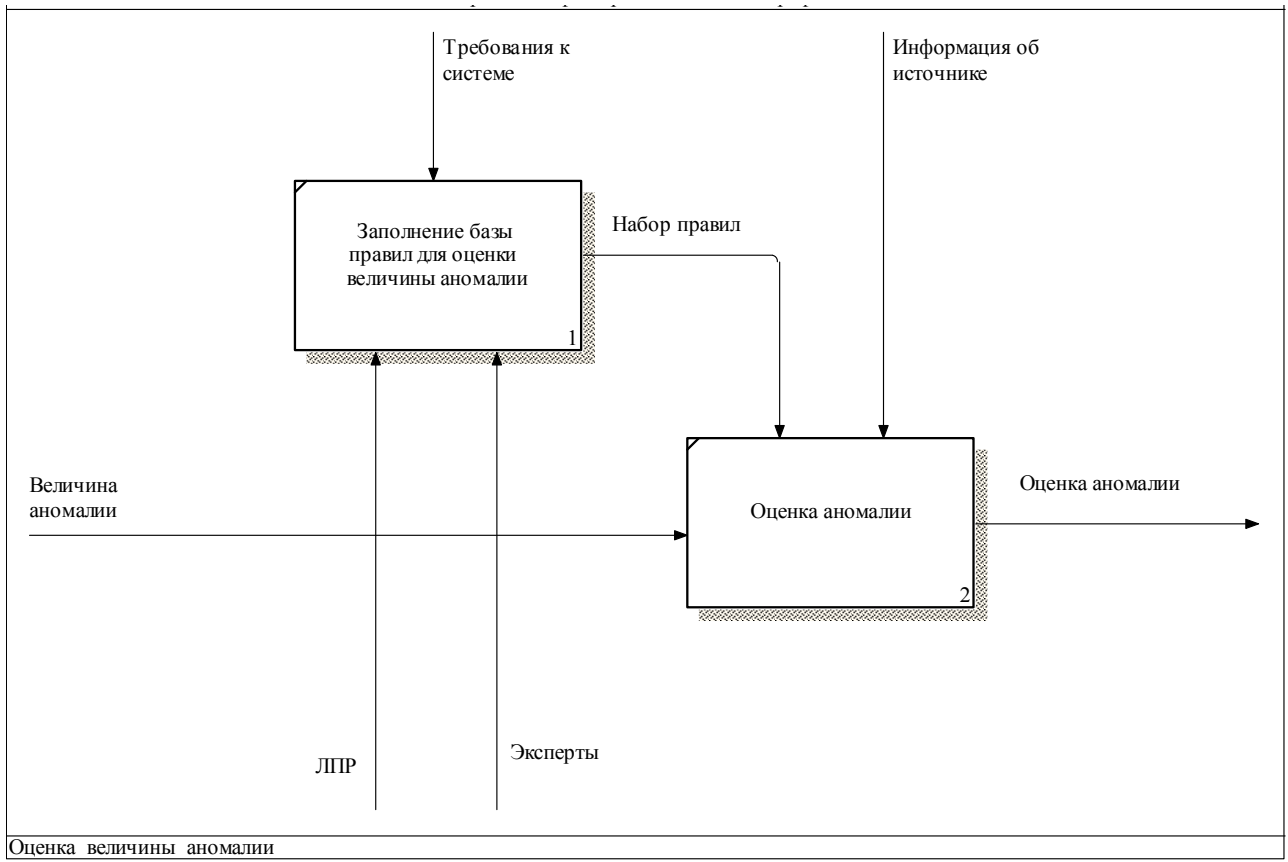

Рис.8.5. Декомпозиция блока «Оценка величины аномалии»

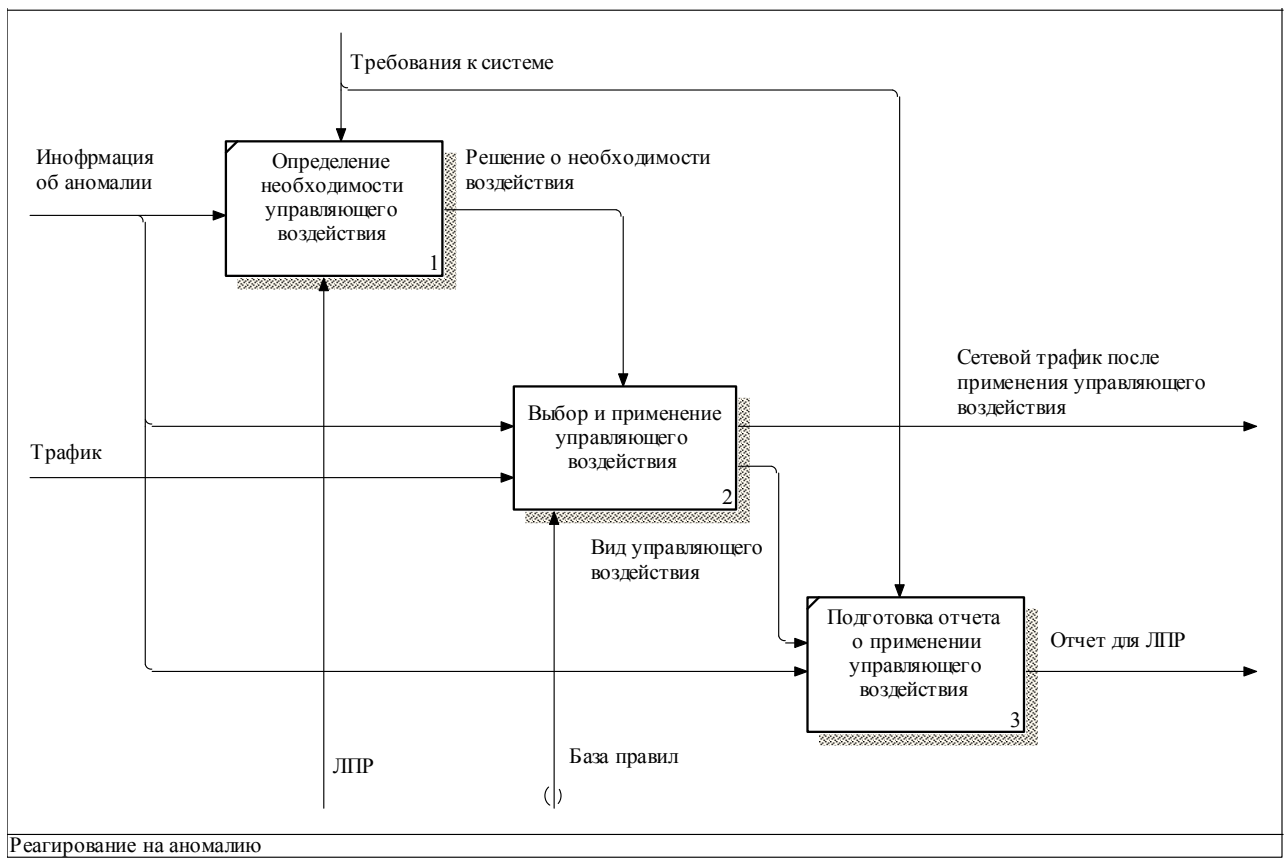

Рис.8.6. Декомпозиция блока «Реагирование на аномалию»

Рассмотрим работу основных функциональных блоков более подробно.

## 8.2. Прогнозирование объема сетевого трафика

Методика прогнозирования объема сетевого трафика предполагает использование следующей пошаговой процедуры.

*Отбор данных.* На первом этапе необходимо определить состав и количество данных, на основании которых будет производиться прогнозирование.

Циклический анализ сильно зависит от однородности данных, иначе структура циклов, скорее всего, будет искажена. Поэтому, при поиске и оценке аномалии должны учитываться резкие изменения в работе сети (например, при подключении большого количество хостов или изменениях в расписании), которые могут повлиять на форму циклов.

Поскольку циклический анализ предполагает работу с рядом данных, необходимо сформировать имеющиеся данные в виде ряда значений, описывающих изменение объема трафика во времени. Для этого необходимо провести дискретизацию объема трафика. Рассмотрим этот процесс на примере (рис. 8.7).

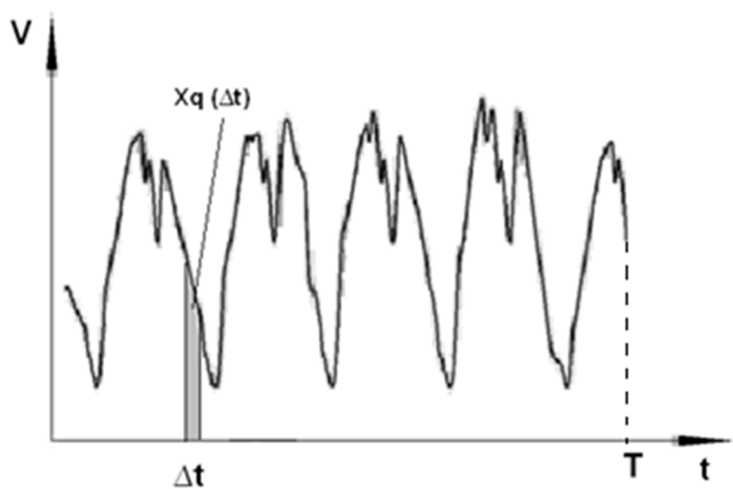

Рис.8.7. Дискретизация объема трафика

На рисунке представлен график, отображающий изменение объема сетевого трафика $V$ от времени $t$. Пусть имеется статистика по трафику, собранная за период времени $T$. Чтобы получить ряд данных, разделим период времени $T$ на целое число $Q$ равных интервалов $\Delta t$:

$$Q = \frac{T}{\Delta t}.$$

Далее для каждого интервала $\Delta t$ сложим объемы сетевых пакетов, попавших в данный интервал времени:

$$X_q(\Delta t) = \sum_{j=1}^{R} V_{\text{пакета}},$$

где: $R$ – количество сетевых пакетов, попавших в интервал $\Delta t$, $q$ – номер интервала, $q = 1, \ldots, Q$; $X_q$ – ряд упорядоченных данных, описывающий изменения объема трафика по времени, с частотой дискретизации $\Delta t$.

*Сглаживание данных.* Определившись с составом и количеством данными, необходимо исключить из трафика случайные колебания. Для этого используется метод сглаживания на основе краткосрочной центрированной скользящей средней.

Количество точек для сглаживания данных возьмем равным $L$ ($L = 3, 5, 7, \ldots$). При вычислении скользящей средней по $L$ точкам, из первоначального ряда данных будет выброшено $L - 1$ точек: $(L-1)/2$ – в начале и столько же в конце ряда. Таким образом, длина нового ряда данных $\bar{X}_k$ равна: $N = Q - (L - 1)$, $k = 1, \ldots, N$:

$$\bar{X}_k = \frac{1}{L} \sum_{j=k}^{k+(L-1)/2} X_j$$

*Поиск возможных циклов.* Устранив случайные колебания, можно приступить к непосредственному поиску циклов. Чтобы определить частотные составляющие рассматриваемого ряда, используем метод спектрального анализа. Математической основой спектрального анализа является преобразование Фурье [150]. Поскольку обрабатываемая статистика сетевого трафика имеет вид цифрового ряда, для определения частотных составляющих необходимо воспользоваться методом дискретного преобразования Фурье. С помощью прямого дискретного преобразования Фурье найдем комплексные амплитуды ряда данных $\bar{X}_k$:

$$Y_n = \sum_{k=1}^{N} \bar{X}_k \, e^{-\frac{2\pi i}{N} nk},$$

где $N$ – количество элементов ряда данных $\bar{X}_k$ и количество компонентов разложения, $i$ – мнимая единица.

Модуль комплексного числа может быть найден как:

$$|Y_n| = \sqrt{Re^2(Y_n) + Im^2(Y_n)}.$$

На основе комплексных амплитуд $Y_k$ вычисляется спектр мощности:

$$R_n = |Y_n|^2 = Re^2(Y_n) + Im^2(Y_n),$$

Изобразим спектр мощности графически (рис. 8.8) [151].

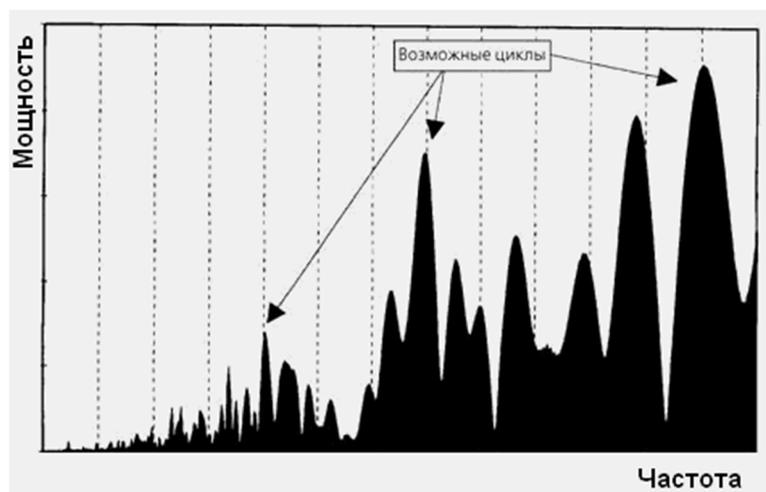

Рис. 8.8. Спектр мощности

Пики с высокими значениями показывают возможные циклы. Значением частоты цикла будет являться индекс $n$, при котором наблюдается высокое значение спектра мощности $R_n$.

Определив возможные циклы и их частоты, рассчитаем обычную (вещественную) амплитуду $A$ и фазу $\varphi$. Пусть найдено $b$ возможных циклов, частоты которых составляют множество $S$, т.е. каждое значение частоты, при которой наблюдается пик в области скопления высоких значений спектра, является элементом множества $S$.

Тогда амплитуды и фазы найденных циклов могут быть вычислены по формулам:

$$A_h = \frac{|S_h|}{N} = \frac{1}{N}\sqrt{Re^2(S_h) + Im^2(S_h)},$$

$$\varphi_h = Arg\,(S_h) = arctg\left(\frac{Im(S_h)}{Re(S_h)}\right),$$

где $h = 1, \ldots, b$; $Arg\,(S_h)$ – аргумент мнимого числа $S_h$.

Функция, описывающая цикл, выглядит как:

$$f_h(t) = A_h \cos(S_h t + \varphi_h).$$

Однако, как уже было сказано, высокое значение спектра мощности позволяет лишь предположить наличие цикла. Поэтому следующим шагом является подтверждение наличия циклов. Для этого необходимо проверить определенное количество критериев.

*Удаление трендовых компонентов в трафике.* Качество проверки циклов на статистическую значимость сильно зависит от существования направленности в данных. Поэтому перед проверкой необходимо провести удаление тренда из данных. Для этого можно применить метод отклонения от скользящего среднего. В данном случае, скользящая средняя будет отражать силы роста в данных, следовательно, ее вычитание из данных удалит и трендовую составляющую.

Таким образом, чтобы удалить тренд в данных необходимо для каждой найденной частоты рассчитать скользящую среднюю для ряда данных $\bar{X}_k$ с количеством точек сглаживания $L = S_h$:

$$\tilde{X}_k = \frac{1}{L} \sum_{j=k}^{k+(L-1)/2} X_j$$

где полученный ряд данных будет короче исходного на $L - 1$ точек: $\tilde{N} = N - (L - 1)$, $k = 1, \ldots, \tilde{N}$.

Далее вычитаем из исходного ряда данных $\bar{X}_k$ полученную скользящую среднюю $\tilde{X}_k$:

$$\bar{\tilde{X}}_k = \bar{X}_k - \tilde{X}_k.$$

Удалив силы роста в данных, можно приступать к проверке найденных циклов на статистическую значимость.

*Проверка циклов на статистическую значимость.* Для проверки на статистическую значимость обычно используются критерий Фишера и тест хи-квадрат.

Результаты теста зависят от количества повторений цикла в данных. Чем больше таких повторений, тем более статистически значим данный цикл.

*Комбинирование и проецирование циклов в будущее.* Прогнозирование трафика происходит на этапе комбинирования и проецирования циклов. Для этого циклы объединяются и на основе полученного результата можно спрогнозировать объем сетевого трафика в будущее. Для проецирования циклы математически комбинируются в одну общую кривую.

Допустим, что $D$ циклов прошли тесты. Эти циклы объединяются в общую кривую, описывающую периодичность в объеме сетевого трафика, найденную на основе данных за период времени $T$:

$$\bar{V}(t) = \sum_{j=1}^{D} f_h(t)\ .$$

Полученная функция может быть экстраполирована, что позволяет получить прогнозируемое значение трафика на период времени $t' \in (T, \bar{T})$ в будущем (рис. 8.9):

$$V_{\text{прогноз}}(t') = \sum_{j=1}^{D} f_h(t')\ ,$$

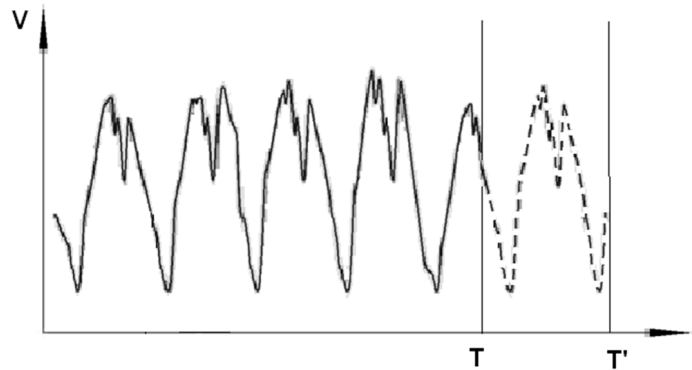

Рис. 8.9. Прогнозирование трафика

Таким образом, основанная на идеях циклического анализа и наличия у сетевого трафика фрактальных свойств математическая модель полностью решает задачу построения прогноза изменения объема трафика (рис.8.2).

## 8.3. Поиск и оценка величины аномалии в объеме сетевого трафика

Рассмотрим следующий блок системы поддержки принятия решений (СППР) об управлении трафиком в сетях общего пользования: «Поиск и оценка аномалии» (рис. 8.4-8.5).

Для определения объема трафика, циркулирующего в сети, используется извлекаемая из сетевых пакетов информация.

На основе сравнения текущих данных с прогнозом производится поиск аномалий. Для этого в единицу времени $t$ сравниваются два значения: $V_\text{реал}$ – величина объема текущего трафика и $V_\text{прогноз}$ – прогнозируемое значение объема. Аномальным будет считать отклонение объема, равное или превышающее заданную величину $\Delta V_\text{крит}$:

$$\Delta V = \left|V_\text{реал} - V_\text{прогноз}\right| \geq \Delta V_\text{крит}$$

Отметим, что прогноз представлен в виде относительно гладкой кривой. В свою очередь, реальный трафик состоит из кратковременных случайных флуктуаций. Если сравнивать эти два ряда данных, возможны ложные определения аномалий (рис. 8.11, а). Чтобы этого избежать, реальный трафик сглаживается методом скользящей средней. При этом окно сглаживания смещается по мере поступления новых данных об объеме трафика (рис. 8.11, б).

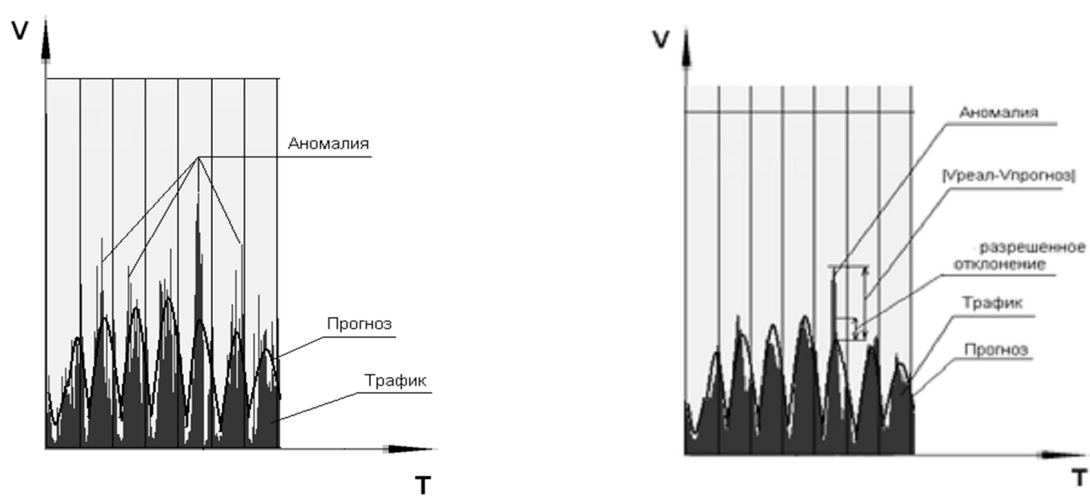

а) Трафик до сглаживания     б) Трафик после сглаживания

Рис. 8.10. Аномалия трафика

В случае если аномалии найдены, результат поиска передается на блок поиска источников аномалий. Определение источников осуществляется на основе информации о сетевых пакетах, циркулирующих в сети.

Далее производится оценка величины аномалии. Кроме самой величины аномалии, в оценке используется информация об источниках аномально высокого объема трафика, а также сведения, полученные от ЛПР и экспертов.

Вся полученная информация об аномалии обобщается и передается для дальнейшего рассмотрения ЛПР.

Оценка величины аномалии происходит на основе типичной для экспертных систем продукционной базы правил (БП). Начальное заполнение БП осуществляется экспертами, каждый из которых, независимо друг от друга, формулирует набор правил, которые затем, после проверки на полноту, независимость (избыточность) и непротиворечивость, образуют базу правил [81]. В дальнейшем ЛПР может корректировать БП, исходя из результатов работы СППР.

Рассмотрим структуру правил, входящих в базу. В наборе правил для оценки уровня аномалии $E$ в качестве входных используются следующие лингвистические переменные с терм-множеством значений фактора {Низкий(Н), Ниже среднего(НС), Средний(С), Выше среднего(ВС), Высокий(В)}:

- величина отклонения $\Delta V$;
- частота возникновения аномалии $M$;
- количество источников аномалий $I$;
- средний объем трафика от одного источника $W$.

Структуру правил можно пояснить следующими типовыми примерами:

➢ ЕСЛИ $\Delta V$=(«Н» или «НС») и $M$=(«Н» или «С») и $I$=(«Н» или «С») и $W$=(«Н» или «С») ТО $E$=«Н»;

➢ ЕСЛИ $\Delta V$=(«ВС» или «В») и $M$=«В» и $I$=(«ВС» или «В») и $W$=«ВС» ТО $E$=«В»;

➢ …

Таким образом, если база правил обладает свойствами полноты, непротиворечивости и независимости, то для любого набора входных данных {$\Delta V$; $M$; $I$; $W$} величина аномалии объема сетевого трафика $E$ может быть однозначно определена с помощью алгоритмов нечеткого вывода, описанных в главе 3 (см. п.3.5).

Зная величину аномалии и ее оценку, можно приступить к выработке решения о выборе и применении управляющих воздействий.

## 8.4. Реагирование на обнаружение аномалии

Рассмотрим следующий блок СППР: «Реагирование на аномалию» (рис.8.6).

Процесс реагирования на аномалию может быть декомпозирован на три основных блока:

- определение необходимости воздействия на трафик;
- выбор вида управляющего воздействия и его применение;
- подготовка отчета.

Решения о необходимости воздействия на трафик, а также выборе вида управляющего воздействия, принимаются на основе сформулированных ЛПР правил, которые в совокупности образуют Базу Правил по Управлению Сетевым Трафиком (БПУСТ). Эта база, как и база правил по оценке величины аномалии, должна обладать свойствами независимости, полноты и непротиворечивости.

Чтобы различать трафик из разных подсетей, необходимо учитывать IP-адрес и маску подсети. Это позволит индивидуально настраивать фильтрацию трафика для каждой подсети. Также должно быть предусмотрено раздельное отслеживание входящего и исходящего трафика.

Поскольку трафики из внешних и внутренних сетей имеют разную информативность и, как следствие, различия при построении модели прогноза, то необходимо различать трафик внешней и трафик внутренней сети.

Таким образом, при формулировании правил должны учитываться следующие характеристики:

- величина отклонения реального трафика от прогнозируемого;
- IP-адрес источника аномалии и маска подсети;
- направление трафика (входящий или исходящий);
- местонахождение источника относительно защищаемой сети (внешний или внутренний).

Примеры правил:

- ЕСЛИ $E$ = «Н» ТО Воздействие = «Отсутствует»;
- ЕСЛИ {$E$=«В» и Источник=«Внутренний» и Направление трафика=«Исходящий» и Привилегия IP-адрес источника= «Средняя»} ТО {Воздействие = «Блокировать» и Время воздействия = 30 минут};
- ЕСЛИ {$E$=«В» и Источник=«Внутренний» и Направление трафика=«Входящий» и Привилегия IP-адрес источника= «Высокая»} ТО {Воздействие = «Подключить дополнительное оборудование» и Время воздействия = «До исчезновения аномалии»};
- ….

После применения любого управляющего воздействия в системе должен быть сформирован отчет для ЛПР и произведена запись в соответствующий журнал аудита сетевой активности.

Таким образом, общая схема управления сетевым трафиком может быть представлена в виде [146]:

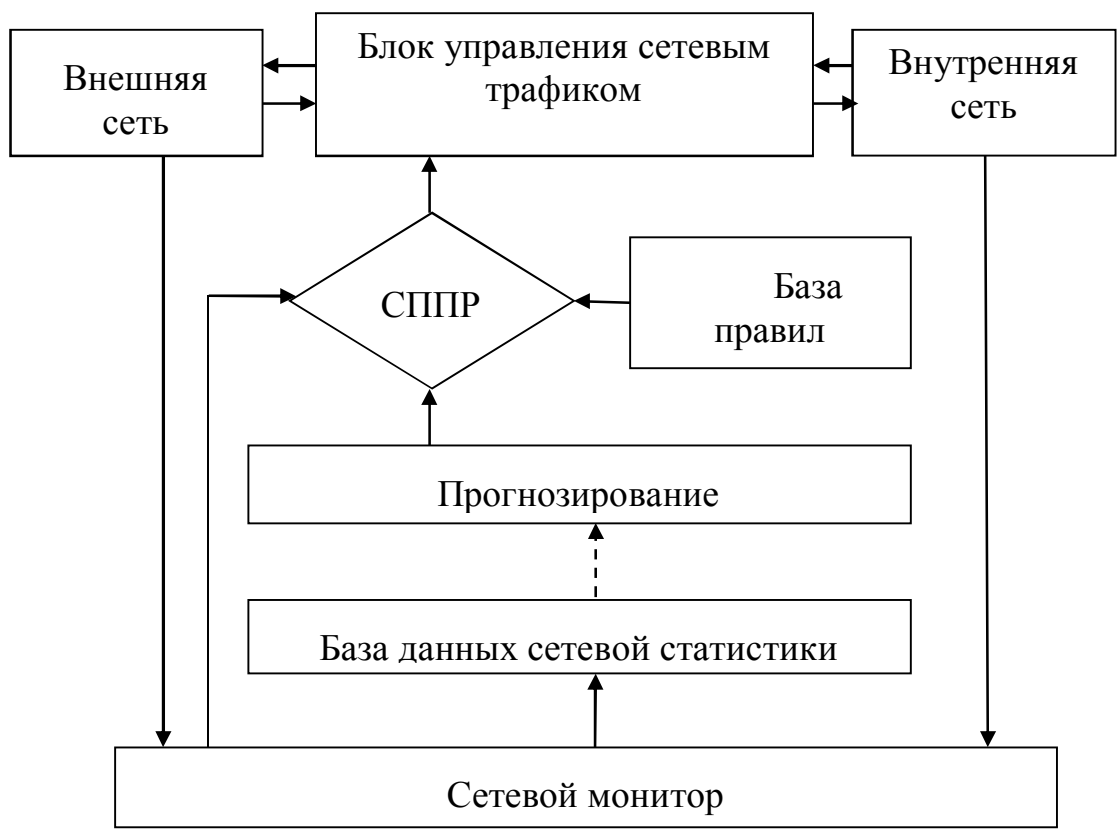

Рис.8.11. Схема управления сетевым трафиком

Для реализации приведенной схемы был разработан программный продукт и предложено техническое решение, содержащее в себе инструменты для сбора, обработки и хранения сетевой статистики, прогнозирования загрузки сети, СППР для поиска и оценки величины аномалии, а также выбора механизмов управления вычислительной сетью и фильтрации IP-пакетов [152-153].

Экспериментальная проверка на реальных сетях показала эффективность предлагаемого решения в сочетании с традиционными, более сфокусированными на сигнатурном анализе, методами обнаружения атак на сетевые ресурсы. Применение разработанной методики позволило увеличить количество обнаруженных аномалий в среднем на 9-13%.

## 8.5. Выводы по главе 8

Описанное в главе решение задачи управления сетевым трафиком представляет собой синтез строго математической модели прогнозирования объема сетевого трафика, основанной на циклическом анализе временных рядов с привлечением методов статистической обработки данных, и модели когнитивной карты, управляемой нечеткими правилами. Подобный синтез позволил построить эффективную модель управления сетевым трафиком, и предложить техническое решение, которое способствует повышению информационной безопасности сетей общего пользования.

Полученные результаты могут быть использованы для обнаружения неисправностей сетевого оборудования, поиска ошибок в настройке программного обеспечения, выявления случайных и преднамеренных деструктивных действий со стороны легитимных пользователей и злоумышленников.

# ЗАКЛЮЧЕНИЕ

Применение информационных технологий немыслимо сегодня без решения вопросов, связанных с информационной безопасностью.

При этом с одной стороны, внимание к данной проблеме постоянно усиливается (существенное повышаются требования к защите информации, принимаются международные стандарты в области информационной безопасности, постоянно растут расходы на обеспечение защиты), с другой – столь же неуклонно растет ущерб, причиняемый владельцам информационных ресурсов.

Исследования в области теории защиты информации сталкиваются с существенными проблемами, связанными с постоянно изменяющимся перечнем возможных угроз безопасности ИС, которая является сложной человеко-машинной системой, со слабой структурированностью и плохой формализуемостью задач данной предметной области, поскольку большая часть параметров являются качественными, не имеющими количественной оценки, а иногда и четкого описания, что вызывает существенные затруднения при их анализе и формализации.

Трудности формализации усугубляются еще и весьма высоким уровнем неопределенности, связанной с непредсказуемостью поведения источников потенциальных угроз.

Все это делает затруднительным применение существующих методик для исследования проблем информационной безопасности и свидетельствует о необходимости комплексного (системного) подхода к исследованию проблемы защиты информации, циркулирующей в ИС.

Указанные обстоятельства приводят к выводу, что интенсификация процессов защиты информации предполагает широкое использование интеллектуальных технологий и неформально-эвристических методов, основанных на экспертных оценках, теории нечетких множеств и когнитивном моделировании.

Однако эффективность указанных методов существенно зависит от представительности выборок, на которых они осуществляются. Кроме того, непрерывное изменение условий защиты, постоянный рост возможностей злоумышленного доступа к защищаемой информации, а также совершенствование методов ее защиты требуют того, чтобы экспертные оценки были не просто перманентными, а практически непрерывными. Этого можно достичь лишь при наличии стройной и целенаправленной системы оценки уровня защищенности информационных активов.

В данной работе предпринята попытка создания на основе системного подхода с использованием методов нечеткого когнитивного моделирования методики комплексной оценки уровня информационной безопасности, позволяющей унифицировать подходы к управлению процессом обеспечения сохранности информационных активов, и показана возможность ее использования в различных задачах, связанных с информационной безопасностью.

Для этого в монографии рассмотрены особенности применения системного подхода в задачах обеспечения информационной безопасности, методы и принципы построения когнитивных моделей. Приведены основные теоретические сведения об использовании лингвистических переменных, принципах построения нечеткой логики и действиях с нечеткими числами. Описана методика оценки уровня информационной безопасности на основе нечеткой логики и нечеткая когнитивная модель управления уровнем ИБ. Рассмотрено применение НКМ при решении различных задач в области защиты информации (оценка повреждений ИС, подготовка кадров, информационная безопасность ВУЗа, анализ сетевого трафика с целью выявления аномалий и др.). Приведено описание технических решений и программных продуктов. Часть задач, имеющих прикладной характер, не вошла в основное содержание монографии и вынесена в приложения.

# СПИСОК ЛИТЕРАТУРЫ

# ПРИЛОЖЕНИЯ

# ОПТИМИЗАЦИЯ ПРОПУСКНОЙ СПОСОБНОСТИ В СИСТЕМАХ ЗАЩИЩЕННОЙ ПЕРЕДАЧИ ДАННЫХ

**Введение**

При решении практических задач, связанных с обеспечением информационной безопасности, часто прибегают к использованию специальных защищенных каналов передачи. Финансовые затраты на поддержание таких «закрытых» каналов могут быть весьма значительными. Поэтому являются весьма актуальными вопросы, связанные с выбором оптимального количества каналов, способного обеспечить различную пропускную способность при случайно возникающей потребности в их использовании.

При этом необходимо сделать обоснованный выбор между достаточным для удовлетворения запросов количеством каналов и обеспечением рентабельности функционирования системы.

**Постановка и решение задачи**

Предположим, разворачивается система, способная обеспечить прохождение информации по некоторому количеству защищенных каналов. Стоимость создания и поддержания канала составляет $S$ условных единиц. Прибыль от предоставления канала пользователю составляет $k*S$, где $k$ – норма рентабельности (отношение прибыли к издержкам на создание и поддержку закрытого канала).

Возникающая потребность в защищенных каналах представляет собой дискретную величину $Z$, распределенную равновероятно в диапазоне от $N_1$ до $N_2$ :

$$P(Z) = 1/(N_2 - N_1); Z = N_1 \ldots N_2$$

Требуется определить оптимальное количество защищенных каналов в разворачиваемой системе для обеспечения максимальной прибыли компании провайдера.

Таким образом, критерием выбора оптимальной стратегии $n$ - количества каналов является величина

$$F(n,Z) = \begin{cases} n(kS) & npu \quad n \leq Z \\ Z(kS) - (n-Z)S & npu \quad n > Z \end{cases}$$

Величину $F$ необходимо максимизировать при условии разрешённости осреднения критерия по случайному фактору. В этом случае приходим к критерию

$$\tilde{F}(n) = \sum_{Z=N_1}^{N_2} F(n,Z)P(Z)$$

Оптимальное значение $n$ можно определить из условия перехода разности $\tilde{F}(n+1) - \tilde{F}(n)$ из положительной области в отрицательную. Запишем выражение для нахождения данной разности:

$$\tilde{F}(n+1) - \tilde{F}(n) =$$
$$= \sum_{Z=N_1}^{n}(Z(kS) - (n+1-Z)S)P(Z) + \sum_{Z=n+1}^{N_2}(n+1)(kS)P(Z) -$$
$$- \sum_{Z=N_1}^{n}(Z(kS) - (n-Z)S)P(Z) + \sum_{Z=n+1}^{N_2}n(kS)P(Z) =$$
$$= 1/(N_2 - N_1)\big[(N_2 - (n+1))kS - ((n+1) - N_1)S\big]$$

Таким образом,

$$n^{opt} = \Omega\left(\frac{N_2(kS) + N_1 S}{(k+1)S}\right) = \Omega\left(\frac{N_2 k + N_1}{k+1}\right),$$

где $\Omega$ - обозначена целая часть числа.

Рассмотрим в качестве примера расчет $n^{opt}$ при $N_1 = 100$; $N_2 = 130$. Соответствующий график приведен на рисунке. По оси ординат отложено количество каналов, которые необходимо задействовать дополнительно к $N_1$ имеющимся.

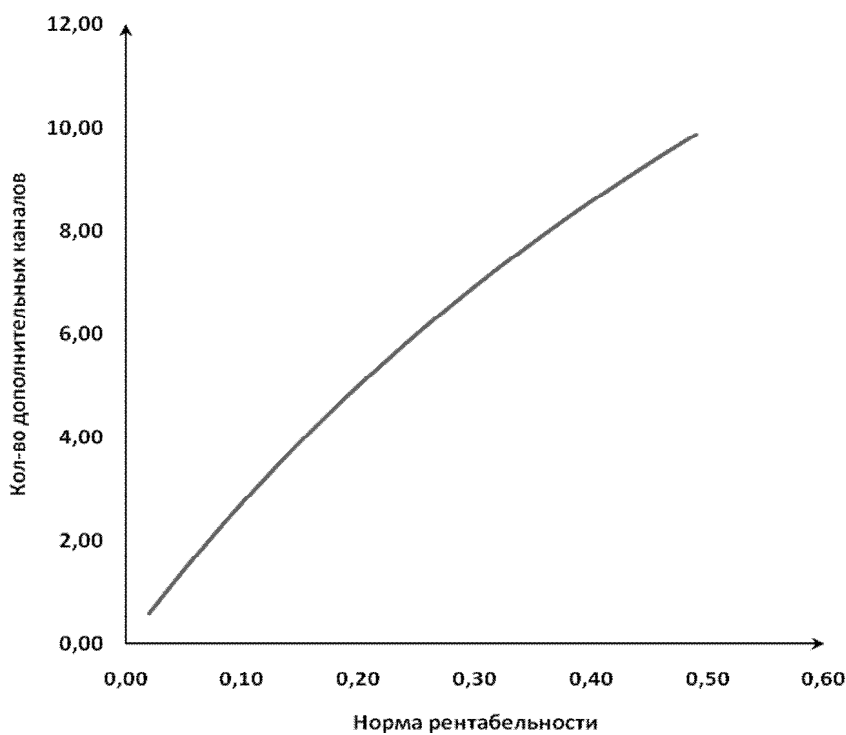

Рис.П.1. Оптимальное количество каналов при различных значениях нормы рентабельности

Из графика видно, что с ростом коэффициента рентабельности оптимальное число каналов нелинейно увеличивается.

**Выводы**

Таким образом, для каждой нормы рентабельности при условии разрешённости осреднения критерия по случайному фактору существует оптимальное (в плане достижения максимума прибыли) количество защищенных каналов передачи информации.

Необходимо заметить, что изложенный в данной работе подход может быть использован не только для расчета оптимальной структуры информационных сетей, но и любой другой транспортной инфраструктуры (нефте-газопроводов, автомобильных и железнодорожных путей), систем коммуникаций, инженерных сетей, аварийных систем жизнеобеспечения, резервных каналов контроля и т.п. В последних двух случаях функцию $F(n,Z)$ необходимо интерпретировать как функцию убытков и искать решение, минимизирующее данную функцию.

# ОЦЕНКА ЭФФЕКТИВНОСТИ ВЫБОРА СИСТЕМЫ ОБЕСПЕЧЕНИЯ БЕЗОПАСНОСТИ ПРИ РАЗЛИЧНОЙ ИНТЕНСИВНОСТИ ПОТОКА УГРОЗ

**Введение**

В современных условиях сложной криминогенной обстановки в РФ и в мире вопросы обеспечения безопасности населения и промышленных объектов приобретают особую актуальность. Особую опасность для промышленных объектов представляют злоумышленные действия физических лиц (нарушителей): террористов, диверсантов, преступников, экстремистов. Результаты их действий не предсказуемы: от хищения имущества до создания чрезвычайной ситуации на объекте (пожар, разрушение, затопление, авария и т.п.).

Одной из эффективных превентивных мер по обеспечению безопасности важных промышленных объектов является создание автоматизированной системы охраны от несанкционированного проникновения физических лиц - системы физической защиты (СФЗ).

Современные СФЗ в корне изменили тактику охраны объектов. В таких системах нет необходимости в организации постовой службы на периметре объекта. Вместо этого создаются дежурные тревожные группы, которые начинают немедленные действия по нейтрализации нарушителей после поступления сигнала тревоги на центральный пульт управления СФЗ. В таких системах сведено до минимума влияние человеческого фактора и достигается высокая эффективность защиты объекта при минимальном количестве личного состава сил охраны.

Однако, приобретение, установка и обслуживание таких систем требует значительных финансовых вложений. Поэтому, предварительная (на стадии проектирования) оценка эффективности развертывания таких систем на конкретном объекте является весьма актуальной задачей.

**Постановка и решение задачи**

Пусть планируется построение системы безопасности, экономическая эффективность которой оценивается как отношение технической эффективности на сумму вложенных в развертывание системы средств (фактически это соотношение «цена-качество»). Техническая эффективность, в свою очередь, определяется как отношение количества предотвращенных атак на объект к их общему количеству.

При традиционном режиме охраны стоимость развертывания и обслуживания системы обеспечения безопасности зависит от плотности потока угроз на охраняемый объект.

Пусть $s_{11}$, $s_{22}$ и $s_{33}$ - стоимости развертывания и обслуживания традиционной системы при разреженном, плотном и сплошном потоке угроз безопасности соответственно.

Как альтернативный вариант рассматривается автоматизированная система охраны, стоимость которой не зависит от плотности потока угроз на систему безопасности и равна

$$s_2 = s_{21} = s_{22} = s_{23}.$$

Вероятности возникновения соответствующих угроз считаются априорно заданными. Обозначим их $P(Z_i)$ – для случаев разреженного ($i = 1$), плотного ($i = 2$) и сплошного ($i = 3$) потоков угроз безопасности.

Таким образом, имеются две альтернативные стратегии: $X_1$ и $X_2$ – развертывания традиционной и автоматической системы обеспечения безопасности соответственно. Возникновение потока угроз с той или иной плотностью выступает в качестве случайного фактора $Z$ с априорно заданной вероятностью. Положим $Z=1$ для случая разреженного потока, $Z=2$ для плотного и $Z=3$ для сплошного потоков.

Тогда, в качестве критерия эффективности для данной задачи выступает общая сумма затрат на создание и обслуживание системы безопасности $S(X,Z)$, которую необходимо минимизировать выбором соответствующей стратегии.

Для достижения наилучшего гарантированного результата необходимо рассмотреть матрицу:

$$S(X,Z) = \begin{pmatrix} s_{11} & s_{12} & s_{13} \\ s_2 & s_2 & s_2 \end{pmatrix},$$

и выбрать такую стратегию, которая соответствует

$$\overline{S} = \min_X \max_Z S(X,Z)$$

Если допускается возможность осреднения критерия по случайному фактору Z, то в качестве оптимальной необходимо выбрать стратегию, отвечающую условию минимизации

$$\tilde{S}(X) = \sum_{i=1}^{3} S(X, Z_i) P(Z_i) \to \min$$

Рассмотрим в качестве примера задачу выбора системы охраны границы контролируемой зоны (КЗ) на одном из судоремонтных заводов. Специфика данной задачи заключается в том, что часть границы КЗ является водной и проходит по акваторию реки.

Вероятность возникновения потоков угроз нарушения периметра охраняемой зоны на этом участке для различных интенсивностей была оценена экспертами как

$$P(Z_1) = 0.15; \quad P(Z_2) = 0.50; \quad P(Z_3) = 0.35$$

При условии принятия стоимости автоматизированной системы $s_2$ за единицу, стоимость традиционных систем при различных потоках угроз для данного предприятия составила соответственно

$$s_{11} = 0.25; \quad s_{12} = 0.85; \quad s_{13} = 1.25 \text{ от } s_2$$

Исходя из этих данных, получаем: $\overline{S} = 1$, что соответствует второй стратегии, т.е. разворачиванию автоматизированной системы охраны;

Таким образом, для гарантированного «закрытия» периметра в данном случае экономически целесообразно применение автоматизированной системы.

**Выводы**

Рассмотренный в данной работе подход может быть применен к оценке эффективности выбора систем обеспечения безопасности при различной интенсивности потока угроз с целью оптимизации затрат на их проектирование, внедрение и обслуживание.

# ФОРМАЛИЗАЦИЯ ЗАДАЧИ РАЗМЕЩЕНИЯ ЭЛЕМЕНТОВ ОХРАННОЙ СИСТЕМЫ В КОНТРОЛИРУЕМОЙ ЗОНЕ

**Введение**

Очень часто при разработке систем охраны объекта от несанкционированного доступа возникает задача оптимального расположения ее элементов вдоль периметра контролируемой зоны. Например, это может быть задача выбора места расположения датчика охранной сигнализации с целью оптимизации его действия, или задача расположения поста охраны на территории контролируемой зоны таким образом, чтобы минимизировать время прибытия оперативной группы к месту нарушения границы. При этом вероятность нарушения считается априорно заданной функцией от координат точки вероятного проникновения $Z$.

**Постановка задачи**

Пусть $G(Z)$ – распределение вероятностей нарушения границы контролируемой зоны. После осреднения критерия эффективности по $Z$, оценка эффективности стратегии $D(Z)$ будет иметь вид:

$$\iint\limits_{X\,Z} F(X,Z)\,dD(Z)\,dG(Z)$$

где $F(X,Z)$ – критерий эффективности, $X$ – точка размещения узла (контролируемый фактор). Соответствующие стратегии, имеющие максимальные оценки эффективности, являются оптимальными.

Для отыскания оптимальных стратегий необходимо решить условные экстремальные задачи вычисления максимина и минимакса некоторого функционала на произведении компактных множеств, что в общем случае является нетривиальной задачей.

**Аналоговая модель для нелинейной границы**

На практике распределение вероятностей проникновения через охраняемый периметр обычно оценивается и задается дискретно в некотором конечном числе точек.

Для решения задачи в случае произвольной нелинейной границы может быть использована следующая аналоговая модель, позволяющая решить задачу размещения элементов без вычислительных процедур.

На планшет наносится схема контролируемой территории, выполненная в нужном масштабе с соблюдением необходимых пропорций. В точках периметра, в которых заданы вероятности нарушений, проделываются отверстия, через которые пропускаются нити.

Над поверхностью планшета нити связываются между собой или крепятся к кольцу, размер которого соответствует элементу в масштабе схемы. На другие концы нитей навешиваются грузы, пропорциональные вероятностям проникновения через соответствующие точки границы.

Под воздействием силы тяжести, действующей на грузы, узел или кольцо, к которому прикреплены нити, займут над поверхностью планшета такое положение, которое и будет соответствовать оптимальному месту расположения элемента охраны на схеме.

Следует заметить, что, сделав «нити» не абсолютно жесткими и меняя их свойства, можно моделировать нелинейное влияние отдельных факторов на принимаемое решение.

**Решение задачи для линейной границы**

Для случая линейной границы задача может быть решена аналитически. Пусть, например, прямолинейный участок охраняемого периметра $AB$ имеет длину $L$. Заданы вероятности нарушения периметра в начальной точке $A = P_A$, в $n$ промежуточных точках $C_i = P_{C_i}$ ($i = 1,\ldots,n$) и в конечной точке $B = 1 - P_A - \sum_{i=1}^{n} P_{C_i}$

Требуется разместить элемент охранной системы (например, датчик или пост) таким образом, чтобы расстояние от него до вероятной точки нарушения было минимальным. В этом случае $X, Z \in E_1$.

Выберем ось $Ox$ вдоль периметра с началом в точке $A$. Пусть точке $B$ соответствует значение координаты равное «1». Тогда в качестве стратегии выступает значение $x$ – координаты расположения элемента охранной системы.

В качестве критерия выберем расстояние от точки $x$ до точки вероятного нарушения z:

$$W(x,z) = |x-z|. \quad (1)$$

Учитывая, что $z$ – случайная величина с распределением $P_z$; $z \in [0;1]$ проведем осреднение критерия по $z$. В результате получим оценку эффективности произвольной стратегии $x$:

$$\overline{W}(x) = P_A W(x,0) + \sum_{i=1}^{n} P_{C_i} W(x,z) + (1 - P_A - \sum_{i=1}^{n} P_{C_i}) W(x,1) \quad (2)$$

С учетом (1) можно записать:

$$\overline{W}(x) = P_A |x| + \sum_{i=1}^{n} P_{C_i} |x - z_{C_i}| + (1 - P_A - \sum_{i=1}^{n} P_{C_i}) |x-1| \quad (3)$$

Далее, рассматривая промежутки изменения $x$:

$$x < 0; \quad 0 \leq x < z_{C_1}; \quad z_{C_{i-1}} \leq x < z_{C_i} \ (i = 2,...,n); \quad z_{C_n} \leq x < 1; \quad x \geq 1 \quad (4)$$

находим $\min_{x} \overline{W}(x)$ и соответствующую ей стратегию (относительную координату) $x$. Абсолютная координата расположения элемента охранной системы вычисляется как произведение длины прямолинейного участка $L$ на относительную координату $x$.

Пусть, например, прямолинейный отрезок границы контролируемой зоны разделен равноотстоящими точками на 4 участка. Тогда точкам соответствуют следующие относительные координаты:

$$z_A = 0; \quad z_{C_i} = i/4 \ (i = 1,...,3); \quad z_B = 1 \quad (5)$$

Пусть в заданных точках имеются следующие априорные вероятности нарушения границы:

$$P_A = 1/2; \quad P_{C_i} = 1/12 \ (i = 1,...,3); \quad P_B = 1 - 1/2 - 3/12 = 1/4 \quad (6)$$

Тогда

$$\overline{W}(x) = \frac{1}{2}|x| + \frac{1}{12} \sum_{i=1}^{3} |x - z_{C_i}| + \frac{1}{4}|x-1| \quad (7)$$

или

$$\overline{W}(x) = \frac{1}{2}|x| + \frac{1}{12}[|x - \frac{1}{4}| + |x - \frac{1}{2}| + |x - \frac{3}{4}|] + \frac{1}{4}|x-1| \quad (8)$$

Анализируем случаи:

$x < 0$: $\overline{W}(x) = -\frac{1}{2}x - \frac{1}{12}[(x-\frac{1}{4})+(x-\frac{1}{2})+(x-\frac{3}{4})] - \frac{1}{4}(x-1) = -x + \frac{3}{8}$. Минимума нет.

$x \geq 1$: $\overline{W}(x) = x - \frac{3}{8}$. Необходимое условие существования экстремума для оценки эффективности не выполняется. Следовательно, минимума на этом промежутке нет.

$0 \leq x < \frac{1}{4}$: $\overline{W}(x) = \frac{3}{8}$. $\overline{W}'(x) = 0$ при любом x $\in$ [0;1/4). Следовательно, в любой точке этого интервала достигается минимум оценки эффективности.

$\frac{1}{4} \leq x < \frac{1}{2}$: $\overline{W}(x) = \frac{1}{6}x + \frac{1}{3}$. Минимума нет.

$\frac{1}{2} \leq x < \frac{3}{4}$: $\overline{W}(x) = \frac{1}{6}x + \frac{1}{4}$. Минимума нет.

$\frac{3}{4} \leq x < 1$: $\overline{W}(x) = \frac{1}{2}x + \frac{1}{8}$. Минимума нет.

**Выводы**

Таким образом, при заданных значениях априорных вероятностей нарушения линейной границы *AB* контролируемой зоны, расположение элемента охраны может быть выбрано в любой точке участка, находящегося в первой четверти прямолинейного отрезка.

При отсутствии возможности расположения элемента в непосредственной близости к охраняемому периметру, он должен располагаться на прямой, проведенной из оптимальной точки перпендикулярно границе. В рассмотренном примере таким месторасположением является полоса, перпендикулярная периметру, с основанием [$AC_1$].

В случае нелинейной границы для решения задачи на практике может быть использована предложенная выше аналоговая модель нахождения оптимального решения, не требующая применения вычислительных процедур и сложных математических выкладок.

# МАРКИРОВКА ПЕЧАТНЫХ КОПИЙ КОНФИДЕНЦИАЛЬНЫХ ЭЛЕКТРОННЫХ ДОКУМЕНТОВ

**Введение**

Несмотря на совершенствование электронных систем документооборота, бумажные носители информации продолжают занимать в данной сфере лидирующее положение.

Документ на бумажном носителе обладает следующими важными свойствами:

информация физически неотделима от носителя. Ее можно воспроизвести или повторить на таком же носителе, но это будет уже копия документа с другим правовым статусом. Подпись под документом существует отдельно от информации, то есть является атрибутом носителя, а не самого документа.

Процесс простановки подписи заключается в начертании автором собственного стилизованного идентификатора на материальном носителе. Авторство собственноручной подписи доказуемо вследствие физиологических особенностей каждого индивидуума, влияющих на процесс воспроизведения подписи.

Нормативные документы, действующие в настоящее время в области информационной безопасности, также включают в себя требования по регистрации и маркировке распечатываемых документов, содержащих конфиденциальную информацию [1].

Таким образом, современная система делопроизводства немыслима без встроенных средств маркировки и идентификации документов, которые позволяют защищать документы от подделки и подтверждают их действительность.

Как правило, на основании реквизитов документа формируется контрольная информация в цифровой форме, контрольная информация преобразуется в штриховой код и штриховой код наносится на бумажный носитель документа.

Легитимность документа, исполненного на бумажном носителе, проверяется с помощью технических средств путем считывания штрихового кода, преобразования его в цифровую форму и сравнения полученных данных с информацией о документе, хранимой в системе делопроизводства.

Описанный алгоритм маркировки и идентификации имеет один существенный недостаток – маркировка в явном виде наносится на документ и открыта для модификации.

Технологии скрытой маркировки основаны в основном на нанесении на маркируемую поверхность специальных составов в виде метки, невидимой при обычном освещении. Метку визуализируют путем освещения источником инфракрасного и/или ультрафиолетового излучения. Типичный пример такого решения приведен в [2].

Данный способ достаточно дорогостоящий, требует специфического оборудования и специальных расходных материалов.

Для улучшения качества маркировки иногда применяется подход, широко используемый в компьютерной стеганографии для охраны интеллектуальной собственности и основанный на информационной избыточности, присущей цветным графическим изображениям. Он связан с неспособностью человеческого зрения в известных пределах различать близкие цвета.

В существующей реализации этого подхода к маркировке электронных документов с помощью манипулирования младшими битами цифрового описания близких цветов, можно осуществить дополнительное информационное насыщение изображений [3-4].

Однако такой подход неприемлем для документов, напечатанных в монохромном режиме. Большая часть документов печатается именно в таком режиме на лазерных принтерах с черным картриджем.

Таким образом, создание эффективных методов и алгоритмов маркировки печатных копий электронных документов, является весьма актуальной задачей.

**Постановка задачи.**

Сформулируем основные требования, которым должна отвечать маркировка бумажной копии конфиденциального документа.

Внедрение в страницы распечатываемых документов дополнительной информации при маркировке должно быть малозаметным. Для ее нанесения и считывания необходимо использовать обычное офисное оборудование (компьютер, принтер, сканер). Дополнительная информация должна позволять установить данные о времени печати, месте печати (названии принтера), о субъекте, осуществившем печать (имя учетной записи и рабочей станции), о количестве страниц в документе. Это, в свою очередь, позволит определить происхождение любого распечатанного экземпляра документа.

Положение отдельных элементов маркера на странице должно быть нерегулярным и зависеть от значения параметра, введенного в программу маркировки. Информация, содержащаяся в маркере документа, также должна быть зашифрована секретным ключом администратора системы.

**Методика формирования маркера**

С учетом сформулированных требований для маркировки конфиденциальных документов на бумажном носителе предлагается следующая схема.

В программе формирования маркера задаются строки N и позиции табуляции P, начиная с которых наносится маркировка, замаскированная под дефекты принтера с целью обеспечения малозаметности. Расчет множителя межстрочного интервала при внедрении маркера осуществляется таким образом, чтобы маркер не попадал на печатаемый текст и вид документа визуально не изменялся.

При этом маркер представляет собой случайным образом расположенные в тексте точечные блоки, содержащие уникальный 64 разрядный код, соответствующий документу – идентификатор печатной копии документа (ИПКД).

Распределение информации внутри маркера может быть следующим. Первый блок из 8 бит отводится под идентификационный номер пользователя. Следующий блок из 8 бит содержит идентификационный номер принтера, на котором печатается документ. 17 бит третьего блока несут в себе информацию о самом документе и степени его конфиденциальности. Четвертый блок из 27 бит содержит информацию о времени печати документа в формате «ДД.ММ.ГГ.чч.мм.». К каждому блоку добавляется по одному контрольному биту для проверки корректности нанесенного кода.

При использовании такой маркировки ИПКД может содержать сведения о 256 различных пользователях и принтерах, работающих в системе, и позволяет установить с точностью до одной минуты время получения печатной копии конфиденциального документа. С использованием данной схемы может быть промаркировано более 32 тысяч различных документов. Для практического применения этого обычно бывает более чем достаточно. При необходимости система может быть легко модифицирована под более высокие требования.

Перед нанесением информация, помещаемая в ИПКД, должна быть зашифрована, например, при помощи гаммирования с секретным ключом администратора системы. Это с одной стороны обеспечит невозможность корректной злоумышленной модификации маркера, с другой – позволит сохранить конфиденциальность информации, включенной в маркер документа.

Схема печати документа с ИПКД показана на рисунке. Формирование и внедрение маркера осуществляется в модуле «Виртуальный принтер» («ВП»), при настройке которого ему в соответствие ставится физическое печатающее устройство. Информация о выбранном принтере включается в состав второго блока маркера.

Кроме этого, входными данными модуля «ВП» являются параметры, определяющие положения блоков маркера на странице, а также ключ, с помощью которого осуществляется шифрование информации. На выходе получается документ с ИПКД, который затем распечатывается на выбранном при настройке модуля «ВП» физическом принтере.

Невозможность «обхода» модуля «ВП» путем назначения другого принтера реализуется средствами администрирования операционной системы.

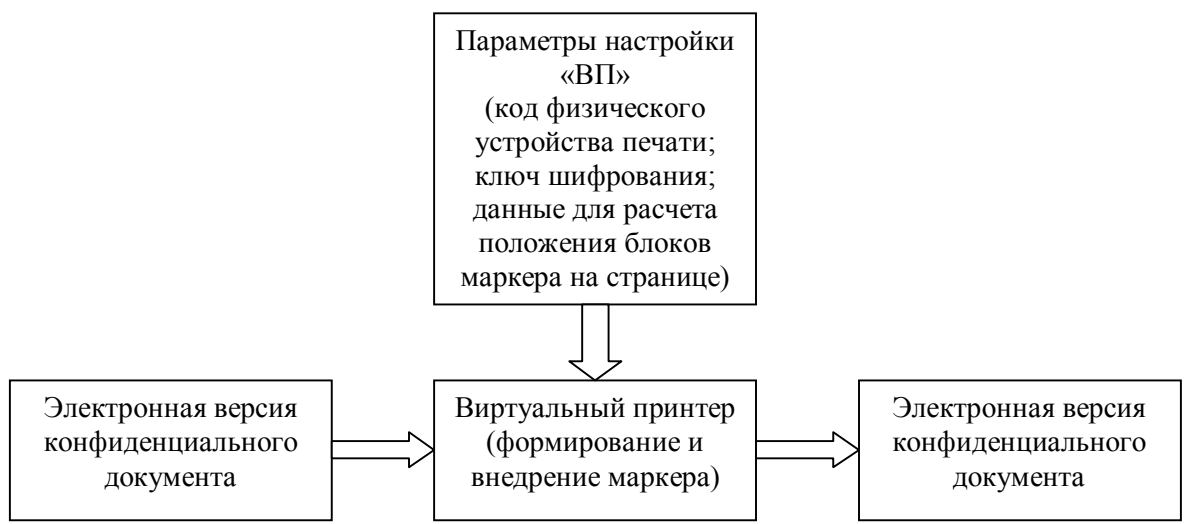

Рис. Схема печати конфиденциального документа с ИПКД

**Заключение**

Таким образом, предложенная методика позволяет формировать и внедрять в печатные копии конфиденциальных электронных документов специальные маркеры, содержащие предварительно зашифрованную информацию о документе, дате и времени печати, о лице, его распечатавшем, а также об оборудовании, на котором была осуществлена печать. Все это усиливает безопасность конфиденциального документооборота на предприятии.

**СПИСОК ЛИТЕРАТУРЫ**

# ИСПОЛЬЗОВАНИЕ СТЕГАНОГРАФИЧЕСКИХ И КРИПТОГРАФИЧЕСКИХ АЛГОРИТМОВ В ЭЛЕКТРОННЫХ УДОСТОВЕРЕНИЯХ ЛИЧНОСТИ

**Введение**

Документы, удостоверяющие личность владельца, всегда играли большую роль в обеспечении общественной и личной безопасности. Развитие технологии изготовления таких документов происходит непрерывно. Вначале объектом совершенствования были степени защиты бланков. Далее, с наступлением цифровой эры, были созданы машиносчитываемые документы, которые были введены в обращение с начала 70-х годов прошлого столетия.

В настоящее время, развитие персональных документов связывают с биотехнологиями. Биометрические документы, удостоверяющие личность, и биометрические системы находятся в последнее время в фокусе внимания государственных структур и бизнеса. Например, создаются паспорта нового поколения. В них, кроме фотографии владельца, дополнительно помещается электронный чип, на который возможно записать дополнительную информацию, в том числе и биометрические параметры. Такие документы вводят США, Германия, Великобритания, Франция, Япония и др. Аналогичный проект по созданию документов нового поколения реализуется и в России [1].

В странах Ближнего Востока распространен другой подход к е-паспортам. Сначала в ОАЭ, а затем и в Иордании население получило идентификационные смарт-карты. Полицейские в этих странах теперь носят с собой специальные ридеры для проверки документов. Причем на карточках нет фотографии владельца. Изображение лица, так же как и отпечатков пальцев, записано в чип [2]. Это связано, в частности с тем, что по религиозным требованиям изображение человека запрещено нормами Ислама.

Документы, удостоверяющие личность также выдаются сотрудникам на предприятиях для организации пропускного режима. Например, все еще широко используются персонифицированные бумажные пропуска, на которых присутствуют фотография владельца и его данные. При этом такой документ легко подделать.

Электронные пропуска основаны на бесконтактных радиочастотных, штрих-кодовых или магнитных картах, картах Виганда, устройствах Touch Memory. Все эти устройства предназначены для идентификации пользователя. Каждый идентификатор характеризуется определенным уникальным двоичным кодом, которому ставится в соответствие информация о правах и привилегиях его владельца. Однако существует ряд недостатков подобного метода организации пропускного режима:

- невозможно точно определить владельца пропуска, т.к. пропуск может быть утерян, украден, передан злоумышленнику;

- необходимо обеспечить уникальность кодов;

- необходимо обеспечить безопасное хранение базы данных (БД) с кодами и сведениями об их владельцах;

- необходимо обеспечить безопасную передачу данных между точкой предъявления пропуска и сервером, на котором находится БД.

Таким образом, существующие идентификаторы, используемые для организации пропускного режима, обладают рядом существенных недостатков, влияющих на безопасность системы. Кроме того, они не могут служить удостоверяющими личность документами без связи с соответствующей БД.

**Постановка и решение задачи**

Основной задачей при изготовлении любого документа, удостоверяющего личность, является обеспечение совокупной (нераздельной) целостности идентифицирующих человека признаков (например, его фотографии) и персональных данных, указываемых в документе (например, фамилия, имя, год рождения и т.д.) и невозможности их подделки или несанкционированного изменения после изготовления документа. Также необходимо свести к минимуму возможность использования чужого документа, который мог быть утерян или выкраден злоумышленником.

Для решения перечисленных задач при изготовлении электронных документов, удостоверяющих личность, представляется целесообразным применение криптографических методов и алгоритмов стеганографии.

Направленные на решение разных задач по обеспечению информационной безопасности, криптографические и стеганографические преобразования, примененные в совокупности, позволяют обеспечить необходимые качества документов, удостоверяющих личность.

Анализ стойкости современных стеганографических алгоритмов показал [3], что создание алгоритма, обладающего достаточной степенью робастности к различным преобразованиям и противостоящего средствам статистического и визуального стегоанализа и разработка на его основе программного продукта для скрытия достаточно большого объема данных методами цифровой стеганографии, является весьма актуальной задачей.

Поэтому был разработан алгоритм цифровой стеганографии с шифрованием данных, позволяющий успешно противостоять визуальной атаке и статистическому стегоанализу, обладающий однозначной функцией восстановления данных, устойчивый к низкочастотной фильтрации и к геометрической атаке типа «обрезание краев», обладающий возможностью встраивания данных в графические файлы, в том числе сжатые стандартом JPEG, обеспечивающий достаточную вместимость (в несжатое изображение размером 1280×1024 можно внедрить до 150 килобайт данных). Кроме того, реализованный на основе алгоритма программный продукт позволил осуществлять дополнительное шифрование информации [4].

При встраивании данных в пространственную область алгоритм оперирует 24-битовыми несжатыми образами с RGB-кодировкой. Встраивание выполняется в канал синего цвета, так как к нему система человеческого зрения наименее чувствительна. Порог чувствительности глаза к изменению освещенности при средних ее значениях составляет $\Delta l = 0.01 - 0.03l$ или 1~3%, где яркость $l(p) = 0,299r(p)+0,587g(p)+0,114b(p)$, $p=(x,y)$ – псевдослучайная позиция, в которой выполняется вложение.

Пусть сообщение S представляет собой последовательность бит длиной N, количество пикселов в области преобразования равно M. Пиксели, в которые выполняется встраивание, равномерно распределяются по всему изображению псевдослучайным образом на основании ключа. Исходное изображение или его часть, ограниченная областью преобразования, разбивается на два типа блоков длины r1 и r2, причем r1 = r2+1. Количество блоков длины r1 и r2 равно n1 и n2 соответственно. При условии, что n1+n2=N и r1*n1+r2*n2=M , значения r1,r2,n1,n2 вычисляются по следующим формулам:

$$r2 = M / N,$$
$$n2 = (r2+1)*N - M,$$
$$r1 = r2+1,$$
$$n1 = N - n2,$$

где «/» - операция целочисленного деления.

Например, если область преобразования содержит M=64 пиксела и длина сообщения N=24 бита, в результате разбиения получится 8 блоков длины 2 и 16 блоков длины 3.

Блоки чередуются псевдослучайным образом на основании ключа. Каждому биту сообщения соответствует свой блок, в котором в соответствии с ключом выбирается пиксел подвергающийся изменению.

Рассмотрим процедуру формирования ключевой последовательности. Пароль, введенный пользователем, с помощью алгоритма хеширования MD5 преобразуется в 16 байтовое слово. С помощью циклической перестановки символов пароля формируется еще 3 таких слова. В результате получается массив 128×4 бита, столбцы которого представляют собой псевдослучайную последовательность чисел длиной 128. Получая остаток от деления этих чисел на длину блока, вычисляем позицию изменения пикселя внутри этого блока.

Цветовая компонента каждого пиксели описывается одним байтом. Изменение происходит по маске (11100011), то есть модификации подлежат 4, 5 или 6 биты. Отклонение интенсивности цвета в данном случае не превышает 6,3%, а общее изменение яркости пикселя не превышает 1%. Такая модификация устойчива против низкочастотной фильтрации изображения, поскольку выходит за рамки шума, возникающего в последних битах цветовой компоненты.

Выбор изменения трех битов вместо одного обусловлен соображениями устойчивости алгоритма к методам статистического стегоанализа и визуальной атаке при просмотре битовых срезов. Обычно в стеганографических системах предлагается изменять какой-то определенный бит изображения, как правило, последний. При использовании контроля четности этого бита изменению подлежит примерно половина отсчетов изображения, что существенно нарушает его статистические свойства.

Модификация одного из трех бит уменьшает объем изменяемых бит в каждом битовом срезе до 1/6, а уменьшение объема скрываемого сообщения с равномерным распределением пикселов хотя бы до 1/3 объема изображения уменьшает объем изменяемых бит в каждом битовом срезе до 1/18. Поэтому применение статистических методов стегоанализа, в частности по критерию хи-квадрат чрезвычайно затруднено, так как алгоритм почти не изменяет статистику первоначального изображения.

Рассмотрим встраивание одного бита S секретной информации. В соответствии с ключом в блоке пикселов выбирается номер пикселя, в который выполняется встраивание. Номер бита определяется на основании значений других пикселов этого же блока, чтобы не прерывать цепочки повторяющихся битов исходного изображения. Например, если все четвертые биты из данного блока равны 0 или 1, то алгоритм оставляет четвертый бит этого пикселя без изменений и переходит к рассмотрению всех пятых битов данного блока и т.д. Если все четвертые, пятые и шестые биты всех пикселов из данного блока равны 0 или 1, предпочтение отдается четвертому для большей помехоустойчивости вложения. После нахождения номера бита, подлежащего модификации, происходит его замена на секретный бит.

Для примера рассмотрим модификацию одного блока исходного изображения. Допустим, блок состоит из пяти пикселов. Так как встраивание выполняется в канал синего цвета, необходимо проанализировать пять байтов (рис. 1).

```
b1 = 1 1 1 1 1 0 1 0
b2 = 1 1 0 1 1 0 1 1
b3 = 1 0 1 1 0 0 0 0
b4 = 1 0 1 1 0 0 0 1
b5 = 1 0 0 1 1 0 0 1
```

Рис 1 Битовое представление пикселов из одного блока.

Как видно из рисунка, последовательность четвертых и шестых битов пикселов из этого блока представляет собой однородную цепочку бит и лишь пятый бит удовлетворяет требованию неоднородности и может быть изменен на секретный бит.

Для обеспечения возможности восстановления сообщения, в изображение встраиваются дополнительные 128 бит, в которых содержится информация о расширении файла сообщения, его размере, относительных координатах области скрытия и используемом алгоритме дополнительного шифрования.

При извлечении сообщения сначала производится извлечение дополнительной информации, определяются координаты области скрытия. Затем изображение разбивается на блоки аналогично процедуре внедрения и в каждом блоке на основании ключа определяется позиция, в которую было произведено встраивание секретного бита. После определения номера бита 8-битной последовательности по критерию неоднородности, значение секретного бита устанавливается равным значению найденного бита изображения.

Используя в качестве «контейнера» стегосистемы файл с электронной фотографией лица, которому выдается документ, а в качестве стегосообщения – предварительно зашифрованные с помощью закрытого ключа ассиметричного алгоритма шифрования персональные данные, можно сформировать содержимое электронного документа, удостоверяющего личность данного лица. После записи на любой подходящий носитель (который, кроме того, может содержать и графическое изображение (фотографию) человека) получаем удостоверение, подлинность которого легко устанавливается после считывания данных и их расшифровки с использованием открытого ключа шифрования.

Полученный по такой схеме документ защищен от атак на его целостность: фотография («контейнер») и персональные данные («стего») являются нераздельными. Без знания специального ключа, часть которого известна только владельцу документа, а часть хранится в программе для считывания данных, невозможно извлечь стегосообщение из контейнера. Поэтому становится весьма затруднительным использование украденного или утерянного документа.

Для подделки или модификации документа требуется знание закрытого ключа шифрования персональных данных. Знание открытого ключа ничего злоумышленнику не дает из-за особенностей ассиметричных криптоалгоритмов.

Кроме того, для идентификации личности по такому электронному документу не требуется связь с какой-либо внешней базой данных. Вся необходимая информация в криптографически защищенном виде хранится в самом документе. В качестве персональных данных, стеганографически «вшитых» в документ, могут выступать и биометрические характеристики личности, что сделает систему еще более надежной.

**Выводы**

Предложенная схема изготовления автономного электронного удостоверения личности на основе стеганографических и криптографических алгоритмов позволит создать на ее основе качественно новую пропускную систему с повышенным уровнем безопасности.

Описанный подход к изготовлению электронных документов может быть применен не только для организации пропускного режима, но и для других документов удостоверяющих личность (например, водительского удостоверения, социальной карты, паспорта и т.п.)

**Литература**

*к.т.н. Ажмухамедов И.М.*
*Колесова Н.А.*

# МЕТОДИКА ФОРМИРОВАНИЯ КЛЮЧЕВОЙ ИНФОРМАЦИИ ИЗ ГРАФИЧЕСКИХ ФАЙЛОВ

**Введение**

В настоящее время с повышением уровня использования вычислительных устройств, проникающих во все сферы жизни человечества, необходимость в криптографической защите данных становится все более актуальной.

Основыми требованиями, которые предъявляются к современным системам шифрования, являются требования по надежности, скорости, простоте реализации и использования. Однако почти все системы шифрования обладают условной надежностью, поскольку могут быть раскрыты при наличии достаточных вычислительных ресурсов. Единственным абсолютно надежным шифром, является одноразовая система шифрования, которую разработали еще в 1917 г. Дж. Моборн и Г. Вернам [1].

Абсолютная надежность одноразовой системы доказана Клодом Шенноном в его работе «Теория связи в секретных системах». Одноразовые системы нераскрываемы, поскольку их шифротекст не содержит достаточной информации для восстановления открытого текста. Однако возможности использования одноразовых систем на практике ограничены. Ключевая последовательность длиной не менее длины сообщения должна передаваться получателю сообщения заранее или отдельно по некоторому секретному каналу, что практически неосуществимо в современных информационных системах, где требуется шифровать многие миллионы символов и обеспечивать засекреченную связь для множества абонентов.

Поэтому блокноты используются редко, в особо критичных случаях. Чаще применененяются алгоритмы симметричного и асимметричного шифрования [2], которые являются условно надежными. Тем не менее при достаточной длине ключа их криптостойкости хватает для подавляющего большинства практических приложений. При этом симметричные алгоритмы шифрования по сравнению с асимметричными являются более эффективными, однако при их использовании возникает проблема выработки и хранения ключевой информации.

**Постановка и решение задачи**

Необходимо разработать методику удобного формирования и хранения ключевой информации для использования в системах защиты.

Для решения задачи в качестве ключевых предлагается использовать последовательности случайных чисел (ПСЧ), полученные из младших бит графических файлов.

Как было показано в [3], младшие биты каждого цветового канала изображения могут образовывать последовательности случайных бит. Причем длина таких последовательностей зависит от размера исходного изображения. Так, например, из изображения размером 640*480 пикселей можно получить три различные случайные числовые последовательности длиной 307200 символов каждая, которые затем могут применяться в качестве ключевых последовательностей или служить источником для их формирования.

Предлагаемых подход по использованию изображений в качестве источника и хранилища ключевой информации обладает рядом преимуществ:

1. Каждый пользователь может самостоятельно создавать свою уникальную базу графических файлов, и в любой момент времени легко извлекать нужный ключ или пароль.

2. В отличие от ПСЧ, получаемых с помощью специальных таблиц, которые несложно найти в Internet, для последовательностей случайных бит, полученных предложенным методом, исключается возможность их компрометации в глобальной сети.

3. Исчезает необходимость запоминать длинную случайную ключевую последовательность из нулей и единиц. Пользователю достаточно лишь

запомнить то или иное изображение (фотографию), что в большинстве случаев не составляет труда.

4. Злоумышленнику сложно догадаться, где и каким образом хранится ключевая информация. И даже при полном доступе к базе изображений пользователя при известной методике получения ключевой последовательности, задача подбора «правильного» ключа (пароля) методом лобовой атаки практически неразрешима. Так например, для изображения размером 227*170 пикселей можно получить три ПСЧ длиной по 38590 бит, каждая из которых в свою очередь может служить источником для формирования $2^{15}$ - $2^{16}$ подпоследовательностей длиной 256 бит. Таким образом, если база картинок пользователя содержит хотя бы 100 изображений, то для реализации «лобовой атаки» злоумышленнику придется перебрать $300*2^{15}$-$300*2^{16}$ или $2^{23}$ - $2^{25}$ вариантов ключей. Соответственно при использовании полноразмерных изображений, данное значение возрастает на порядки величин.

Проверка степени случайности числовой последовательности, состоящей из младших бит изображения, может быть осуществлена с помощью методики, описанной в [4].

Согласно данной методике, влияние результатов различных проверок последовательности бит на общий уровень ее качества может быть представлено в виде ориентированного трехуровневого графа *G*, имеющего одну корневую вершину и не содержащего петель и горизонтальных ребер в пределах одного уровня иерархии:

$$G = <\{F_i\};\{D_{ij}\}>,$$

где $\{F_i\}$ – множество вершин графа; $\{D_{ij}\}$ – множество дуг, соединяющих *i*-ю и *j*-ю вершины; $K_0$ – корневая вершина, отвечающая интегральному критерию.

При этом дуги расположены так, что началу дуги соответствует вершина нижнего уровня иерархии (ранга), а концу дуги – вершина ранга, на единицу меньшего.

На втором (нижнем) уровне расположены тесты $T_i$, используемые для проверки различных характеристик числовой последовательности. На уровень выше находятся обозначенные через $K_i$ основные характеристики случайности: равномерность, стохастичность и независимость. И, наконец, корневой вершине нулевого уровня соответствует комплексный критерий $K_0$ оценки качества проверяемой последовательности.

Далее на полученный граф необходимо наложить систему весов или отношений предпочтения одних критериев над другими по степени их влияния на заданный элемент следующего уровня иерархии.

Для комплексной оценки качества последовательности бит производится агрегирование данных, собранных в рамках иерархии $G$.

При этом агрегирование совершается по направлению дуг графа иерархии, где при переходе со второго уровня на первый применяется аддитивная свертка, а при переходе с первого уровня на нулевой – мультипликативная.

Таким образом, пройдя последовательно снизу вверх по всем уровням иерархии $G$ можно путем комплексного агрегирования данных выработать суждение о качественном уровне показателя на каждой ступени иерархии (вплоть до $K_0$).

**Расчетный пример**

С целью исследования принципиальной возможности такого подхода к формированию ключевой информации было протестировано 60 числовых последовательностей, полученных из 20-ти изображений. В качестве исходных изображений для повышение быстродействия расчетов были взяты фотографии невысокого разрешения (227*170 пикселей) и сформированы последовательности из младших бит длиной 38590 символов каждая.

По результатам тестирования из 20-ти исходных изображений было выделено 24 последовательности случайных бит, каждая из которых может самостоятельно использоваться в качестве ключевой последовательности или служить источником для формирования подключей меньшей длины.

В таблице приведено процентное распределение случайных, неслучайных и неопределенных последовательностей бит для каждого цветового канала. Видно, что распределение бит в G-канале более случайно, чем в других цветовых каналах.

Таблица

Процентное соотношение случайных, неслучайных и неопределенных последовательностей, полученных из RGB-каналов пикселей изображений

|  | Случайные, % | Неопределенные, % | Неслучайные, % |
|---|---|---|---|
| Последовательности бит R-канала | 35 | 35 | 30 |
| Последовательности бит G-канала | 60 | 25 | 15 |
| Последовательности бит B-канала | 25 | 10 | 65 |

**Заключение**

Таким образом, предложенная методика оценки качества последовательностей случайных чисел позволяет отбирать графические файлы, которые могут использоваться в качестве источника и хранилища ключевых последовательностей.

Предложенный подход к получению критически значимых для обеспечения безопасности информации данных обладает рядом существенных преимуществ, что позволяет усилить безопасность компьютерных сетей и систем.

**СПИСОК ЛИТЕРАТУРЫ**

*к.т.н. И.М. Ажмухамедов,*
*Е.В.Кульчихин,*
*А.О.Кайдалова*

## ОБ ОЦЕНКЕ СТЕПЕНИ ОПАСНОСТИ ПРОЦЕССОВ, ИСПОЛНЯЕМЫХ В ВЫЧИСЛИТЕЛЬНОЙ СИСТЕМЕ

**Введение**

В настоящее время одним из основных методов поиска вредоносных программ является метод, основанный на сравнении синтаксических сигнатур, при котором программа-антивирус, просматривая очередной файл или пакет, выделяет его уникальную сигнатуру, после чего обращается к словарю с известными угрозами. Если в словаре присутствует сигнатура программы, то антивирус идентифицирует её как опасную и пытается блокировать или удалить.

При использовании этого метода необходимо периодически пополнять словарь известных вирусов новыми сигнатурами. Плюс данного подхода состоит в том, что вирусы можно обнаружить сразу же после их появления на компьютере, при этом количество ошибок первого и второго рода будет минимально. Но у данного подхода есть и минусы:

• невозможно выявить какие-либо новые (неизвестные) вредоносные программы;

• данный подход бесполезен при заражении полиморфическими вирусами;

• необходимо регулярно обновлять базы сигнатур;

• необходимо кропотливо вручную анализировать файлы с целью выявления степени их вредоносности и уникальности сигнатуры.

Существует другой подход выявления вредоносных программ - эвристический анализ. Он широко применяется компаниями-лидерами по производству антивирусных программ. Но для эвристического анализа характерна высокая степень ошибок первого и второго рода [1].

Еще один подход основан на анализе поведения программ и применения так называемых блокираторов. Преимущества блокираторов состоят в том, что они в меньшей степени, чем классические подходы замедляют работу системы, позволяют откатывать изменения, совершенные в операционной системе, вернув её к незараженному состоянию [2].

**Постановка задачи и подходы к ее решению**

Для улучшения характеристик поведенческих блокираторов необходимо создать методику оценки степени опасности исполняемых операционной системой процессов.

Определим понятие критичной области вычислительной системы (КОВС) как области, изменения в которой могут существенно повлиять на безопасность функционирования системы. В качестве КОВС могут выступать, например, раздел реестра, область дискового пространства или оперативной памяти и т.д. Степень критичности $j$-й КОВС для вычислительной системы в целом обозначим $R_j$.

Процесс P, вызывая с определенной частотой $\Omega_{ij}$ множество функций (элементарных базовых операций) $F=\{f_i\}$, оказывает воздействие на множество КОВС $A=\{A_j\}$. Данное воздействие можно формализовать с помощью матрицы $D=\{d_{ij}\}$, где $d_{ij}=0$ если $i$-я функция процесса P не влияет на $j$-ю критическую область. В противном случае $d_{ij}$ характеризует степень опасности проведения $i$-ой функцией изменений в $j$-ой КОВС.

Множество значений $\Omega_{ij}$ частот вызова $i$-й функции для воздействия на $j$-ю КОВС образует матрицу частот $\Omega$.

Множество функций F, которые могут быть вызваны процессом P, также как и множество A областей КОВС - ограничено.

Следовательно, работу вычислительной системы с точки зрения оценки степени опасности выполняемого в ней процесса P можно представить в виде следующего кортежа, который представляет собой общую математическую модель решаемой задачи:

$G = \langle P, A, R, F, D, \Omega \rangle$.

В каждый квант времени *t* (один такт работы вычислительной системы) процесс P вызывает с учетом частот $\Omega$ множество функций F, обладающих степенями опасности D при проведении соответствующих изменений в критических областях A, которые в свою очередь имеют степень критичности R для функционирования вычислительной системы в целом.

Разработка методики оценки степени опасности выполняемых в вычислительной системе процессов предусматривает решение следующих задач:

• анализ предметной области с целью выявления основных достоинств и недостатков в подходах, используемых в настоящее время при эвристическом анализе;

• составление обобщенной математической модели для решения поставленной задачи;

• содержательное наполнение математической модели (определение множества процессов P, множества КОВС и степеней их критичности R, множества функций F, заполнение матрицы опасности воздействий D и матрицы частот $\Omega$);

• составление методики оценки опасности процесса и его программная реализация;

• верификация разработанного алгоритма и реализованного на его основе программного продукта;

• оценка валидности разработанной методики на примере применения в реальных вычислительных средах.

Функциональная диаграмма процесса создания методики была выполнена в стандарте IDEF0 с декомпозицией некоторых блоков в нотации IDEF3 и представлена на рис.1-3.

**Выводы.**

Предложенная в работе обобщенная математическая модель анализа опасности выполняемых в вычислительной системе процессов позволяет перейти к ее содержательному наполнению. В дальнейшем это позволит разработать методику оценки опасности процессов и приступить к ее программной реализации.

СПИСОК ИСПОЛЬЗОВАННЫХ ИСТОЧНИКОВ

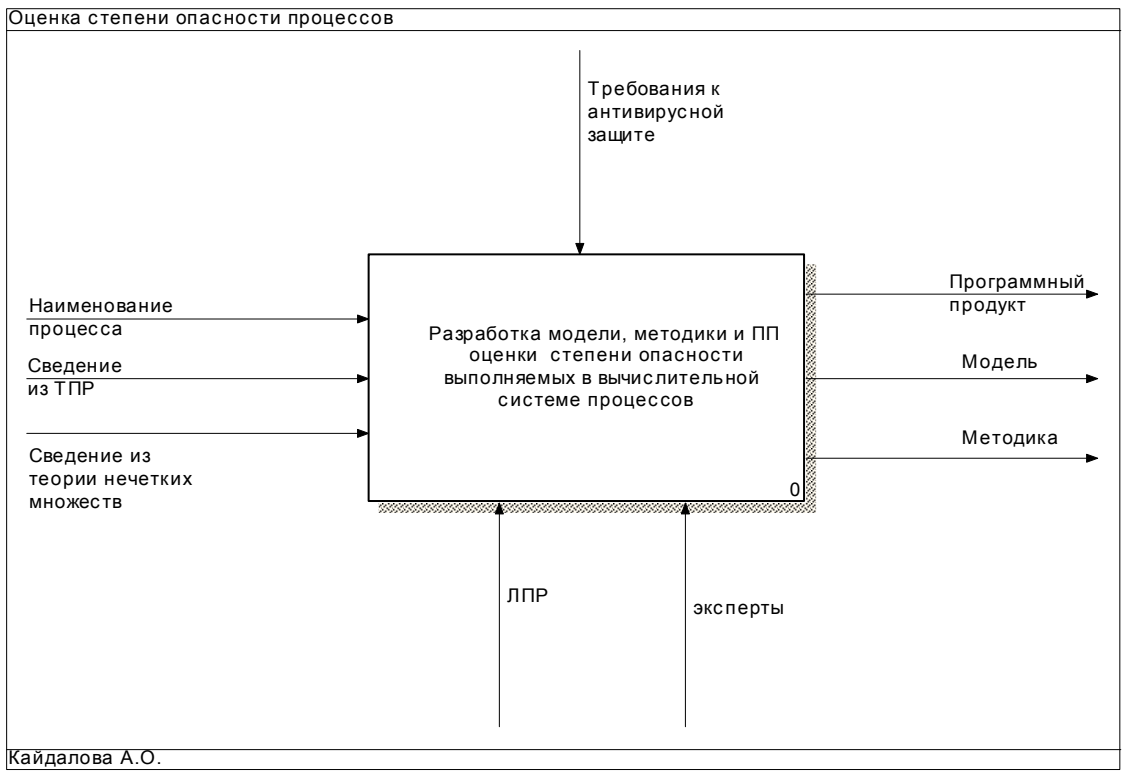

Рис.1. Концептуальная модель

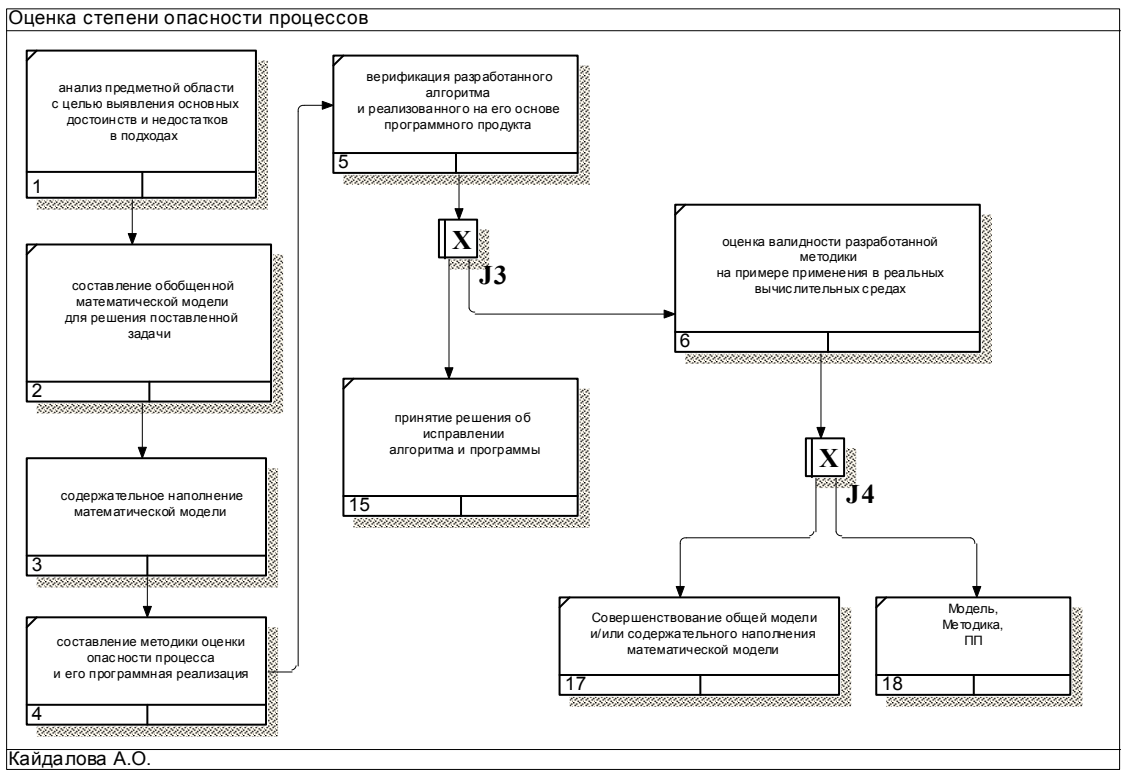

Рис.2. Декомпозиция концептуальной диаграммы в нотации IDEF3.

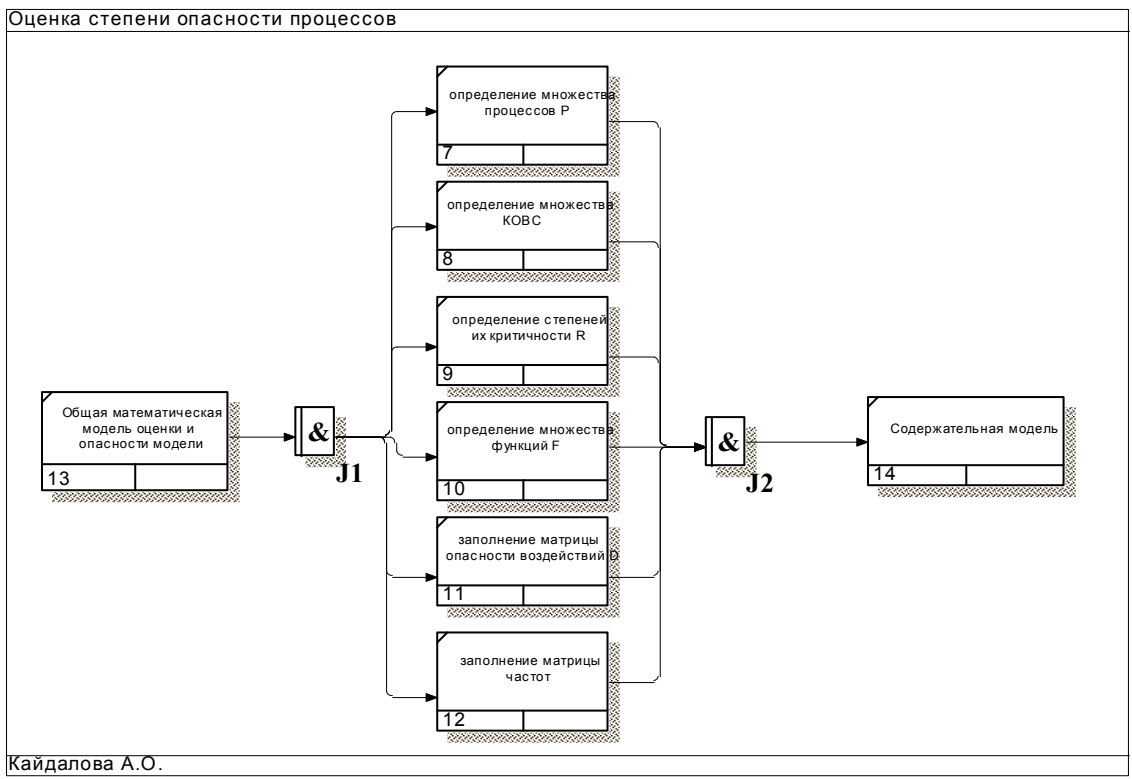

Рис.3. Декомпозиция работ по содержательному наполнению общей математической модели